\documentclass[useAMS,usenatbib,twocolumn, fleqn]{mn2e} \usepackage{natbib,
graphics, epsfig, color}

\usepackage{times} 
\usepackage{amsmath} 
\usepackage{amssymb}
\usepackage{textcomp} 
\usepackage{verbatim}
\usepackage{hyperref}
\newcommand{\sm}{\small}
\newcommand{\dint}{\,{\rm d}} 
\newcommand{\hmsun}{\,h^{-1}\,{\rm M_\odot}} 
\newcommand{\msun}{{\,\rm M_\odot}}
\newcommand{\kms}{\,{\rm km}\,{\rm s}^{-1}}

\newcommand{\erg}{\,{\rm erg}}
\newcommand{\ergs}{\,{\rm erg}\,{\rm s}^{-1}}
\newcommand{\yr}{\,{\rm yr}}
\newcommand{\Gyr}{\,{\rm Gyr}}
\newcommand{\Myr}{\,{\rm Myr}}
\newcommand{\kpc}{\,{\rm kpc}}

\newcommand{\hkpc}{\,h^{-1}\,{\rm kpc}}
\newcommand{\hmpc}{\,h^{-1}\,{\rm Mpc}}

\urldef\myhomepage\url{http://www.cfa.harvard.edu/itc/research/arepogal/}
\title[A model for cosmological simulations of galaxy formation physics] 
      {A model for cosmological simulations of galaxy formation physics} 
       \author[M. Vogelsberger et al.]
      {\parbox{20cm}{ 
        Mark Vogelsberger$^{1}$\thanks{Hubble Fellow, mvogelsb@cfa.harvard.edu}, Shy Genel$^{1}$, Debora Sijacki$^{1,2}$, Paul Torrey$^{1}$, Volker Springel$^{3,4}$, \\and Lars Hernquist$^{1}$}\vspace{0.3cm}\\ 
        $^1$ Harvard-Smithsonian Center for Astrophysics, 60 Garden Street, Cambridge, MA, 02138, USA\\  
	$^2$ Kavli Institute for Cosmology, Cambridge and Institute of Astronomy, Madingley Road, Cambridge, CB3 0HA\\
	$^3$ Heidelberg Institute for Theoretical Studies, Schloss-Wolfsbrunnenweg 35, 69118 Heidelberg, Germany\\ 
        $^4$ Zentrum f\"{u}r Astronomie der Universit\"{a}t Heidelberg, ARI, M\"onchhofstr. 12-14, 69120 Heidelberg, Germany\\}

\begin{document}

\maketitle

\begin{abstract}

We present a new comprehensive model of the physics of galaxy formation
designed for large-scale hydrodynamical simulations of structure formation
using the moving mesh code {\sm AREPO}. Our model
includes primordial and metal line cooling with self-shielding corrections,
stellar evolution and feedback processes, gas recycling, chemical enrichment, a
novel subgrid model for the metal loading of outflows, black hole (BH) seeding,
BH growth and merging procedures, quasar- and radio-mode feedback, and a
prescription for radiative electro-magnetic (EM) feedback from active galactic
nuclei (AGN).  Our stellar evolution and chemical enrichment scheme follows
nine elements (H, He, C, N, O, Ne, Mg, Si, Fe) independently. Stellar
feedback is realised through kinetic outflows. The metal mass loading of outflows can be
adjusted independently of the wind mass loading.  This is required to
simultaneously reproduce the stellar mass content of low mass haloes and their
gas oxygen abundances.  Radiative EM AGN feedback is implemented assuming an
average spectral energy distribution and a luminosity-dependent scaling of
obscuration effects.  This form of feedback suppresses star formation more
efficiently than continuous thermal quasar-mode feedback alone, but is less
efficient than mechanical radio-mode feedback in regulating star formation in
massive haloes.   We contrast simulation predictions for different variants of our
galaxy formation model with key observations, allowing us to constrain the
importance of different modes of feedback and their uncertain efficiency
parameters.  We identify a fiducial best match model and show that it
reproduces, among other things, the cosmic star formation history, the stellar
mass function, the stellar mass -- halo mass relation, g-, r-, i-, z-band SDSS
galaxy luminosity functions, and the Tully-Fisher relation.  We can achieve
this success only if we invoke very strong forms of stellar and AGN feedback
such that star formation is adequately reduced in both low and high mass systems.
In particular, the strength of radio-mode feedback needs to be increased
significantly compared to previous studies to suppress efficient cooling in
massive, metal-enriched haloes.

\end{abstract}


\section{Introduction}\label{sec:intro}

In the standard model of cosmology, the dominant mass contribution to
the Universe is in the form of cold dark matter (DM). Together with a
large dark energy (DE) component \citep[][]{Riess1999,
  Perlmutter1999}, this is the defining feature of the concordance
$\Lambda$ cold dark matter ($\Lambda$CDM) cosmogony.  Early models of
structure formation have predicted the condensation of baryons via
radiative cooling at the centres of a population of hierarchically
growing DM haloes, forming galaxies \citep[][]{Silk1977, Rees1977,
  White1978, Blumenthal1984}.  Galaxy formation theory aims to explain
the observed galaxy population of our Universe through a
self-consistent model which requires as input only the initial
conditions left behind after the Big Bang, as inferred from the cosmic
microwave background (CMB).  Such a model needs to correctly predict
the evolution of all constituents of the Universe: DM, DE {\em and}
baryons. Despite the fact that it is still unknown what DM and DE are,
numerical N-body simulations of these ``dark components''
have reached a high degree of sophistication over the past decade,
making possible accurate calculations far into the non-linear
regime~\citep[e.g.,][]{SpringelWhite2005, Springel2008,
  BoylanKolchin2010, Klypin2011, Wu2013}.

However, to make direct connections with the large amount of available
and upcoming observations of the {\it visible} universe, simulations
must also account for baryonic physics.  Especially in the era of
``precision cosmology'' based on large surveys like SDSS, LSST, etc., it
is crucial to have predictive, accurate and reliable galaxy formation
models to test our understanding of structure formation by confronting
these models with a plethora of data. Including baryons in
cosmological models is also a necessity for predicting the
back-reaction of baryons on the dark matter distribution, an effect
that is particularly important for observational searches of dark
matter and for concluding whether the $\Lambda$CDM model is viable on
small scales~\citep[][]{Vogelsberger2009, Ling2010}.

There are currently two main approaches for computing cosmological
models of galaxy formation and evolution: so-called semi-analytic
models and self-consistent hydrodynamical simulations.
Semi-analytical models \citep[e.g.,][]{White1991, Kauffmann1993,
  Baugh2006, Croton2006, Guo2011, Benson2012} account for baryonic
physics by adding galaxy formation as a post-processing step on top of
the output of N-body simulations or onto Monte-Carlo realisations of
dark matter merger history trees.  Specifically, this approach uses
simple analytical prescription to describe the baryonic physics that
shape galaxy populations. Typically, the free parameters of the models
are tuned to reproduce certain key observations, like the local
stellar mass function, and are then used to predict other
characteristics of the observed galaxy population. By contrasting
those predictions to observations the model can then be refined.  While
this approach allows a fast parameter space exploration of the
relevant physics at play, its predictive power for certain observables
is limited.  For example, semi-analytic models do not permit a
detailed prediction for the gas properties of the intergalactic medium
(IGM), let alone the circumgalactic medium (CGM) in and around galaxy
haloes.  Moreover, since the gas dynamics in the semi-analytic
approach is approximated only in a very crude way, it is difficult to
correctly identify the primary physics driving the outcome.

The most general way to overcome these limitations is to couple the
gas physics to the ``dark components'' and perform fully
self-consistent hydrodynamical simulations in cosmological volumes.
This approach is significantly more expensive from a computational
point of view than semi-analytic models.  In addition, modelling the
required physics in a numerically meaningful way is also challenging.
The difficulties arise from two different considerations: (i) the
underlying numerical technique must solve the hydrodynamical equations
efficiently and accurately, and offer a sufficiently large dynamic
range to describe both cosmological and galactic scales; and (ii) the
relevant baryonic processes must be implemented in a physically and
numerically meaningful way.  Especially the latter point should not be
viewed as an after-thought -- owing to inevitable technical limitations,
brute force implementations of the physics are often ill-posed
numerically, non-convergent, and may produce spurious results.  These
limitations can be overcome only by employing sub-grid prescriptions
that provide numerical closure, albeit at the cost of some modelling
uncertainty.  Such sub-resolution models aim to link the unresolved
small-scale physics within galaxies to the scales that can actually be
resolved in large-scale cosmological simulations. We stress that any numerical
model always requires closure at some scale, since cosmological simulations
cannot resolve all scales. Schemes without closure naturally lack convergence.
This can be seen for example in the recent comparison paper by 
\cite{Scannapieco2012}
and in more detail in 
\cite{Marinacci2013}, 
where different numerical implementations were compared.
Fig. 21 (bottom left) in \cite{Marinacci2013} demonstrates that the convergence properties of a numerically well-posed scheme
(as presented below) are significantly better than those other methods. We note that this does not
mean, that smaller scales can and should not be resolved in future simulations. But we stress
that this has to be done in a numerically meaningful way such that convergence can be achieved.

Many previous studies have argued that accurate schemes for solving
the inviscid Euler equations are crucial for reliably modelling galaxy
formation~\citep[][]{Frenk1999, Agertz2007, Mitchell2009,
  Sijacki2012}.  Recently, we have corroborated this point explicitly
through a detailed comparison campaign between simulations carried out
with the moving-mesh code {\small AREPO}~\citep[][]{Springel2010} and with smoothed particle
hydrodynamics (SPH) with the {\small GADGET}~\citep[last described in][]{Springel2005} code. The differences we identified are sufficiently
large to qualitatively change the way galaxies accrete their gas \citep[][]{Nelson2013}, and
to quantitatively modify the strength of required feedback
processes. We stress that these results were obtained with ``standard SPH'' as implemented
in the {\small GADGET} code. 
We hence consider it essential to eliminate numerical
artifacts and errors in galaxy formation simulations as far as
possible.  Without doing so, the physical models used to describe
effects associated with star formation, black hole growth, and
feedback will be corrupted by numerical limitations.  Furthermore, 
galaxy formation simulations should ideally be performed at the
highest possible resolution in order to faithfully model also small
galaxies and to limit the amount of relevant physics modelled as
``sub-grid''.  This clearly requires a very good scaling behaviour of
the employed numerical scheme to allow the use of state-of-the-art
supercomputers.

Over the last several years, a large body of results based on
hydrodynamical simulations of galaxy formation have been
obtained~\citep[e.g.,][]{Ocvirk2008, Crain2009, Schaye2010, Dave2011,
  McCarthy2012, Puchwein2013, Kannan2013}. A detailed investigation of
the impact of various physical mechanisms on cosmic gas can be found
in~\cite{Schaye2010} as well as in other recent studies
\citep[e.g.][]{Haas2012a, Haas2012b}. A consensus from these works is
that feedback processes are crucial for any successful model of galaxy
formation.  For example, \citet{Dave2011} focused on the impact of
stellar winds on the faint-end of the stellar mass
function. \cite{Puchwein2013} tried to match the stellar mass function
and the stellar mass -- halo mass relation through stellar and active
galactic nuclei (AGN) feedback, but their simulations did not include
effects like metal line cooling and gas recycling. Other studies
focused specifically on low mass haloes and showed that specific
forms of stellar feedback may be successful in producing the correct
amount of stellar mass in these systems~\citep[e.g,][]{McCarthy2012,
  Kannan2013}.

We note that these cosmological simulations of the formation of
representative galaxy populations are markedly different from studies
where the computational power is focused on a single
galaxy~\citep[e.g.,][]{Governato2010, Agertz2011, Guedes2011, Genel2012, 
  Scannapieco2012, Aumer2013}. In the latter case, a very high spatial
and mass resolution can be achieved, allowing one to resolve the
physics to smaller scales. However, one does not obtain a statistical
sample of objects to compare to observations.  Moreover, the impact of
numerical artifacts inherent to some methods such as SPH varies
significantly from one halo to
another~\citep[e.g.,][]{Vogelsberger2012, Sales2012}.  This means that
physics models calibrated against individual objects will be corrupted
by numerical errors in unpredictable ways, rendering their
applicability to galaxy populations problematic.  These considerations
limit the ability of such ``zoom-in'' simulations by themselves to
constrain galaxy formation models.  Nevertheless, when used in
conjunction with large-scale simulations, the zoom-in procedure will
be essential for constructing more reliable sub-resolution models for
characterising entire galaxy populations.

Modern cosmological hydrodynamical simulations typically include a
sub-resolution model for star formation~\citep[e.g.,][]{Ascasibar2002,
  Springel2003a, Schaye2008, Dubois2008, Few2012}, radiative
cooling~\citep[e.g.,][]{Katz1996,Wiersma2009a}, and chemical
enrichment~\citep[e.g.,][]{Steinmetz1994,Mosconi2001,Lia2002,Springel2003a,
  Kobayashi2004, Scannapieco2005, Tornatore2007, Oppenheimer2008,
  Wiersma2009b,Few2012}.  Furthermore, some small-scale simulations
also include magnetic fields~\citep[e.g.,][]{Teyssier2006, Dolag2009, Collins2010, Pakmor2011, Pakmor2012},
radiative transfer~\citep[e.g.,][]{Abel2002, Petkova2010, Cantalupo2011}, cosmic ray
physics~\citep[][]{Jubelgas2008} and thermal
conduction~\citep{Jubelgas2004, Dolag2004}.  Most importantly,
hydrodynamic simulations of galaxy formation usually include some form
of stellar feedback~\citep[e.g.,][]{Dekel1986,
  Navarro1993,Mihos1994,Gerritsen1997,Thacker2000,Kay2002,Kawata2003,SommerLarsen2003,Springel2003a,Brook2004,Oppenheimer2006,Stinson2006,Dubois2008,DallaVecchia2008,Okamoto2010,
  Piontek2011,DallaVecchia2012, Stinson2013}, and more recently also
AGN feedback~\citep[e.g.,][]{Springel2005, Kawata2005, DiMatteo2005,
  Thacker2006, Sijacki2006, Sijacki2007, DiMatteo2008, Okamoto2008, Booth2009,
  Kurosawa2009,Teyssier2011, Debuhr2011, Dubois2012}.

In a recent series of papers
\citep{Vogelsberger2012,Keres2012,Sijacki2012,Torrey2012,Nelson2013, Bird2013}
we have demonstrated that a moving mesh approach as implemented in the
{\sm AREPO} code~\citep[][]{Springel2010} provides a promising
numerical method for performing galaxy formation calculations.
Indeed, such a scheme combines most of the advantages of previous
Eulerian and pseudo-Lagrangian techniques used in the field while
avoiding many of their weaknesses.  We therefore concluded in our
previous work that the quasi-Lagrangian finite volume scheme of
{\small AREPO} is well-suited for reliably solving the hydrodynamical
equations posed by the galaxy formation problem.  However, the
simulations presented in~\cite{Vogelsberger2012} accounted for only a
rather limited set of physical processes in order to simplify
comparisons with earlier numerical approaches.  Most importantly, they
did not include any form of explicit stellar feedback and also did not
account for feedback from active galactic nuclei (AGN). But it has
become clear over the last decade that both forms of feedback are
crucial for shaping, for example, the stellar mass function.  The
stellar mass function may be viewed as a convolution of the underlying
DM halo mass function with the efficiency with which stars form in
these haloes~\citep[][]{Springel2003b}.  Stellar feedback regulates
star formation in low mass haloes, since their potential 
wells are sufficiently shallow such that energy supplied through supernovae can remove gas
efficiently from the galaxy, yielding
galactic winds and outflows.  For more massive systems, this channel
of feedback becomes inefficient. However, as was realised already many
years ago, AGN feedback from black holes (BH) can supply sufficient
energy to quench star formation in these more massive systems. A
combination of both feedback channels is the leading theoretical
conjecture to explain the observed stellar mass function in the local
Universe.

Besides these feedback mechanisms other physical processes are known
to be important but were not included in our first moving-mesh
simulations of galaxy formation presented
in~\cite{Vogelsberger2012}. In particular, these simulations
considered cooling processes only through a primordial mixture of
hydrogen and helium. But evolving and dying stars release significant
amounts of heavier elements, which leads to an important source of
additional atomic line cooling once these metals are mixed with
ambient gas.  Stellar evolution processes also lead to gas recycling
such that baryonic matter is not completely locked up in stars in a
permanent fashion; rather it is returned at some rate to the gas phase,
even by old stellar populations. Both metal line cooling and gas
recycling elevate the supply of cold gas to galaxies and hence tend to
increase the star formation rates in these systems.

In this paper, we present a new implementation of galaxy formation
physics in the moving code {\sm AREPO}, suitable for large--scale
hydrodynamical simulations of structure formation. Our goal is to
account for all primary physical processes that are known to be
crucial for shaping the galaxy population.  While we aim to be
reasonably comprehensive in this endeavour, we note that it is clear
that we cannot be fully complete in this first installment of our
model. For example, we neglect some ``non-standard'' physics that has
been proposed to influence galaxy formation, such as cosmic rays and
magnetic fields.  Besides presenting our implementation we also
demonstrate how different feedback effects influence properties of
galaxies and how the simulated galaxy population compares to current
observations. We also identify the physical parameters of a fiducial
reference model that leads to reasonable agreement with many key
observables.

This paper is organised as follows. In Section~2, we present our
galaxy formation model.  Analysis and post-processing techniques that
we have implemented in {\sm AREPO} are described in Section~3.  The
first cosmological simulations using the newly added physics, and a
comparison to different observational datasets, are discussed in
Section~4.  Here we also explore different feedback settings and
introduce a fiducial reference model which provides a good fit to a
set of important key observations.  Finally, we give our conclusions
and summary in Section~5 of the paper.

\section{Galaxy formation model}\label{sec:galmod}

In the following subsections we describe the physics implementation of our
galaxy formation model. We first give a very brief description of our numerical
scheme for solving the hydrodynamical equations.  We then present our
sub-resolution model for star formation followed by a discussion of our stellar
evolution model. Next we discuss our approach for handling gas cooling.
Finally, the second half of this section covers our methods for including black
holes and feedback processes (stellar and AGN).

\subsection{Hydrodynamical method}\label{sec:hydro_method}

Most previous hydrodynamical simulations of galaxy formation have
employed the smoothed particle hydrodynamics (SPH) technique
\citep[][]{Lucy1977, Gingold1977, Monaghan1992, Monaghan2005}, where
gas is discretised into a set of particles for which appropriate
equations of motion can be derived.  The SPH method is well-suited for
cosmological applications owing to its pseudo-Lagrangian character
that automatically adjusts its resolution when structures collapse
gravitationally.  Furthermore, SPH manifestly conserves energy,
momentum, mass, entropy {\em and} angular momentum, 
a feature unmatched by
competing techniques.\footnote{We note, however, that simultaneous
conservation of entropy (for adiabatic flows) and energy are not
automatically guaranteed if SPH smoothing lengths are
adaptive \citep[][]{Hernquist1993}.  This can be achieved, in
general, only if the
equations of motion are derived using a variational principle
\citep[][]{Springel2002, Hopkins2013}, which is not the case
for many popular SPH codes.}
Other methods for solving the equations of
hydrodynamics in a cosmological context have been employed as well,
most of which are descendants of classic Eulerian methods
\citep[][]{Stone1992} implemented as adaptive mesh refinement (AMR)
codes \citep[][]{Berger1989, Teyssier2002, OShea2004}. Several studies
have however pointed out that these widely used schemes can lead to
significant differences in the results for galaxy formation
\citep[e.g.,][]{Frenk1999, Agertz2007, Mitchell2009}.

In this paper, we use a new numerical approach where a moving-mesh is
used to solve the hydrodynamical equations \citep[the AREPO
  code,][]{Springel2010}, based on a quasi-Lagrangian finite volume
method.  The primary motivation is similar to earlier
implementations of moving meshes by \citet{Gnedin1995} and
\citet{Pen1998}, but our method avoids the mesh-twisting problems that
troubled these first attempts.  To this end, {\sm AREPO} employs an
unstructured Voronoi tessellation of the computational domain. The
mesh-generating points of this tessellation are allowed to move
freely, offering significant flexibility for representing the geometry
of the flow. This mesh is then used to solve the equations of ideal
hydrodynamics with a finite volume approach using a second-order
unsplit Godunov scheme with an exact Riemann solver.  This technique
avoids several of the weaknesses of the SPH and AMR methods while
retaining their most important advantages \citep[see the detailed
  discussion in][]{Springel2010, Vogelsberger2012, Keres2012,
  Torrey2012, Sijacki2012}. We note that a number of SPH extensions
  were proposed, which solve some of the 
defects of standard SPH \citep[e.g.,][]{Price2008, Wadsley2008, Cullen2010, Abel2011, Read2012, Saitoh2013}.

\subsection{Star formation}\label{sec:starformation}

Owing to the limited resolution in large-scale structure simulations,
we must treat the star-forming interstellar medium (ISM) in a
sub-resolution fashion.  Here we follow~\cite{Springel2003a} and model
the star-forming dense ISM gas using an effective equation of state
(eEOS), where stars form stochastically above a gas density of
$\rho_\mathrm{sfr}$ with a star formation time scale of
$t_\mathrm{sfr}$.  Such an approach for simulations with an unresolved
ISM structure is very common~\citep[e.g.,][]{Ascasibar2002,
  Springel2003a, Springel2005c, Robertson2006, Schaye2008, Dubois2008,
  Hopkins2008, Hopkins2009, Hopkins2009b, Few2012}.  The idea behind
the use of an eEOS is that the ISM is believed to be governed by
small-scale effects like supersonic turbulence, thermal instability,
thermal conduction, and molecular cloud formation and evaporation
processes. The combination of these processes is believed to quickly
establish a self-regulated equilibrium state of the ISM. In that
regime, the average ISM temperature may be approximated as a function
of density only.

We introduce a few modifications of the original~\cite{Springel2003a}
implementation. First, to seamlessly connect to our stellar evolution
model (see below) we use a~\cite{Chabrier2003} initial mass function
(IMF) instead of the Salpeter IMF. Note that this also implies a
larger supernovae type II (SNII) energy input in the ISM per formed
solar mass of stars.  Second, we take stellar mass loss processes into
account when deriving the star formation rate based on the cold gas
fraction in the ISM.  Third, we prevent cells from being star-forming
(both in terms of their star-formation rate, and of the effective
equation of state) if their temperature is above the temperature
corresponding to the effective equation of state at their
density. This prevents spurious star-formation in diffuse hot gaseous
haloes. Fourth, for determining the temperature of star-forming gas,
we interpolate between the full~\cite{Springel2003a} eEOS and an
isothermal EOS at $10^4\,\mathrm{K}$ with an interpolation parameter
${q_\mathrm{EOS}=0.3}$, which avoids an over-pressurisation of the
ISM~\citep[][]{Springel2005, Hopkins2010}.  We did not incorporate
metal dependent density thresholds nor did we take into account metal
cooling within the two-phase model. Although adopting this would be
straightforward, these effects do not lead to any significant changes
in the functionality of our model, hence we ignore them here in order
to keep the model as simple as possible. The adopted parameter choices
for our star formation model are
$\rho_\mathrm{sfr}=0.13\,\mathrm{cm}^{-3}$ and
$t_\mathrm{sfr}=2.2\Gyr$, in the notation of \citet{Springel2003a}.
Our ISM sub-resolution model alone, by construction, does not drive galactic winds.
We will discuss this below in more detail and present a model for the explicit generation
of galactic outflows.

We note that once the numerical resolution is sufficient to resolve
all the relevant ISM phases and processes, an eEOS is in principle not
necessary anymore. In this case, a direct modelling of the relevant
physical ISM processes, like momentum/energy injection and radiation
pressure associated with star formation, is expected to lead to a
turbulent and self-regulated quasi-equilibrium structure of the
ISM. Impressive progress in this direction has recently been achieved
\citep[e.g.][]{Hopkins2012b, Hopkins2012a}.  However, we point out
that the resolution requirements for such an approach are
significantly beyond what is currently feasible in large-scale
cosmological simulations, hence this is applicable only in ``zoom-in''
simulations of individual galaxies of low to moderate mass or in
idealised simulations of isolated galaxies \citep[][]{Hopkins2012d,
  Hopkins2013b, Hopkins2013c}.  In a forthcoming study~(Marinacci et
al., in prep), we will present a much more detailed ISM prescription
on top of the other physics implementations discussed below.

\subsection{Stellar evolution and chemical enrichment}\label{sec:stellar_evolution_and_enrichment}
Elemental abundance patterns are an important diagnostic tool for
galaxy formation. Cosmological simulations track the evolution of
baryons as they fall into galaxies, form stars, and are recycled back
into the ISM. Including estimates for the recycling of material from
aging stellar populations is important because it plays a central role
in determining the evolution of a galaxy's gas content and because the
metal-rich ejecta from evolved stars determine its heavy element
content.

Handling the mass and metal return from massive stars can be well-approximated
by assuming instantaneous massive star evolution, which is implemented by
lowering the effective star formation rate while the metal content of nearby
gas is increased to account for metal
production~\citep[e.g.,][]{Steinmetz1994,Springel2003a}.  This instantaneous
recycling approach is justified for massive stars because they have relatively
short lifetimes ($\lesssim 10^7\yr$).  However, an instantaneous recycling
model cannot be accurately applied to intermediate mass stars, as the timescale
for stellar evolution, and therefore the timescale for mass return, is
comparable to or larger than galactic dynamical times.  Most notably,
asymptotic giant branch (AGB) stars return a substantial fraction of their mass
to the ISM over expected lifetimes that can exceed $10^9\yr$.  Similarly, Type
Ia supernovae explode with substantial delay times and, although they do not
return a very large amount of mass to the ISM, they produce substantial
quantities of iron.  This gives rise to a situation where both the mass return
rate and the elemental composition of the returned material will change as a
function of time, depending on which mechanisms are operating.

A more general enrichment approach can be developed by tracking the
mass and metal return of a stellar population to the ISM as a function
of time~\citep[e.g.,][]{Mosconi2001,Lia2002,Kobayashi2004,
  Scannapieco2005, Tornatore2007, Oppenheimer2008, Wiersma2009b,
  Few2012}.  In our model, we integrate the time evolution of stellar particles
to determine their mass loss and chemical enrichment as a continuous
function of time.  We use information from stellar evolution
calculations to determine the expected main sequence lifetime, mass
return fraction, and heavy element production for a wide range of
initial stellar masses and metallicities.  As a result, our model
calculates not only the appropriate time-delayed mass loss rate, but
also the chemical composition of the mass loss to the ISM.  In the
following, we briefly outline our scheme, which is similar to the
implementation presented in~\citet[][]{Wiersma2009b}.

\subsubsection{Stellar evolution}\label{sec:stellar_evolution}
We assume that each star particle in our simulations is a single-age
stellar population (SSP) that represents a collection of discrete
stars all born at the same time $t_0$. The distribution of stellar
masses contained in each SSP is described by an initial mass function
(IMF) $\Phi (m)$ such that the integral of the IMF is equal to the
mass of a star particle at its birth time,
\begin{equation} 
M_{*} (t=t_0) = \int_{0}^{\infty} \!\!\!\!\!\! m \, \Phi(m) \dint m.  
\label{eq:MtotIMF} 
\end{equation}
We further assume that the post main sequence evolution of stars is
instantaneous. This assumption is justified because all stars have
post main sequence evolutionary timescales that are typically
$\lesssim 1/10$ of their total lifetime.  All stars within each SSP
have thus a well-defined lifetime after which they instantly return
some fraction of their mass and metals to the ISM.  We keep track of
the expected lifetimes using the stellar lifetime function $\tau
(m,Z)$, which specifies the lifetime of stars on the main sequence as
a function of their initial mass $m$ and metallicity $Z$. We can
invert this function to obtain the inverted lifetime function
$\mathcal M(t=t_0+\tau,Z)$, which gives the mass of stars that are
moving off the main sequence at time $t$.  Using the inverted lifetime
function in conjunction with a chosen IMF, we can find the mass of
stars evolving off the main sequence during any time step $\Delta t$,
\begin{equation} 
\Delta M(t, \Delta t,Z) = \int_{\mathcal M(t+\Delta t,Z)}^{\mathcal M(t,Z)} \!\!\!\!\!\!\!\!\! m  \, \Phi(m) \dint m.
\label{eqn:MainSequenceMassLoss} 
\end{equation}
In our fiducial model, we choose a~\cite{Chabrier2003} IMF of the form 
\begin{equation}
\Phi(m) = \left\{
  \begin{array}{l l}
    \!A \, m^{-1} \, \exp\left(-\frac{\log(m/m_c)^2}{2 \sigma^2}\right) & \text{, $m \leq 1\,\msun$ }\\ & \\ 
    \!B \, m^{-2.3}                                                     & \text{, $ m > 1\,\msun$ },\\
    \end{array} \right.
\end{equation}
where $m_c = 0.079$, $\sigma = 0.69$, and $A=0.852464$ and
$B=0.237912$ are normalisation coefficients constrained to yield a
continuous function at the transition mass scale.  We note that our
IMF choice influences the total mass returned to the
ISM~\citep[e.g.,][]{Leitner2011} through the relative fraction of
stars at the high and low mass ends. Using a different IMF would
involve a trivial change to our model.  For convenience, we have
chosen the~\cite{Chabrier2003} IMF because it provides a reasonable
fit to observational data both in our Galaxy and in nearby early-type
galaxies for which detailed dynamical data are
available~\citep[e.g.,][]{Cappellari2006}. For the normalisation of
the IMF according to equation~(\ref{eq:MtotIMF}) we apply a lower mass
limit of $0.1\msun$ and an upper limit of $100\msun$. We note
that there is also the possibility of a varying IMF \citep[e.g.][]{Conroy2012}.

We adopt the lifetime function from~\cite{Portinari1998}, which gives
the expected stellar lifetimes as a function of initial stellar mass
and metallicity. These lifetimes are calculated as the sum of the
hydrogen and helium burning timescales for stars as a function of mass
and metallicity.

\subsubsection{Mass and metal return}\label{sec:mass_return}

To calculate the stellar mass loss of SSPs, we define the stellar
recycling fraction $f_\mathrm{rec} (m, Z)$, which indicates the total
fraction of a star's initial mass that is recycled to the ISM over its
entire lifetime.  The dominant source of recycled material will change
as a function of stellar mass $m$ and metallicity $Z$.  While the most
massive stars ($m > 13\,\msun$) will return most of their mass in core
collapse supernovae, less massive stars return their mass
via AGB winds.  Regardless of the source, we can calculate the amount
of mass returned to the ISM $\Delta M_\mathrm{rec}$ during a time step
in our simulation as
\begin{equation} 
\Delta M_\mathrm{rec}(t, \Delta t, Z) = \int_{\mathcal M(t+\Delta t)}^{\mathcal M(t)} \!\!\!\!\!\!\!\!\! m \, f_\mathrm{rec}(m, Z) \, \Phi(m) \dint m.
\label{eqn:MainSequenceMassReturn} 
\end{equation}
Besides the total mass return we also track the return and production
of individual chemical elements. If we denote the initial mass
fraction of each element as $Z_i$, such that $\sum_i Z_i = 1$, where
$i$ is a sum over all elements (including hydrogen, helium, and all
metals), then the mass of element $i$ that is ejected from an
unenriched stellar population during a simulation time step is given
by
\begin{equation} \Delta M_i(t, \Delta t, Z) = Z_i \int_{\mathcal M(t+\Delta
t)}^{\mathcal M(t)} \!\!\!\!\!\!\!\!\! m \, f_\mathrm{rec}(m, Z) \, \Phi(m) \dint m.
\label{eqn:MainSequenceMassReturn2} \end{equation}
In the absence of chemical enrichment, the composition of the returned material
is, by construction, identical to the initial metallicity of the star,
independent of our choice for the returned mass fraction or IMF.

To include chemical enrichment, we must have some knowledge about the
production or destruction of each element. We achieve this by using elemental
mass yields $y_i (m, Z)$ that specify the amount of mass created or destroyed
for each element $i$, and each initial stellar mass $m$ and initial
metallicity $Z$.  By definition, the yield for each element is 
\begin{equation}
y_i (m, Z) =  M_{i,\mathrm{enrich}} (m, Z) - m \, Z_i \, f_\mathrm{rec}(m, Z),
\label{eqn:BasicYields}
\end{equation}
where $M_{i,\mathrm{enrich}} (m,Z)$ is the total mass of element $i$ that is
returned to the ISM. We note that the purpose of the mass yield is to track the
transfer of mass  between elements, not to create or destroy mass. As such, the
production of one element must come at the expense of the destruction of
another element such that $\sum_i y_i = 0$.  We tabulate the elemental mass
yields and incorporate them into our simulations via lookup tables. Using
equations~(\ref{eqn:MainSequenceMassReturn}) and~(\ref{eqn:BasicYields}) we
calculate the mass return as 
\begin{equation}
\Delta M_i(t, \Delta t, Z) \!=\!\! \int_{\mathcal M(t+\Delta t)}^{\mathcal M(t)} \!\!\!\!\!\!\!\!\!\left( y_i + m \; Z_i \, f_\mathrm{rec}(m,Z)\right)\, \Phi(m) \dint m.
\label{eqn:MainSequenceElementMassReturn}
\end{equation}
Equation~(\ref{eqn:MainSequenceElementMassReturn}) is used to
determine the mass return for each element at each time step so that
we can physically transfer the proper elemental abundances from
stellar particles to ISM cells. In our fiducial model we track
nine chemical elements: H, He, C, N, O, Ne, Mg, Si, Fe.

We use the elemental mass yields for AGB stars
from~\citet{Karakas2010}, which are calculated by dynamically evolving
the thermally pulsating AGB stars and then inferring the
nucleosynthetic yields using a full reaction network over a wide range
in metallicity ($0.0001 < Z < 0.02$) and initial stellar masses
($1\msun < M < 6\msun$). We adopt the elemental mass yields for core
collapse supernova from~\cite{Portinari1998}, which are calculated
using the Padova stellar evolutionary tracks in conjunction with the
explosive nucleosynthesis of~\cite{Woosley1995}. We also extract
$f_\mathrm{rec}(m, Z)$ from these references.

\subsubsection{Supernovae rates}\label{sec:typeIa}
The majority of a stellar population's mass loss and metal production
comes from AGB stars or core collapse supernovae.  However, Type
Ia supernovae are also important to consider because they produce a
substantial amount of iron. Besides, the Type Ia supernova rate is
crucial for cosmological studies.

The progenitor systems of Type Ia supernovae are still
uncertain.  
The two most commonly discussed progenitor
channels are the single degenerate scenario, where a Type Ia supernova results
from a white dwarf accreting material from a companion star, and the double
degenerate scenario, where a Type Ia supernova results from the merging of two
white dwarf stars.
In the standard single degenerate scenario, 
nearly all Type Ia progenitors will have the same mass and will be
comprised mostly of carbon and oxygen.  Since the conditions leading
to a Type Ia event are comparatively well-posed in this picture, the
chemical yields expected in each Type Ia supernova have been
calculated~\citep[e.g.,][]{Thielemann2003}.

What remains nevertheless highly uncertain even in this scenario is the rate we
expect for Type Ia events after the birth of an SSP.  In contrast, determining
the expected number and rate of AGB stars and core collapse supernovae is
relatively straightforward once we have made a choice for the IMF and lifetime
functions. This is because we expect that all stars of a given mass will end
their lives in the same way (e.g., we assume all $20\,\msun$ stars end as core
collapse supernovae).  The same simple logic cannot be applied to Type Ia
supernovae because of the uncertainty that exists about their progenitor
systems and the long time delay between the birth of a star and its eventual
explosion in a Type Ia event.
Our lack of a precise knowledge about the true form of the
IMF, stellar binary fractions, and Type Ia progenitor systems 
prevents us from implementing a first-principles based model for Type Ia supernova rates.

Instead, a simple parameterisation of the Type Ia rate based on the
delay time distribution (DTD) of Type Ia
events~\citep[e.g.,][]{Dahlen2004,Strolger2004,Greggio2005,
  Mannucci2006,Matteucci2006} is often adopted. In this formalism, the
global Type Ia rate is determined by a convolution of the star
formation rate with the DTD,
\begin{equation}
\dot N_\mathrm{Ia} (t) = \int_{0}^{t} \!\! \Psi(t^\prime) g(t-t^\prime) \dint t^\prime,
\end{equation}
where $\Psi(t)$ is the star formation rate, and $g(t)$ is the DTD.  For a
single SSP, the  star formation history is a delta function centred on the
stellar population's birth time, which gives a simplified Type Ia rate $\dot
N_\mathrm{Ia} (t) = g(t-t_0)$.  Once the DTD is specified, we can calculate the
number of SNIa events for $t>t_0$ in a given time step by 
\begin{equation}
N_\mathrm{Ia} (t,\Delta t) = \int_t^{t+\Delta t} \!\!\!\!\! \dot N_\mathrm{Ia}
(t^\prime) \dint t^\prime = \int_t^{t+\Delta t} \!\!\!\!\!
g(t^\prime - t_0) \dint t^\prime.
\end{equation}
Note that this does not require any explicit assumptions about the
progenitors of Type Ia events, the form of the IMF, or the stellar
binary fraction. All this information is implicitly contained in the
DTD. Based on the number of SNIa events in a time step for a given
SSP, we can calculate the returned mass and elements simply by
multiplying this number by the corresponding yields per SNIa, as
discussed above. Unfortunately, the exact form of the DTD is still
poorly constrained. In the following, we will use a DTD model which 
consists of a power law in time,
\begin{equation}
g(t) = \left\{
  \begin{array}{l l}
    0 & \quad \mathrm{if\;} t< \tau_{8\msun} \\
    N_0 \left(\frac{t}{\tau_{8\msun}}\right)^{-s} \frac{s-1}{\tau_{8\msun}}& \quad \mathrm{if\;} t \ge \tau_{8\msun}, \\
  \end{array}
  \right.
\end{equation}
where $s$ is the power law index and $\tau_{8 M_\odot}$ is an offset time
between the birth of the SSP and the first expected Ia event.  We take $s=1.12$
as the fiducial value~\citep{Maoz2012}, which is consistent with theoretical
expectations that relate the Type Ia rate to the loss of energy and angular
momentum to gravitational radiation in a binary
system~\citep[e.g.,][]{Greggio2005}.  Furthermore, we take $\tau_{8 M_\odot} =
40 \Myr$, corresponding to the main sequence lifetime of $\sim 8\msun$ stars,
which we assume to be the upper mass limit for progenitors of Ia events.  For
the normalisation we adopt $N_0 = 1.3 \times 10^{-3} \mathrm{[SN \; M_\odot
^{-1}]}$~\citep[][]{Dahlen2004, Maoz2012}.

With the Type Ia supernova rate specified, determining the mass return
and metal production is easily done using the yield calculations
of~\cite{Thielemann2003} and \citet{Travaglio2004}.  Mass and metals
produced in Type Ia events are then returned to the ISM in the same
fashion as for core collapse supernovae and AGB winds. We have
checked that the exact choice of the DTD (e.g., power law vs. exponential) 
does not affect most of our results in any significant way. The reason
for this is that the SNIa mainly affect the iron production, but they
are not a significant source of stellar mass loss, which is dominated
by stars in the AGB phase and SNII events.

We note that the number of SNII events in a given time step can be calculated
based on the IMF
\begin{equation}
N_\mathrm{SNII}(t,\Delta t) = \int_{\max[\mathcal M(t+\Delta t), M_\mathrm{SNII, min}]}^{\min[\mathcal M(t), M_\mathrm{SNII, max}]} \!\!\!\!\!\!\!\!\!\!\!\!\!\!\!\!\!\!\!\!\!\!\!\!\!\!\!\!\! \Phi(m) \dint m,
\label{eqn:SNIInumber}
\end{equation}
which is non-zero only if the integration interval is positive. In the
following we set $M_{\rm SNII, min}=6\msun$ and $M_{\rm SNII, max}=100\msun$ as
the mass range for stars ending in SNII.

\subsubsection{Gas enrichment}\label{sec:gas_enrich}

For each active stellar particle we calculate the total mass, total metal mass
and element return during its current time step following the scheme described
above.  These masses are then returned to the nearby Voronoi cells of the
stellar particle, where we enforce momentum and energy conservation. We
distribute the mass and metals over a set of nearest neighbouring cells using a
top hat kernel enclosing a mass of approximately $N_{\rm enrich} \times m_{\rm
target}$, where $m_{\rm target}$ is the cell target mass according to our
(de-)refinement scheme (see below).  For convergence studies we normally hold this
``enrichment mass'' fixed if we change resolution, which means that we scale
$N_{\rm enrich}$ such that the change of the target mass with resolution is
compensated. However, we find that using a constant $N_{\rm enrich}$ produces
almost identical results.

\subsection{Cooling and heating}\label{sec:cooling}

Energy loss via radiative cooling is a key physical process for galaxy
formation.  Early cosmological hydrodynamical
simulations~\citep[e.g.,][]{Katz1992, Katz1996} included cooling due to a
primordial mixture of hydrogen and helium, where two-body processes
(collisional excitation, collisional ionisation, recombination,
dielectric recombination and free-free emission) and inverse Compton
cooling off the CMB~\citep[e.g.,][]{Ikeuchi1986} were considered.
Furthermore, energy input due to photo-ionisation was injected as heat
into the gas. Typically these simulations considered a spatially
uniform, time-varying UVB radiation field.  The presence of metals in
the ISM and IGM due to enrichment processes changes the cooling rates
by increasing the number of possible transitions.  A primordial
treatment is then not accurate anymore for such an enriched gas. 

For high-temperature gas, \citet[][]{Sutherland1993} published cooling
rate tables for $14$ heavy elements over a range of metallicities,
using solar relative abundances.  Their rates assume collisional
ionisation equilibrium and do not include effects due to background
radiation fields.  However, it is important to include background
radiation when calculating the cooling rates, because it affects both
the thermal and ionisation state of the
plasma~\citep[e.g.,][]{Efstathiou1992,Gnedin2012} leading to a
suppression of the cooling rate and increase of heating
rates. \cite{Smith2008} introduced a table-based method to include
metal line cooling, accounting for photo-ionisation due to the UVB.
Their technique involves the use of the 
photo-ionisation code {\sm CLOUDY} to construct a lookup table of
metal cooling rates. This approach was also followed in other studies;
e.g.~\citet{Wiersma2009a} even considered the cooling contribution
from individual elements.

We adopt a spatially uniform time-dependent UVB and also implement
metal-line cooling in {\sm AREPO} based on {\sm CLOUDY} cooling
tables. We assume ionisation equilibrium such that we can calculate
cooling rates a priori for a given portion of dust-free and optically
thin gas, for a given background radiation field.  We still carry out
a self-consistent calculation of the primordial cooling
following~\cite{Katz1996}, on top of which we add the cooling
contribution of the metals. In principle, the metal-line cooling can
be implemented on an element-by-element basis once the main coolants
are tracked by the enrichment scheme. \cite{Wiersma2009a} showed that
it is sufficient to follow $11$ elements (H, He, C, N, O, Ne, Mg, Si,
S, Ca, Fe) to recover the main metal line cooling contribution.
Although most of these elements are explicitly followed by our
chemical enrichment model, we decided not to implement cooling on an
element-by-element basis in our default setup because: (i) individual
element yields have large uncertainties, and element-wise cooling
would directly couple those into the hydrodynamical evolution via the
cooling contribution; (ii) departures of the relative abundances from
solar lead only to small corrections compared to the overall
photo-ionisation modification~\citep[see][for example]{Wiersma2009a};
and (iii) the computational overhead for evaluating the cooling rate
scales directly with the number of elements used in the cooling. We
base our implementation of metal-line cooling therefore on the rates
for a solar composition gas, and these rates are scaled linearly with
the total metallicity $Z$.

The total net cooling rate in the simulation is then evaluated based on 
\begin{align} 
&\Lambda(T, \rho, z, Z) =  \nonumber \\  
&\Lambda_{\rm p}(T, \rho, z)  + \frac{Z}{Z_\odot}\Lambda_{\rm m}(T, \rho,z, Z_\odot) + \Lambda_{\rm
C}(T, \rho, z).  
\label{eq:coolrate} 
\end{align}
Here $\Lambda_{\rm p}$ is the net cooling contribution due to
primordial species (${\rm H}$, ${\rm H}^{+}$, ${\rm He}$, ${\rm
He}^{+}$, ${\rm He}^{++}$), $\Lambda_{\rm m}$ is the net cooling
contribution due to metals, and $\Lambda_{\rm C}$ represents Compton
cooling off the CMB. The primordial cooling and heating is calculated
directly from ionisation equations using the cooling, recombination
and collisional ionisation rates from~\cite{Cen1992,
Katz1996}. Photoionisation rates, which affect abundances and inject
energy into the gas, are calculated based on the UVB intensity
of~\cite{Faucher2009}. This ionising background has contributions
from both quasars and star-forming galaxies, with the latter dominating at
approximately $z>3$. In addition to being compatible with recent
luminosity functions, the model was calibrated to satisfy the measured
mean transmission of the Ly-alpha forest at intermediate redshifts
$z=2-4.2$ \citep[][]{Faucher2008b, Faucher2008}, published HeII to HI
column density ratios, HeII reionisation by $z\sim 3$ 
\citep[][]{McQuinn2009}, and complete HI
reionisation by $z=6$. We use the spectra from the same background as
an input to {\sm CLOUDY} to calculate matching metal line cooling
rates $\Lambda_{\rm m}$. 

For simplicity we neglect the metal
contribution to the mean molecular weight $\mu$; i.e. we calculate
temperatures based on the molecular weight corresponding to the
ionisation state of the primordial gas composition. This is
well-justified since in a fully ionised metal enriched plasma of
metallicity $Z$ the mean molecular weight is approximately given by
$\mu \cong 4 /(8 X \, + \, 3 Y + 2 Z)$, such that even at super-solar
composition the metal contribution leads only to minor changes in
$\mu$~\citep[see also][]{Smith2008}.  We also neglect the metal
contribution to the free electron density, which is also justified
since those will only be relevant for $Z \gg
Z_\odot$~\citep[e.g.,][]{Wiersma2009a}. For solving the primordial
network we use the advected H, He mass fractions as an input.
Together with the spatially uniform but temporally varying UVB this
gives us the primordial abundances and a self-consistent temperature
based on $\mu$.  With this temperature we calculate $\Lambda_{\rm p}$,
and look up $\Lambda_{\rm m}$ in the pre-calculated {\sm CLOUDY} tables
to arrive at the total cooling rate through
Equation~(\ref{eq:coolrate}).

For the construction of the lookup tables, we tabulated metal line net
cooling rates on a grid in $\log(T/{\rm K})$, $\log(n_{\rm H}/{\rm
  cm}^{-3})$ and redshift: $1<\log(T/{\rm K})<9$ at $200$ equally
spaced grid points, $-8<\log(n_{\rm H}/{\rm cm}^{-3})<2$ at $50$
equally spaced grid points, and $0<z<10$ at $50$ grid points.  During
the simulation, we linearly interpolate the metal line cooling rates
in $[z, \log(n_{\rm H}/{\rm cm}^{-3}), \log(T/{\rm K})]$ space.

\subsubsection{Self-shielding}\label{sec:selfshielding}

So far we described our treatment of cooling and heating in the
presence of a spatially uniform UVB radiation under the assumption
that the gas is optically thin to the radiation. However, this
approximation breaks down at gas densities $\rho\gtrsim10^{-3}\,{\rm
  cm}^{-3}$, depending on the details of the radiation field. Above
such densities, the gas absorbs the radiation to such a degree that the
radiation field is attenuated compared to the optically thin
case. This modified radiation field in turn affects the heating rate
and ionisation state, and hence the cooling rate, differently than the
unattenuated radiation field. As a result, the equilibrium gas
temperature ends up being different, and hence the dynamical evolution
can be affected.  

In order to approximately account for this effect, we implemented a
simple prescription for the self-shielding of gas from the UVB
radiation to be used on-the-fly, rather than only in post-processing
as has been common practice~\citep[e.g.,][]{Bird2011}.  Our
implementation is derived from the results of radiation transfer
simulations by~\cite{Rahmati2013}, who post-processed cosmological
simulations using the radiation transfer code
{\sm TRAPHIC}~\citep[][]{Pawlink2008, Pawlink2011}, and quantified the
self-shielding of the intergalactic gas as a function of cosmological
epoch and gas density. We implemented their Equation (A1) with the
parameters given in their Table A1 both for the primordial network
that is directly followed in our code, and for the {\sm CLOUDY} calculations
we use to generate the lookup table.  Specifically, the ionisation and
heating rates entering into the primordial network are suppressed by 
\begin{equation}
(1-f) \left[1+\left(\frac{n_\mathrm{H}}{n_\mathrm{0}}\right)^\beta\right]^{\alpha_\mathrm{1}} + f \left[1+\frac{n_\mathrm{H}}{n_\mathrm{0}}\right]^{\alpha_\mathrm{2}},
\end{equation}
and the same factor is used to suppress the normalisation of the input
radiation field that is given to the {\sm CLOUDY} code, making the approximation that
its spectrum is unchanged. The numerical values of the various parameters in
the self-shielding formula ($\alpha_\mathrm{1}$, $\alpha_\mathrm{2}$, $\beta$,
$f$, $n_\mathrm{0}$) are interpolated linearly in redshift $z$ between the
values provided in Table A1 of~\cite{Rahmati2013}, up to $z=6$, above which we
assume zero self-shielding. Finally, we note that the radiation field is not
considered for star-forming gas, which is placed on the effective equation of
state of the star-formation model.

\subsection{Stellar feedback}\label{sec:stellar_feedback}

Although our subgrid ISM model implicitly invokes thermal SNII
feedback, this is not sufficient to avoid the well-known overcooling
problem~\citep[e.g.,][]{Springel2003a}. Large-scale cosmological
simulations therefore typically resort to some subgrid model for galactic winds
and outflows, capable of efficiently removing baryonic material from
the star-forming phase.  Two approaches are typically followed in a
regime of limited resolution: injection of SN energy in kinetic
form~\citep[e.g.,][]{Navarro1993,Mihos1994,Kay2002,Springel2003a,Oppenheimer2006,Dubois2008,DallaVecchia2008,Okamoto2010,
  Hopkins2012} or suppression of radiative cooling after injection of
thermal energy~\citep[e.g.,][]{Gerritsen1997, Thacker2000,
  Kawata2003,SommerLarsen2003, Brook2004, Stinson2006,
  Piontek2011,DallaVecchia2012, Stinson2013}.  

The kinetic wind models are typically characterised by a mass loading
factor $\eta_{\rm w}$, which gives the ratio of the wind mass flux and
star formation rate, and an initial wind velocity $v_{\rm w}$.  In the
original kinetic wind scheme of~\cite{Springel2003a}, star-forming SPH
particles are converted into wind particles stochastically.  In order
to allow a precise specification of the wind mass flux, the
implementation ensured that wind particles can leave the dense ISM gas
without entraining additional star-forming particles. In order to
technically realise this sub-grid prescription, hydrodynamical
interactions were temporarily disregarded for newly launched wind
particles until they ``recouple'' just outside of the star-forming
phase, based on a density threshold criterion.  The mass loading and
wind velocity in~\cite{Springel2003a} were chosen to be constant. More
recent studies propose variable wind models, however, where the wind
velocity is not held fixed. For example, \cite{Oppenheimer2006} argue
for momentum driven winds, where the wind speed scales as the galaxy
velocity dispersion and the mass loading is inversely proportional to
that~\citep[see also][]{Murray2005}. Such an approach appears to be
supported by observations.  \cite{Martin2005} and
\citet{Weiner2009} found that the final wind velocity scales
approximately linearly with the circular velocity.  \cite{Okamoto2010}
use the local velocity dispersion of dark matter to prescribe the wind
velocity, although they consider energy driven winds, which are generated
locally around the stellar particles similar to the approach
of~\cite{DallaVecchia2008}, which uses constant local winds.
\cite{Puchwein2013} reported a good match to the observed stellar mass
function with an energy-driven variable wind model, where the wind
velocity is derived from the escape velocity of the host halo.

We have implemented two alternative wind schemes in {\sm AREPO}.  Our
basic approaches for modelling winds and outflows are similar to those
presented in previous work~\citep{Springel2003a, Oppenheimer2006,
  Oppenheimer2008, DallaVecchia2008, Okamoto2010, Puchwein2013}, but
our technical implementation is slightly different to account for our
mesh-based hydro scheme.  Also, we note that the overcooling problem
tends to be more severe in {\sm AREPO} compared to previous SPH
studies~\citep[see][for details]{Vogelsberger2012, Keres2012}. The
origin of this lies in part in the treatment of subsonic turbulence
and in spurious viscous heating in SPH, which leads to an artificial
heating of gaseous haloes in SPH simulations~\citep{Vogelsberger2012,
  Bauer2012}. This effect is strongest for massive haloes, and we will
show below that our simulations require stronger AGN feedback because
of this (see below).

\subsubsection{Non-local SNII feedback}\label{sec:nonlocal_feedback}

In our first stellar feedback implementation, which we call
``non-local'', we launch winds directly from the star-forming ISM gas,
which is part of the two-phase medium. This ties the wind rate
directly to the star formation rate of star-forming gas. For
large-scale simulations, the time steps are typically not much shorter
than the SNII delay time; i.e. it is a valid approximation to neglect
locality for SNII driven winds in that case.

We realise the generation of the wind at a technical level by turning
gas cells (or a fraction of their mass) for a short period of time
into massive wind particles which interact gravitationally but do not
couple to the hydrodynamical calculation.  For these non-local winds,
we probabilistically select a star-forming ISM gas cell, which we then
convert into a wind particle. This particle is then allowed to travel,
decoupled from hydrodynamical forces, until it reaches a certain
density threshold or a maximum travel time has elapsed. Once either of
these criteria is fulfilled, we dissolve the particle and deposit its
mass, momentum, thermal energy and all tracked metals into the gas
cell in which it is currently located. The net effect of this
procedure mimics the sub-grid wind prescription of
\cite{Springel2003a} and facilitates a precise control of the wind
parameters independent of numerical resolution. We note that \cite{DallaVecchia2008}
demonstrated that decoupling wind particles can have an effect on the galaxy properties.
However, a decoupled scheme gives better control over mass loadings and typically
converges better, which is why we favour such an approach.

We set the mass loading and wind velocity based on the local DM
velocity dispersion~\citep[similar to][]{Oppenheimer2008, Okamoto2010}:
to this end, each gas cell calculates the local one-dimensional DM
velocity dispersion ($\sigma_{\rm DM}^{1D}$) at its current location,
and the wind velocity is then set to
\begin{equation}
v_{\rm w} = \kappa_{\rm w} \, \sigma_{\rm DM}^{1D},
\end{equation}
where $\kappa_{\rm w}$ is a dimensionless model parameter.
\cite{Okamoto2010} found that the local velocity dispersion measured
at the position of star forming gas strongly correlates with the
maximum of the circular velocity profile of its host halo ($V_{\rm
  max}\simeq 1.45\,\sigma_{\rm DM}^{1D}$), which motivates such a wind
velocity scaling.

Once the wind velocity is determined, the wind mass, given by the mass loading $\eta_{\rm w}$,
is determined by combining momentum and energy driven wind contributions 
\begin{equation}
\eta_{\rm w} = \frac{1}{v_{\rm w}^2} \left(\! {\rm egy}_{\rm w} + \sqrt{{\rm egy}_{\rm w}^2 + v_{\rm w}^2 \, {\rm mom}_{\rm w}^2}\right), 
\label{eq:windmassloading}
\end{equation}
where ${\rm mom}_{\rm w}$ is the specific wind momentum, ${\rm
egy}_{\rm w}$ is specific energy available for wind generation (available SNII
energy per formed stellar mass).
Both ${\rm mom}_{\rm w}$
and ${\rm egy}_{\rm w}$ are free parameters, which are motivated by the
underlying driving mechanism of the winds.
Equation~\ref{eq:windmassloading} gives the asymptotic mass loading scalings $\eta_{\rm w} \propto v_{\rm
w}^{-2}$ for energy-driven, and $\eta_{\rm w} \propto v_{\rm w}^{-1}$ for
momentum-driven winds. This therefore represents
an average mass loading with energy and momentum contributions, where the
relative importance is regulated by ${\rm mom}_{\rm w}$ and ${\rm egy}_{\rm
w}$. For example, purely energy driven winds would be represented by ${\rm
egy}_{\rm w}=1.73\times 10^{-2}\, E_{\rm SNII, 51}\, 10^{51} \erg
\msun^{-1}$, where $E_{\rm SNII, 51}$ denotes the available energy per SNII in
units of $10^{51} \erg$ and $1.73\times 10^{-2}$ is the number of SNII per
stellar mass formed for our adopted stellar evolution model.

For the non-local ISM driven winds, star formation and the generation
of wind particles are two different mechanisms that drain gas mass
simultaneously from the ISM. It is therefore desirable to have a
common treatment of the generation of stellar and/or wind particles in
this case. Our solution is based on a unified equation for
star-formation and wind generation
\begin{equation}
\frac{d}{\dint t}(M_\star + M_{\rm w}) = -\dot{M}=(1+\eta_{\rm w})\Psi=(1+\eta_{\rm w})\frac{M}{t_{\rm SF}},
\label{eq:sf_winds}
\end{equation}
where $M$, $M_\star$ and $M_{\rm w}$ are the mass of a gas cell $i$, its
associated stellar and wind mass, respectively. 
The quantity $\Psi$ denotes the star
formation rate which is determined by the star formation time scale $t_{\rm
SF}$.  The solution of Equation~(\ref{eq:sf_winds}) specifies the amount of
stellar and wind material created during a time step of size $\Delta t$ between
$t$ and $t+\Delta t$
\begin{align}
\Delta(M_\star + M_{\rm w}) &= (M_\star + M_{\rm w})(t+\Delta t) - (M_\star + M_{\rm w})(t)  \nonumber \\
			    &= M(t) - M(t+\Delta t) \nonumber \\
                            &= M(t) \left( 1-e^{-(1+\eta_{\rm w})\Delta t/t_{\rm SF}} \right).
\end{align}
At every time step $\Delta t$, for each active gas cell, we first make a
probabilistic decision whether star-formation or wind-launching is
treated in that step. This is done by drawing a uniformly distributed
random number $x\in U(0,1)$. Then, star-formation is treated if
$x<1/(1+\eta_{\rm w})$, and wind-launching otherwise. This procedure
ensures that the expectation value for the mass-loading factor of the
wind; i.e. the ratio between the wind-launching rate and the
star-formation rate, is exactly $\eta_{\rm w}$. To avoid forming very
large numbers of stellar or wind particles,
Equation~(\ref{eq:sf_winds}) is implemented
stochastically~\citep[e.g.,][]{Springel2003a}. This is done by
assigning a probability
\begin{equation}
p^{\rm gas\,\rightarrow\,stars/wind}=\frac{M_i}{M_{\star}}(1-e^{-(1+\eta_{\rm w})\Delta t/t_{\rm SF}}),
\label{eq:star_form}
\end{equation}
to generate a star or wind particle of mass $M_{\star}$. We note that it is
guaranteed that the correct amount of stellar mass (wind mass) is created
because star-formation (wind-formation) is treated in only $1/(1+\eta_{\rm w})$
($\eta_{\rm w}/(1+\eta_{\rm w})$) of the time steps by the random decision
described above. In Equation~(\ref{eq:star_form}), the star particle mass
$M_{\star }$ is determined in the following way: if $M<2\,m_\mathrm{target}$, we
set $M_{\star}=M$ and the full cell is converted into a star particle, while if
$M \ge 2\,m_\mathrm{target}$,  and the cell
only spawns a star particle of mass $m_\mathrm{target}$. As discussed above,
$m_\mathrm{target}$ is the mean gas cell mass in the initial conditions, which
plays the role of target gas cell mass in our (de-)refinement scheme. This
procedure ensures that we avoid large mass variations among the baryonic cells
and particles.

For choosing the wind direction, we implemented two commonly used
approaches: isotropic and bipolar winds. For the isotropic case, each
wind particle is kicked in a completely random direction. In the
bipolar case, each wind particle is kicked, with a random sign, in the
direction ${\bf v}\times\nabla\phi$~\citep{Springel2003a} in the rest
frame of the corresponding FoF group. Note that both schemes do not
add total momentum on average. In the following we will only use the
bipolar implementation.

\subsubsection{Local SNII feedback}\label{sec:local_feedback}

In the wind scheme described above, wind particles are launched from
the ISM and are therefore not spawned directly from stellar
particles. Although slightly inconsistent with our stellar evolution
and enrichment model, where we use a delayed SNII return, this model
works very well for cosmological applications because of the relevant
time scales, as argued above. Nevertheless it is desirable to have the
possibility of using a more local SNII feedback implementation for
comparison. For example, \cite{Okamoto2010} use an energy-driven
local stochastic scheme, where wind particles are launched in the
neighbourhood of stellar particles. This approach is similar to the
implementation of~\cite{DallaVecchia2008} with the exception of the
way in which ``excess energy/mass loading'' is handled. Another variation
of this approach was presented in~\cite{DallaVecchia2012}.

To explore the effect of non-locality we implemented a wind creation
scheme similar to that of~\cite{Okamoto2010} for purely energy driven
winds, whose velocity scales with the local DM velocity dispersion.
Specifically, we modified our enrichment routines to also distribute
SNII energy among the gas cells that receive metals and mass. During
the enrichment step, each cell keeps track of how much energy it
receives from stellar particles such that once the enrichment is
finished, some cells have received the SNII energy $\Delta E_{\rm
  SNII}$. Each of these cells is then assigned a probability $p=\Delta
E_{\rm SNII} / (0.5 M v_{\rm w}^2)$, where $M$ is the mass of the gas
cell and $v_{\rm w}=\kappa_{\rm w} \sigma$ is the wind velocity
assigned to this particle. Based on this probability it is then
decided whether this cell should be converted to a wind particle. Once
a cell is converted, it is treated exactly the same way as in the
non-local case described above. We note that we do not require gas
cells to be above the density threshold for star formation for this to
happen.

To make contact with our non-local scheme, we can derive an effective mass
loading.  For this we follow~\cite{Okamoto2010} and assume the instantaneous
recycling approximation ($\dot{M}_{\rm w}=2 \dot{E}_{\rm w}/v_{\rm w}^2$, where
$\dot{E}_{\rm w}$ is the available energy flux for the wind generation) to
derive
\begin{equation}
\eta_{\rm w} = E_{\rm SNII, 51}  \left(\frac{v_{\rm w} = \kappa_{\rm w} \times \sigma_{\rm DM}^{1D}}{1307\kms}\right)^{-2}, 
\end{equation}
where $E_{\rm SNII, 51}$ is the energy per SNII in units of
$10^{51}\erg$ and we assumed $1.73\times 10^{-2}$ SNII per formed
stellar mass, which are the appropriate values for
a~\cite{Chabrier2003} IMF with the SNII mass limits discussed above.

\subsubsection{Wind metal loading}\label{sec:wind_metal_loading}

Wind models with an energy or momentum scaling are the two most
commonly applied wind parameterisations in the literature, and both
rely on mass loading factors that decrease with galaxy mass.  In fact,
the mass loading factors associated with low mass galaxies can often
be well in excess of unity, naturally indicating that low mass
galaxies do not efficiently hold onto their gaseous content.
Moreover, such high mass loading factors appear to be necessary for
obtaining a reasonable shape for the galaxy stellar mass
function~\citep[e.g.,][]{Dave2011, Puchwein2013}.  However,
calculations of the metal content of low-mass galaxies based on their
observed gas fractions and metallicities indicate that low mass
galaxies nevertheless retain a substantial fraction of their
metals~\citep[][]{Zahid2012}.  Thus, there is a certain tension
between the need to efficiently expel a lot of material to reduce the
star formation efficiency in low mass galaxies, and the need to retain
a sizable fraction of the metals in these same low-mass galaxies.

The origin of this tension lies in our construction of the sub-grid
wind model.  Traditionally, it is assumed that the wind particles have
the same metallicity as the ambient ISM from where they are
launched~\citep[e.g.,][]{Springel2003a, Dave2011}.  However, while the
actual wind material will be partially composed of the material from
the site where the wind is launched, a substantial fraction of the
wind's mass will be entrained as the wind vents out of the galaxy and
blows through low density regions of the ISM.  This suggests a picture
where the metallicity of the wind material will almost certainly be
different -- and likely lower -- than the metallicity of the
ISM-region from where the wind originated.

To address this point, we introduce a wind metal loading factor
$\gamma_\mathrm{w}$, which is set independently from the wind mass loading
factor, $\eta_\mathrm{w}$.  The wind metal loading factor defines the
relationship between the metallicity of newly created wind particles
$Z_\mathrm{w}$, and the metallicity $Z_\mathrm{ISM}$ of the ambient ISM, 
\begin{equation} 
Z_\mathrm{w} = \gamma_\mathrm{w} \, Z_\mathrm{ISM}.
\end{equation} 
The case $\gamma_\mathrm{w}= 1$ corresponds to the traditionally
assumed scenario of ``fully metal-loaded'' winds. The opposite extreme
is described by $\gamma_\mathrm{w}= 0$, which assumes that winds carry
no metals along with them. In this case, a created wind particle
deposits all its metals in the surrounding ISM cells before being
kicked, thereby ensuring conservation of total metal mass.

We found that the traditional $\gamma_\mathrm{w}=1$ leads to an
under-production of the gas-phase metallicity and an over enrichment
of halo gas, especially in low mass galaxies.  Both of these problems
directly result from an overly efficient ejection of metals from
galaxies via winds.  Not only does this lead to a poor match to the
observationally constrained mass-metallicity relation, but the over
enrichment of halo gas also leads to enhanced metal-line cooling. The
other extreme, when using $\gamma_\mathrm{w}=0$, leads to overly
enriched galaxy gas-phase metallicities compared to the
mass-metallicity relations, suggesting that the correct behaviour is
bracketed by these two extremes.

We note that wind metal loadings were recently discussed
in~\cite{Peeples2011}, where it was argued that winds should actually
be ``super-enriched'' such that their metallicity should be higher
than the of the surrounding ISM gas. We note however that this
conclusion was reached based on the very low mass loading factors
discussed in~\cite{Peeples2011}, which then automatically require
highly enriched winds to match the observed mass--metallicity
relation.  Such low mass loadings are however incompatible with
results from cosmological simulations; the latter require rather much
higher mass loadings for low mass systems to match the stellar mass
function (see below).

\subsection{Black hole growth and AGN feedback}\label{sec:BH_growth_and_AGN}

As with star formation, it is currently computationally impossible to
resolve the details of accretion flows around central galactic black
holes (BHs) in large-scale simulations of galaxy formation. While the
gravitational radius of supermassive black holes starts to be resolved
in zoom-in simulations, the Schwarzschild radius of these BHs is still
many orders of magnitude smaller than the spatial resolution limits of
these calculations.  Describing BH growth and related feedback
therefore requires a sub-resolution model \citep[e.g.][]{Springel2005,
  Kawata2005, DiMatteo2005, Thacker2006, Sijacki2007, DiMatteo2008,
  Okamoto2008, Booth2009, Kurosawa2009,Teyssier2011, Debuhr2011,
  Dubois2012}.  Our BH and AGN feedback implementation follows closely
previous studies~\citep[][]{Springel2005, DiMatteo2005, Sijacki2007,
  DiMatteo2008}, with modifications to adopt the scheme to the moving
mesh and finite volume technique.  In what follows, BHs are represented
by collisionless, massive sink particles that are created at early
. times, and subsequently grow in mass by gas accretion or BH
mergers. We include three different forms of back-reaction of the
accretion: thermal, mechanical and electro-magnetic feedback.  Thermal
and mechanical feedback have already been considered in some previous
large-scale cosmological simulations, whereas electro-magnetic
radiative feedback has typically been neglected (except for a few cases, see below). We here present a
novel phenomenological treatment of this form of feedback.

\subsubsection{Black hole seeding}\label{sec:BH_seeding}

Our seeding strategy for BHs in cosmological simulations follows previous work.
We regularly run a FoF group finder and assign seed BHs to massive enough FoF
groups~\citep[see][for details]{Sijacki2007, DiMatteo2008}. We set the seeding mass
threshold for FoF groups to $5\times 10^{10}\hmsun$. For a FoF group that is
selected to acquire a seed BH, we turn the gas cell with the highest density in
the group into a BH sink particle.  The dynamical mass of the particle is set
to the cell mass. This dynamical mass is typically significantly higher than the real
expected seed mass of the BH, which we set in the following to $10^{5}\hmsun$,
as in many previous studies. To address this problem, we
follow~\cite{Springel2005} and track the BH mass as a separate subgrid
variable.  This internal BH mass smoothly increases according to the estimated
BH accretion rate $\dot{M}_{\rm BH}$ (see below), such that the value of this
internal mass represents the BH mass for the case where accretion can be fully
resolved and mass is not discretised. 

We allow the BH sink particle to drain mass continuously from the primary cell
of the BH particle.  Specifically, during each time step, an active BH particle
removes the mass ${\Delta M = (1-\epsilon_\mathrm{r}) \, \dot{M}_{\rm BH} \,
\Delta t}$ from its primary cell, where $\Delta t$ is the current time step and
$\epsilon_\mathrm{r}$ radiative efficiency (see below). BH accretion rates are
typically very small compared to the actual cell masses of the primary cell
such that $\Delta M$ is usually significantly smaller than the mass of the
primary cell.  In the rare event that this is not the case, we added a ``bucket
mechanism'' which collects $\Delta M$ in an accretion mass bucket to be then
handled with a small time delay.  During subsequent time steps, this bucket is
then emptied such that we always accrete the correct amount of mass.
Specifically, if $\Delta M$ is larger than $90\%$ of the primary cell mass, we
remove these $90\%$ of the cell mass and put the remaining mass of $\Delta M$
into the bucket. During the next accretion event of the BH we then try to
remove the current $\Delta M$ from the cell plus the mass that is currently in
the bucket.  Here again, we allow at maximum a removal of $90\%$ of the cell
mass.  This procedure guarantees a continuous accretion scheme, where the
internal and dynamical masses of the BH particles grow at the same rate, which
is different from the original stochastic accretion implementation
used in previous SPH studies of this model~\citep[][]{Springel2005}. We note
that the draining scheme does not change the real dynamical mass of BH sink
particles until the internal mass is equal to or larger than the initially
assigned dynamical mass.  Also, only at that point do we actually really drain
mass from the primary cell, thereby maintaining mass conservation during the
whole evolution.

\subsubsection{Black hole growth}\label{sec:BH_growth}

BH sink particles grow in mass by accreting surrounding gas or through
BH mergers.  In our model, BH accretion is described using a
Bondi-Hoyle-Lyttleton based Eddington-limited rate
\begin{equation}
\dot M_\mathrm{BH} = \min\left[\frac{4 \pi \alpha G^2 M_\mathrm{BH}^2 \rho}{(c_s^2 + v_\mathrm{BH}^2)^{3/2}}, \dot M_\mathrm{Edd}\right],  
\label{eqnBHaccr}
\end{equation}
where $\rho$ and $c_s$ are density and sound speed of the surrounding
gas, respectively, and $v_\mathrm{BH}$ is the BH velocity relative to
the gas. $\dot M_\mathrm{Edd}$ denotes the Eddington accretion rate of the BH. In the following we will usually use a repositioning scheme
for BH sink particles that ties them to the minimum of the
gravitational potential, in which case we will neglect the relative
gas velocity term ($v_\mathrm{BH}$) in the accretion rate.  The
repositioning ensures that BHs stay at the centre of their haloes (FoF
groups), which is important to guarantee their correct growth
rate. Note that because of the relatively coarse mass resolution
available for dark matter and stars, the BHs would not by themselves
sink to the centre on the correct timescale.  But the repositioning
implies that the velocity of the BH sink particle is unphysical, such
that we do not consider it in the accretion rate. Due to our feedback
implementation the sound speed ($c_s$) near BHs is typically
significantly higher than the relative BH -- gas velocity such that
this does not introduce any relevant effects.  As in previous
studies we multiply the theoretical Bondi-Hoyle-Lyttleton accretion
rate by a factor $\alpha$ to approximately account for the
volume-average of the Bondi-rates for the cold and hot phases of the
sub-grid ISM model.

However, when no star-forming gas is present in the immediate vicinity 
of the BH (and the quasar accretion is hence in a ``low state''), this
prescription is expected to significantly overestimate the true accretion
rate.  In particular, this will happen when the BH has grown to a large
mass and is embedded in comparatively low density background gas.
Then the residual BH accretion we estimate with equation
(\ref{eqnBHaccr}), together with our quasar-mode feedback scheme (see
below), can create a hot, low-density bubble around the black hole,
with a pressure that matches the background gas pressure $P_{\rm ext}$
at the centre of the halo.  The formation of such a bubble with
density well below the star-formation density threshold would clearly 
be an unrealistic artifact of our subgrid model.  We address this
as follows. 

First, we note that if a quasar-heated bubble in
a quasi-stationary state forms, we expect that the cooling losses in
the bubble will approximately balance the injected feedback energy from the
quasar mode
\begin{equation}
\Lambda(T)\,\rho\,M_{\rm fb} \simeq  \epsilon_\mathrm{f}  \, \epsilon_\mathrm{r} \,  \frac{4 \pi \alpha G^2 M_\mathrm{BH}^2 \rho}{c_s^3}\, c^2.
\end{equation}
Here $M_{\rm fb}$ is the gas mass in the bubble (which is equal to the amount
of material that receives the feedback energy), and $\Lambda(T)$ is the cooling
function. The product $\epsilon_\mathrm{f}  \, \epsilon_\mathrm{r}$ specifies
the quasar-mode feedback strength (see below). Neglecting metallicity effects
for the moment, this equation describes an equilibrium temperature $T_{\rm eq}$
for the bubble (or equivalently a thermal energy $u_{\rm eq}$ per unit mass)
which depends only on the black hole mass, because the density dependence drops
out.  We use this temperature to define a reference pressure
\begin{equation}
P_{\rm ref} = (\gamma -1 ) \, \rho_{\rm sfr} \, u_{\rm eq},
\end{equation}
where $\rho_{\rm sfr}$ is the star-formation threshold. Note that this
definition makes $P_{\rm ref}$ effectively a function of the black
hole mass alone. We now compare $P_{\rm ref}$ for each BH to its
actual surrounding gas pressure $P_{\rm ext}$, which each BH sink
particle measures for its current location. If we have $P_{\rm ext} <
P_{\rm ref}$ then the external gas pressure is not able to compress
the gas around the BH against the quasar-mode feedback to a density
exceeding the star-formation threshold. In this case our
multiplication factor for $\alpha$ is not really meaningful. We
compensate this by lowering the accretion rate estimate by the factor
$(P_{\rm ext}/ P_{\rm ref})^2$. In the regime of very hot gas (where
$\Lambda(T) \sim T^{1/2}$), this will lower the bubble temperature
approximately by the factor $P_{\rm ext}/ P_{\rm ref}$, and increase
its density by the inverse of this factor, while the actual accretion
rate is hardly changed. We have found that this scheme reliably
prevents the formation of unphysically large bubbles around BHs in the
``low state'', whereas the self-regulated growth of the black holes and
their final masses is unaffected.

BHs do not only grow through gas accretion, but also when they merge with other BHs.
BHs merge once they are within their ``feedback radius'' (see below).
Here we also do not consider the relative velocities of BHs for the same reason
that we neglect the velocity term in the accretion rate.

\begin{figure*}
\centering
\includegraphics[width=0.49\textwidth]{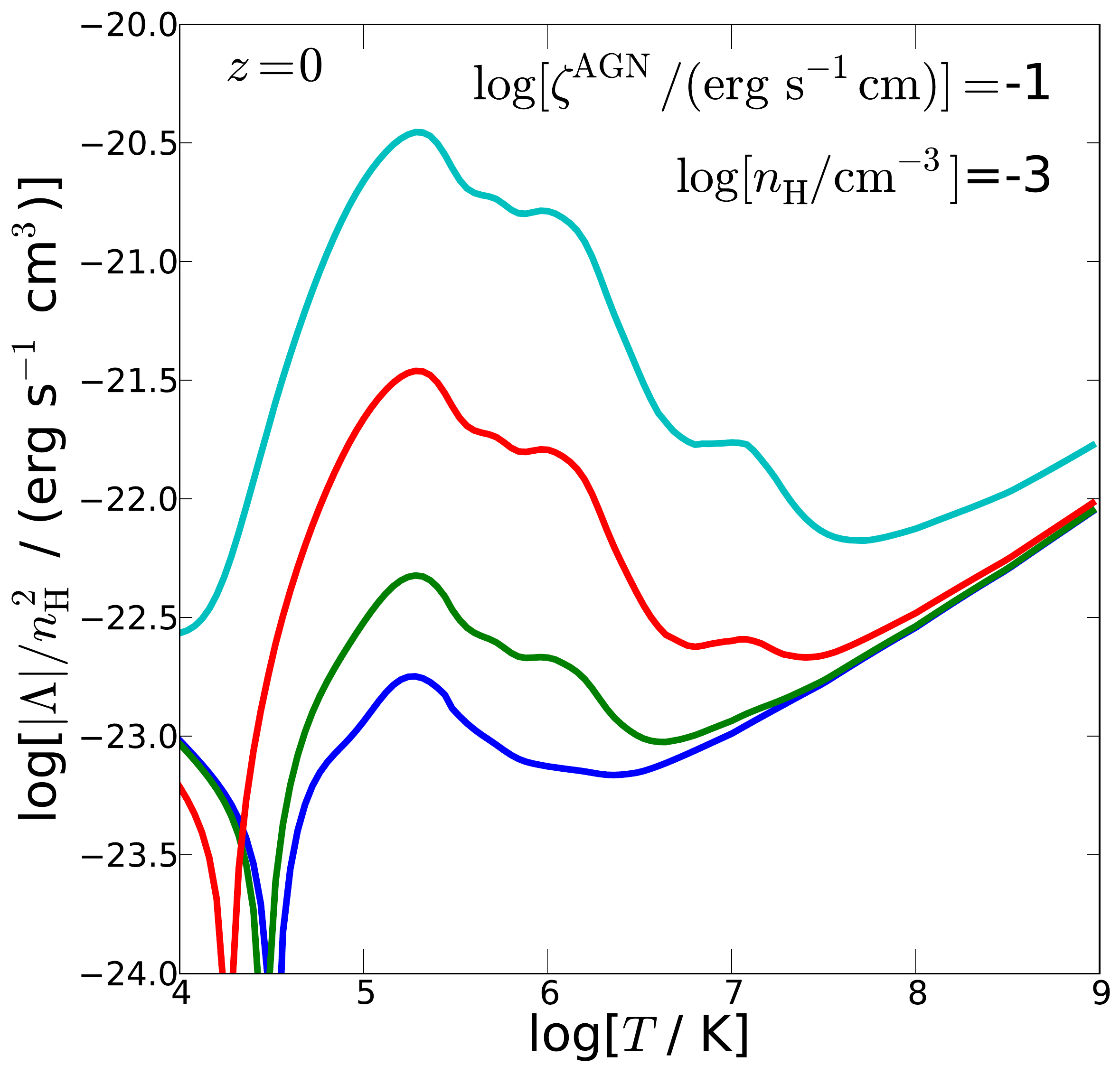}
\includegraphics[width=0.49\textwidth]{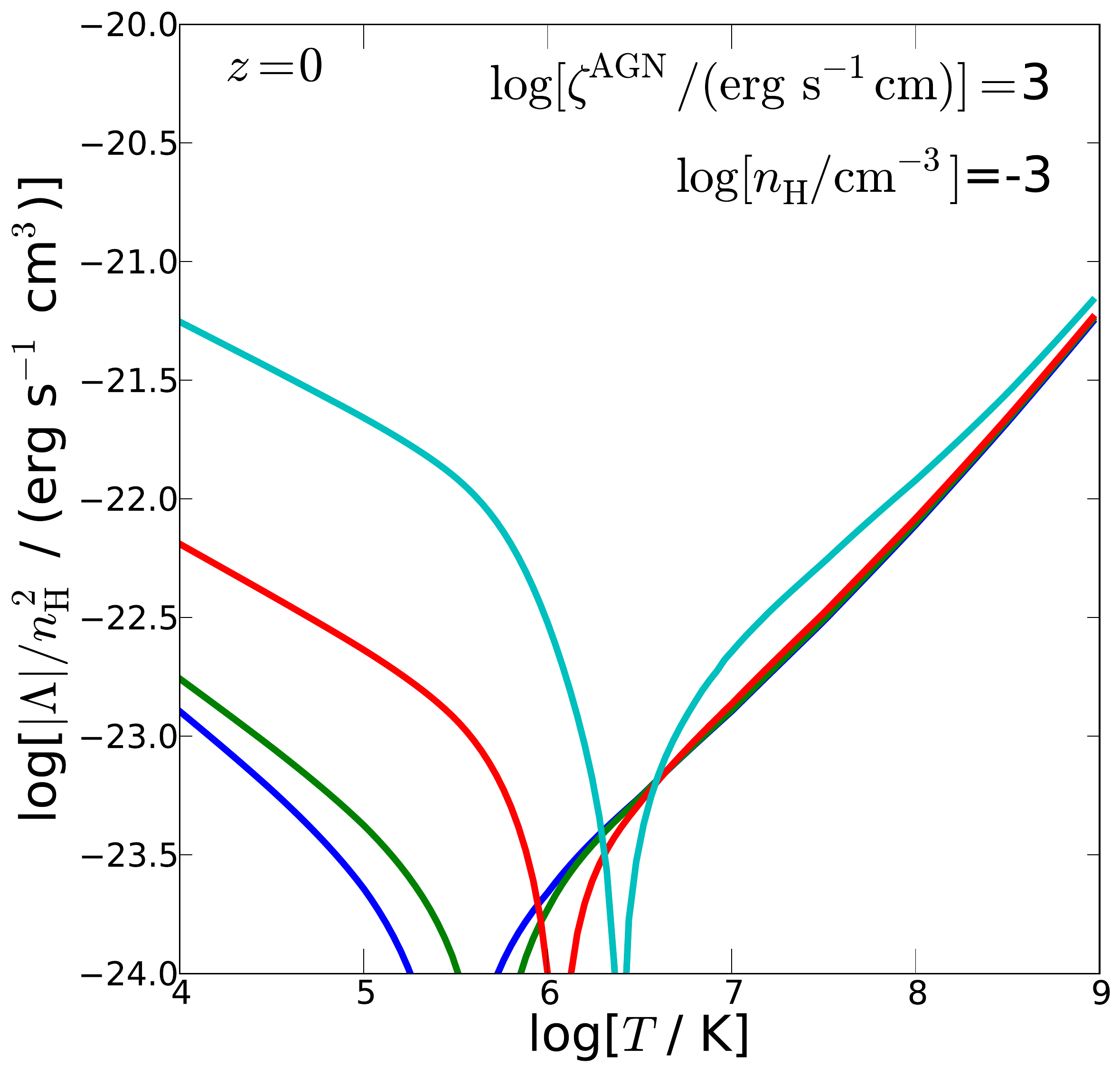}
\caption{Influence of the AGN radiation field (``electro-magnetic
  feedback'') on the net cooling rate for gases of different
  metallicities, in the range $\log[{\rm Z}/{\rm Z}_\odot]=-2.0,
  -1.0, 0.0, 1.0$ (lowest to highest curve). The AGN ionisation
  parameter is negligibly small in the left panel ($\zeta^{\rm
    AGN}_{\rm thresh}=10^{-1}\,{\rm erg}\,{\rm s}^{-1}{\rm cm}$), but
  substantial in the right panel ($\zeta^{\rm AGN}_{\rm
    thresh}=10^{3}\,{\rm erg}\,{\rm s}^{-1}{\rm cm}$). Clearly, a high
  ionisation parameter leads to a strong cooling suppression and
  significant additional heating, as seen in the net cooling rates in
  the right panel. We note that such large ionisation parameters
  require high BH accretion rates close to Eddington. Therefore, halo
  gas is exposed to such a radiation field typically only a rather
  short period of cosmic time during quasar activity. For lower BH
  accretion rates the radiation field is essentially too small to
  produce large ionisation parameters for a significant fraction of
  the halo gas.}
\label{fig:cooling_test_AGN} 
\end{figure*}

\subsubsection{Quasar- and radio-mode feedback}\label{sec:BH_feedback}

For AGN feedback we use a two-state model as used in previous
studies~\citep[see][for details]{Springel2005, Sijacki2007}. This model
distinguishes between quasar-mode feedback (high BH accretion rates) and
radio-mode feedback (low BH accretion rates).  Quasar-mode AGN feedback is
modelled by assuming that a fraction ($\epsilon_\mathrm{f}$) of the radiative
energy released by the accreted gas couples thermally to nearby gas within a
radius that contains a pre-calculated mass scale.  Significant quasar activity
requires high densities of relatively cold gas around the black hole, supplied
through large-scale inflows during galaxy mergers.  We will in the following
assume a radiative efficiency of $\epsilon_\mathrm{r}\sim
0.1-0.2$~\citep{Shakura1973,YuTremaine2002}.  For quasar-mode AGN feedback we
follow the same approach as for the enrichment scheme presented above and
distribute energy over a number of BH neighbour cells such that a pre-defined
mass is enclosed within the thermal ``feedback radius''.  We set the mass over
which to spread the BH thermal energy to the same mass that is also used for
the enrichment scheme, which again is typically set to a multiple of the target
mass ($m_{\rm target}$) of our (de-)refinement scheme. For convergence studies
we hold this mass fixed if we change resolution; i.e.~the feedback energy is
always distributed over a fixed physical mass, following the same strategy as
for the enrichment scheme.

For low activity states of the BH, we consider a form of mechanical
radio-mode AGN feedback following~\cite{Sijacki2007}. Quasar- and
radio-mode feedback are distinguished based on the BH accretion
rate. For BH accretion rates below a fraction $\chi_\mathrm{radio}$ of
the Eddington rate, we assume that feedback operates in radio-mode,
where AGN jets inflate hot, buoyantly rising bubbles in the
surrounding halo atmosphere. The duty cycle of bubble injection,
energy content of the bubbles as well as their initial size are
estimated based on the BH accretion rate and the current BH mass. The
duty cycle is coupled to the mass growth of the BH such that energy is
released once the BH has increased its mass by a factor of
$\delta_\mathrm{BH}$. For the the bubble properties we assume the
scaling relations of~\cite{Sijacki2007}, with
$R_\mathrm{bub,0}=50\kpc$, $E_\mathrm{bub,0}=10^{60}\erg$ and
$\rho_\mathrm{ICM,0}=10^{4}\msun\kpc^{-3}$, which we will keep fixed
in the following. For the scaling of the distance over which bubbles
are placed we use a normalisation value of $100\kpc$, which we will
also keep fixed in the following.  We assume the radio-mode feedback
efficiency provided by the bubbles to be $\epsilon_\mathrm{m} \times
\epsilon_\mathrm{r}$ of the accreted rest mass energy. 

\subsubsection{Radiative AGN feedback}\label{sec:EM_feedback}

So far we discussed only thermal and mechanical AGN feedback, which are
the most commonly employed feedback channels in BH simulations.
However, near strong ionising radiation sources like AGN the radiation field
is very different from the spatially uniform UVB assumed for the net
cooling rate calculation above. It is well-known that this can alter the
photo-ionisation and photo-heating rates of nearby
plasma~\citep[e.g.,][]{Rees1986, Efstathiou1992, Hambrick2009,
  Hambrick2011, Gnedin2012}.  For AGN, this provides a channel of
non-thermal/non-mechanical feedback~\citep[e.g.,][] {Sazonov2005,
  Ciotti2007, Hambrick2011, Kim2011, Gnedin2012, Choi2012}. Fully accounting for
this effect in detail requires polychromatic radiative transfer which
is too expensive to be applied to large-scale galaxy formation
simulations.  We will try here to take radiative electro-magnetic
feedback into account using an approximate approach. We assume a
universal and time-independent AGN SED parameterised as~\citep[see][for
  more details]{Korista1997}
\begin{equation}
f^{\rm AGN}(\nu) \!\!=\! \nu^{\alpha_{\rm UV}}\!\!\exp\!\left(\!-\frac{h \nu}{k T_{\rm BB}}\!\right) 
\!\exp\!\left(\!-\frac{10^{-2}{\rm Ryd}}{h \nu}\!\right)\! + a\nu^{\alpha_{\rm X}},
\end{equation}
where the big bump component is exponentially cut off at a temperature $T_{\rm
BB}$, and the UV-bump is suppressed exponentially in the infrared at a
temperature $k T_{\rm IR} = 0.01\,{\rm Ryd}$. The $\alpha_{\rm X}$ and
$\alpha_{\rm UV}$ parameters describe the slope of the X-ray component and the
low-energy slope of the big bump continuum, respectively. The ratio of X-ray to
UV, $\alpha_{\rm 0X}$,
\begin{equation}
\frac{f^{\rm AGN}(2\,{\rm keV})}{f^{\rm AGN}(2500\,{\rm \AA})} = \left(\frac{2\,{\rm keV}}{2500\,{\rm \AA}}\right)^{\alpha_{\rm 0X}} \!\!\!\!\!= 403.3^{\alpha_{\rm 0X}},
\end{equation}
is set through a corresponding choice of the normalisation $a$. In the
following we will assume a fixed SED with $T_{\rm BB}=10^6$~K,
$\alpha_{\rm 0X}=-1.4$, $\alpha_{\rm UV}=-0.5$ and $\alpha_{\rm
  X}=-1$~\citep[][]{Zamorani1981,Francis1993,Elivs1994} as our default
reference model.  We note that this radiative feedback is most
effective for accretion rates close to Eddington, and it is therefore
a reasonable choice to assume a fixed SED~\citep[][]{Sazonov2005}.

Based on the AGN SED, we can tabulate metal line cooling and heating rates for
different AGN bolometric intensities. For that we superpose the redshift
dependent UVB with the AGN radiation field and create similar net metal line
cooling rate tables as for the UVB described above.  Primordial cooling is
changed by calculating the photo-ionisation rates
\begin{align}
\Gamma_i \!&=\! \int_{\nu_i}^{\infty} \!\frac{4\pi (J^{\rm UVB}(\nu) + J^{\rm AGN}(\nu))}{h \nu} \sigma_i(\nu) \dint \nu \nonumber \\
           &= \Gamma^{\rm UVB}_i + \Gamma^{\rm AGN}_i, 
\end{align}
and photo-heating rates
\begin{align}
\epsilon_i \!&=\! \int_{\nu_i}^{\infty} \!\frac{4\pi (J^{\rm UVB}(\nu) + J^{\rm AGN}(\nu))}{h \nu} \sigma_i(\nu) (h \nu - h\nu_i)\dint\nu \nonumber \\
             &= \epsilon^{\rm UVB}_i + \epsilon^{\rm AGN}_i. 
\end{align}
Here $\Gamma_i$ and $\epsilon_i$ are the photo-ionisation and -heating rates,
respectively, for ${\rm H}^0$, ${\rm He}^+$ and ${\rm He}^0$. 
The quantities $\nu_i$ and
$\sigma_i(\nu)$ are the threshold frequencies and cross sections for
photo-ionisation. We note that the superposition of both radiation fields (UVB
plus AGN) is a valid approximation since near BHs the local AGN radiation field
is significantly stronger than the overall smooth and uniform UVB. The addition
of $J^{\rm UVB}(\nu)$ has in that case no impact on the cooling and heating
rates. We further note that $\Gamma^{\rm AGN}_i$ and $\epsilon^{\rm AGN}_i$
need to be calculated only for one fixed bolometric luminosity since we can
simply scale them linearly with bolometric intensity and add them to
$\Gamma^{\rm UVB}_i$ and $\epsilon^{\rm UVB}_i$, respectively. We assume that
the gas is optically thin to AGN radiation.  To improve on that assumption we
take into account a simple approximation for AGN obscuration as a function of
the AGN bolometric luminosity $L_{\rm bol}^{\rm AGN}$. We employ a power-law
parameterisation~\citep{Hopkins2007}, which we apply to the full bolometric
luminosity
\begin{equation}
L_{\rm bol}^{\rm AGN, obs} = \omega_{\rm 1}\left(\frac{L_{\rm bol}^{\rm AGN}}{10^{46}\ergs}\right)^{\omega_{\rm 2}},
\end{equation}
where we adopt $\omega_{\rm 1}=0.3$ and $\omega_{\rm 2}=0.07$ as our
default choice. With this obscuration scheme, we arrive at a rather
conservative estimate for the radiation intensity impinging the halo
gas. In addition, we assume that star-forming gas is optically thick
to the AGN radiation and therefore neglect this effect for gas above
the SF density threshold. To incorporate the impact of AGN radiation in the simulation, we assign to each
gas cell a bolometric intensity based on the obscured bolometric luminosities
$L_{\rm bol}^{\rm AGN, obs}$ of all BHs and the cell's distance to the BHs.  

To
capture both radiatively efficient and inefficient accretion we change our
default radiative efficiency of $\epsilon_{\rm r}$ to
\begin{equation}
\tilde\epsilon_{\rm r} = \epsilon_{\rm r} \, \frac{2\,x}{1+x}, \qquad x = \frac{1}{\chi_\mathrm{radio}} \, \frac{\dot M_\mathrm{BH}}{\dot M_\mathrm{Edd}}, \\ 
\end{equation}
if the BH accretion rate falls below $\chi_\mathrm{radio}$ of the corresponding
Eddington rate~\citep[see also][]{Ciotti2009}; i.e. if $x<1$. This leads to a
continuous transition from the ADAF~\citep[][]{Narayan1994} to the radiatively
efficient accretion regime, where we assume, as discussed above, a constant
efficiency of $\epsilon_{\rm r}$. We note that we use this efficiency scaling
only for the luminosity calculation for the AGN and not for the AGN thermal
and mechanical feedback discussed earlier. To be consistent with the quasar-
and radio-mode AGN feedback described above we will only allow an energy
fraction $(1-\epsilon_{\rm f})$ or $(1-\epsilon_{\rm m})$ to go into
electro-magnetic feedback for quasar-mode or radio-mode feedback, respectively.
We note that this correction and the smooth $\epsilon_{\rm r}$ parameterisation
are not crucial since radiative feedback is anyways only significant for BHs in
the quasar-mode phase with very high accretion rates.

In Figure~\ref{fig:cooling_test_AGN}, we show net cooling rates for gases of
different metallicities which are exposed to different AGN radiation sources
characterised by their ionisation parameter $\zeta^{\rm AGN}=L_{\rm bol}^{\rm
AGN}/(r^2 n_{\rm H})$, where $L_{\rm bol}^{\rm AGN}=(1-\epsilon_\mathrm{f}) \,
\tilde\epsilon_\mathrm{r} \, \dot{M}_{\rm BH} c^2$ is the bolometric luminosity
and $r$ is the distance from the plasma cell to the AGN radiation source. We
note that $J^{\rm AGN}\equiv L_{\rm bol}^{\rm AGN}/r^2$ is the bolometric
intensity that we tabulate in the cooling table to look up the metal line
cooling.  As Figure~\ref{fig:cooling_test_AGN} demonstrates, a simple
$Z$-scaling of the rates does not work in this regime.  We hence need to
tabulate metal-line cooling rates also as a function of $Z$. Specifically, we
extend the grid of net cooling rates by $[\log(Z/Z_\odot), \log(4 \pi J^{\rm AGN}/(\ergs\,{\rm
cm}^{-2}))]$, with $-5 < \log(4 \pi J^{\rm AGN}/(\ergs\,{\rm cm}^{-2})) < 5$
sampled equally spaced at $11$ bins, and six different metallicities
$\log(Z/Z_\odot) = -4, -3, -2, -1, 0, 1$. At each redshift, we also add grid
points for a pure UVB which we use for heating and cooling if there is no
nearby AGN.  Although Figure~\ref{fig:cooling_test_AGN} clearly demonstrates
that the AGN radiation field can strongly influence net cooling rates, we note
that this occurs only for a short interval of cosmic time for each BH once it
is in its maximum accretion phase with an accretion rate close to Eddington.
This implies that the impact of electro-magnetic AGN feedback is strongest
during quasar activity, but is limited when AGN enter the low accretion
radio-mode state.

The intensity assignment is performed by looping over all BHs and
calculating their bolometric luminosities based on the current mass
accretion rates. We first calculate a search radius for each BH. For
this we derive the BH's bolometric obscured luminosity and solve for
the radius within which the ionisation parameter is larger than
$\zeta^{\rm AGN}_{\rm thresh}=10^{-3}\,{\rm erg}\,{\rm s}^{-1}{\rm
  cm}$, assuming gas with mean hydrogen number density.  We then
calculate for all cells within this radius the incoming AGN intensity.
To avoid excessively large search radii we impose an upper limit of three
times the virial radius to this search radius. We have checked that
our results are not sensitive to this radius once it is chosen to be
of the
order of the virial radius or larger.  Overlapping AGN contributions
of several BHs are handled by summing up their bolometric intensities
at the cell's position.  Once we know all local intensities we can
calculate the net cooling rate for each cell.  We do not include
speed-of-light delay effects in the propagation from the AGN, which is
a reasonable approximation since we limit the sphere of influence of
the radiation field as described above, and do not allow the radiation
field to propagate over arbitrarily large 
distances through the simulation volume.

\section{Analysis techniques}\label{sec:analysis}

\subsection{Tracer particles}\label{sec:tracer}

The pseudo-Lagrangian nature of SPH is convenient for approximately
tracing mass elements in simulations.  (Although we emphasise that
this 'advantage' of SPH comes at the expense of an inaccurate 
integration of the mass continuity equation
\citep[see, e.g.,][]{Vogelsberger2012}.)
On the other hand, the Eulerian
nature of finite volume schemes does not directly allow this because
the mass continuity equation is integrated correctly on the
resolution scale. 
One approach to address this ``disadvantage'' of grid codes
is to use an additional passive
tracer particle species which is advected along with the gas by
inferring a tracer's velocity by interpolating the calculated
hydrodynamic velocity field.

{\sm AREPO} employs a quasi-Lagrangian scheme; i.e. it is neither Eulerian nor
strictly Lagrangian since mass can be advected through cell boundaries as in an
ordinary finite volume approach.  (We note here that SPH is not actually
Lagrangian either, at best ``pseudo-Lagrangian'', as the remapping that
implicitly occurs when local neighbourhood relations of particles change is
ignored in this approach.)  This quasi-Lagrangian nature of {\sm AREPO} implies that
tracing the evolution of ``closed-box'' fluid elements is not
possible a priori, similar to the situation in other finite volume schemes.
We have therefore realised two different passive tracer techniques in 
{\sm AREPO}
\citep[][]{Genel2013}.
First, we implemented the velocity
field approach, which ties the tracers to the reconstructed and interpolated
fluid velocity field.  Second, we developed a new scheme, which determines the
exchange of tracers between cells in a Monte Carlo fashion based on the actual
mass exchanges between them, which is given by the Riemann solver. We briefly
describe the main characteristics of these schemes in the following two
subsections.

\subsubsection{Velocity field tracer particles}\label{sec:vel_tracer}

Velocity field tracer particles move according to the piece-wise
linear reconstruction of the velocity field. At each time step, we
make a look up of the closest mesh-generating point for every active
tracer particle, which provides the Voronoi cell that this tracer 
resides in. We
then use the fluid velocity field gradients of that cell to
interpolate the velocity field to the tracer position. The gradient
information is readily available for the cells, because it is already
calculated for the MUSCL-Hancock step in the finite volume
solver. Once a new velocity is assigned to each active tracer particle
we drift them according to their individual time steps. Tracer
particles inherit the time step from the cells they fall in, which
makes their time integration adaptive and consistent with the dynamics
of the underlying cell timestep hierarchy.

Velocity field tracer particles try to follow the stream lines of the
underlying flow field and are essentially noise-free. Moving them is
computationally very cheap, the only moderate computational effort lies in the
required closest cell lookup, which however can be realised very efficiently
through a tree search with an initial search radius guessed based on the
nearest cell distance of the last search. However, we have found that this
tracer particle approach does not follow the gas mass properly. Since the fluid is evolved based on the local solutions to the Riemann
problems across each cell interface, simply using the reconstructed velocity
field to advect the tracers does not lead to consistent evolution between the
tracers and the fluid. Moreover, the fluid is averaged in the cell and
reconstructed after it is evolved, while those steps are not performed on the
tracers.

The inconsistency between the flow of the fluid and that of the tracers leads
in cosmological simulations to large biases~\citep[][]{Genel2013}. For example, the tracer density
profiles of haloes deviate significantly from the actual gas density profiles.
This effect was already found in other studies, but was not interpreted
appropriately. For example, \cite{Price2010} studied turbulence with the
velocity field tracer implementation of the FLASH code and found that tracer
particles tend to clump on sub-resolution scales in turbulence simulations.

\subsubsection{Monte Carlo tracers}\label{sec:mc_tracer}

Instead of relating the tracer evolution directly to the velocity field, we can
also link them to the mass exchange between cells. The basic idea of this
approach, presented and studied in \cite{Genel2013}, is to attach a population
of $N_i \ge 0$ Monte Carlo tracers to a computational Voronoi cell $i$. Based
on the finite volume update solution for this cell we know the mass fluxes
through each cell face during the time step.  We can then probabilistically
sample the transfer of tracer particles from one cell to the other. This
results in a Monte Carlo sampling of the underlying gas mass fluxes over the
computational domain, which does, by construction, not suffer from the bias
effect.  Such a scheme can be easily inlined in the finite volume calculations,
which loop over all Voronoi cell faces. This face list is constructed by the
tessellation engine. During the interface loop we keep track of the current
number of tracers per cell, and of the current total mass in each cell.  We
only consider outgoing mass fluxes $\Delta M_{i, j} < 0$ from cell $i$ to cell
$j$.  Furthermore, we keep track only of the reduced mass $\widetilde{M}_i$ of
each cell, which is updated for each outgoing flux, but not for in-going fluxes
since the tracer particle exchange is done for all outgoing fluxes starting
from a given cell. In-going fluxes into cell $i$ are symmetrically treated by
the outgoing fluxes of cell $j$. The probability for a tracer to leave cell $i$
and go into cell $j$ is then given by $p_{i, j}^{\rm flux}  =  \Delta M_{i,
j}/\widetilde{M}_i$.  To decide whether a tracer should leave a cell we draw a
random number $x_\alpha \in U(0,1)$ for each tracer $\alpha \leq N_i$ of cell
$i$.  The tracer is put into cell $j$ if $x_\alpha < p_{i, j}^{\rm flux}$.

In such a Monte Carlo based approach, tracers have no phase-space coordinates
within the cell, corresponding to the assumption that they are always uniformly
mixed within a cell.  We therefore store them simply in a globally distributed
linked list where each list entry has a tracer ID, a set of tracked fluid
properties, and a pointer to the next and the previous tracer in the list.
Each fluid cell then needs only a pointer to the first tracer associated with
it.  Tracer exchanges between cells can be implemented as operations on this
global linked list.

Another advantage of Monte Carlo tracers is that any sort of mass
transfer in or out of the gas cells can be modelled probabilistically in
that scheme. This can be used, for example, to follow mesh
(de-)refinement operations, but also for treating the formation of
stars, BHs or wind particles. We exploit this in our implementation
and have added, in addition to the tracers associated with gas cells,
star, wind and BH-tracers to keep track of {\em all} baryonic
mass exchanges. For example, this allows us to quantify precisely
(albeit with some Monte Carlo noise that can be reduced by averaging
over many cells) what fraction of gas in a cell has previously been
part of a star or a wind, or exactly which gas was swallowed by BHs.

\subsection{On the fly volume rendering}\label{sec:rendering}

Efficient visualisation of complex simulation datasets becomes increasingly
important since it can provide insights into the physical mechanisms at play.
Both for SPH- and AMR-based simulations, a large number of visualisation
toolkits exist. For example, {\sm SPLASH}~\citep[][]{Price2007} or {\sm
SPLOTCH}~\citep{Dolag2011} provide easy-to-use tools to visualise various
aspects of astrophysical SPH simulation data. AMR data can be efficiently
visualised by many different software packages, including Mayavi, Visit,
Paraview, and yt~\citep[][]{Turk2011}. Tools for visualising an unstructured
Voronoi mesh like the one used by {\small AREPO} are much less common and
hardly exist in comparison.

Since {\sm AREPO} contains already all the required Voronoi
mesh infrastructure it is natural to add a volume rendering engine
directly to the code. This has the advantage that the rendering can be
done at a very high time frequency on-the-fly and on very large
datasets due to the implemented parallelism. To this end, we
have added a simple volume rendering algorithm to {\sm AREPO}.
Specifically, we have implemented an image-space ray-casting scheme,
which follows the propagation of individual rays through the rendered
volume. Given a predefined camera path, this allows for orthogonal and
perspective projections.  Our scheme performs the ray-casting
operation in frequent time intervals, where a predefined number of
rays is integrated from a near field plane to a far field plane, using
the front-to-back recursive rendering equation
\begin{align} {\bf C}^{\rm acc}_{n+1} &= {\bf C}^{\rm acc}_{n} + (1-\alpha^{\rm
acc}_{n}) \, {\bf C}^{\rm cell}_{n} \, \alpha^{\rm cell}_{n} \nonumber \\
\alpha^{\rm acc}_{n+1} &= \alpha^{\rm acc}_{n} + (1-\alpha^{\rm acc}_{n}) \,
\alpha^{\rm cell}_{n}, \end{align}
where ${\bf C}^{\rm acc}_{n+1}$ is the accumulated ray colour vector,
${\bf C}^{\rm cell}_{n}$ is the colour vector assigned to the current
Voronoi cell based on some set of transfer functions, $\alpha^{\rm
  acc}_{n+1} \leq 1$ is the accumulated ray opacity, and $\alpha^{\rm
  cell}_{n}$ is the opacity assigned to the current Voronoi cell
specified through some transfer function. We note that this recursive
render equation is simply a discretised version of an
emission-absorption model which ignores any scattering. Since we
perform a front-to-back ray casting, we need to keep track of the
accumulated opacity during the $\alpha$ compositing. Starting from the
near plane allows one to terminate the ray-casting process once the
opacity for a given ray has reached a value of $\lesssim 1$ . Such
early ray termination leads to a speed-up of the rendering process
without affecting the results.

The actual ray traversing is done by exploiting mesh connectivity
information which is available through the Voronoi mesh implementation
in {\sm AREPO}. We also use this connectivity for a smoothing
operation that we optionally apply during the render process.
Especially in regions of low density, the number of Voronoi cells
sampling the gas distribution can be quite low. Visually, this can
then lead to discrete transitions from one cell to the other, which
can show up in the rendered images, similar to the familiar block
structure showing up in many images of AMR simulations.  To reduce
this effect, we have implemented a smoothing procedure, which averages
cell values based on the values of the nearest cells which are
connected to the current ray casting cell.  Specifically, we first
make one loop over all neighbouring cells and calculate the maximum
distance of their mesh generating points to the current ray location
in the primary cell. We then take this radius as a smoothing length
for a top hat kernel based cell value estimate, where we weight the
cell values that are going to be rendered according to the cell volume.

\subsection{Stellar population synthesis models}\label{sec:steller_synth}

To compare our galaxy formation model predictions with observations,
we need to transform stellar information of our simulated galaxies
into photometric properties.  Stellar population synthesis
models~\citep[e.g.,][]{Starburst99,BC03,Pegase} provide a convenient
way to associate the stars in our simulations with observable spectra
or broad band luminosities.  We include this in our simulations by
assigning to each star particle a series of broad-band luminosities based
on the~\cite{BC03} catalogues after taking into account the star
particle's age, mass, and metallicity.  Currently, we choose to
include the U-, V-, B-, K-, g-, r-, i-, and z-bands, however, this
could be extended easily to include any other spectral information
which is accurately tabulated in the stellar population synthesis
models.

The assigned stellar luminosities are not currently used in any part
of the dynamical evolution, nor do we include radiative transfer or
dust attenuation.  However, adding together the luminosity
contributions from all star particles in a given subhalo allows us to
construct estimates of the galaxy luminosities in several bands for
direct comparison against observations.  We construct broad-band
luminosities for all galaxies, and store the values in the group
catalogues to simplify basic post-processing analysis.

\section{Cosmological simulations}\label{sec:cosmo_sims}

\subsection{Simulations}\label{sec:sims}

We will present in this section the first cosmological {\sm AREPO} simulations
including the newly added physics described above.  In the following, we adopt
the cosmological parameters $\Omega_{m0}=0.27$, $\Omega_{\Lambda0}=0.73$,
$\Omega_{b0}=0.0456$, $\sigma_8=0.81$, and $H_0=100\,h\,{\rm km}\,{\rm s}^{\rm
-1}\,{\rm Mpc}^{\rm -1}= 70.4\,{\rm km}\,{\rm s}^{\rm -1}\,{\rm Mpc}^{\rm -1}$
($h=0.704$). We create realisations of this cosmology in periodic boxes with a
side length of $25\hmpc$. Initial conditions are generated at $z=127$ based on
a linear power spectrum made by CAMB, with gas particles/cells added to the
initial conditions by splitting each original particle into a dark matter and
gas particle/cell pair, displacing them with respect to each other such that
two interleaved grids are formed, keeping the centre-of-mass of each pair
fixed. 

Our fiducial parameters for the different physical processes are
summarised in Table~\ref{table:fiducial}. These parameters are
physically plausible and were chosen such that they provide a
reasonable fit to most key observables at $z=0$, as we discuss below.
Most importantly, the feedback parameters of our fiducial model are
chosen such that they reasonably well reproduce the stellar mass
function and the stellar mass -- halo mass relation. We also explore a
few modifications of this fiducial model, where we focus on
differences in the stellar and AGN feedback processes.  All our
simulations are summarised in Table~\ref{table:cosmo_sims}.  We
specify two Plummer-equivalent gravitational softening length
values. For DM particles, we use a fixed comoving softening length
(second value).  For baryonic collisionless particles (stars and BHs),
we further assume a maximum physical softening length (first value),
which limits the growth of the physical softening length. Gas cells
use an adaptive softening length tied to the cell radius, limited by a
floor. This floor is set to the same value as for the other baryonic
particles (i.e.~stars and BHs).  In the table, we also specify the
simulation volume and the mass resolution for DM and baryons.
Finally, the last column describes the physics characteristics of the
specific model with respect to our fiducial model. We employ a
(de-)refinement scheme which keeps the cell masses close to a
specified target mass $m_\mathrm{target}$, which results in a nearly constant number of
cells. This scheme is identical to that used in
\cite{Vogelsberger2012}, and we also follow the same mesh
regularisation strategy.

\begin{table*}
\begin{tabular}{lll}
\hline
variable                                                                     & fiducial value & description \\
\hline
\hline
stellar feedback (non-local energy-driven)                                   &                & \\
\hline
$\kappa_\mathrm{w}$                                                          & $3.7$          & wind velocity relative to local DM 1D velocity dispersion \\
${\rm egy}_{\rm w}/{\rm egy}_{\rm w}^0$                                      & $1.09$         & available SNII energy per formed stellar mass in units of $[{\rm egy}_{\rm w}^0]$\\
${\rm mom}_{\rm w}/\kms$                                                     & $0$            & specific wind momentum in units of $[\kms]$\\
\hline
AGN feedback (quasar-mode)                                                   &                & \\
\hline
$\epsilon_\mathrm{f}$                                                        & $0.05$         & quasar mode feedback energy fraction \\
$\epsilon_\mathrm{r}$                                                        & $0.2$          & radiative efficiency for Bondi accretion \\
\hline
AGN feedback (radio-mode)                                                    &                & \\
\hline
$\chi_\mathrm{radio}$                                                        & $0.05$         & accretion rate threshold for radio-mode in units of $[\dot M_\mathrm{Edd}]$ \\
$\delta_\mathrm{BH}$                                                         & $1.15$         & duty cycle of radio-mode \\
$\epsilon_\mathrm{m}$                                                        & $0.35$         & radio-mode feedback energy fraction \\
\hline
AGN feedback (electro-magnetic)                                              &                & \\
\hline
$\omega_1, \omega_2$                                                         & $0.3, 0.07$    & AGN obscuration parameterisation \\
\hline
wind metal loading                                                           &                & \\
\hline
$\gamma_\mathrm{w}$                                                          & $0.4$          & metal loading of wind particles \\
\hline
\end{tabular}
\caption{Fiducial model parameters. We assume purely energy-driven,
  non-local winds for stellar feedback. AGN feedback consists of three
  components: quasar-mode (thermal), radio-mode (mechanical) and
  radiative (electro-magnetic) feedback.  For SNII winds, we specify
  the mass loading parameter ${\rm egy}_{\rm w}$ (in units of ${\rm
    egy}_{\rm w}^0 = 1.73\times 10^{-2}\, 10^{51} \erg \msun^{-1}$; i.e.~$E_{\rm SNII, 51}=1$)
  and the wind velocity $\kappa_{\rm w}$ (in units of the local 1D DM
  velocity dispersion). For AGN feedback, $\epsilon_\mathrm{r}$
  indicates the radiative efficiency, $\epsilon_\mathrm{f}$ the
  fraction of the bolometric luminosity that is thermally coupled to
  nearby gas as a form of quasar-mode feedback and
  $\epsilon_\mathrm{m}$ the energy fraction that goes into bubbles
  through radio-mode feedback once the BH accretion rate drops below
  $\chi_\mathrm{radio}$ of the Eddington rate.  Electro-magnetic AGN
  feedback is specified by a fixed SED ($T_{\rm BB}=10^6$~K,
  $\alpha_{\rm 0X}=-1.4$, $\alpha_{\rm UV}=-0.5$ and $\alpha_{\rm
    X}=-1$) and the obscuration factors $\omega_{\rm 1,2}$.  The
  feedback parameters are chosen such that they are physically
  plausible and provide a good match to key observations.}
\label{table:fiducial}
\end{table*}

\begin{table*}
\begin{tabular}{llllll}
\hline
name                            & volume            & cells/particles  & $\epsilon$       & $m_\mathrm{DM}/m_\mathrm{target}$                      &physics\\
                                & [$(\hmpc)^3$]     &                  & [$\hkpc$]        & [$\hmsun$]                                             &\\
\hline
\hline
L25n512                         & $25^3$            & $2\times512^3$   & $0.5/1.0$        & $7.33 \times 10^{6}/ 1.56 \times 10^{6}$         &fiducial\\
L25n256                         & $25^3$            & $2\times256^3$   & $1.0/2.0$        & $5.86 \times 10^{7}/ 1.25 \times 10^{7}$         &fiducial\\
L25n128                         & $25^3$            & $2\times128^3$   & $2.0/4.0$        & $4.69 \times 10^{8}/ 1.00 \times 10^{8}$         &fiducial\\
\hline
stronger winds                  & $25^3$            & $2\times256^3$   & $1.0/2.0$        & $5.86 \times 10^{7}/ 1.25 \times 10^{7}$         &${\rm egy}_{\rm w}/{\rm egy}_{\rm w}^0=2.18$\\
weaker winds                    & $25^3$            & $2\times256^3$   & $1.0/2.0$        & $5.86 \times 10^{7}/ 1.25 \times 10^{7}$         &${\rm egy}_{\rm w}/{\rm egy}_{\rm w}^0=0.545$\\
faster winds                    & $25^3$            & $2\times256^3$   & $1.0/2.0$        & $5.86 \times 10^{7}/ 1.25 \times 10^{7}$         &$\kappa_\mathrm{w}=7.4$\\
slower winds                    & $25^3$            & $2\times256^3$   & $1.0/2.0$        & $5.86 \times 10^{7}/ 1.25 \times 10^{7}$         &$\kappa_\mathrm{w}=1.85$\\
\hline
stronger radio                  & $25^3$            & $2\times256^3$   & $1.0/2.0$        & $5.86 \times 10^{7}/ 1.25 \times 10^{7}$         &$\epsilon_\mathrm{m}=0.7$\\
weaker radio                    & $25^3$            & $2\times256^3$   & $1.0/2.0$        & $5.86 \times 10^{7}/ 1.25 \times 10^{7}$         &$\epsilon_\mathrm{m}=0.175$\\
higher radio threshold          & $25^3$            & $2\times256^3$   & $1.0/2.0$        & $5.86 \times 10^{7}/ 1.25 \times 10^{7}$         &$\chi_\mathrm{radio}=0.1$\\
lower radio threshold           & $25^3$            & $2\times256^3$   & $1.0/2.0$        & $5.86 \times 10^{7}/ 1.25 \times 10^{7}$         &$\chi_\mathrm{radio}=0.025$\\
\hline
no feedback                     & $25^3$            & $2\times256^3$   & $1.0/2.0$        & $5.86 \times 10^{7}/ 1.25 \times 10^{7}$         &no stellar/AGN feedback\\
\hline
\end{tabular}
\caption{Summary of the different cosmological simulations. The L25n128,
L25n256, L25n512 simulations employ the fiducial physics parameters listed in
Table~\ref{table:fiducial} and explore convergence by changing the mass
resolution by a factor of $64$ in total. The remaining simulations consider
variations of certain physical processes at the intermediate resolution
level.
Parameters that are varied are indicated in the last column. Here we increase
or decrease stellar and AGN feedback parameters by a factor of two. The ``no
feedback'' simulation does not include stellar/AGN feedback (except for the
implicit ISM pressurisation).}
\label{table:cosmo_sims}
\end{table*}

\begin{figure*}
\centering
\includegraphics[width=0.32\textwidth]{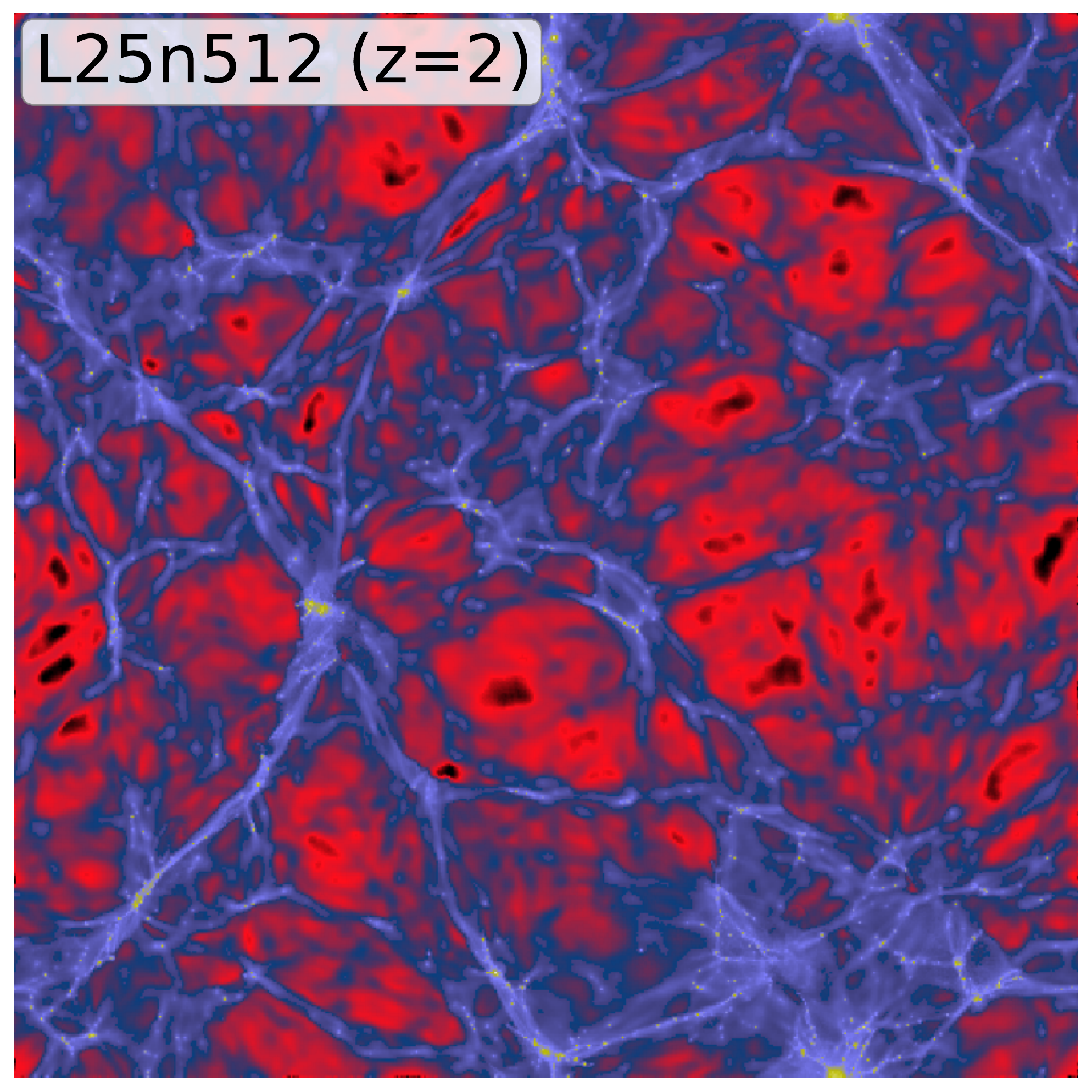}
\includegraphics[width=0.32\textwidth]{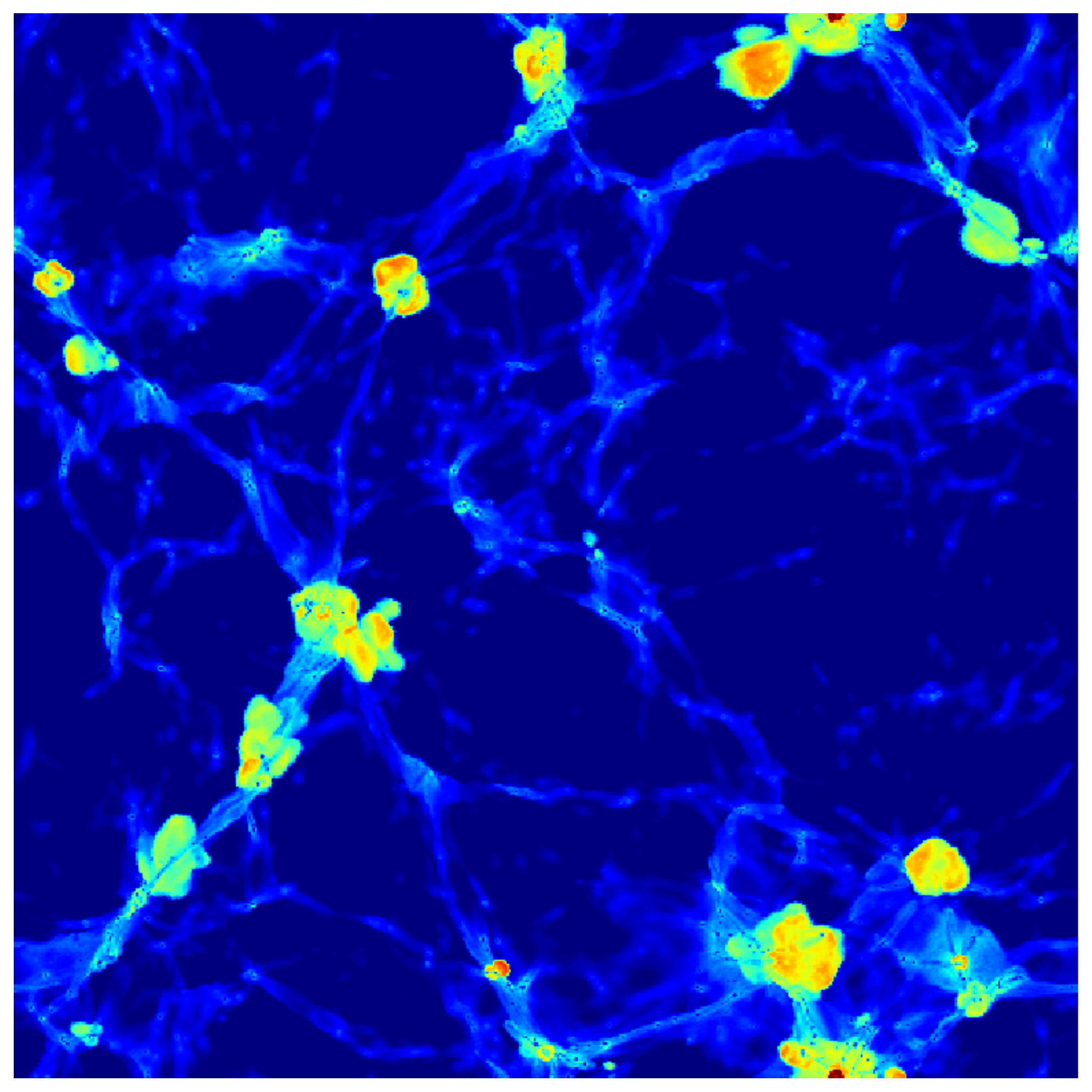}
\includegraphics[width=0.32\textwidth]{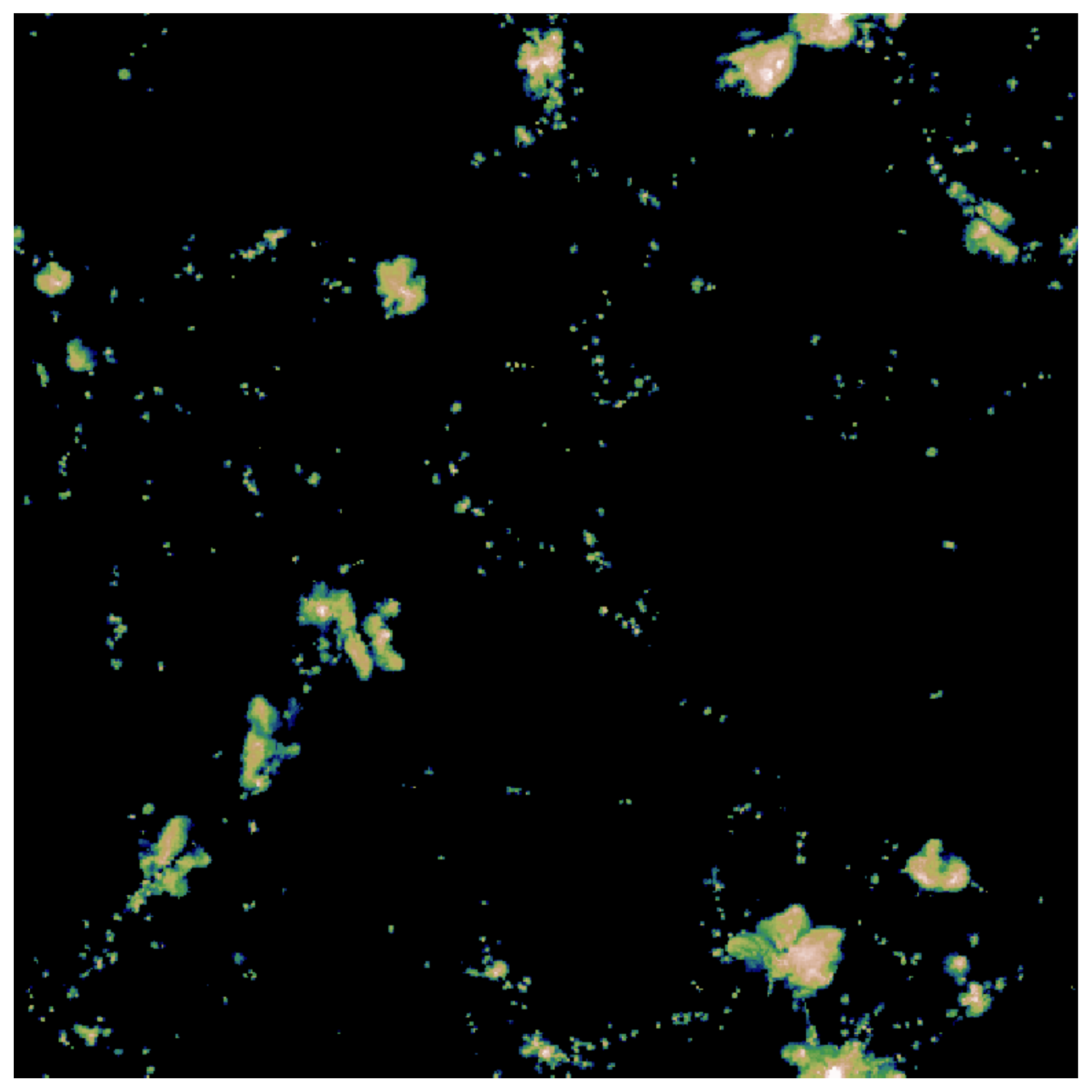}
\includegraphics[width=0.32\textwidth]{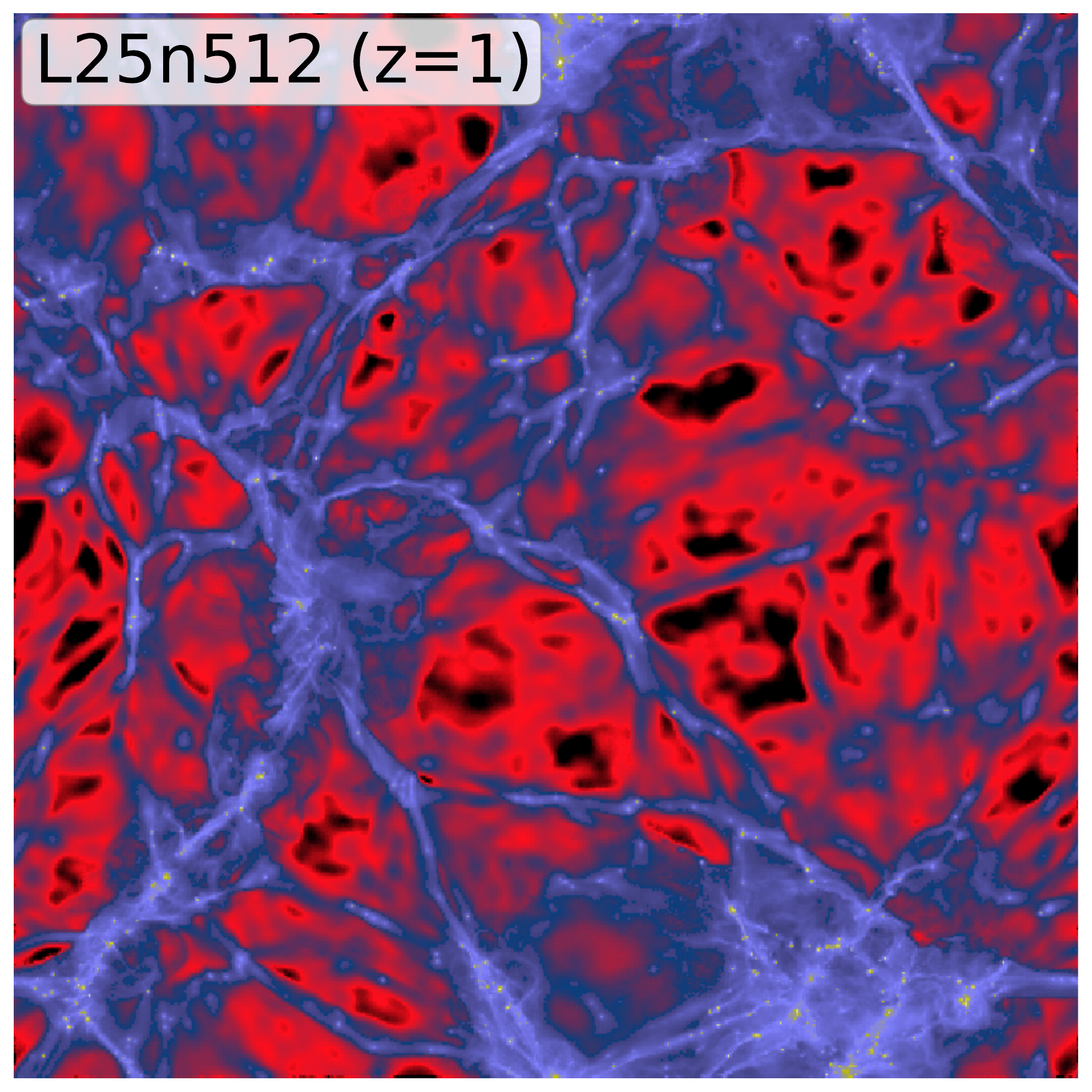}
\includegraphics[width=0.32\textwidth]{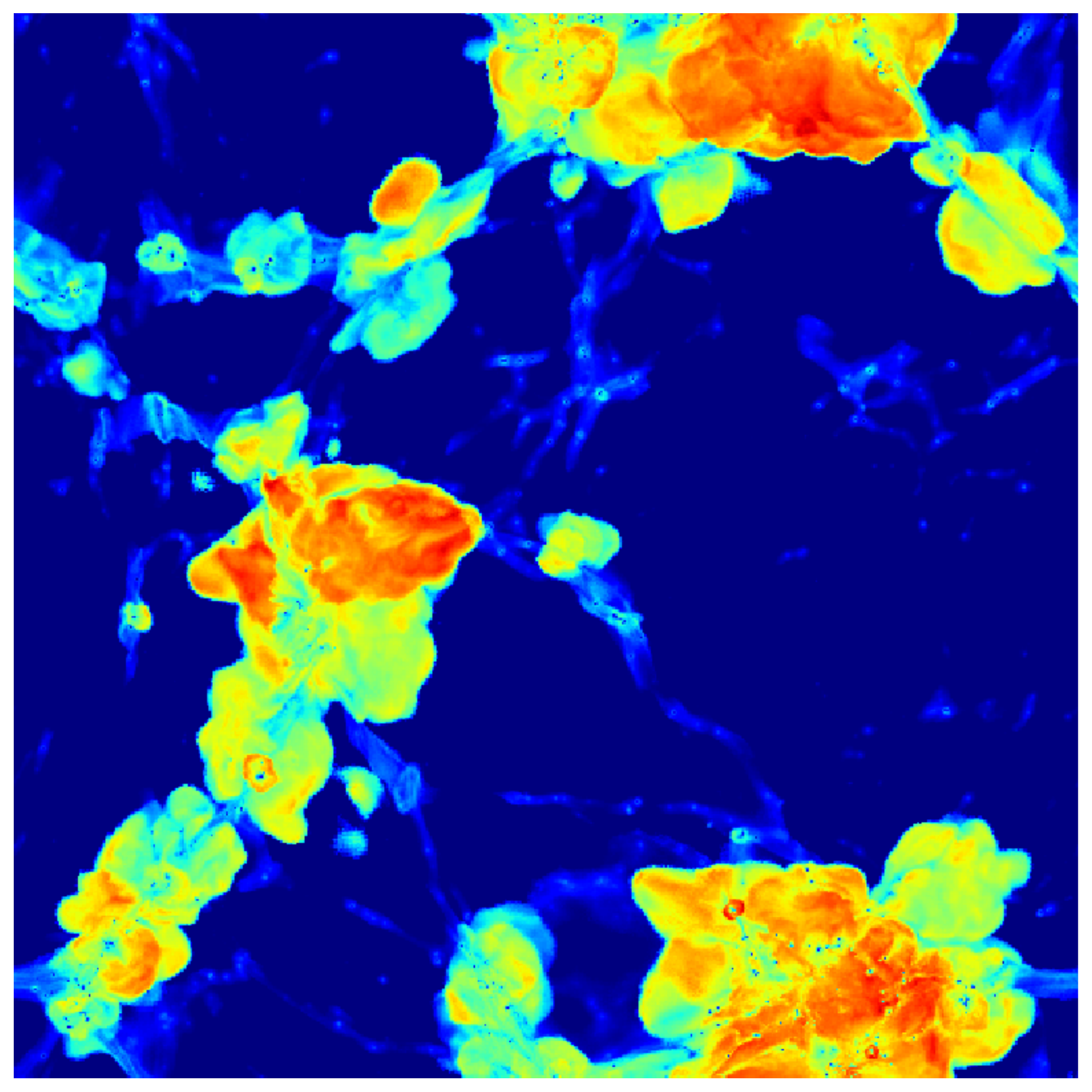}
\includegraphics[width=0.32\textwidth]{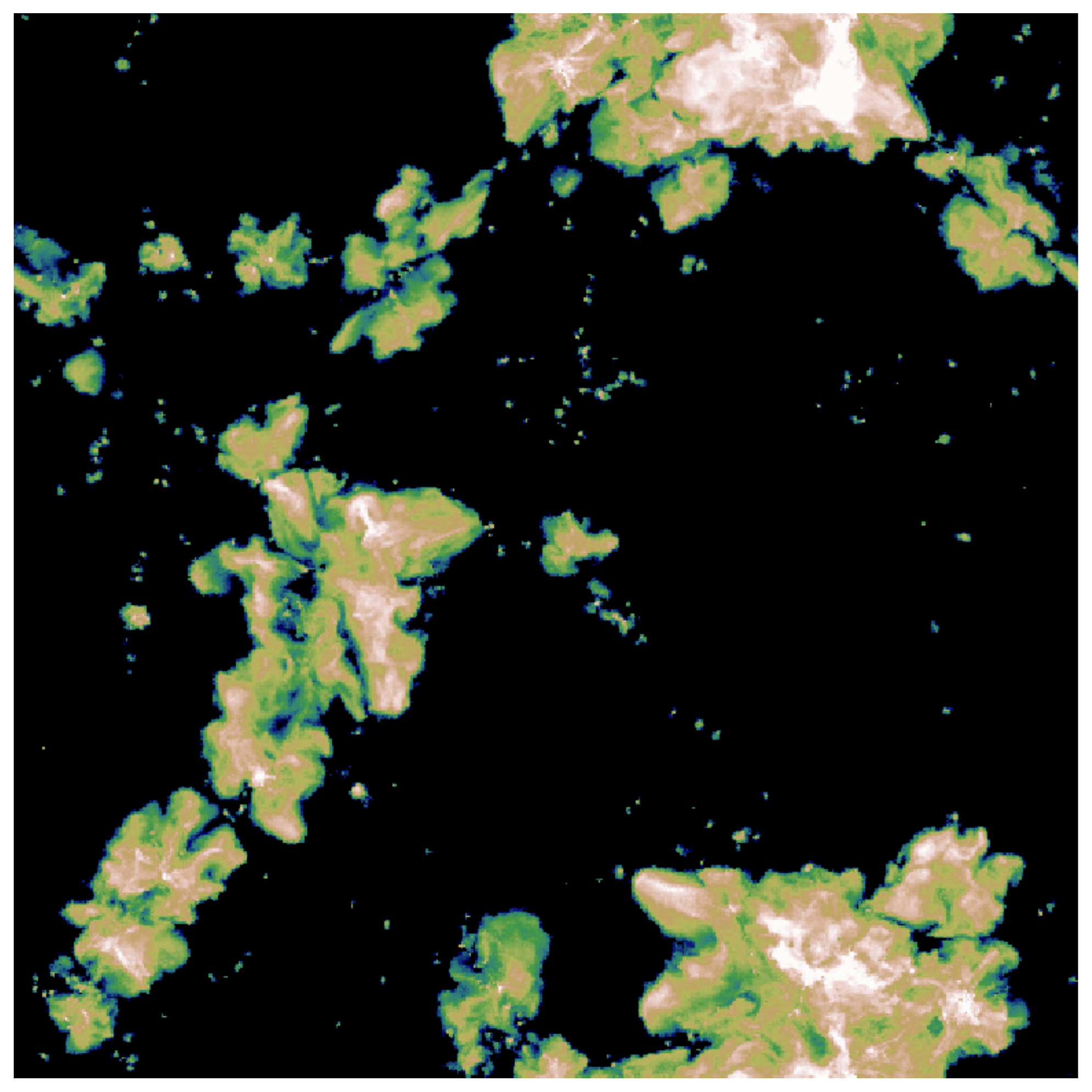}
\includegraphics[width=0.32\textwidth]{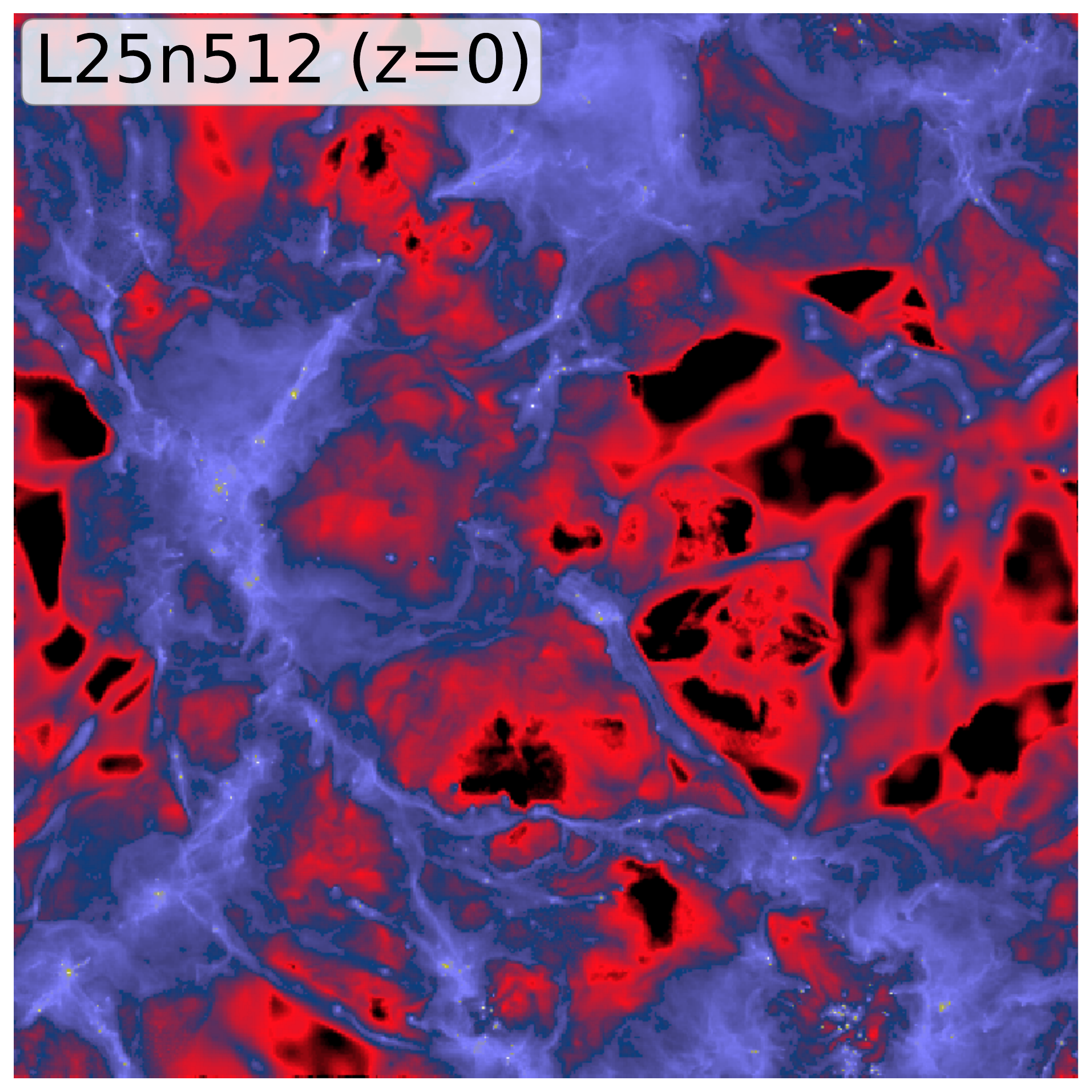}
\includegraphics[width=0.32\textwidth]{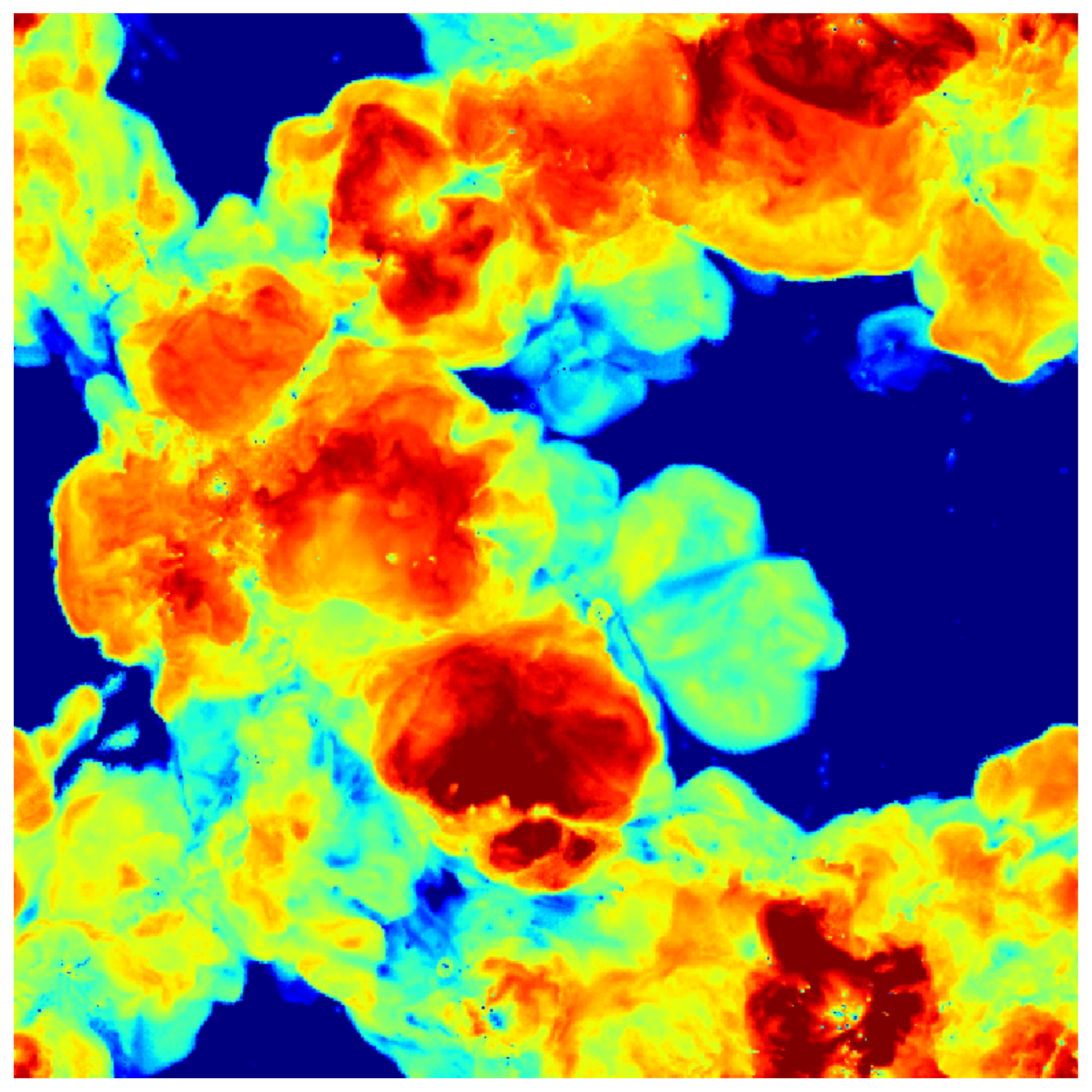}
\includegraphics[width=0.32\textwidth]{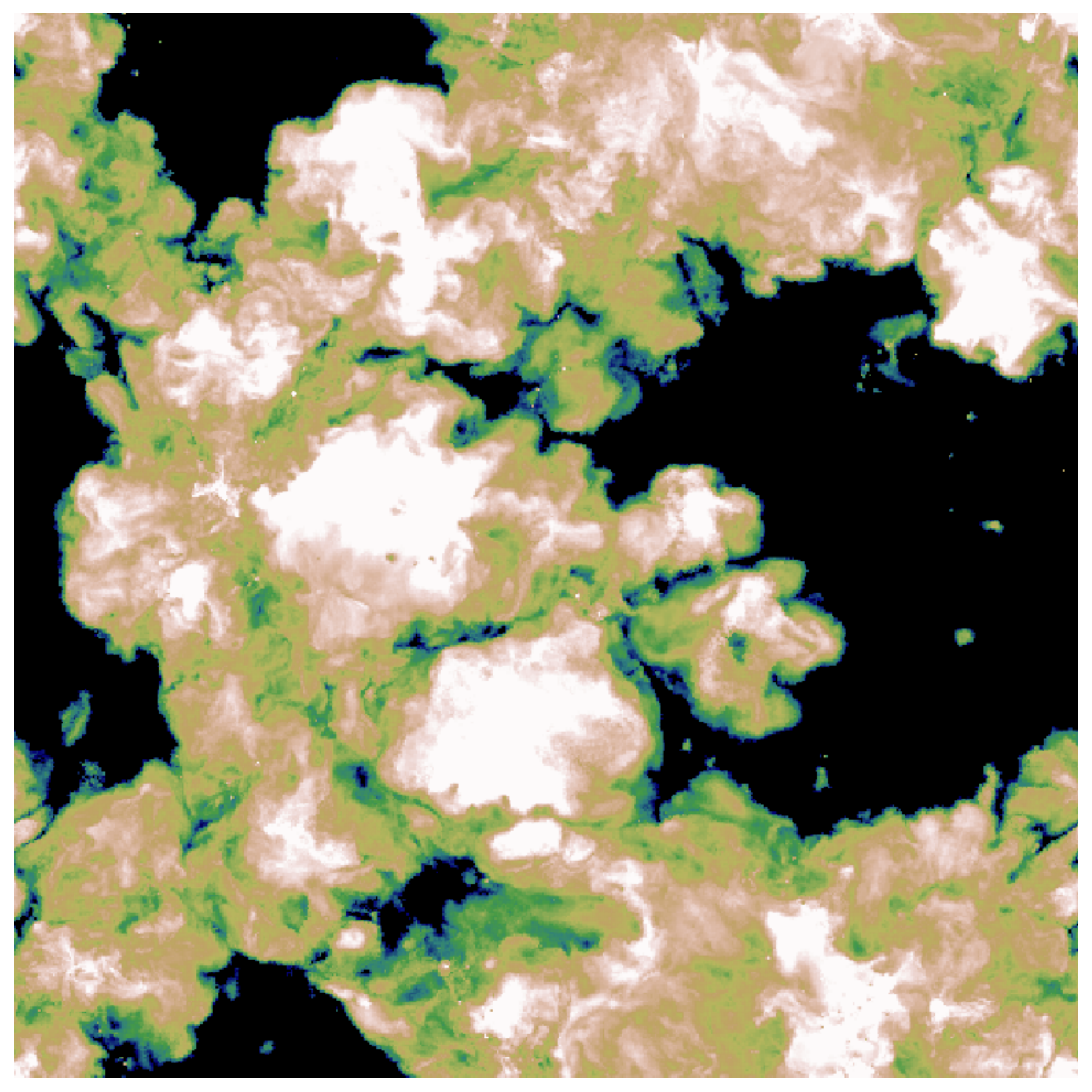}
\caption{Gas density (left panel), gas temperature (middle panel) and
  gas metallicity (right panel) projections of the L25n512 simulation.
  Each panel is $25\hmpc$ on a side and has a thickness of
  $1\hmpc$. We show the fields at three different redshifts $z=0$,
  $1$, and $2$. At $z=2$ some haloes show outflows generated mainly by
  strong winds through stellar SNII feedback. The more dramatic
  heating effects at late times ($z=0$, $1$) are largely caused by
  strong radio-mode AGN feedback. Both stellar and AGN feedback 
  lead to a significant enrichment of the IGM as can be seen in the
  metallicity projections. Strong AGN feedback also alters the density
  structure of the gas at $z=0$ (see left panel).}
\label{fig:cosmo_box}
\end{figure*}

\begin{figure*}
\centering
\includegraphics[width=0.32\textwidth]{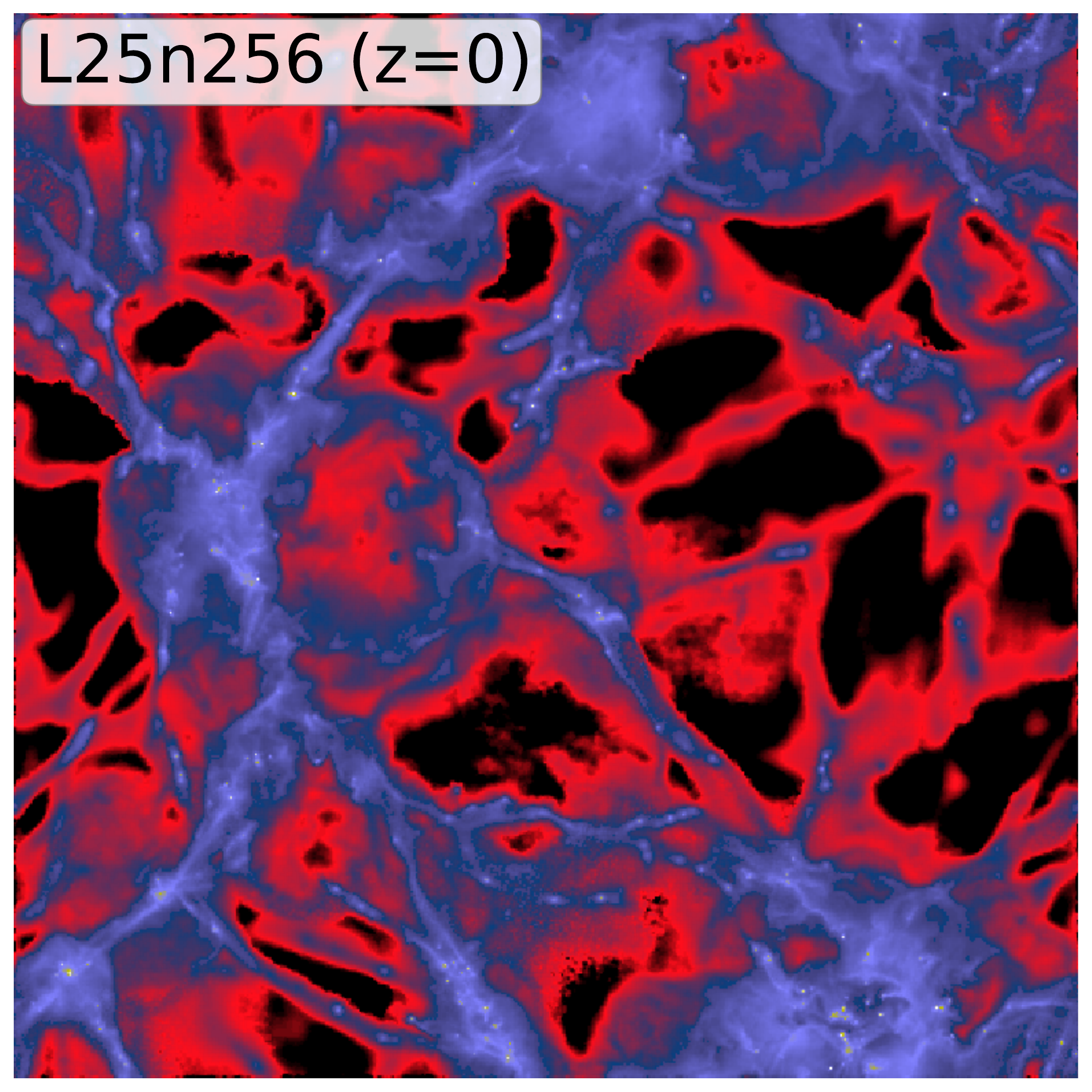}
\includegraphics[width=0.32\textwidth]{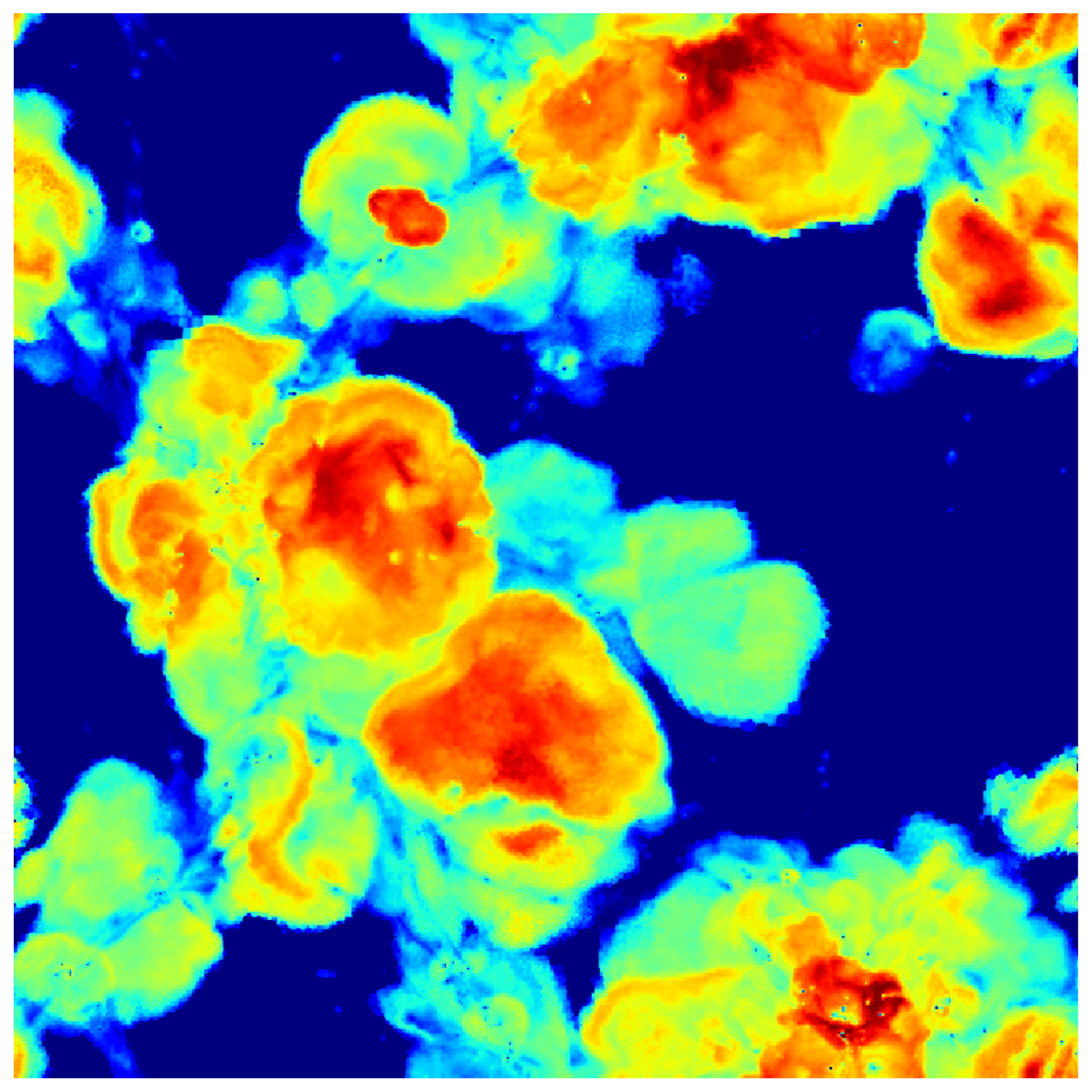}
\includegraphics[width=0.32\textwidth]{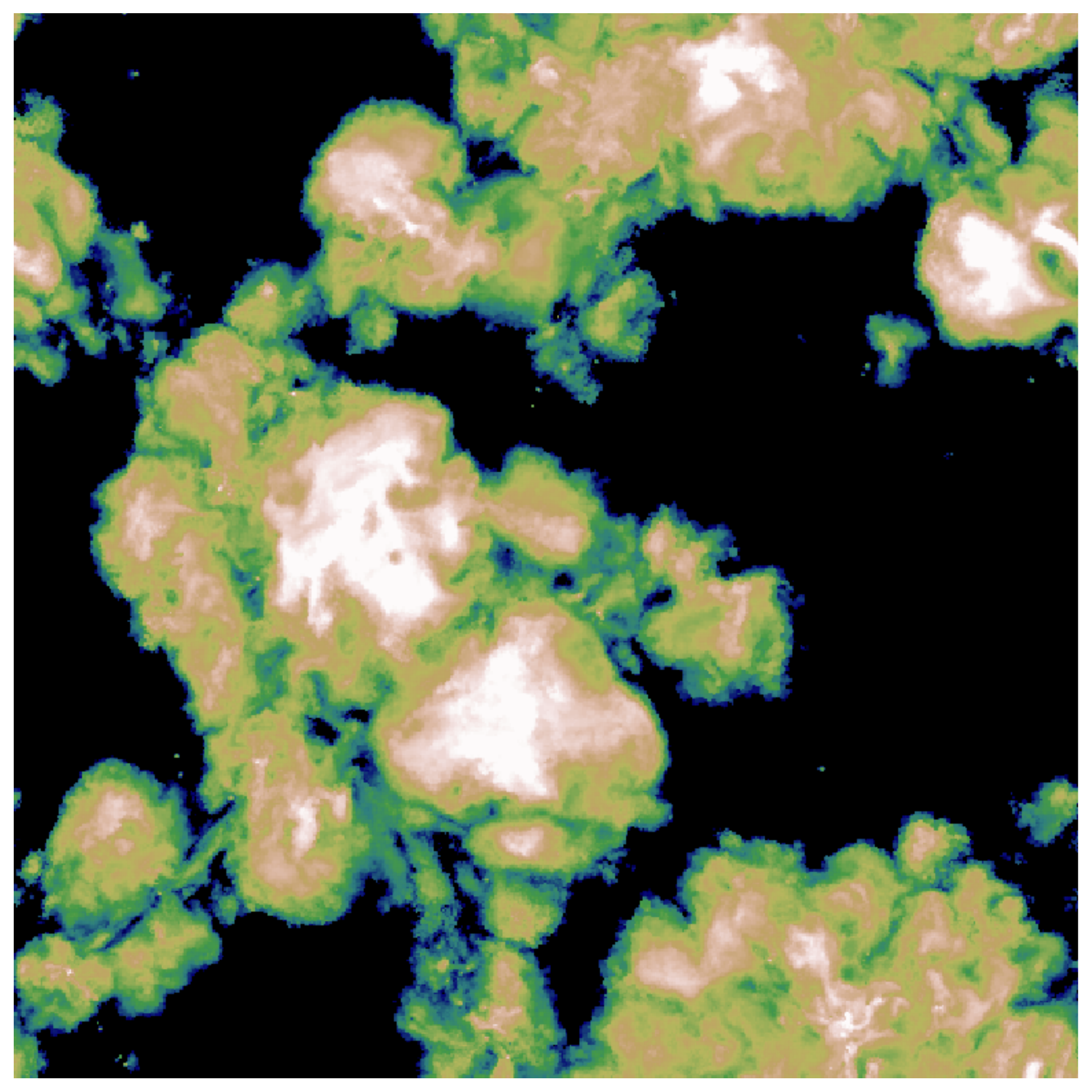}
\includegraphics[width=0.32\textwidth]{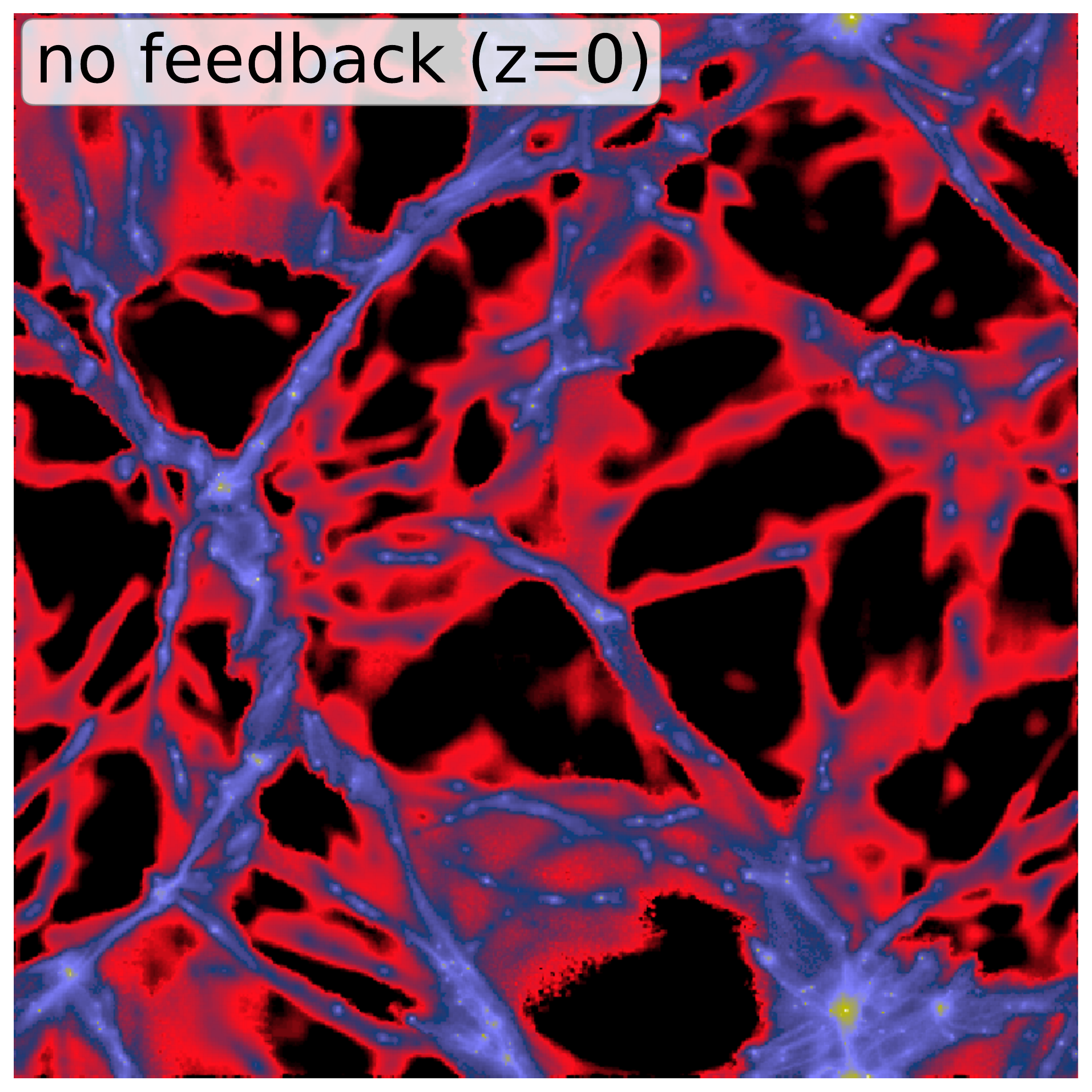}
\includegraphics[width=0.32\textwidth]{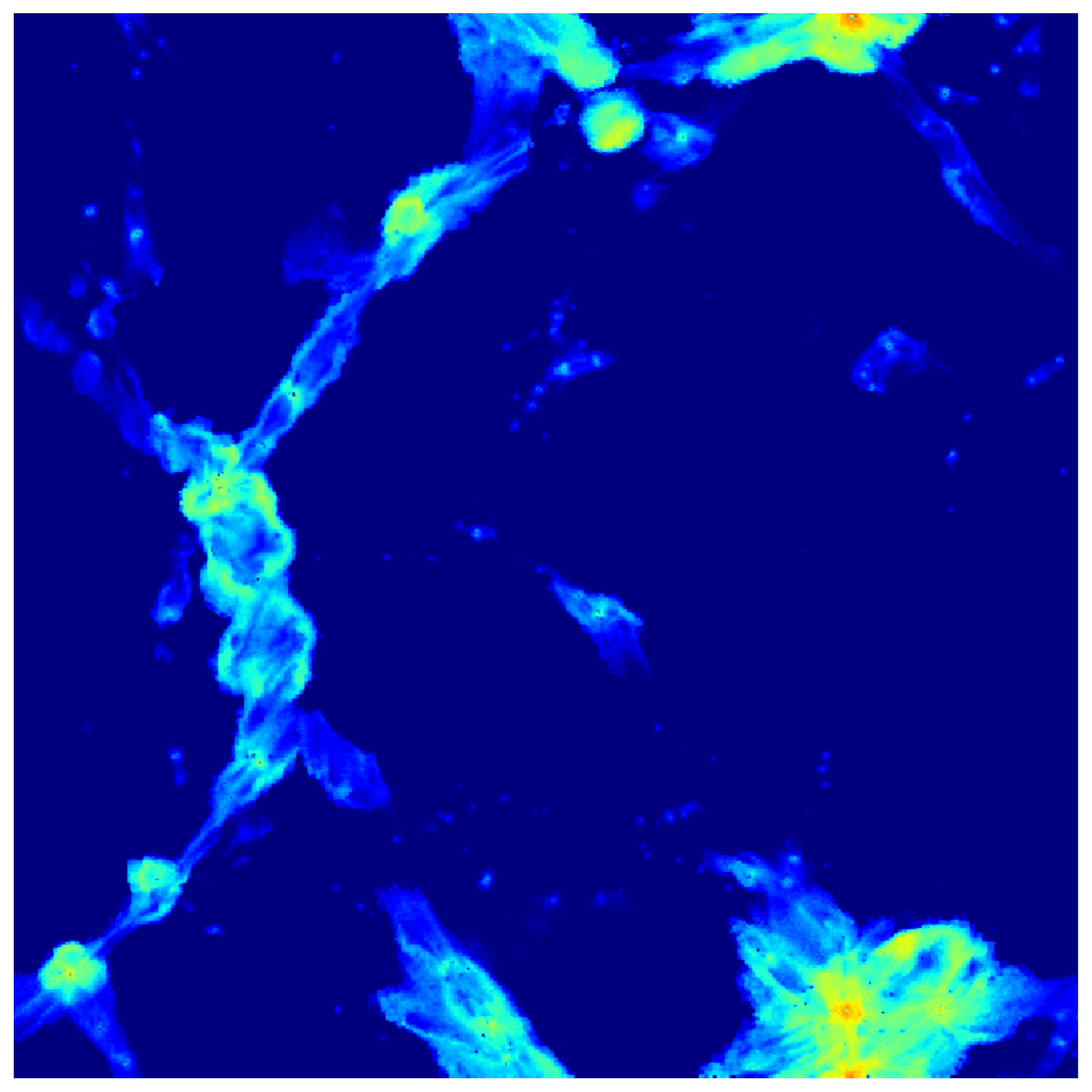}
\includegraphics[width=0.32\textwidth]{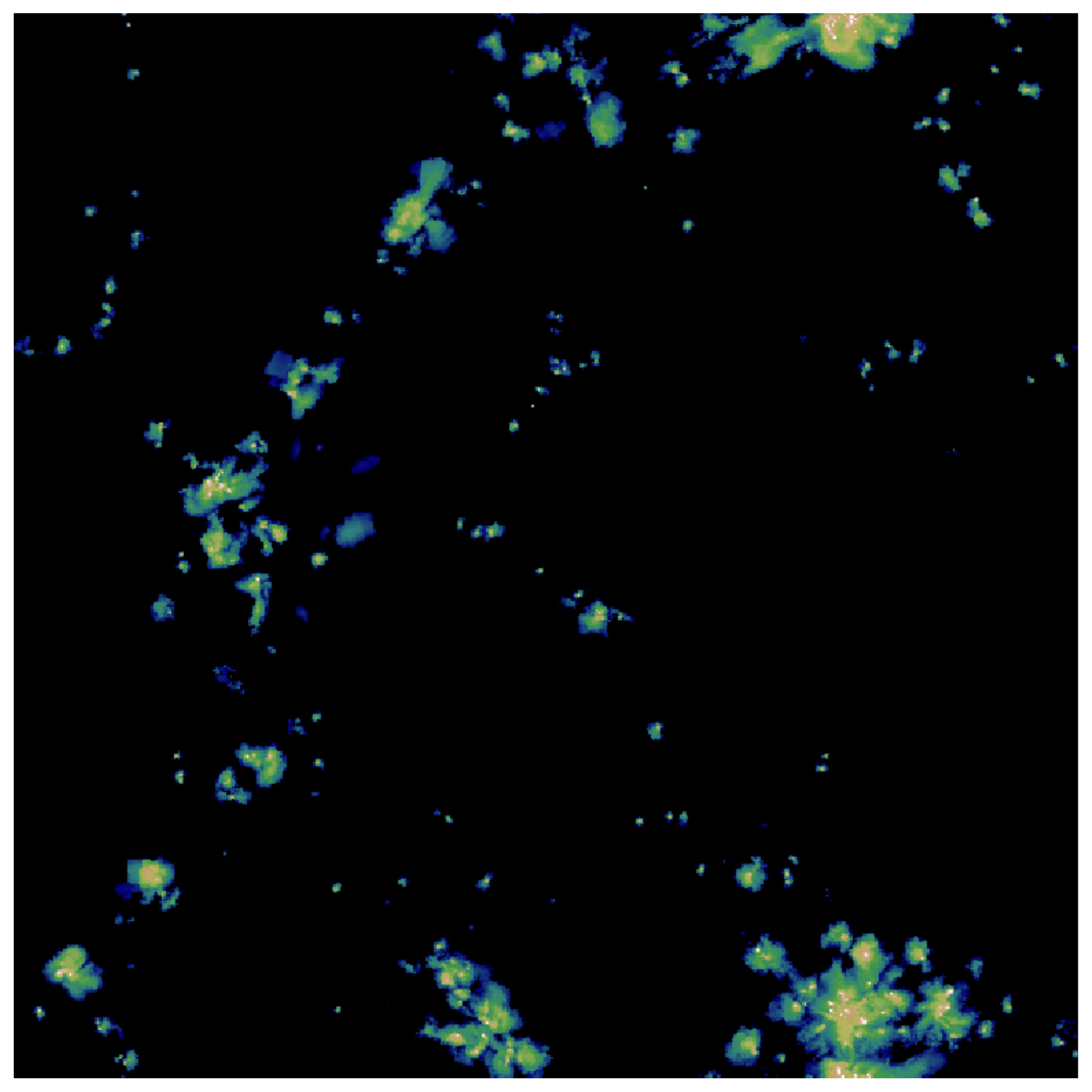}
\caption{Gas density (left panel), gas temperature (middle panel) and
  gas metallicity (right panel) projections of the L25n256 simulation
  (top panels) and ``no feedback'' simulation (bottom panels) at
  $z=0$.  Each panel is $25\hmpc$ on a side and has a thickness of
  $1\hmpc$ (same as Figure~\ref{fig:cosmo_box}). Clearly, the addition
  of more efficient cooling and stellar/AGN feedback strongly affects
  the thermodynamic properties of the gas. Most dramatically, the
  metal distribution is altered through the additional feedback
  enriching the IGM. We note that the colour scale used here is
  different from the one in Figure~\ref{fig:cosmo_box} as it has been
  adapted to the different simulation resolution in both figures.}
\label{fig:cosmo_box_no_feedback}
\end{figure*}

Table~\ref{table:cosmo_sims} includes three simulations with the fiducial setup
presented in Table~\ref{table:fiducial}: L25n128, L25n256, L25n512. These three
simulations serve as a resolution study of our fiducial physics setup to
estimate convergence. The other simulations explore modifications of the
fiducial setup.  Each of these modified runs typically varies only one of the
feedback parameters of our fiducial setup and explores it at the intermediate
resolution equivalent to the L25n256 simulation, which is sufficient to obtain
reasonably converged results, and computationally efficient enough to explore
different feedback settings quickly.  Each of the modified parameters is
increased or decreased by a factor of two compared to the fiducial value.  We
explore here only a rather limited number of parameters.  For example, we do
not change the quasar-mode feedback strength since we are mainly interested in
the stellar properties of haloes and quasar-mode feedback mainly regulates
BH growth (see below).  Furthermore, we do not alter the wind model; i.e.~we
only vary the parameterisation of the non-local energy-driven winds presented
in Table~\ref{table:fiducial}, but we do not explore, for example, local or
momentum-driven winds. For the sake of brevity, we here also do not explore a
large number of unrealistic setups like models without metal-line cooling or
without stellar mass loss.  While such test simulations can be interesting for
studying the specific implications of these physical effects, our goal is here
to bracket the feedback parameters required to obtain an acceptable physical
match to observations; hence we explore only models with complete physics in
this subsection. However, we include one simulation (``no feedback''), which
does not include stellar and AGN feedback except for the implicit ISM
pressurisation through SNII feedback as implemented in our effective EOS for
star forming ISM gas. We consider this ``no feedback'' simulation to
demonstrate that such a model gives a very poor fit to observations,
highlighting the difference with our fiducial model which
includes the full feedback physics. We stress that this ``no feedback'' simulation
contains a weak form of stellar feedback in form of the ISM pressurisation. 
To understand the importance of the different physical processes in our model
we will present unrealistic simulations in Section~\ref{sec:disent} below.
There we deactivate certain physical processes to understand better how they
shape our $z=0$ results. Clearly, such simulations are not expected to match
observational constraints, which is why we do not discuss them here.

We will show below that the fiducial parameters presented in
Table~\ref{table:fiducial} produce a galaxy population in good agreement with
various observational probes at $z=0$ in the local Universe. We stress that
some of these feedback parameters differ significantly from those previously
used in SPH simulations with similar physics models. Most importantly, the AGN
radio-mode feedback is more energetic than in our previous simulations
\citep[e.g.,][]{Sijacki2007, Puchwein2008}. This can be seen directly by the
value for the radio-mode feedback energy fraction, which is about a factor of
two larger than in previous SPH studies. Also, the accretion rate threshold for
switching between radio- and quasar-mode feedback is five times larger than in
previous simulations. This means that more of the BH accretion energy goes into
radio-mode feedback compared to previous work, where $\chi_\mathrm{radio}=0.01$
was typically adopted. As a result, more feedback energy is deposited displaced
from the central BH in a bursty fashion rather than smoothly heating the
central region around the BH. This automatically leads to a more efficient
suppression of star formation in massive haloes.  We found that this
combination of an increased radio-mode feedback factor and a higher accretion
rate threshold for the quasar mode are necessary to reproduce the observed
stellar masses in massive haloes.  We also note that the amount of supernova
energy for stellar feedback is quite high in our fiducial setup. In fact, it is
$1.09$ times larger than what is nominally assumed for SNII type events
($10^{51}\erg$). We argue that this is still physically plausible for a couple
of reasons. First, $10^{51}\erg$ is just a canonical value which is in fact
uncertain.  Second, recent work has shown that accounting for early stellar
feedback seems crucial for regulating star formation in low mass
systems~\citep[][]{Stinson2013, Kannan2013}.  In our model, we include this
contribution in the wind energy budget for launching kinetic wind particles.

\begin{figure*}
\centering
\includegraphics[width=0.975\textwidth]{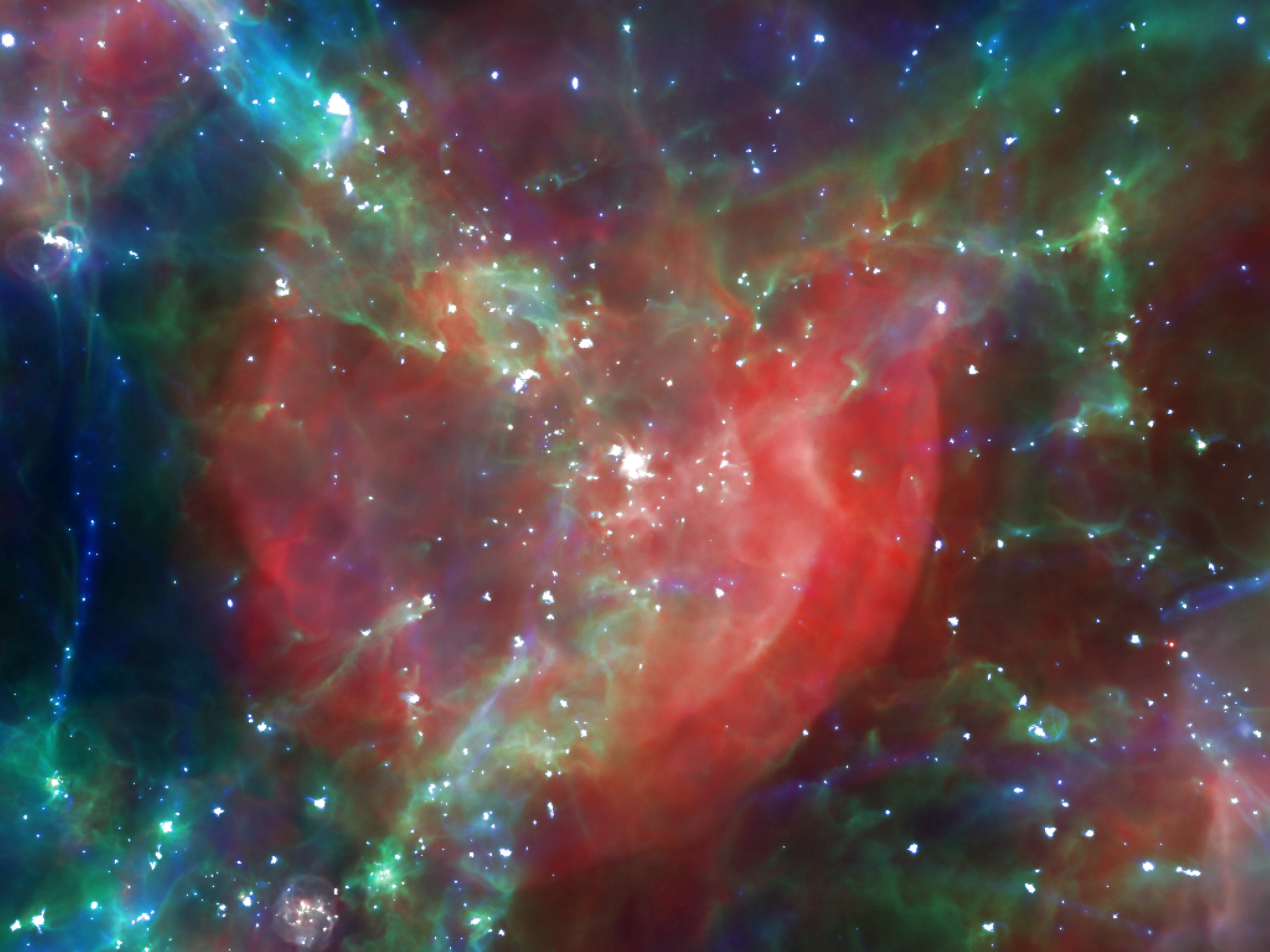}
\caption{Rendered gas temperature field of a large region of the simulation
volume of L25n512 at $z=0$. The render frustum has a near field plane distance
of $2\hmpc$ and the far plane is at $20\hmpc$. The width of the near field
plane is $2\hmpc$ with a height of $1.5\hmpc$.  Hot regions are shown in red
and cold regions in blue. Galaxies are shown in bright white. Part of the
cosmic web and hot halo atmosphere of one of the most massive haloes in the
simulation volume can be seen.}
\label{fig:render_temp}
\end{figure*}

\begin{figure*}
\centering
\includegraphics[width=0.975\textwidth]{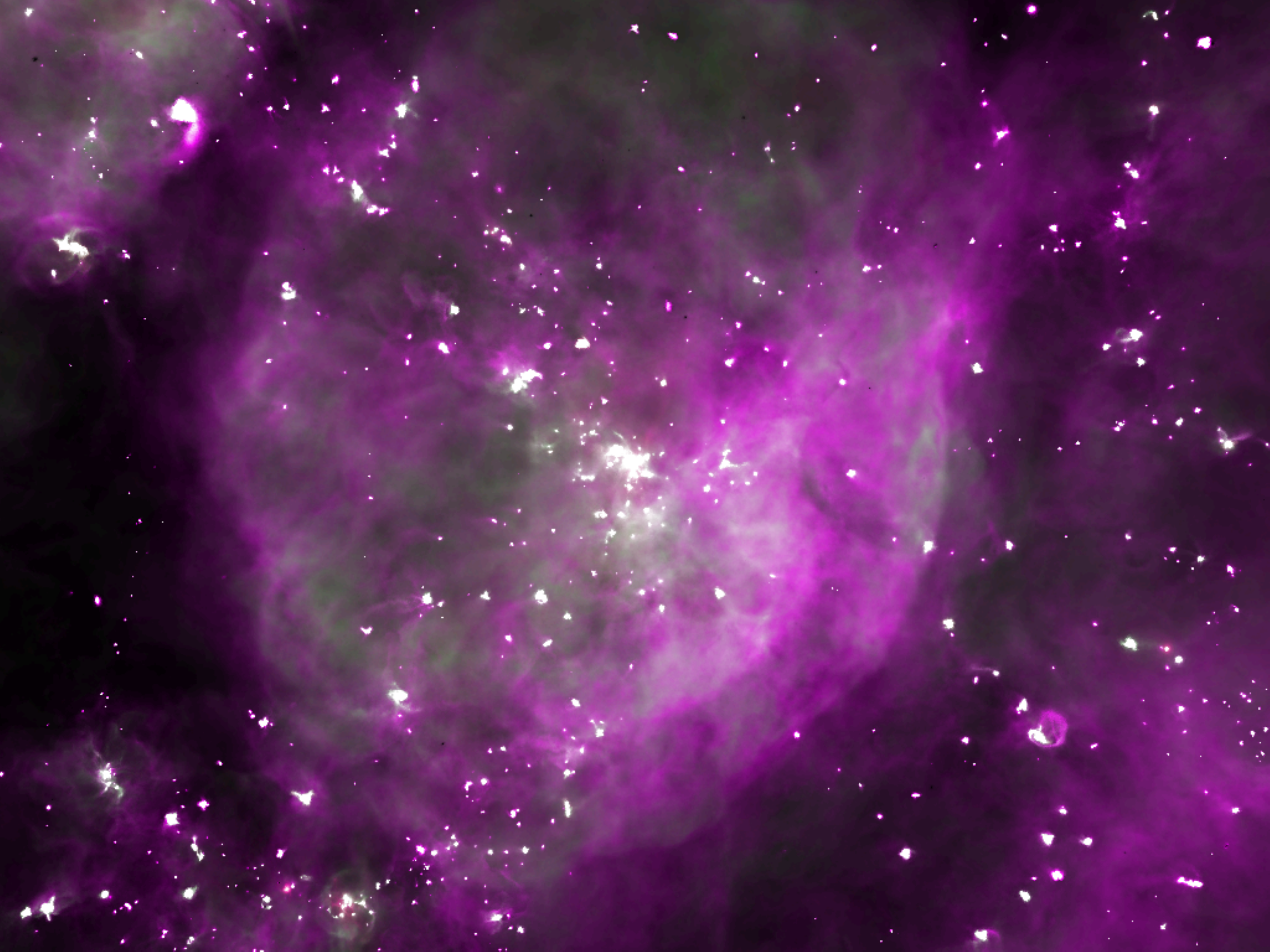}
\caption{Rendered gas metallicity field of a large region of the simulation
volume of L25n512 at $z=0$. The render frustum has a near field plane distance
of $2\hmpc$ and the far plane is at $20\hmpc$. The width of the near field
plane is $2\hmpc$ with a height of $1.5\hmpc$. The hot halo gas shown in
Figure~\ref{fig:render_temp} is significantly enriched.}
\label{fig:render_met}
\end{figure*}

The revised settings of the feedback model are required to match the
$z=0$ observations reasonably well. Less effective feedback leads to a
significant overproduction of stellar mass at both the faint and
bright ends. The need for even stronger feedback than in SPH was
already pointed out in~\cite{Vogelsberger2012}, where it was
demonstrated that {\sm AREPO} shows enhanced gas cooling compared to
previous SPH simulations, which we primarily attribute to spurious
viscous heating effects in SPH and an inaccurate treatment of subsonic
turbulence~\citep[][]{Bauer2012}.  This effect is strongest for high
mass systems where a quasi-hydrostatic hot atmosphere can form. SPH
simulations show here effectively a form of ``numerical quenching'',
reducing the amount of AGN feedback required to regulate the amount of
stars forming in these systems. Since {\sm AREPO} does not suffer from
this problem \citep[][]{Vogelsberger2012}, we are not helped by
``numerical quenching'' and instead need to offset the enhanced
cooling in large haloes at late times {\it physically}
by stronger forms of stellar and
AGN feedback. We note that this form of quenching can also be seen in 
\cite{Springel2003a,Springel2003b}, where the cosmic SFR density could be reproduced without the inclusion of AGN feedback.
This is possible because numerical quenching prevents star formation in massive haloes.

\subsection{Properties of the simulated galaxy population}\label{sec:sim_props}

In the following subsections we present some first comparisons of our
simulation results to observational data, where we mainly focus on the
local Universe ($z=0$). This serves primarily as a validation of our
galaxy formation model and an identification of acceptable parameter
settings. An initial study of the redshift evolution of the predicted
galaxy properties is given in \cite{Torrey2013}, while an
in-depth analysis will be provided in forthcoming studies based on
larger and higher resolution simulations.

We identify haloes, subhaloes and galaxies based on the {\sm SUBFIND}
algorithm~\citep[][]{Springel2001, Dolag2009b}. In the following, we often refer
to the stellar galaxy mass of haloes and subhaloes. For
definiteness, we define this mass as the total gravitationally bound
stellar mass that is contained within twice the stellar half mass
radius of each {\rm SUBFIND} (sub)halo.  This definition of stellar
mass does not significantly differ from the total stellar mass for low
mass systems, but it excludes some of the intracluster light stars for
massive systems. We have checked this definition against surface
brightness cuts in different bands and find it to give very similar
results as such more detailed methods of excluding intracluster light.

\subsubsection{General gas structure}\label{sec:general_gas_struct}

A first qualitative view of the simulations is presented in
Figure~\ref{fig:cosmo_box}, where we show gas density (left panels),
gas temperature (middle panels) and gas metallicity (right panels)
projections of the highest resolution L25n512 simulation with our
fiducial physics setup at three redshifts $z=2,1,0$ (top to
bottom). Each panel is $25\hmpc$ on a side and has a projection thickness of
$1\hmpc$. At $z=2$, some haloes show outflows generated mainly by
winds associated with stellar feedback leading to IGM enrichment. The
more dramatic heating effects at late times are largely caused by
strong radio-mode AGN feedback in massive systems. Both stellar and
AGN feedback lead to a significant enrichment of the IGM, as can
be seen in the right panels.  Furthermore, the AGN feedback
also alters the density structure of the gas at $z=0$.  

To demonstrate the impact of stellar and AGN feedback more clearly, we compare
in Figure~\ref{fig:cosmo_box_no_feedback} the L25n256 simulation with the ``no
feedback'' simulation, which does not include stellar and AGN feedback. We note
that the colour scales in Figure~\ref{fig:cosmo_box} and
Figure~\ref{fig:cosmo_box_no_feedback} are not the same and have 
been adapted for each
resolution.  Figure~\ref{fig:cosmo_box_no_feedback} clearly demonstrates that
stellar and AGN feedback strongly affect the thermodynamic properties of the
gas as can be seen from the density and temperature maps, which differ 
significantly
between the two runs.  Furthermore, the metal distribution is very different.
In fact, the simulation without any feedback does not lead to any significant
IGM enrichment as metals are locked up efficiently in galaxies at the
centres of haloes.

To give a better impression of the three-dimensional distribution of gas in the
simulation volume we show in Figure~\ref{fig:render_temp} a volume-rendered gas
temperature field of a large fraction of the L25n512 simulation volume\footnote{Volume-rendering movies 
and high-resolution images are available for download at the website \myhomepage}. Here,
the newly implemented, inlined volume renderer has been used, as described
above. We do not employ the smoothing procedure for this rendering. The render
frustum has a near field plane distance of $2\hmpc$ and the far plane is at
$20\hmpc$.  The width of the near field plane is $2\hmpc$ with a height of
$1.5\hmpc$. Hot regions are coloured in red, and cold regions in blue.
Galaxies are shown in bright white. Parts of the cosmic web and the hot halo
atmosphere of one of the most massive haloes in the simulation volume can be
seen. In Figure~\ref{fig:render_met}, we show the same region of the simulation
volume but render the gas metallicity instead. The hot gas atmosphere of the
cluster is clearly significantly enriched. We note that the filaments, which
are visible in the temperature rendering, are not clearly visible in
Figure~\ref{fig:render_met}.

\begin{figure*}
\centering
\includegraphics[width=0.49\textwidth]{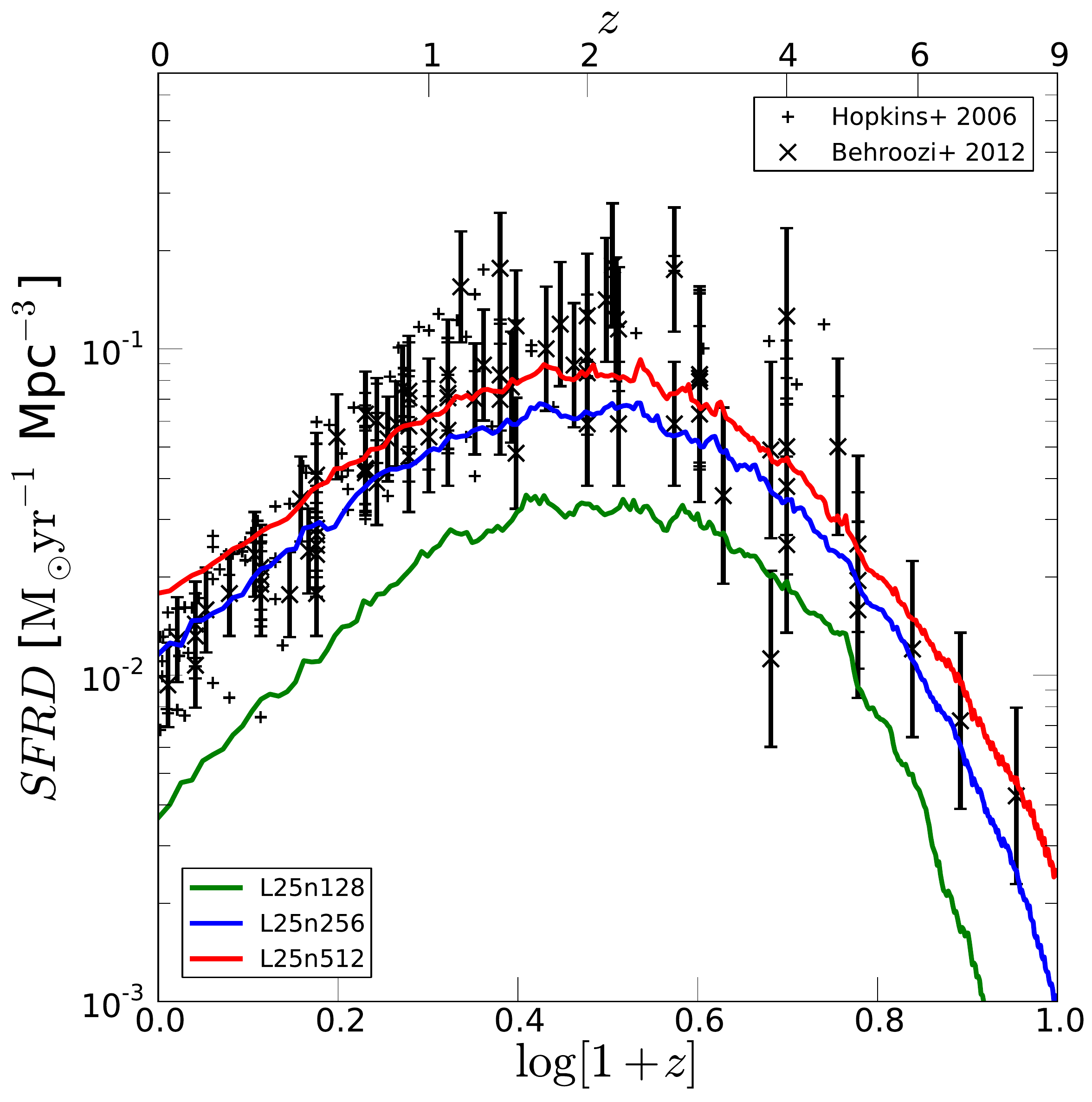}
\includegraphics[width=0.49\textwidth]{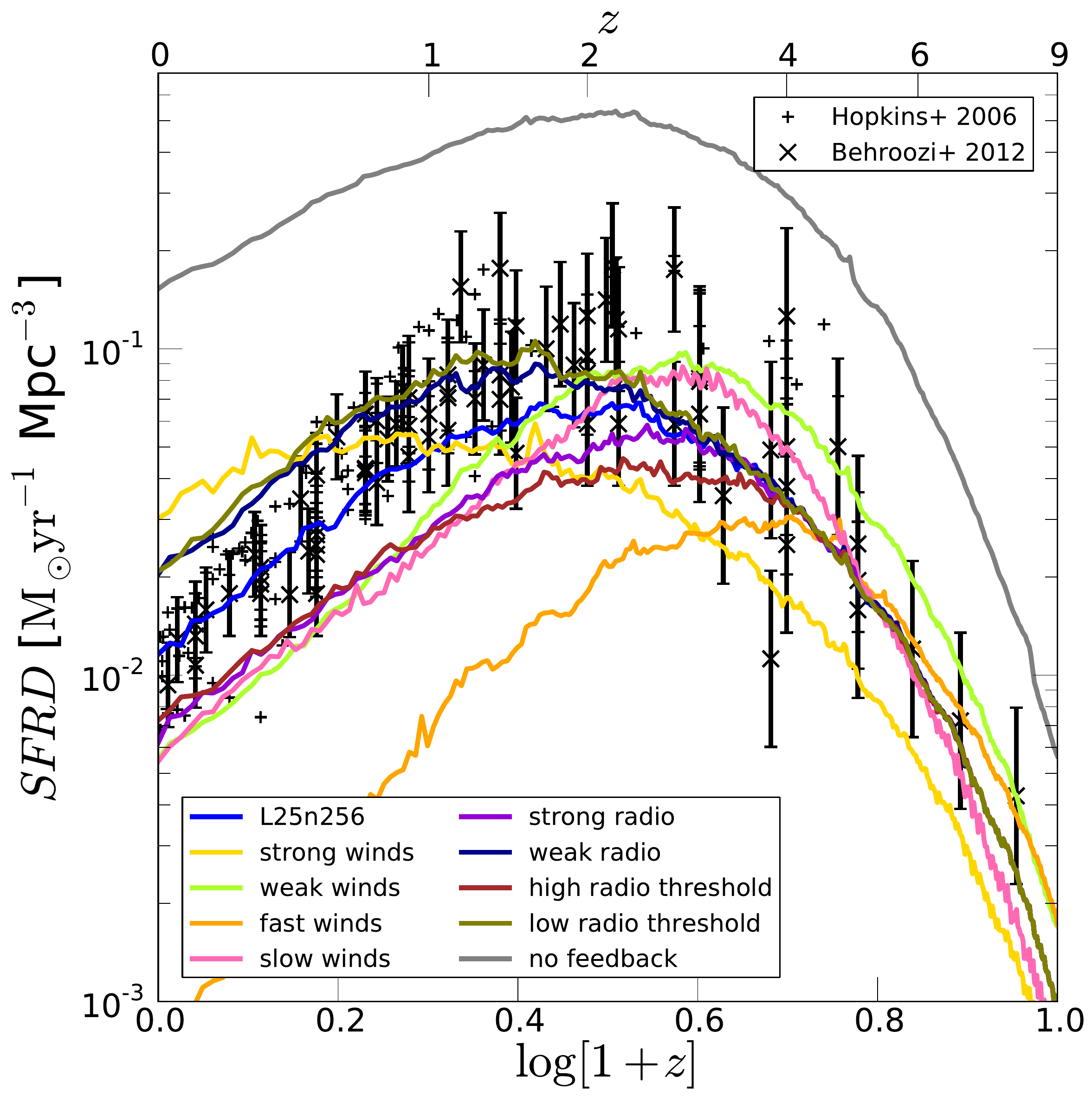}
\caption{Cosmic SFR density as a function of redshift (left panel: resolution
  study; right panel: different feedback models). Observational data
  is taken from~\protect\cite{Hopkins2006} and
  \citet{Behroozi2012}. The left panel demonstrates that the overall
  shape of the SFR evolution is well converged, with a minor
  normalisation change toward higher SFRs with increasing resolution,
  but the peak location does not change with resolution.  Doubling the
  wind velocity (``faster winds'') leads to a strong suppression of the
  overall star formation rate.  Increasing the overall wind energy
  (``stronger winds'') leads to a clear overproduction of stars at late
  times. Variations in the AGN radio-mode feedback do not change the
  SFR density nearly as much as changes in the stellar
  feedback. Stronger radio-mode feedback leads to less star formation
  at late times. The same is true for an increased radio mode
  threshold, where more AGN feedback energy is channelled into the
  radio-mode. Since this mode is more bursty, it is more efficient at
  reducing star formation compared to the continuous quasar-mode AGN
  feedback, which mainly regulates the growth of BHs.}
\label{fig:cosmo_sfr}
\end{figure*}

\subsubsection{Evolution of the cosmic SFR density}\label{sec:sfr}

Observationally it is now well-established that the cosmic SFR density
increased significantly from $z\sim10$ to $z\sim2$ and reached a peak
value at around $z \sim 2$ followed by a rapid
decline~\citep[][]{Lilly1996, Madau1998, Schiminovich2005,
  HopkinsBeacom2006, Bouwens2008}.  In Figure~\ref{fig:cosmo_sfr}, we
show the simulated cosmic SFR density as a function of redshift. To
compare our models against observational constraints, we have included
a set of cosmic SFR density measurements from the literature, as
compiled in~\cite{Hopkins2006} and~\cite{Behroozi2012}.  We show a
convergence study in the left panel, and an exploration of the
different feedback models summarised in Table~\ref{table:cosmo_sims}
in the right panel.  We will in the following subsections follow the
same format and always explore numerical convergence and physical
model variations side-by-side.

The left panel of Fig.~\ref{fig:cosmo_sfr} demonstrates that the shape
of the SFR evolution is well preserved for all three resolutions, with
a minor normalisation change towards higher SFRs with increasing
resolution.  The magnitude of this normalisation offset between our
two highest resolution runs is $\lesssim 1.5$, which we consider to be
reasonably well converged. The overall shape of the cosmic SFR density
is largely determined by stellar feedback at high redshifts and
radio-mode AGN feedback towards lower redshifts. We find that the
rapid decline of the SFR towards lower redshifts can only be achieved
through strong radio-mode AGN feedback~\citep[see also][for
  example]{Schaye2010, vandeVoort2011,Bower2012}.  In fact, in the absence
of AGN feedback, but with the addition of stellar feedback, the
cosmic SFR density would continue to rise beyond $z=0$, since gas
recycling and metal line cooling can support a large amount of star
formation at late times~\citep[see also][]{Schaye2010}.  We note in
particular that the location of the peak of the SFR density does not
change significantly with numerical resolution.  The intermediate
resolution run L25n256 is already sufficient to give a reasonably
reliable estimate of the SFR in our simulation volume. We can hence
study the effects of different physics parameterisations by running
simulations at this intermediate resolution, as listed in
Table~\ref{table:cosmo_sims}.

The comparison among these different feedback models is shown in the
right panel of Figure~\ref{fig:cosmo_sfr}.  Our fiducial model
(L25n256) provides clearly the best match to the observational data
compared to any of the other model variations presented in
Table~\ref{table:cosmo_sims}. The figure also highlights that the
impact of changes in the wind model can be rather dramatic.  For
example, doubling the wind velocity (``faster winds'') leads to a strong
suppression of the overall SFR, especially at late times. Such fast
winds lead to a significant reduction of star formation also for
massive haloes as we show below~\citep[see also][for a similar
  finding]{Schaye2010}.  Increasing the overall wind energy (``strong
winds'') leads to a clear overproduction of stars at late times,
because a very substantial fraction of the gas that is blown out of
galaxies at early times falls back in at late times -- fueling
further star formation along a ``wind accretion'' channel~\citep[see
  also][]{Oppenheimer2010}.  The peak of the SFR density is also
sensitive to the details of the wind and AGN parameterisation.
Interestingly, the changes of the SFR density with respect to
variations in the AGN radio-mode are less dramatic.  As expected,
``stronger radio'' feedback leads to more efficient suppression of
star formation at late times. Likewise, ``weaker radio'' mode feedback
increases star formation at late times. Similar trends can be seen for
changes in the radio threshold. A ``higher radio threshold'' puts more
energy into the radio-mode and is therefore more efficient in
suppressing star formation in massive systems at late times. A ``lower
radio threshold'' on the other hand puts more AGN feedback energy into
the quasar-mode, which is less efficient in suppressing star
formation.  Consequently, such a simulation produces more stars at
late times. In fact, we find that quasar-mode feedback is mainly
responsible in establishing the BH scaling relations, but does not
significantly affect star formation in massive systems.  The ``no
feedback'' simulation, as expected, strongly overproduces the amount
of stars at all times.  However, the peak of the SFR occurs still at
about the right time even for this simulation.

\subsubsection{Stellar mass -- halo mass relation}\label{sec:shm_ratio}

\begin{figure*}
\centering
\includegraphics[width=0.475\textwidth]{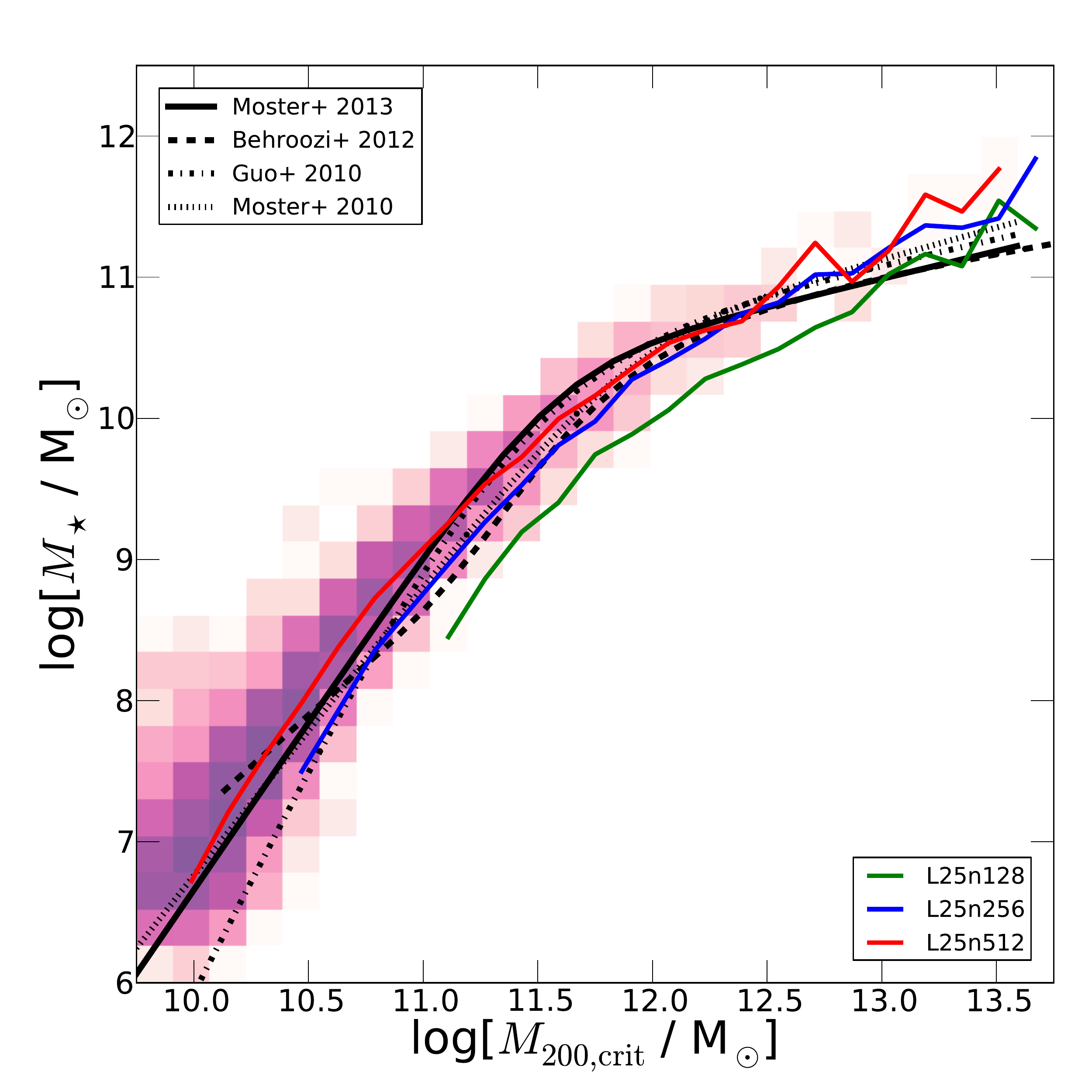}
\includegraphics[width=0.475\textwidth]{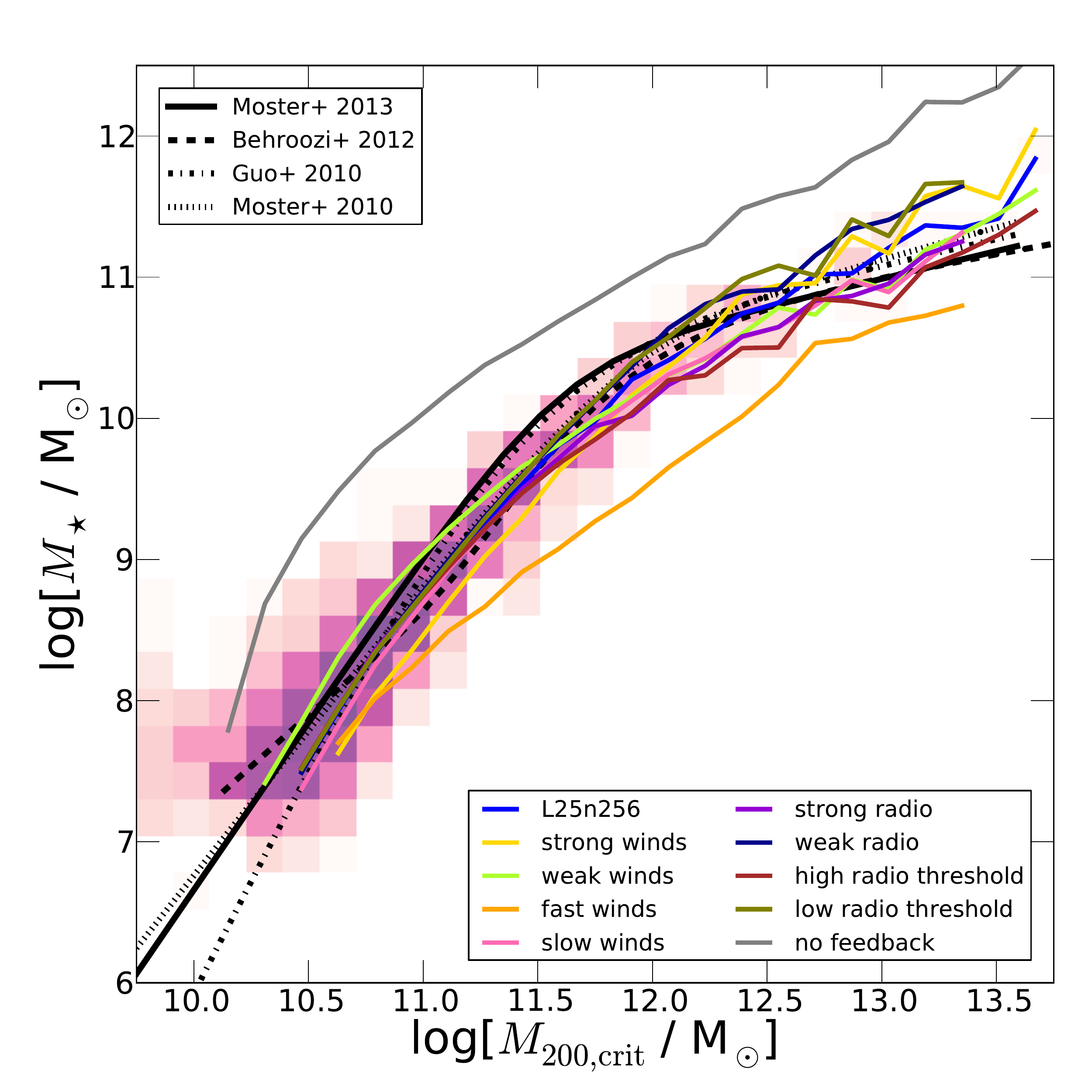}
\caption{Stellar mass of central galaxies as a function of total halo
  mass ($M_\mathrm{200,crit}$) at $z=0$ (left panel: resolution study;
  right panel: various feedback models). Different black lines show
  abundance matching results
  \protect\citep{Moster2010,Guo2010,Behroozi2012,Moster2013}, which are extrapolated beyond the constrained regime. Solid
  coloured lines mark the median relations of the simulations, whereas
  the two-dimensional histograms indicate the distribution of the
  L25n512 (left panel) and L25n256 (right panel) results.  The
  left panel demonstrates that the stellar content of low mass
  galaxies (i.e. $M_\star < 10^9 \msun$) is not yet fully
  converged. But at somewhat higher galaxy masses (i.e.  $M_\star>10^9
  \msun$), we find convergence in the stellar mass -- halo mass
  relationship.  The overall shape of the relation and the turnover
  mass agree reasonably well with the results derived based on
  abundance matching techniques.  The right panel demonstrates that
  both stellar feedback at the faint end and AGN feedback at the
  massive end shape the stellar mass content.  The wind speed has the
  most dramatic impact on the stellar mass content of haloes. ``Faster
  winds'' reduce the amount of stellar mass substantially over a large
  range of halo masses.  ``Weaker winds'' clearly do not suppress star
  formation efficiently enough at the faint end. The massive end is most
  sensitive to changes in the radio-mode accretion rate threshold,
  where more radio-mode AGN feedback leads to more efficient quenching
  of massive systems. The same effect can be achieved through a larger
  radio-mode feedback strength.}
\label{fig:cosmo_dm_mass__vs__stellar_mass_over_dm_mass}
\end{figure*}

The cosmic SFR density can be viewed as a convolution of the halo mass
function and the amount of stars formed in a given halo
\citep{Springel2003a}. It is therefore natural to examine the
average amount of stars formed as a function of halo mass.  Over the
last years, so-called abundance matching results have established the
required relationship between the stellar mass content of haloes and
their total mass~\citep[e.g.,][]{Conroy2006, Conroy2009, Moster2010,
  Guo2010, Behroozi2012, Moster2013} in order to achieve consistency
between the observed stellar mass function and the halo mass function
of the $\Lambda$CDM cosmology.  In
Figure~\ref{fig:cosmo_dm_mass__vs__stellar_mass_over_dm_mass}, we
present the stellar mass -- halo mass relation of our simulations at
$z=0$ and compare it to the four abundance matching results
of~\cite{Moster2010, Guo2010, Behroozi2012, Moster2013}. Specifically,
we compare against the stellar mass -- halo mass relation
parameterisation
\begin{equation}
M_\star = M_\mathrm{200,crit} \times a \left[\left(\frac{M_\mathrm{200,crit}}{10^b\msun}\right)^{c} + \left(\frac{M_\mathrm{200,crit}}{10^b\msun}\right)^{d}\right],
\end{equation}
with the coefficients taken from \cite{Moster2010}, \cite{Guo2010},
and \cite{Moster2013}. For \cite{Behroozi2012} we plot instead the
provided tabulated
data. Figure~\ref{fig:cosmo_dm_mass__vs__stellar_mass_over_dm_mass}
shows the stellar mass of central galaxies as a function of their halo
mass $M_\mathrm{200,crit}$, where the stellar mass is measured within
twice the stellar half-mass radius as discussed above.  Again, the
left panel of
Fig.~\ref{fig:cosmo_dm_mass__vs__stellar_mass_over_dm_mass} shows a
resolution study while the right panel explores the various physics
settings.  Solid lines mark the median relations, whereas the
two-dimensional coloured histograms indicate the distribution of the
L25n512 result (left panel) and L25n256 result (right panel).

\begin{figure*}
\centering
\includegraphics[width=0.475\textwidth]{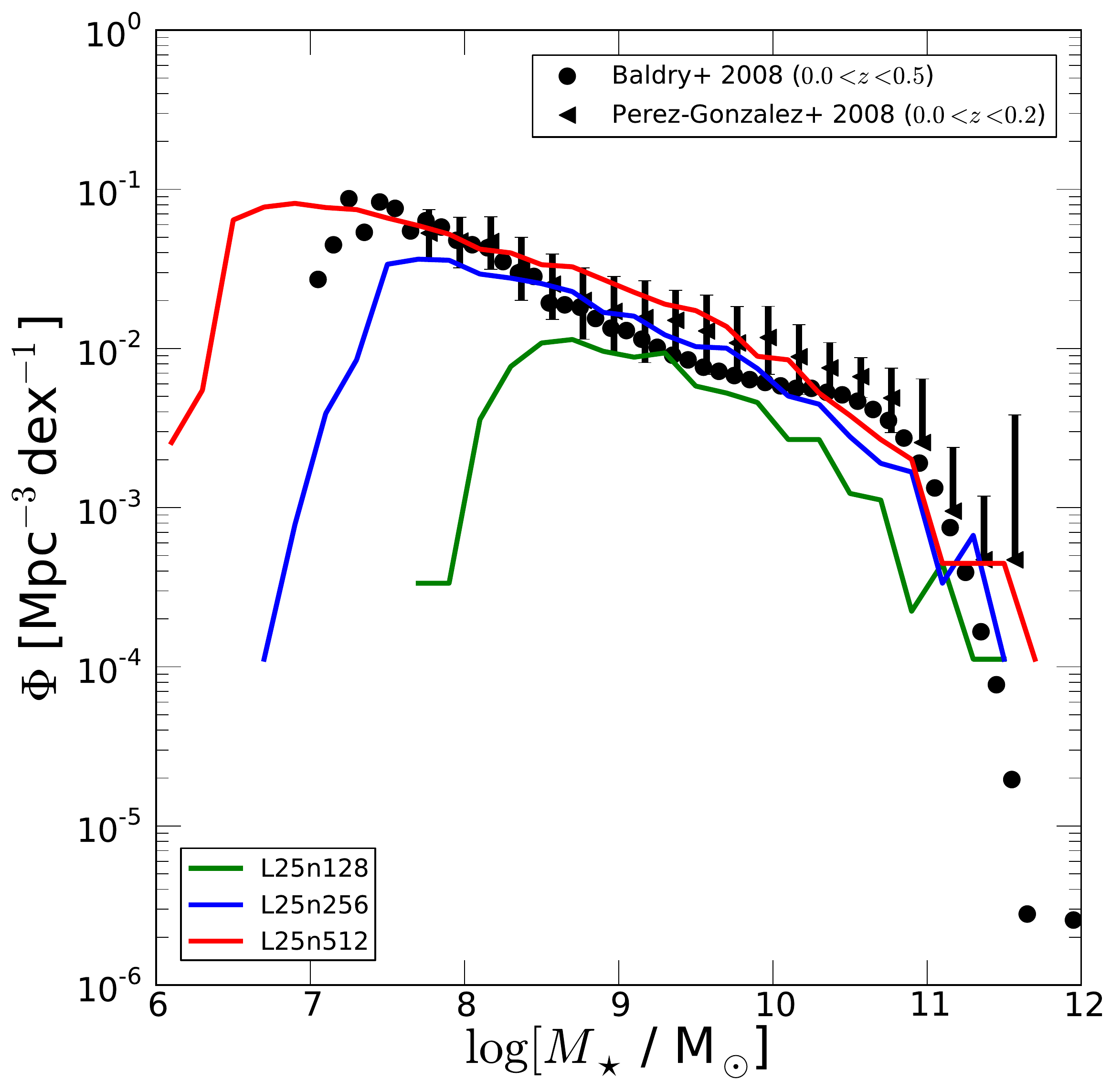}
\includegraphics[width=0.475\textwidth]{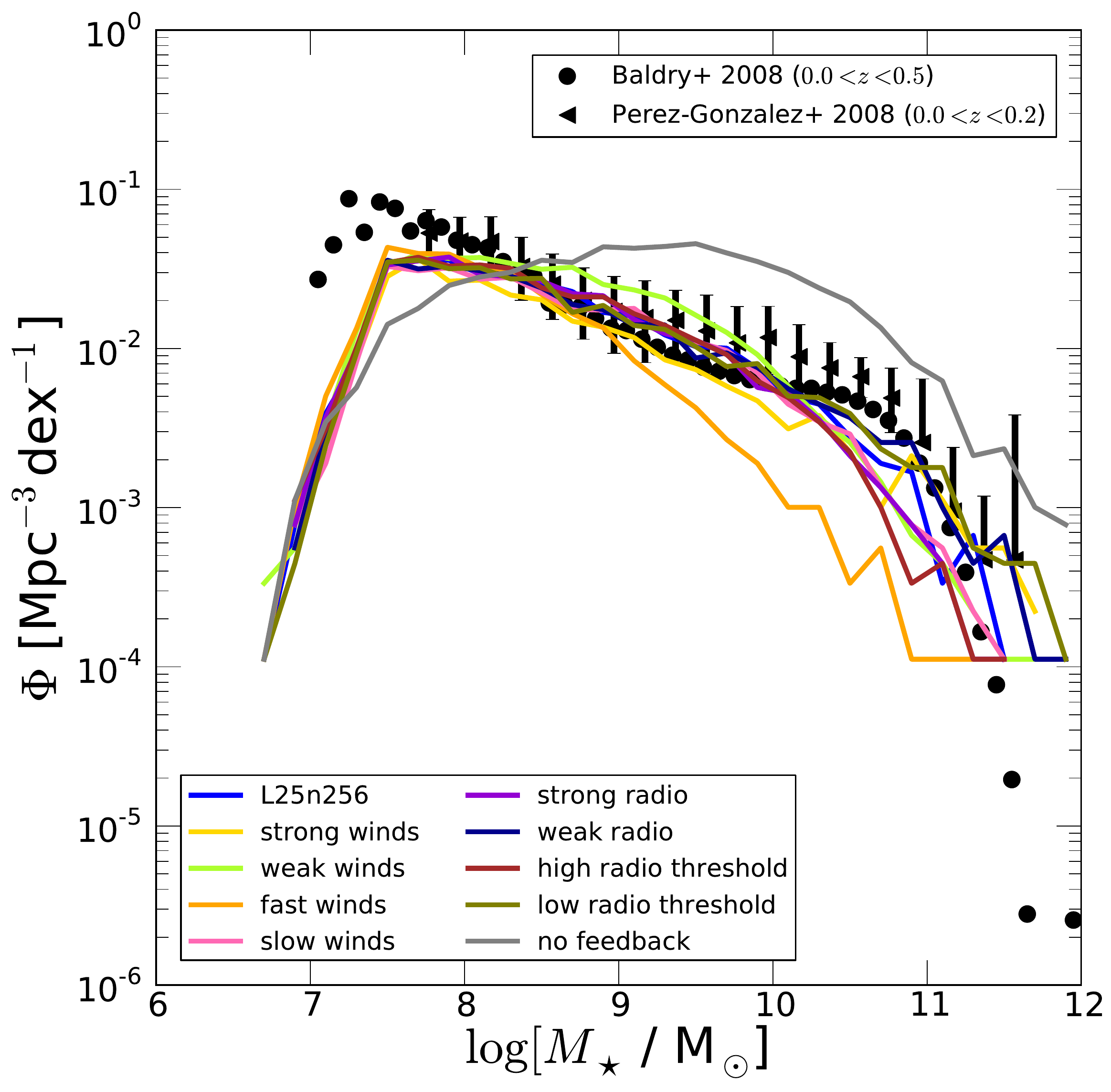}
\caption{
Stellar mass function at $z=0$ including all subhaloes (left
  panel: resolution study; right panel: different feedback
  models). Observational data points are taken
  from~\protect\cite{Bladry2008} and \citet{PerezGonzalez2008}.  The
  left panel shows that the stellar mass function is reasonably
  converged for higher mass systems (i.e.  $M_*>10^9\msun$).  In the
  right panel the ``faster wind'' simulation leads to a strong reduction
  of stellar mass in agreement with similar findings for other
  diagnostics like the SFR density and the stellar mass -- halo mass
  relation.  The ``weaker winds'' simulation overproduces the number of
  galaxies at the faint end significantly, whereas the ``strong
  winds'' suppress star formation too much in these systems.  A weaker
  radio-mode feedback leads to a less sharp drop of the stellar mass
  function towards the massive end. The same is true for a lower
  radio-mode accretion threshold since more energy goes then into the
  non-bursty quasar-mode AGN feedback, which is less effective at
  suppressing star formation.
}
\label{fig:cosmo_stellar_mass_function}
\end{figure*}

The left panel of
Figure~\ref{fig:cosmo_dm_mass__vs__stellar_mass_over_dm_mass}
demonstrates that low mass galaxies (i.e. $M_\star < 10^9 \msun$)
experience a factor of $\lesssim 2$ increase in their stellar masses
as we increase the resolution.  Although we do not expect similarly
sized changes to the stellar masses of these galaxies with further
increased resolution, the stellar masses of these low mass galaxies
are clearly not yet fully converged.  Galaxies with somewhat higher
masses (i.e.  $M_\star>10^9 \msun$) are better resolved and show
reasonable convergence in their stellar mass -- halo mass
relationship.  As a result, the overall shape of the relation and the
turnover mass are quite well converged and agree with the results
derived based on abundance matching techniques. We note that the break
in the stellar mass to halo mass ratio around
$M_\mathrm{200,crit}\sim10^{12}\msun$ can be reproduced only through a
combination of stellar and AGN feedback. In fact, the main challenge
in reproducing the abundance matching results is to correctly match
the change of slope towards more massive systems. Reproducing the
faint-end or massive-end slope alone is typically easier to
achieve. Our simulation volume is too small to properly sample massive
systems, but based on selected test calculations of clusters of
galaxies using a zoom-in procedure, we have verified that our feedback
model also reduces the stellar masses of these more massive systems to the observed levels 

The right panel of
Figure~\ref{fig:cosmo_dm_mass__vs__stellar_mass_over_dm_mass}
demonstrates that both stellar feedback at the faint end and AGN
feedback at the massive end are needed to produce a stellar mass --
halo mass relation that matches the derived
relation from abundance matching reasonably well.
In agreement with our findings for
the cosmic SFR density, the wind speed has the most dramatic impact on
the stellar mass content of haloes, with ``faster winds'' being capable of
substantially suppressing the stellar mass content of haloes over a
wide mass range, also affecting more massive systems. ``Weaker winds''
in contrast clearly do not reduce star formation efficiently enough at
the faint end.  This can be seen from the ``weaker wind'' curve which
significantly overproduces the stellar mass for low mass systems. On
the other hand, ``stronger winds'' lead to an excessive suppression of
star formation towards the low mass end, and also an undershoot around
the turnover point of the observed relation.  The massive end is very
sensitive to changes in the radio-mode accretion rate
threshold. Lowering this value, and thereby putting more AGN feedback
energy into the quasar-mode, leads to an overproduction of stars in
massive systems as can be seen from the ``lower radio threshold''
curve. The same is true if we decrease the radio feedback factor by a
factor of two (``weaker radio'' curve).  These findings agree with the
conclusions drawn from the cosmic SFR density plots.  As for the
cosmic SFR density, the fiducial L25n256 model provides the best fit
to the observational data. The ``no feedback'' simulation clearly
overproduces stars at all halo masses.  Furthermore, the amount of
stars scales nearly linearly with the halo mass in this case (above
$M_\mathrm{200,crit}\sim 10^{10.5}\msun$), and there is clearly no
turnover towards higher masses. This implies that radio-mode AGN feedback is
crucial for the quenching of star formation at the massive
end~\citep[as has been previously found by][for
  example]{Croton2006,Bower2006,Cattaneo2006,Somerville2008,Puchwein2008,McCarthy2010,
  Teyssier2011,Dubois2013}.

We stress again that {\sm AREPO} does not experience the ``numerical
quenching'' of cooling in large haloes present in SPH simulations. This
artificial quenching that occurs mainly in massive haloes helps to
reduce the amount of star formation in these systems
\citep[see][]{Vogelsberger2012}, which implies that even without
explicit AGN feedback the stellar masses of massive haloes are pushed
towards the observed relationship. Even though this is helpful in this
sense, it is an undesirable effect as it is purely of numerical
origin.  In fact, it will then lead one to adopt physically
incorrect settings for the radio-mode AGN feedback parameters.

\subsubsection{Stellar mass function}\label{sec:stellar_mass_function}

The shape of the stellar mass function is determined by the underlying
halo mass function convolved with the efficiency at which stars can
form in these haloes. As shown in the previous subsection, star
formation in our simulations is most efficient for haloes with masses
around $M_\mathrm{200,crit}\sim10^{12}\msun$, in agreement with
abundance matching results.  Star formation in lower mass haloes is
suppressed through stellar feedback, whereas star formation in more
massive systems is quenched by AGN feedback, as found in previous
semi-analytic models~\citep[mainly,][]{Bower2006,Croton2006} and more
recently in self-consistent hydrodynamical
simulations~\citep[e.g.,][]{Puchwein2013}.  Especially, the
exponential suppression at higher masses requires strong AGN feedback.

\begin{figure*}
\centering
\includegraphics[width=0.475\textwidth]{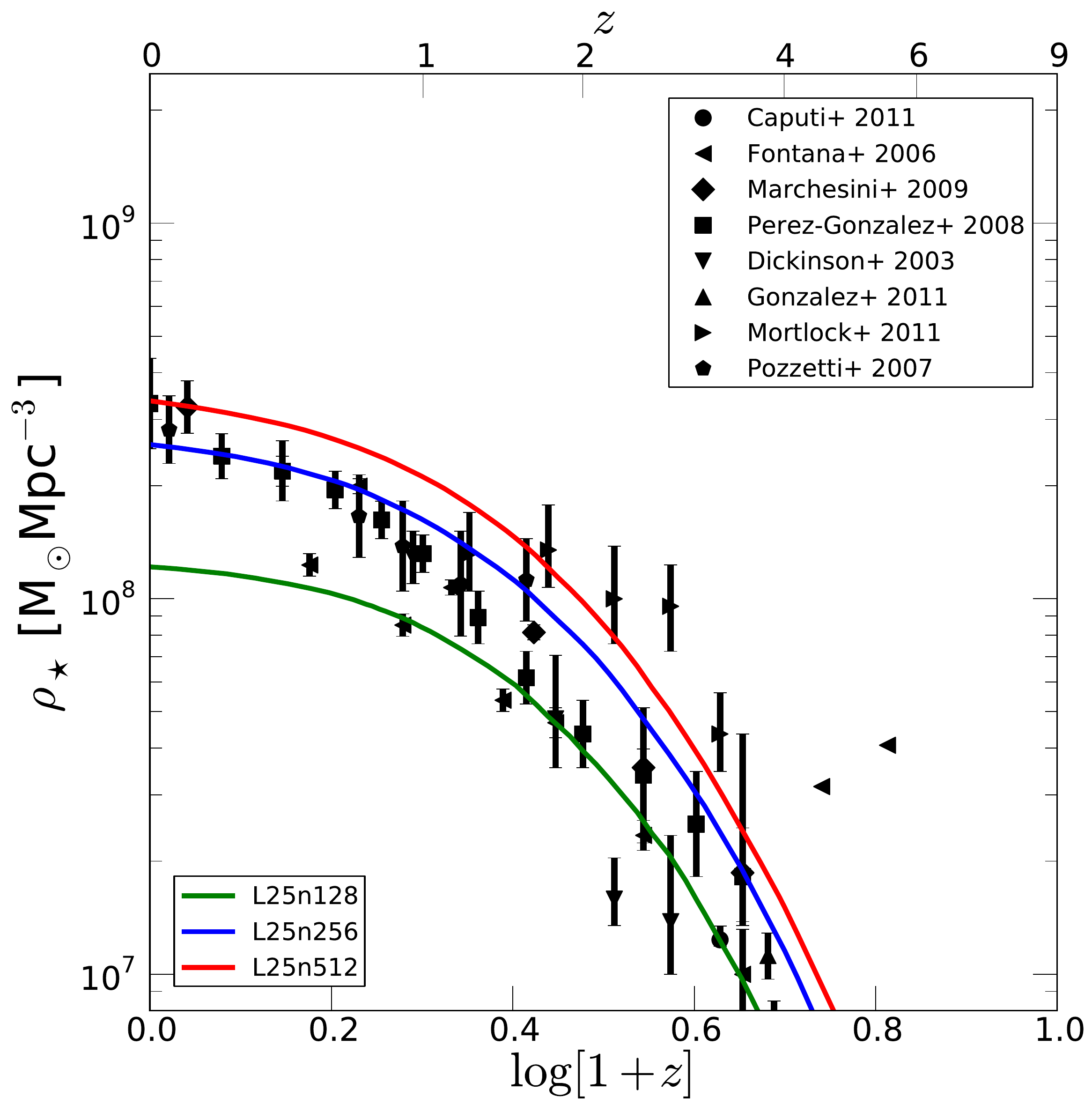}
\includegraphics[width=0.475\textwidth]{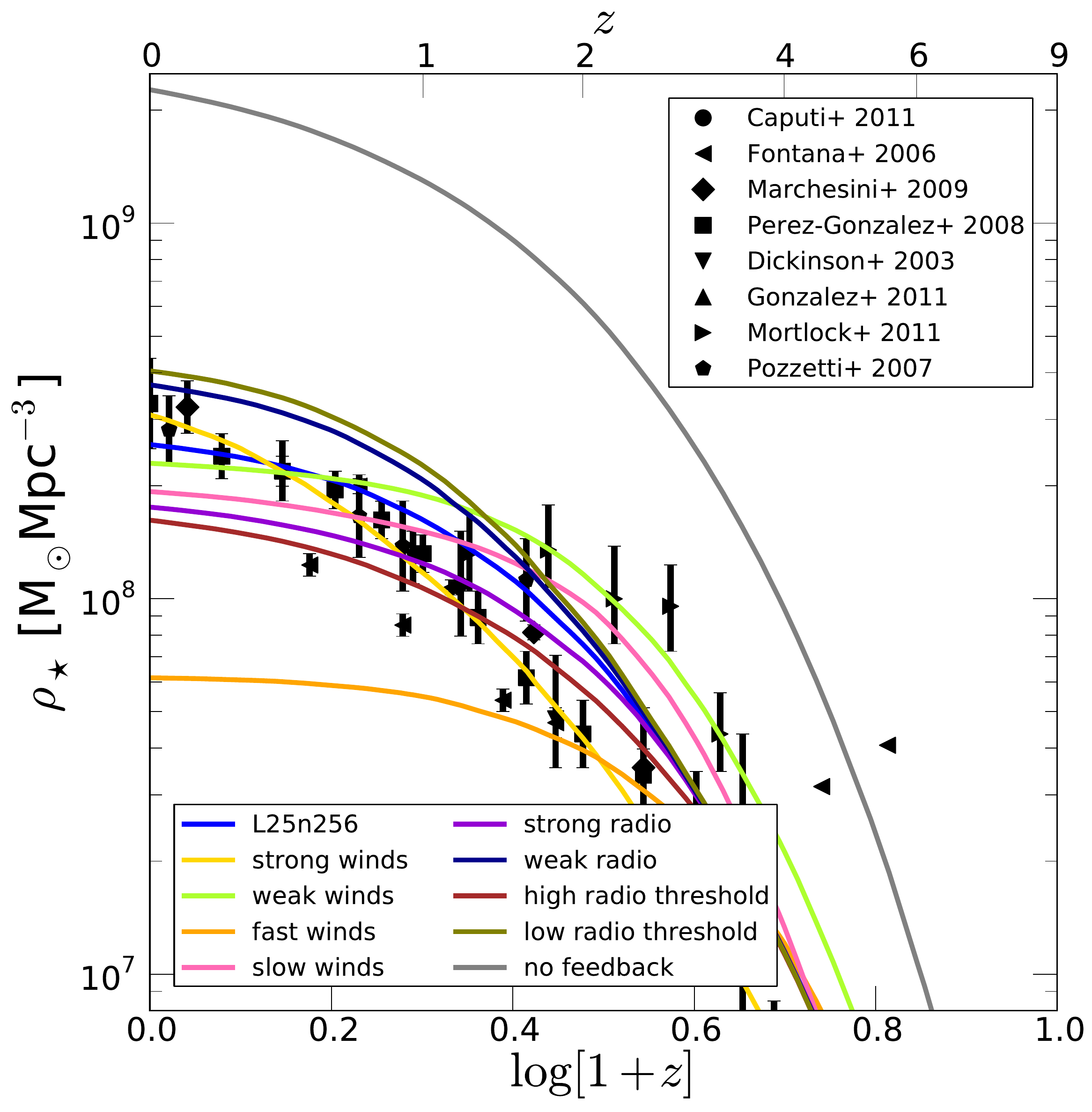}
\caption{Stellar mass density as a function of redshift for the various
models (left panel: resolution study; right panel: different feedback models).
We compare the simulation results to observational data taken
from~\protect\cite{Dickinson2003, Fontana2006, Pozzetti2007, PerezGonzalez2008,
Marchesini2009, Mortlock2011, Gonzalez2011, Caputi2011}. The convergence is
reasonable although the L25n512 simulation is clearly not yet fully converged
(left panel).  But both L25n256 and L25n512 have a stellar mass density
which is consistent with the observations. The ``no feedback'' simulation
strongly overproduces the amount of stellar mass, whereas the ``faster winds''
strongly underproduces it. All the other feedback variations are quite close
to the fiducial model, which provides the best fit to the data. Changes
in the radio-mode AGN feedback only affect the late time evolution of the
stellar mass density.}
\label{fig:cosmo_stellar_mass_density}
\end{figure*}

In Figure~\ref{fig:cosmo_stellar_mass_function}, we show the simulated
$z=0$ stellar mass function and compare it to observational data
of~\cite{Bladry2008} and \citet{PerezGonzalez2008}, with correction
factors to account for our Chabrier IMF.  Here we include all
subhaloes and treat the stellar component of each as a galaxy
contributing to the stellar mass function. As discussed above the
stellar mass is measured within twice the stellar half-mass radius.
The left panel of Figure~\ref{fig:cosmo_stellar_mass_function} shows
that the number density of galaxies at any given stellar mass
increases marginally with resolution.  This is directly tied to the
increase in stellar mass for low mass haloes, as discussed in the
previous subsection.  Consistently with our previous discussion, we find
that the stellar mass function shows reasonable convergence for higher
mass systems (i.e. $M_*>10^9\msun$). But increasing the resolution
allows an examination of progressively smaller galaxy masses.  With
our highest resolution simulation, we are able to compare to the
observed stellar mass functions over quite a large range of masses.
Although both functions differ in their detailed shape, we find that
our simulated stellar mass functions are remarkably similar to the
observations.  This includes the slope of the low mass end of the
stellar mass function, which is shaped by our stellar wind model, as
well as the sharp drop towards the massive end, which is due to the
radio-mode AGN feedback. We emphasise that the energy scaling of our
stellar wind model is crucial to produce a very flat stellar mass
function towards lower mass systems. Note that the mass loading for
low mass systems can achieve very high values if an energy-scaling is
adopted, and it grows more rapidly towards small galaxy sizes than for
a momentum-scaling. A similar conclusion was reached
by~\cite{Puchwein2013} who also argued for energy-driven winds to
obtain a sufficiently flat stellar mass function for low mass
systems \citep[but see ][for arguments in favour of momentum-driven
  winds]{Oppenheimer2010}.

In the right panel of Figure~\ref{fig:cosmo_stellar_mass_function}, we
explore variations around our fiducial model. The ``faster wind''
simulation leads to a strong reduction of stellar mass in agreement
with similar findings for other diagnostics presented above. Such
winds are capable of suppressing star formation even in more massive
haloes.  The other models behave largely very similarly.  The ``weaker
winds'' simulation overproduces the number of galaxies at the faint
end significantly, whereas ``stronger winds'' suppress star formation
too much in these systems.  The ``stronger winds'' also lead to an
undershoot of the stellar mass function around the cut-off scale.
These findings are consistent with the ones presented above for the
stellar mass -- halo mass relationship.  A ``weaker radio'' mode
feedback leads to a less sharp drop of the stellar mass function
towards the massive end. The same is true for a lower radio-mode
accretion threshold since in that case more energy goes then into the
non-bursty quasar-mode AGN feedback. The ``no feedback'' simulation
overproduces the number of galaxies at all stellar masses. Here the
stellar mass function closely follows the shape of the halo mass
function since most of the baryons in a halo can efficiently cool and form
stars if not moderated by feedback. Note that the simulation without
feedback does not show the strong exponential drop towards high mass
galaxies masses. As for the stellar mass -- halo mass relation, we
hence argue that a quenching of star formation at the massive halo end
requires AGN feedback.

\subsubsection{Stellar mass density}\label{sec:stellar_mass_density}

Integrating the cosmic SFR density leads to the overall stellar mass
density in the Universe. In
Figure~\ref{fig:cosmo_stellar_mass_density} we show the stellar mass
density of the total simulation volume as a function of redshift,
comparing it to estimates based on observational
data~\citep{Dickinson2003, Fontana2006, Pozzetti2007,
  PerezGonzalez2008, Marchesini2009, Mortlock2011, Gonzalez2011,
  Caputi2011}. We note that we here include all stellar mass formed in
the simulation volume; i.e.~we do not restrict ourselves to stellar
mass associated with (sub)haloes or that has formed within a certain
radial range in a halo. The reason for this is to make this diagnostic
consistent with the cosmic SFR density (see
Figure~\ref{fig:cosmo_sfr}), which also includes all reservoirs of
stellar mass formed without restriction.  

\begin{figure*} \centering
\includegraphics[width=0.475\textwidth]{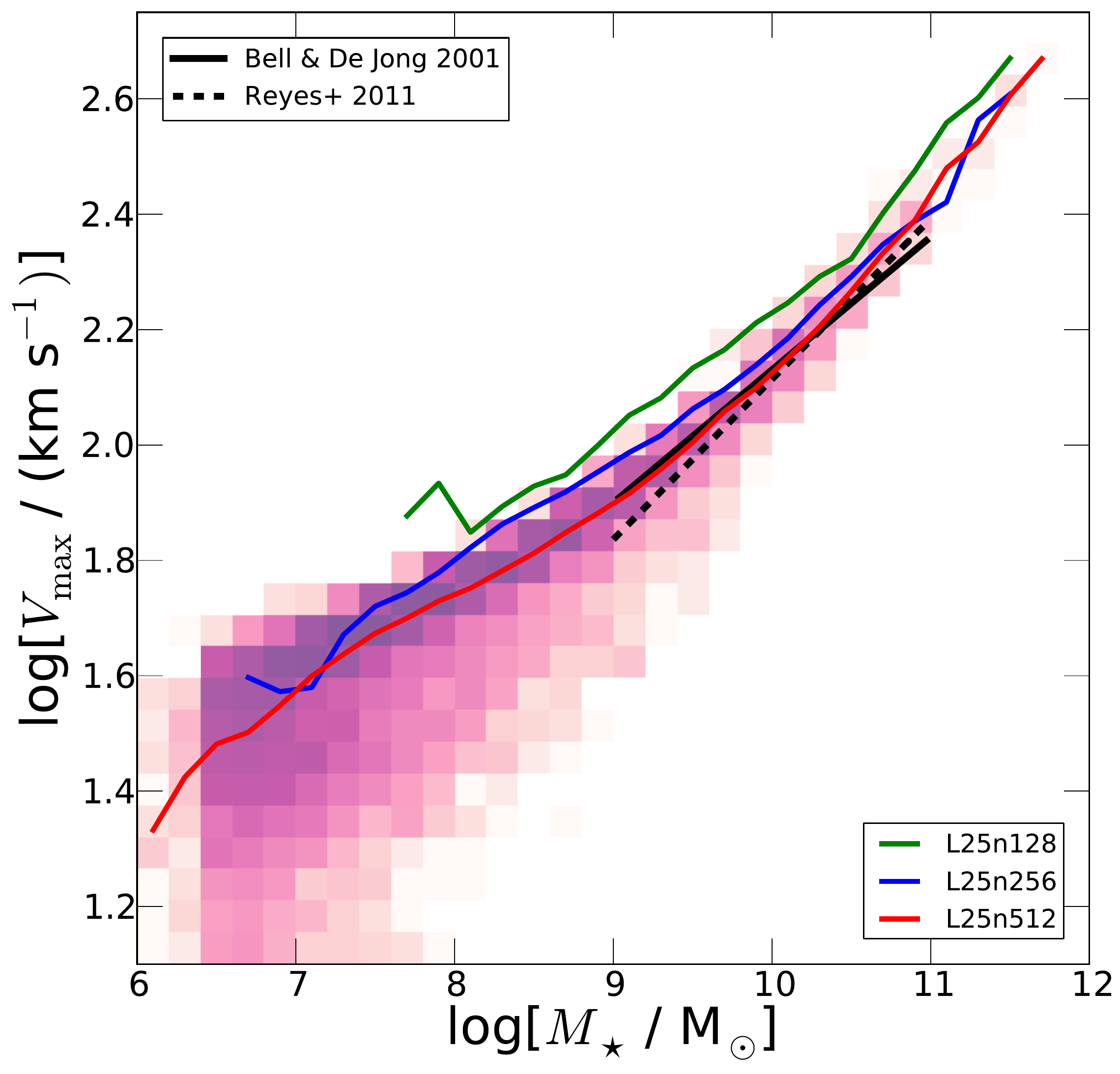}
\includegraphics[width=0.475\textwidth]{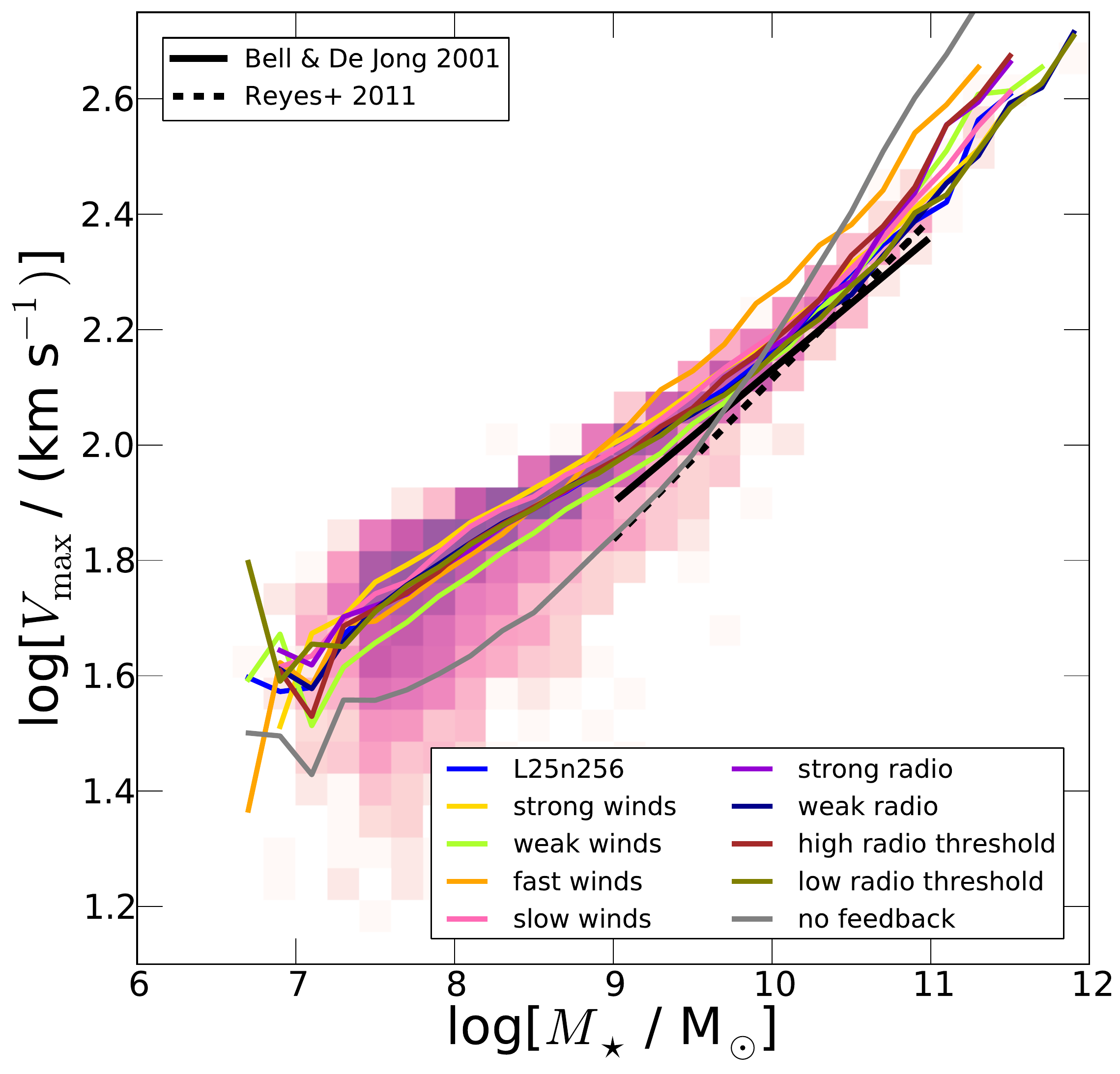}
\caption{Tully-Fisher relation at $z=0$ (left panel: resolution study; right
panel: different feedback models). Observational data is taken
from~\citet{Bell2001} and~\citet{Reyes2011}. Solid lines mark the simulation
median relations, whereas the two-dimensional histograms indicate the
distribution of the L25n512 result (left panel) and L25n256 result (right
panel).  The runs converge toward a relation which is very similar to the
marked observed relation with the slope and normalisation of the simulated
Tully-Fisher relation falling almost directly on top of the two marked
observational Tully-Fisher relations.  Most variations of the feedback model
parameter choices lead to only minor changes in the resulting Tully-Fisher
relation.  The exception is the ``faster wind'' model which gives rise to the
largest change in the Tully-Fisher relation by increasing the rotational
velocity at a fixed stellar mass.  This result is a consequence of the
efficiency with which the fast wind model suppresses star formation.}
\label{fig:cosmo_tully_fisher}
\end{figure*}

Our fiducial model reproduces the observations
reasonably well at all redshifts. As already seen above, the
``faster winds'' suppress star formation too strongly and therefore also
significantly underpredict the total stellar mass density at nearly all
redshifts but especially towards lower redshifts. Changes in the
radio-mode lead, as expected, to differences in the late time
behaviour since the radio-mode becomes dominant only at late times
once more massive haloes with lower BH accretion rates form. For the
radio-mode we also find the anticipated behaviour: a ``stronger
radio'' mode reduces the total stellar mass density at late times,
whereas a ``weaker radio'' mode leads to an increase.  A similar trend
can be seen for variations of the radio-mode accretion threshold.  A
higher threshold suppresses late time star formation compared to our
fiducial model, which is in agreement with the findings above. The
``no feedback'' simulation strongly overproduces the amount of stars.

The left panel of Figure~\ref{fig:cosmo_stellar_mass_density} demonstrates that
the convergence of the overall amount of stellar mass is reasonable although
the L25n512 simulation is clearly not yet fully converged. But both L25n256 and
L25n512 have a stellar mass density at $z=0$ which is consistent with the
observations. The non-converged L25n128 simulation clearly fails in this
respect and has far too few stars at $z=0$ since it does not have adequate mass
resolution. We note that such a low resolution does not even resolve the halo
mass function properly for our simulation volume. It is therefore not
surprising that the stellar mass is substantially underpredicted in that case. 

\subsubsection{Tully-Fisher relation}\label{sec:tullyfisher}

The stellar mass and the circular velocity of disk galaxies are
strongly correlated through the Tully-Fisher
relation~\citep[][]{Tully1977}.  Reproducing the Tully-Fisher
relation, which ties together galactic mass, concentration, and
angular momentum, is an important goal of any galaxy formation
model~\citep[e.g.,][]{Steinmetz1999, SommerLarsen2003, 
Robertson2004, Croft2009,
  Agertz2011, Scannapieco2012,Hummels2012}. In
Figure~\ref{fig:cosmo_tully_fisher}, we compare the $z=0$ Tully-Fisher
relation of simulated galaxies with observational fits of the form
\begin{equation}
\log\!\left(\frac{V_\mathrm{max}}{\kms}\right) \! = \log\!\left(\frac{V_\mathrm{max,10}}{\kms}\right) \! + b  \log\!\left(\frac{M_\star}{10^{10}\msun}\right) ,
\end{equation}
where we take the coefficients from~\cite{Bell2001}
and~\citet{Reyes2011}.  We note that \cite{Reyes2011} provide a fit to
$V_\mathrm{80}$, which is the rotation velocity measured at the radius
containing $80\%$ of the i-band galaxy light. However, as
\cite{McCarthy2012} point out, this does generally not differ 
significantly
from our diagnostic so we decided to compare to their results as well.
Solid lines mark the median relations of the simulation, whereas the
two-dimensional coloured histograms indicate the distribution of the
L25n512 result (left panel) and L25n256 result (right panel). As
before we measure the stellar mass within twice the stellar half-mass
radius.  For the plotted circular velocity, we measure the total mass
within that radius and calculate the associated circular
velocity~\citep[see also][]{Scannapieco2012}. We have experimented
also with other measures of the circular velocity but found only very
small variations. In particular, using $V_\mathrm{max}$ of the subhaloes
gives similar results.

\begin{figure*}
\centering
\includegraphics[width=0.475\textwidth]{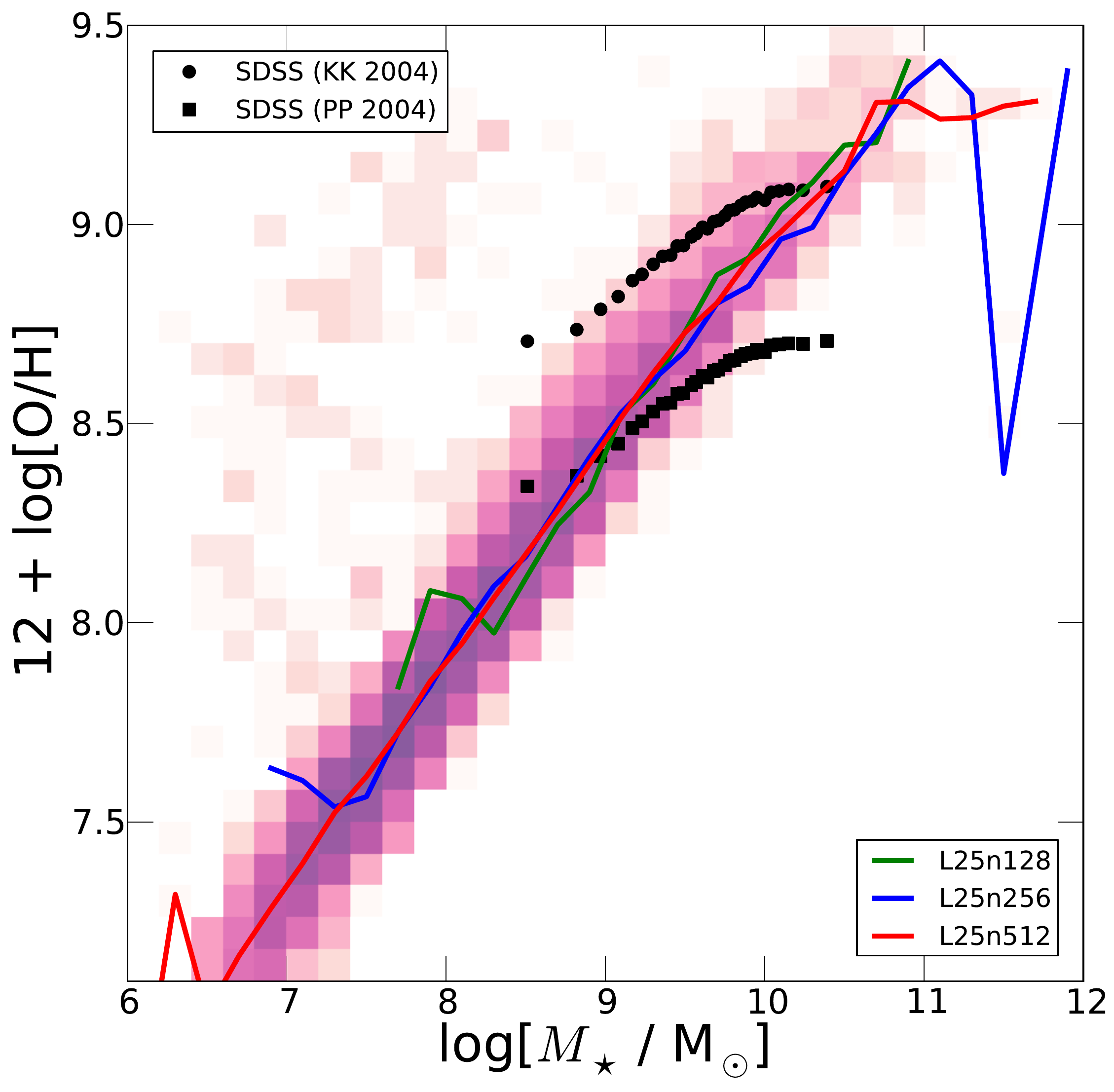}
\includegraphics[width=0.475\textwidth]{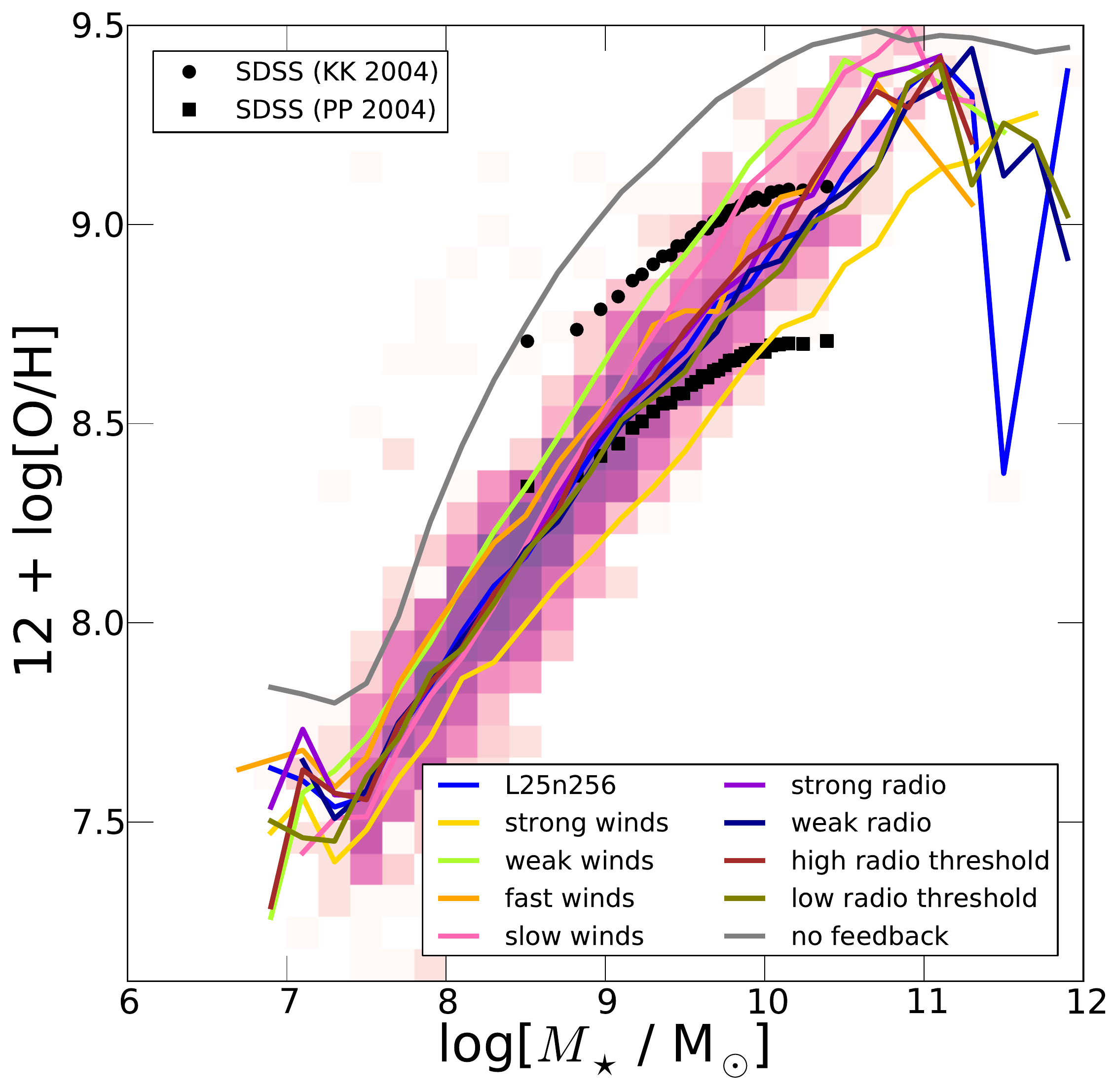}
\caption{Mass--metallicity relation as a function of stellar mass
  (left panel: resolution study; right panel: different feedback
  models). Observational data is taken from SDSS
  DR7~\protect\citep{Abazajian2009}. Solid lines mark the simulation
  median relations, whereas the two-dimensional coloured histograms
  indicate the distribution of the L25n512 (left panel) and L25n256
  (right panel) results.  The metallicity values are well-converged,
  but the relationship which they converge to does not perfectly match
  the observations.  Specifically, the observed mass metallicity
  slope is shallower than the simulated slope, and there is no apparent
  turn-over in the simulated mass metallicity relation at high masses.
  However, the simulated mass metallicity relation falls between the
  PP04 and KK04 relations indicating that galaxies are retaining a
  reasonable fraction of their metals.  The right panel shows that the
  simulated mass metallicity relation is quite sensitive to the
  adopted physics parameters since feedback variations strongly affect
  the amount of metals retained in galaxies and expelled to haloes.}
\label{fig:cosmo_stellar_mass__vs__oxygen_over_H}
\end{figure*}

Examining the different resolutions shown in the left panel of
Figure~\ref{fig:cosmo_tully_fisher}, we find that there is good
convergence for high mass systems, consistent with our previous
findings.  Furthermore our model converges towards a relation which is
very similar to the observed relation, with the slope and
normalisation of the simulated Tully-Fisher relation falling almost
directly on top of the two observational Tully-Fisher
determinations. We note that we did not select the galaxies in
Figure~\ref{fig:cosmo_tully_fisher} according to their type.  This
means that especially at the more massive end our result will be
influenced by a certain fraction of elliptical galaxies, implying that
we should not expect perfect agreement with the observations in this
regime.

In the right panel of Figure~\ref{fig:cosmo_tully_fisher}, we find
that most variations in the feedback parameters do not induce
significant changes in the resulting Tully-Fisher relation.  The
exception is the ``faster wind'' model which gives rise to the largest
change in the Tully-Fisher relation by increasing the rotational
velocity at a fixed stellar mass.  This result is a consequence of the
efficiency with which the fast wind model suppresses star formation,
as seen previously in
Figure~\ref{fig:cosmo_dm_mass__vs__stellar_mass_over_dm_mass}.  Aside
from the ``faster wind'' model, we find that our feedback models are
able to produce reasonable normalisations and slopes for the
Tully-Fisher relation independent of the detailed feedback parameter
choices. We therefore conclude that the Tully-Fisher relation is not
very sensitive to the details of our physics parameterisation. In fact,
among all the observables considered in this paper the Tully-Fisher
relation is the most stable against variations of the underlying
physics. However, the ``no feedback'' simulation clearly fails
to reproduce this key observable.

\subsubsection{Mass-metallicity relation}\label{sec:massmetal_rel}

Our stellar evolution and enrichment implementation allows us to study
chemical abundances of galaxies in detail and confront those with
observations.  In
Figure~\ref{fig:cosmo_stellar_mass__vs__oxygen_over_H}, we show the
$z=0$ mass metallicity relation and compare it to SDSS DR7
data~\citep[][]{Abazajian2009}. Two distinct mass metallicity
relations are shown. Although both use the SDSS data, one is the
KK04~\citep{KK04} nebular emission line diagnostic relation, while the
other is the PP04~\citep{PP04} diagnostic. To facilitate this
comparison, we have taken the metallicity values as tabulated
in~\cite{Zahid2012} using the KK04 diagnostic and converted them to
the PP04 diagnostic using the empirical relations
of~\cite{KewleyEllison2008}.  The main reason for this procedure is to
include the relevant information about the known uncertainty in the
normalisation of nebular emission line metallicity indicators, and to
avoid comparing our models to only one diagnostic method. We calculate
the metallicity value for each simulated galaxy by finding the average
oxygen to hydrogen abundance, weighted by the star formation rates of
gas and excluding non-star forming gas from this calculation in order
to provide a more straightforward comparison to observations, where
the nebular emission lines naturally probe star-forming regions.

In the left panel of 
Figure.~\ref{fig:cosmo_stellar_mass__vs__oxygen_over_H}, we find that the
metallicity values are well-converged for our runs with varying
resolution.  In fact, the convergence is slightly better than in most
of the other relations which we examine in this paper. However, the
relationship which they converge to does not perfectly reproduce the
observations.  There are two main features that make the
simulation results distinct from the observations. The first is that
the observed mass metallicity slope is shallower than the simulated
slope -- regardless of whether we adopt the PP04 or KK04
diagnostic. The second issue is that there is no apparent turn-over in
the simulated mass metallicity relation at high masses, unlike
observed. Although there is a flattening visible for the two highest
resolutions (L25n256, L25n512), this occurs at too large stellar
masses and at too high metallicities to account for this observational
signature.  However, despite these shortcomings, the simulated mass
metallicity relation falls between the PP04 and KK04
relations for most systems, which indicates that the simulated 
galaxies retain a
reasonable fraction of their metals, in acceptable agreement with
observations, on average (a similar level agreement was found in \cite{Dave2011b}).

\begin{figure*}
\centering
\includegraphics[width=0.475\textwidth]{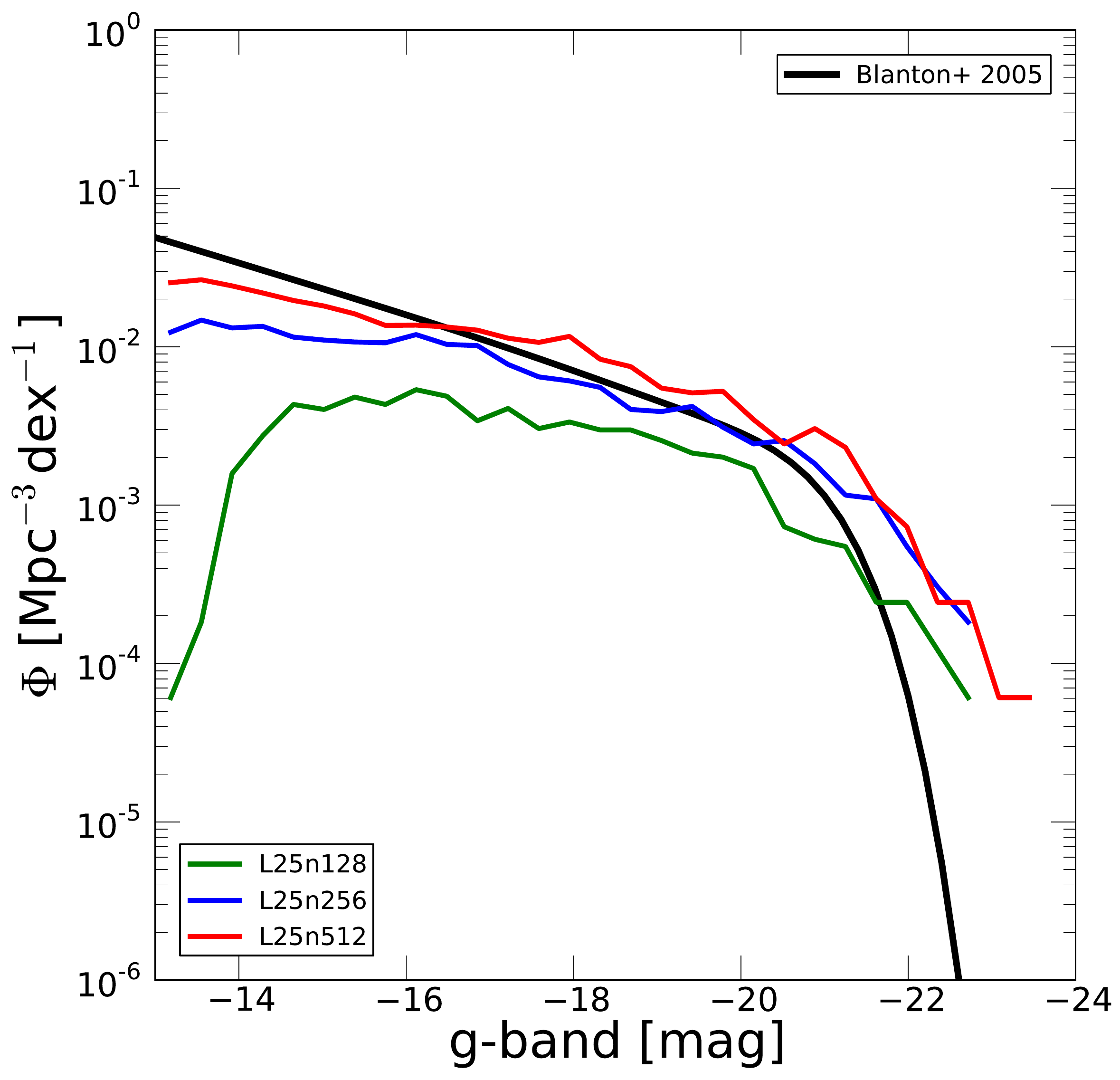}
\includegraphics[width=0.475\textwidth]{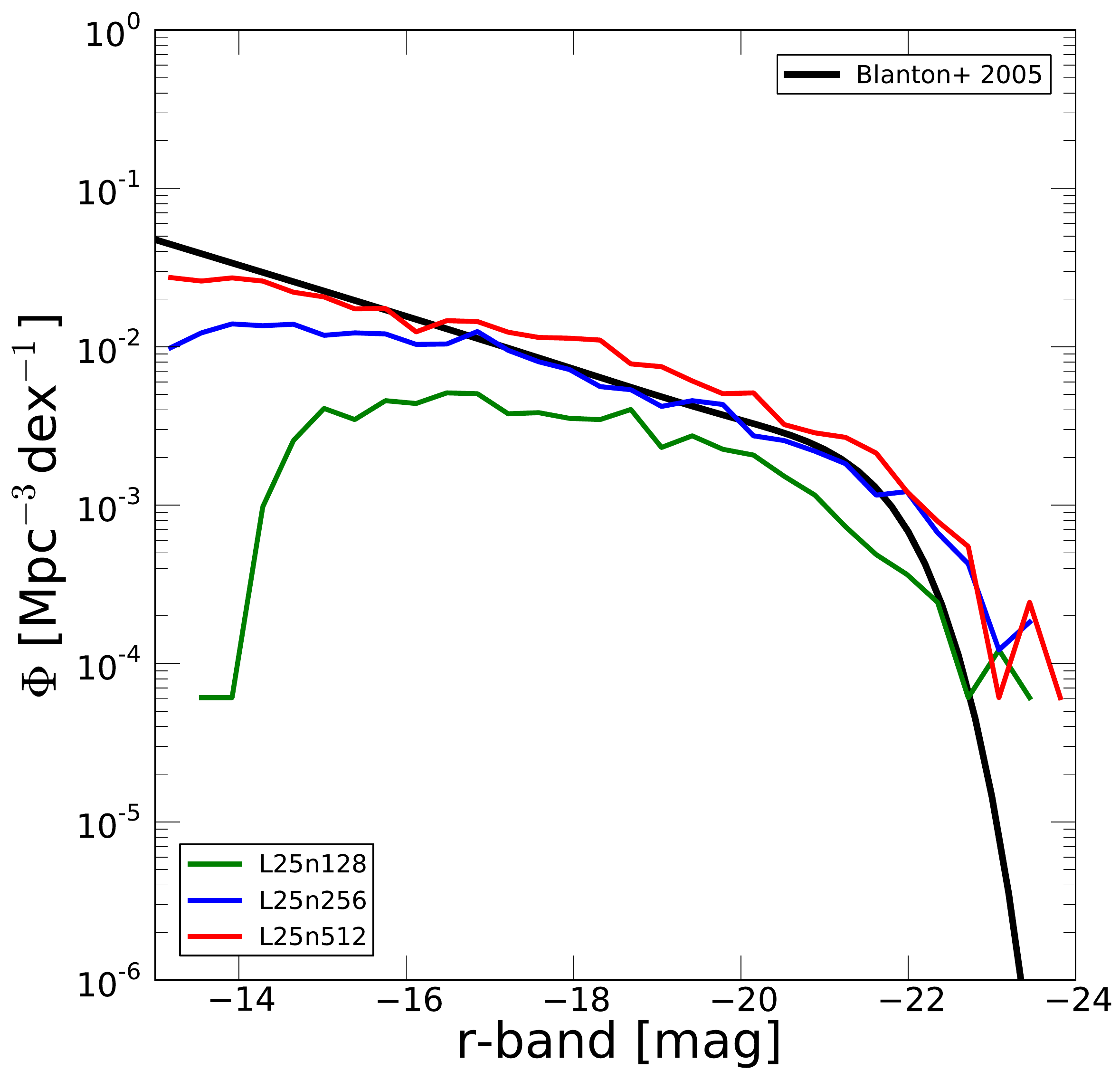}
\includegraphics[width=0.475\textwidth]{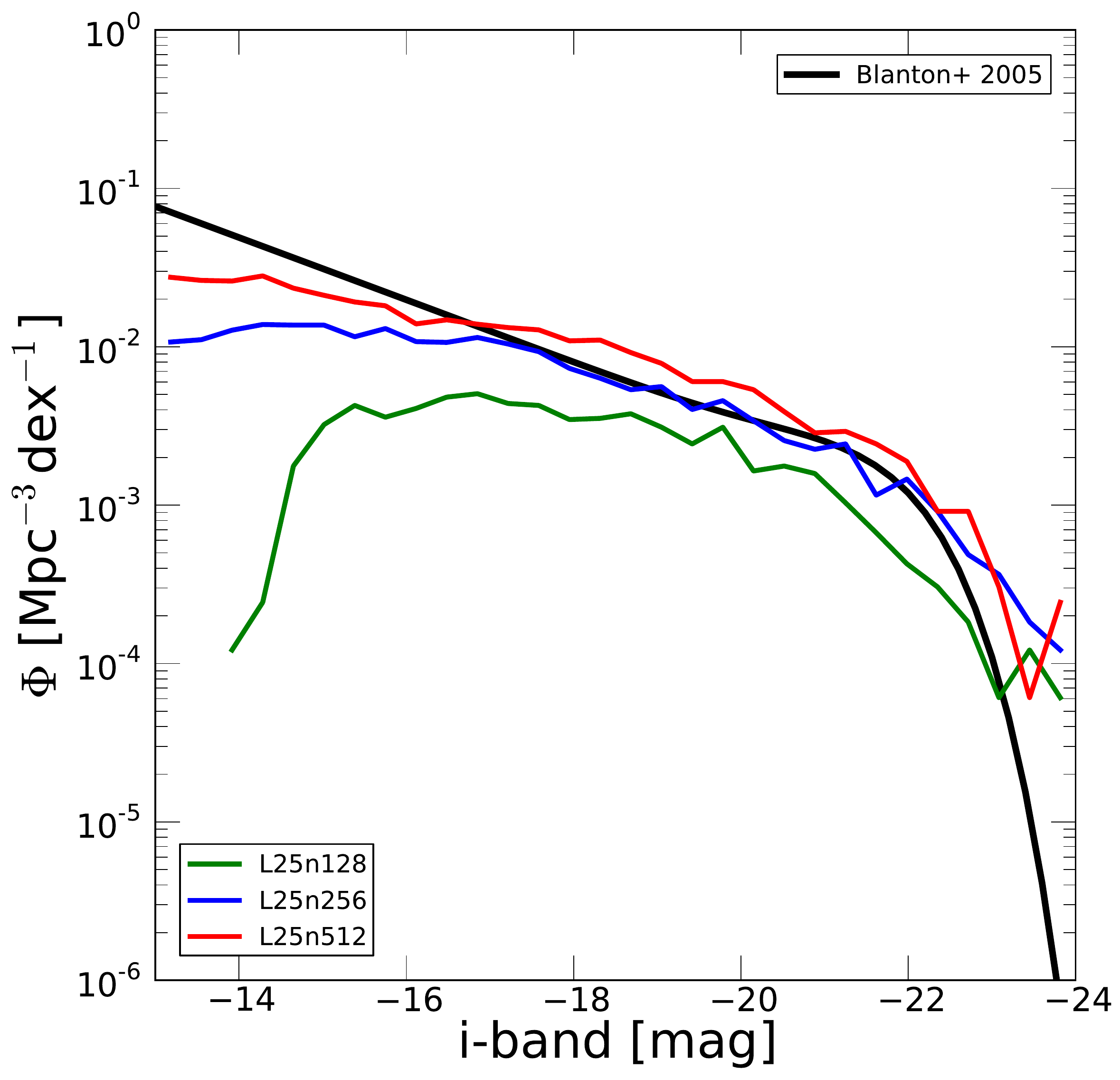}
\includegraphics[width=0.475\textwidth]{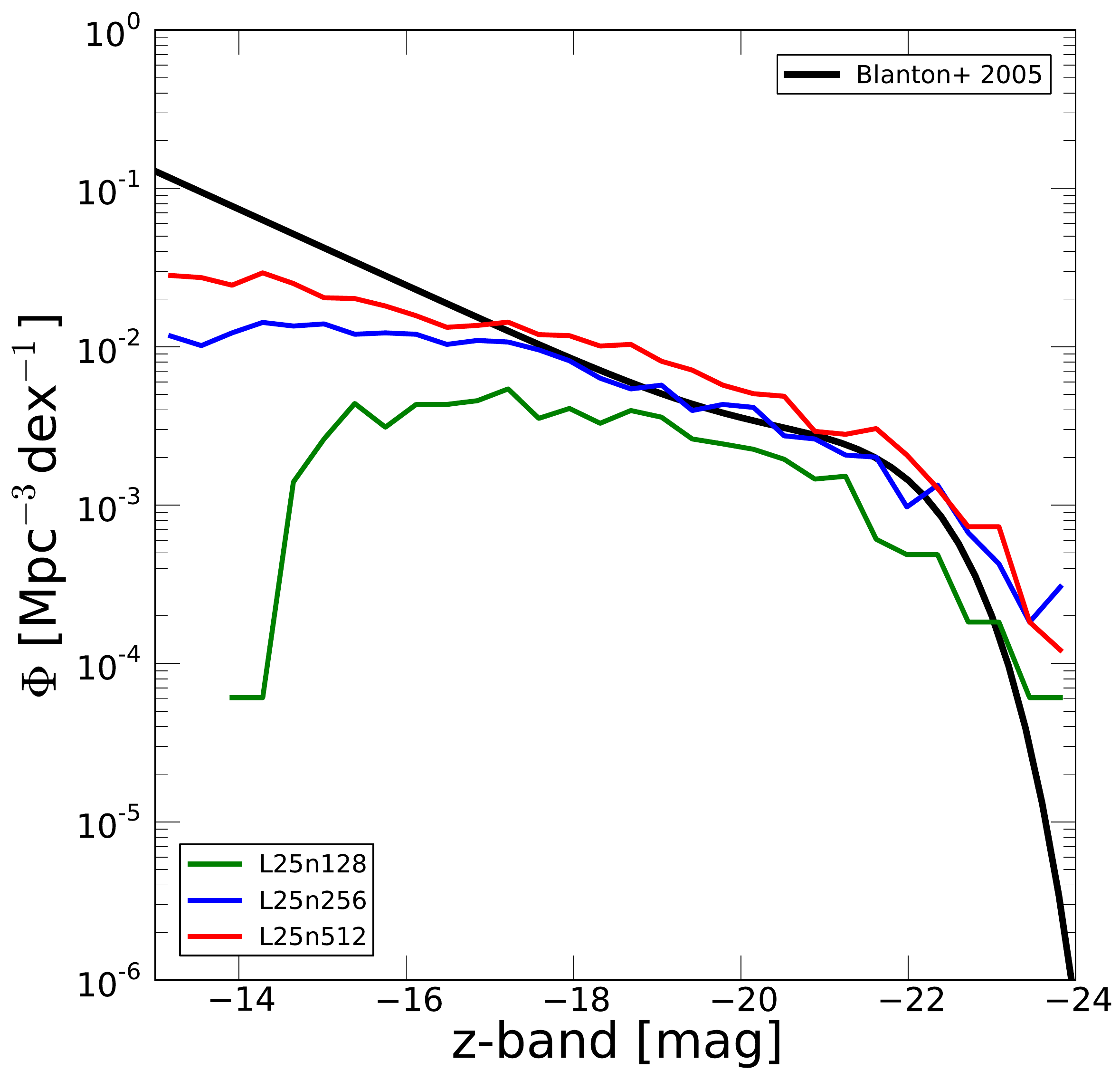}
\caption{SDSS-band (g, r, i, z) luminosity functions for our simulated
  galaxy population (resolution study). We compare to double
  Schechter-fits of the observational data, taken
  from~\protect\cite{Blanton2005}. We find a reasonable agreement of
  L25n256 and L25n512 with these fits. The faint-end slope agrees very
  well with the slope inferred from observational data, although the
  highest resolution simulations have a slightly too high
  normalisation.  Also, the exponential suppression is reasonably well
  reproduced but starts at slightly too bright galaxies. This
  discrepancy is largest for the g-band luminosity function.}
\label{fig:cosmo_luminosity_function_0}
\end{figure*}

We note that this agreement has become possible only through the
independent treatment of mass and metal loading of stellar winds. As
described above, we have adopted a wind metal loading factor
($\gamma_\mathrm{w}$), which determines the metallicity of the wind
material relative to the local ISM metallicity from where the wind is
launched.  Without such a scheme our galaxies would retain too few
metals due to the large wind mass loadings required to match the
observed stellar masses of lower mass systems.  This creates an
interesting tension between the need for high wind efficiencies to
reduce the buildup of stellar mass in low mass galaxies, and the need
for low mass galaxies to be relatively efficient at retaining their
metal content~\citep{Zahid2012}.  The solution to this tension likely
either lies in: (i) fundamentally reducing the accretion efficiency of
low mass galaxies via some feedback mechanism that is able to disrupt
and heat the IGM around these systems, or (ii) in a more detailed wind
model that self-consistently handles the entrainment of low
metallicity material as winds propagate out of
galaxies~\citep[e.g.,][]{Hopkins2012}.

\begin{figure*}
\centering
\includegraphics[width=0.475\textwidth]{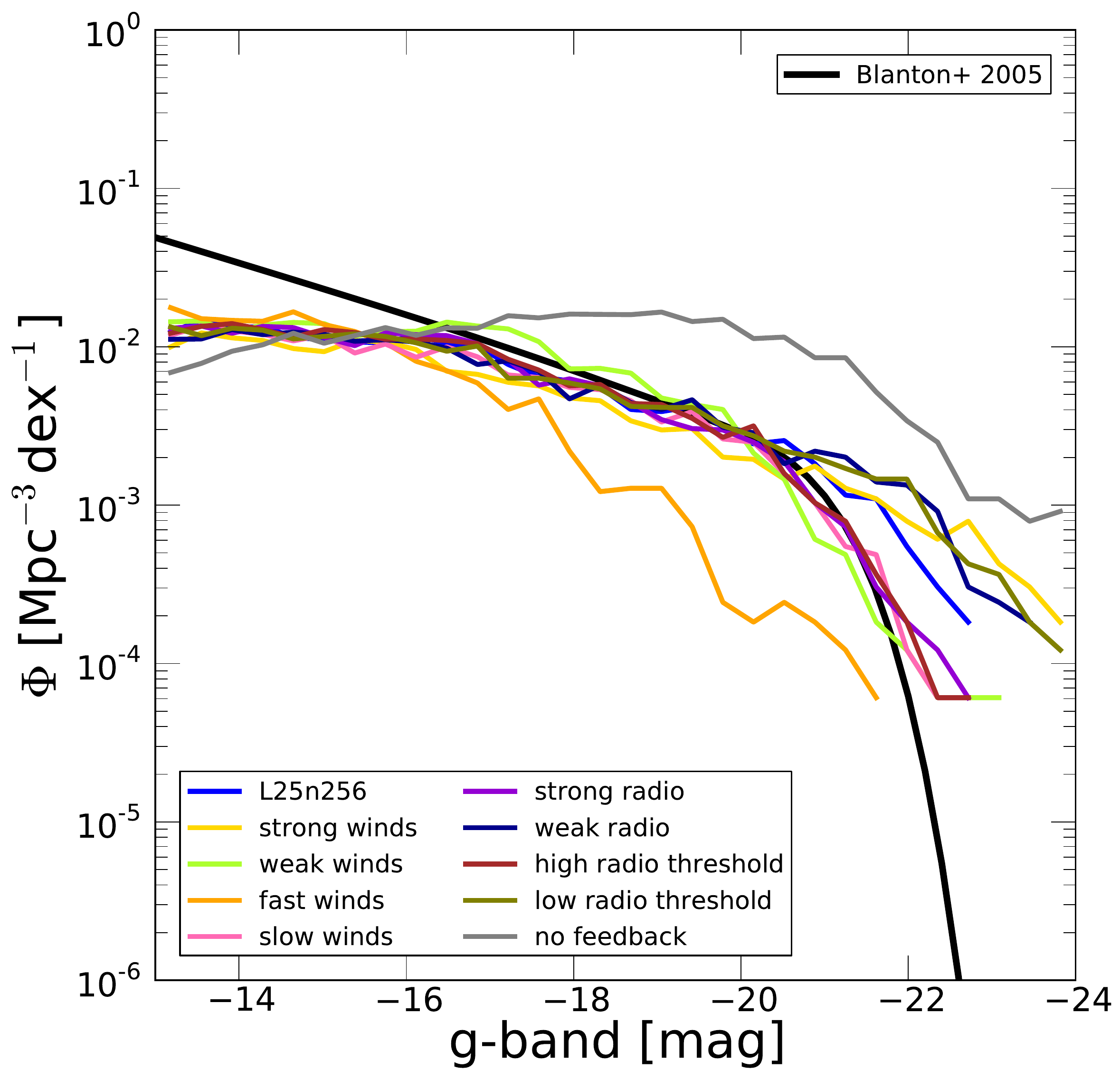}
\includegraphics[width=0.475\textwidth]{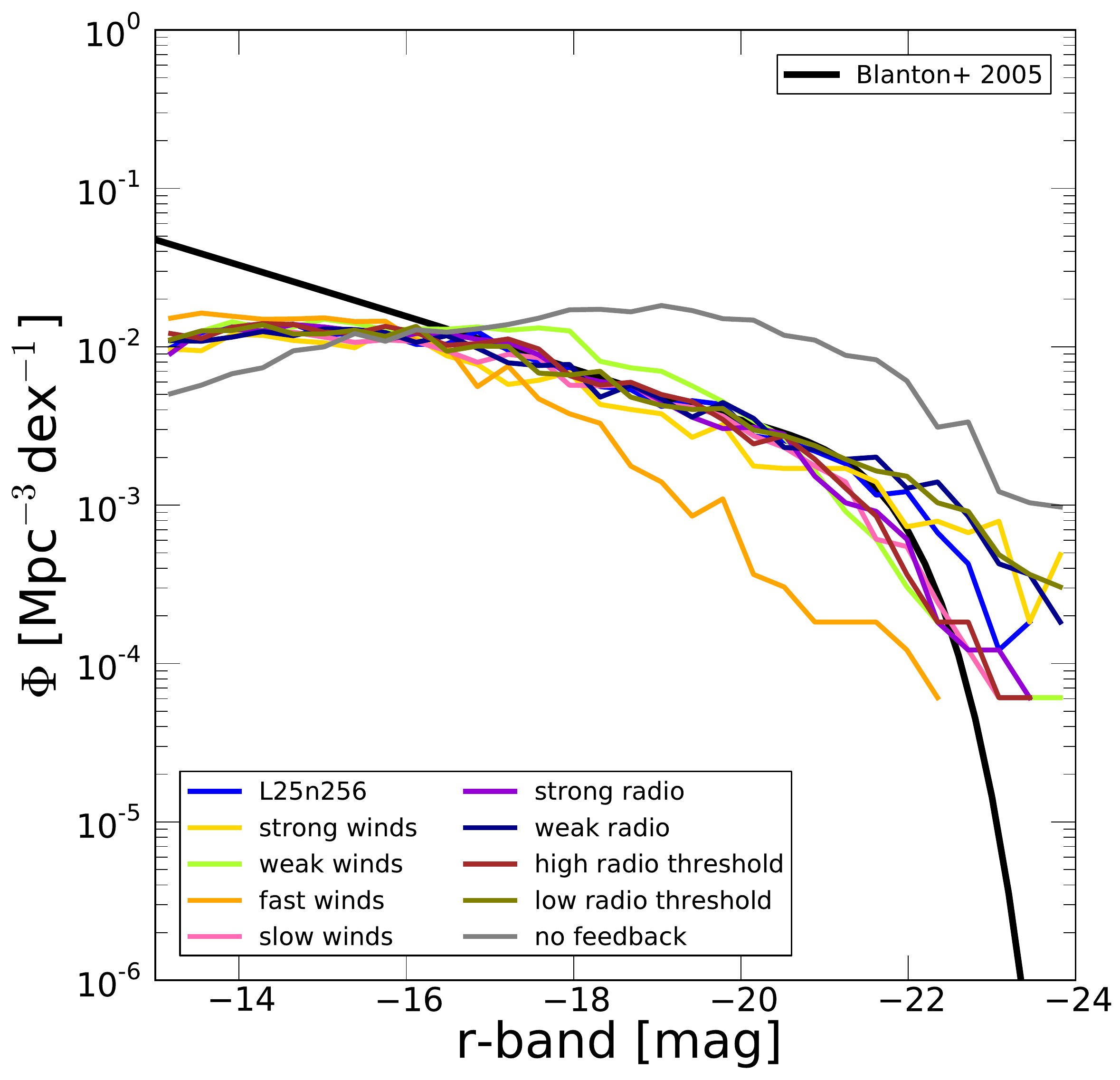}
\includegraphics[width=0.475\textwidth]{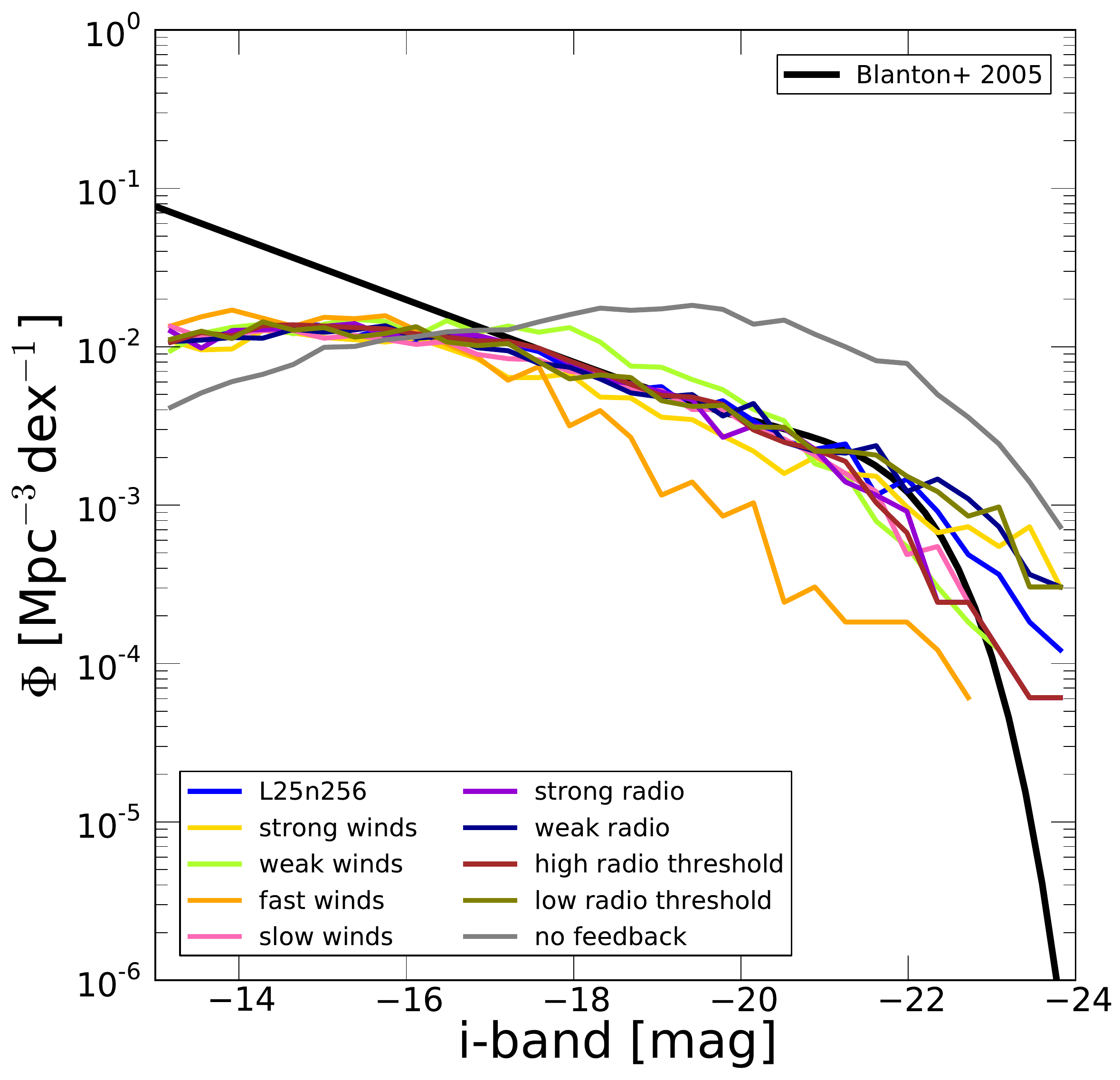}
\includegraphics[width=0.475\textwidth]{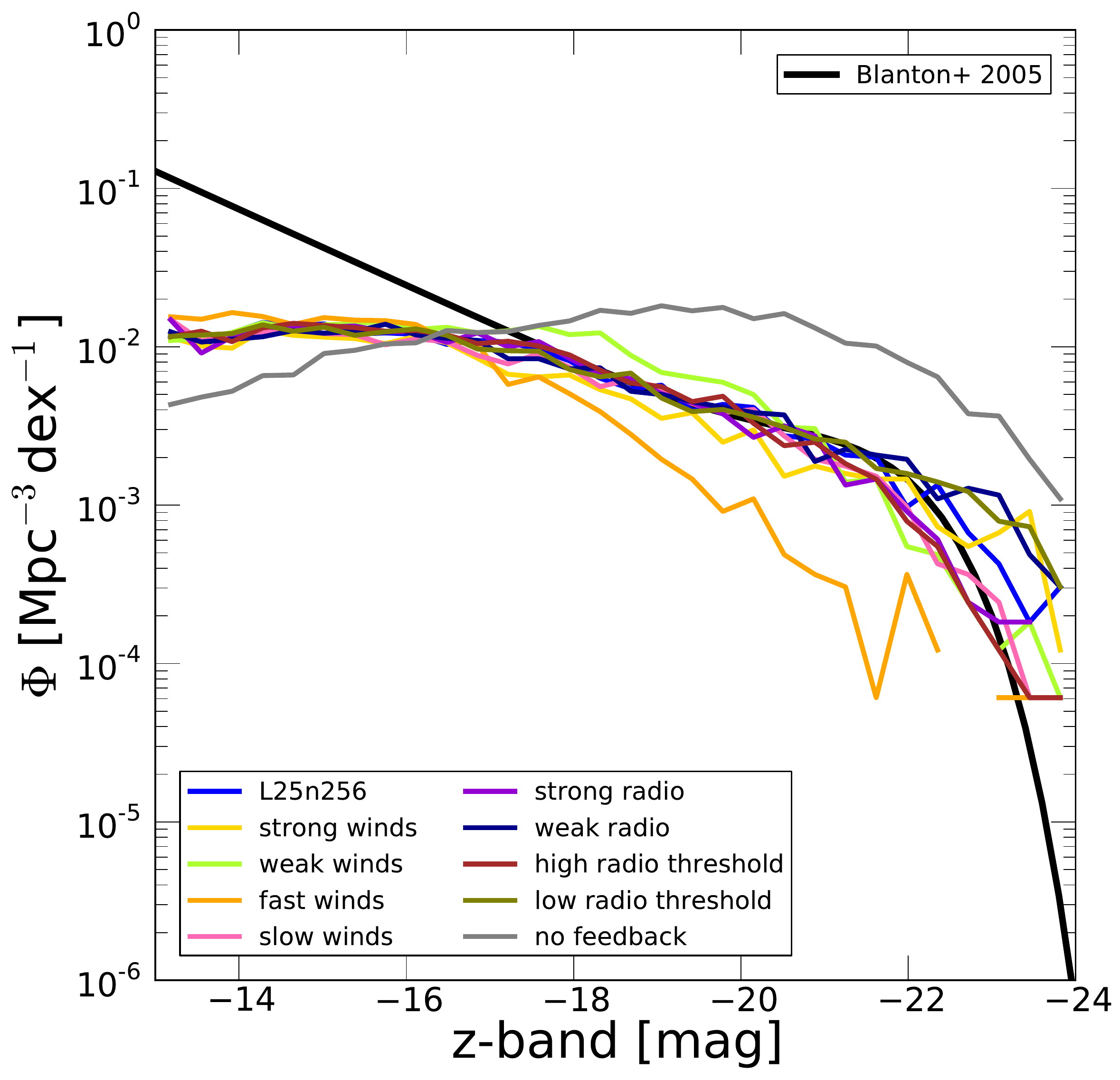}
\caption{SDSS-band luminosity (g, r, i, z) functions for our simulated
  galaxy population (different feedback models). We compare to
  observational double Schechter-fits taken
  from~\protect\cite{Blanton2005}. Our fiducial model reproduces the
  observational fit reasonably well. Models with faster winds clearly
  give the poorest agreement.  Weakening the radio-mode AGN feedback
  leads to a significant overshoot of the luminosity function at the
  bright end. ``Weaker winds'' lead to an overproduction of faint
  galaxies. The ``no feedback'' simulation clearly fails dramatically
  in reproducing the observed luminosity function. Interestingly, a
  model with a higher radio-mode threshold leads to a better agreement
  for the bright end in all bands.}
\label{fig:cosmo_luminosity_function_1}
\end{figure*}

An interpretation of the differences between the simulated mass
metallicity relation and the observed relation is guided by the right
panel of Figure~\ref{fig:cosmo_stellar_mass__vs__oxygen_over_H}.  We
find that the simulated mass metallicity relation is quite sensitive
to the adopted feedback physics parameters.  The reason for this
behaviour is that the level of central galactic metallicity is easily
reduced if we encourage mixing between the central galaxy gas and the
halo gas, which is mediated by AGN or stellar feedback. Our winds,
which by construction transport highly enriched disk gas into the
halo, are very efficient at mixing metals.  Thus, when we use a ``weaker
wind'' prescription we find that the average metal content of the
central dense galaxy gas goes up.  On the other hand, if we employ
``stronger winds'' we find that the average metal content goes down.
The need to regulate the growth of low mass galaxies successfully
while at the same time avoiding overly depleting galaxies of their
metal content prompted us to implement a specific wind metal loading,
as described above. We note that variations in the AGN radio-mode
alone are not able to remove the metallicity tension at the high mass
end. Most of the simulations overshoot the amount of metals in massive
galaxies by $0.3-0.5\,\mathrm{dex}$.

\begin{figure*}
\centering
\includegraphics[width=0.475\textwidth]{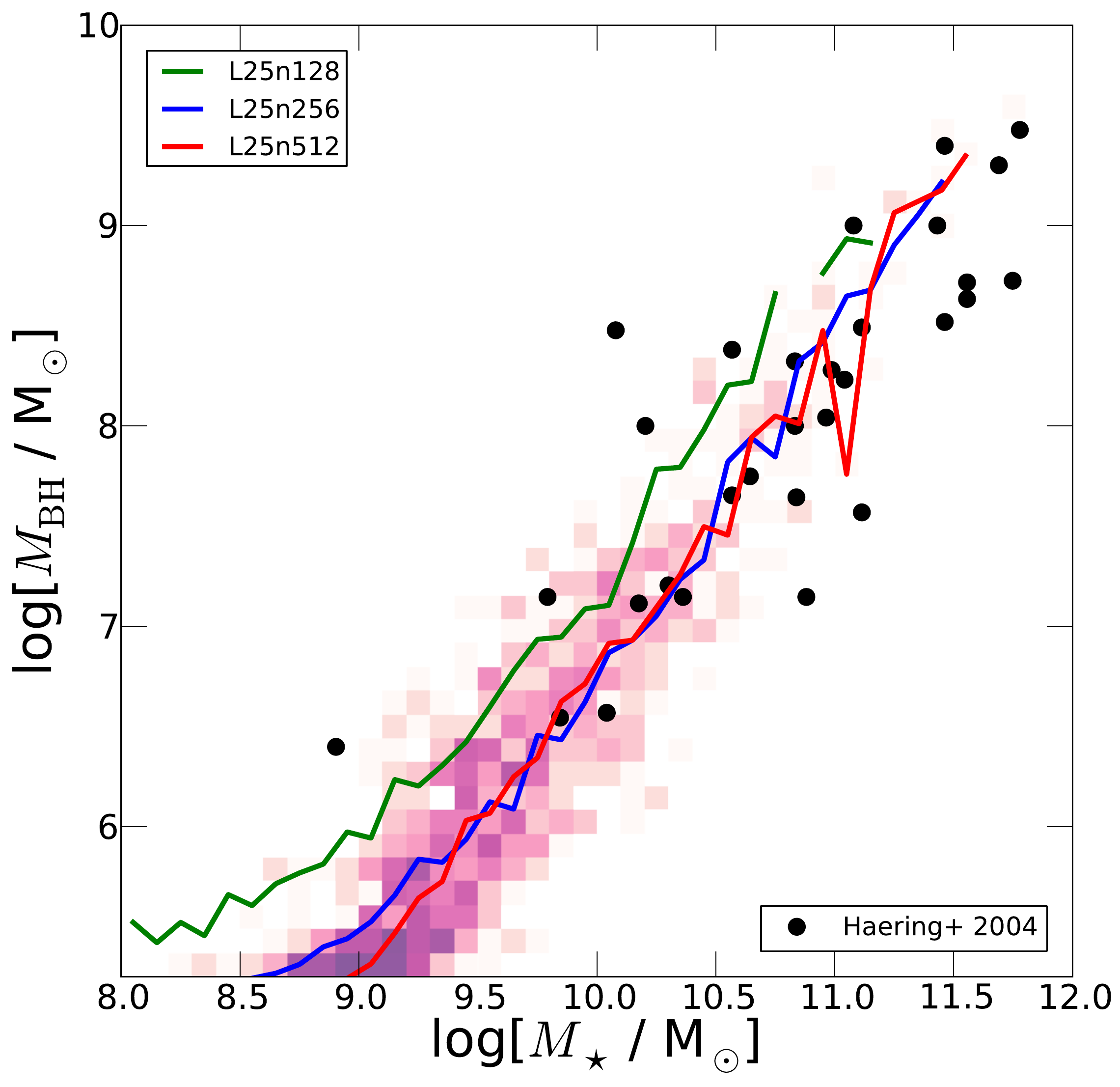}
\includegraphics[width=0.475\textwidth]{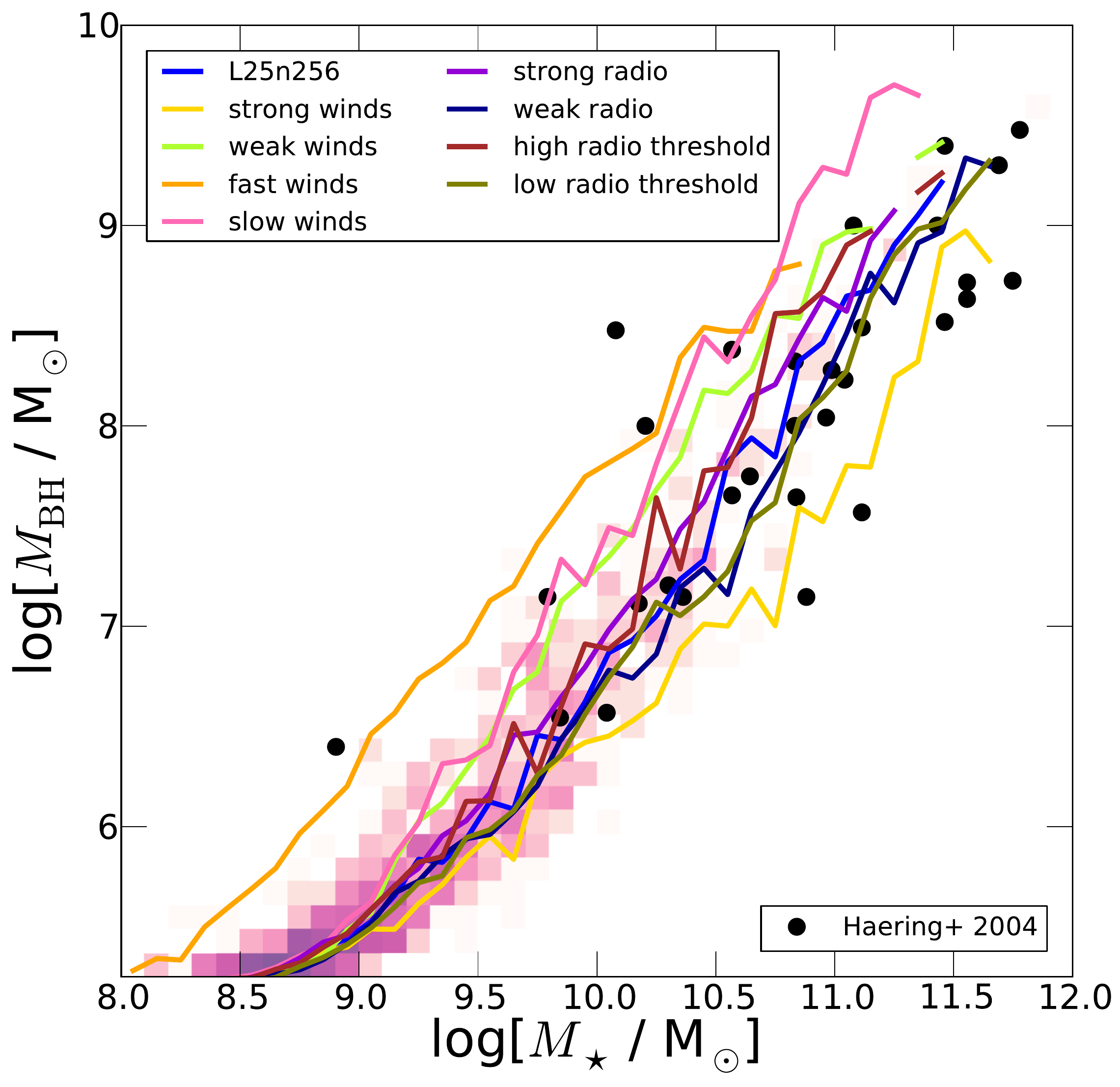}
\caption{Black hole mass -- stellar mass relation (left panel:
  resolution study; right panel: different feedback
  models). Observational data is taken
  from~\protect\cite{Haering2004}. The BH mass -- stellar mass
  relation is well-converged and agrees with the observations both in
  slope and in normalisation.  Also the amount of scatter is in
  reasonable agreement with the observational data.  Normalisation
  changes in this relation are mainly due to changes in the stellar
  mass caused by the different feedback choices.  For example, ``fast
  winds'' strongly reduce the amount of stars forming in haloes, as
  demonstrated above.  We note that the right panel does not include
  the ``no feedback'' simulation since this run did not include any
  BHs.}
\label{fig:stellar_mass__vs__BH_mass}
\end{figure*}

\subsubsection{Luminosity function}\label{sec:luminosity_function}

We can use the stellar population synthesis models described above to
derive luminosities of individual galaxies in various bands.  We here
confront the resulting galaxy luminosity functions with SDSS
observations.  We note that we neglect the impact of dust attenuation,
and simply define the luminosity of each galaxy as the sum of the luminosities of all its
star particles. In Figure~\ref{fig:cosmo_luminosity_function_0} we
show the g-, r-, i-, and z-band luminosities (other bands are compared to observations in \cite{Torrey2013}; there we also include a crude model for dust attenuation) and compare them to
double Schechter-fits
\begin{align}
\Phi(M) &=  0.4 \ln 10 \dint M  \exp\left(-10^{-0.4(M-M_{\ast})}\right) \\
        & \left[\phi_{\ast,1}  10^{-0.4 \left( M-M_{\ast} \right)(\alpha_1+1)} \!+\!  \phi_{\ast,2}  10^{-0.4 \left( M-M_{\ast} \right)(\alpha_2+1)} \right], \nonumber
\end{align}
from SDSS, where we take the fitting parameters $M_{\ast}$,
$\phi_{\ast,1}$, $\phi_{\ast,2}$, $\alpha_1$, and $\alpha_2$
from~\cite{Blanton2005}. We find a reasonable agreement of L25n256 and
L25n512 with these fits.  Especially the faint-end slope agrees well
with the Schechter-fit slope inferred from observational data,
although the highest resolution simulation has a slightly too high
normalisation.  Also the exponential suppression at the bright end is
reasonably well reproduced, although for most bands the suppression
starts at slightly too bright galaxies. We will demonstrate in \cite{Torrey2013} that we also find good agreement of the simulated
B-band luminosity function compared to observations at various
redshifts.  Interestingly, the g-band luminosity function shows the
worst agreement with the observations at the bright end. For all other
bands the exponential drop is significantly better
reproduced. Furthermore, we find that the simulated luminosity
function is well converged even at the faint end. We caution however
that this comparison does not account for dust attenuation.

We compare the luminosity functions of the different feedback models
to the observational fits in
Figure~\ref{fig:cosmo_luminosity_function_1}. Our fiducial L25n256
model gives a good fit to the observed luminosity function, whereas
a model with ``faster winds'' clearly gives the worst agreement.
Weakening the radio-mode AGN feedback leads to a significant overshoot
of the luminosity function at the bright end. ``Weaker winds'' cause 
an overproduction of faint galaxies. Finally, the ``no feedback''
clearly fails dramatically in reproducing the observed luminosity
function, as it did for the stellar mass function. Interestingly, the
``higher radio threshold'' simulation produces luminosity functions
which are in better agreement with the observations than our fiducial
model. This is true for all bands, where the exponential drop at the
bright end is systematically in better agreement with the observational
data for the ``higher radio threshold'' simulation. Also the model with
``stronger radio-mode'' feedback leads to a better agreement of the
luminosity functions in all bands for the bright end. However,
although these two models agree better with the observed luminosity
functions, their stellar mass functions do not agree as well with the
observations as our model with the fiducial feedback settings.

\subsubsection{Black hole mass -- stellar mass relation}\label{sec:BH_mass_rel}

The BH mass growth is mainly regulated by quasar-mode feedback which together
with the BH accretion yields a tightly self-regulated feedback loop. As a
result, a BH mass -- stellar mass relation is expected to be produced through
these two processes \citep[see also][]{Springel2005, DiMatteo2005}. Here we
compare the BH mass -- stellar mass relation to observational data
of~\cite{Haering2004}. As the left panel
Figure~\ref{fig:stellar_mass__vs__BH_mass} demonstrates, the implemented
quasar-mode model leads to the correct growth of BHs in our simulation. Most
important, we reproduce the correct slope of the relation. This demonstrates
that the modifications of the original \cite{Springel2005} model described
above, do not affect the BH mass -- stellar mass relation in any significant
way. Reproducing this relation is also crucial to inject the correct amount of
radio-mode AGN feedback generated by these BHs and to create the appropriate
radiation field used for the radiative (EM) AGN feedback.  We note that we do
not plot the stellar bulge mass in Figure~\ref{fig:stellar_mass__vs__BH_mass}
but rather the stellar mass within twice the stellar half mass radius, as
discussed above. Correcting for the bulge mass does however not affect the
conclusions of this subsection. We will in forthcoming work study this relation
in more detail using extracted bulge masses (Sijacki et al, in prep).

The left panel of Figure~\ref{fig:stellar_mass__vs__BH_mass} also shows
that our BH mass -- stellar mass relation is well converged and agrees
with the observations both in slope and in normalisation. Also, the
amount of scatter is in reasonable agreement with the observational
data. The right panel of Figure~\ref{fig:stellar_mass__vs__BH_mass}
demonstrates that the slope of this relation does not change
significantly if we alter the stellar or radio-mode AGN feedback. We
note that this plot does not include the ``no feedback'' simulation
since this run did not include BHs due to the absence of any BH growth regulating
feedback mechanism.  Normalisation offsets in this relation are mainly
due to the changes in the stellar masses caused by the different
feedback choices.  For example, ``faster winds'' strongly reduce the
amount of stars forming in haloes. This implies that the low mass
stellar systems end up with too massive BHs at their centres, as can
be seen in the right panel. Overall we find that our fiducial model
reproduces the BH mass -- stellar mass relation reasonably well.

\subsection{Disentangling the impact of different physical
processes}\label{sec:disent}

So far we have explored feedback variations around our fiducial model, varying
the strength of stellar and AGN feedback. In this subsection, we briefly
discuss the impact of the different physical effects on the cosmic SFR density,
the stellar mass function, the stellar mass -- halo mass relation, and the
stellar mass density.  Most of the simulations presented here are unrealistic
in the sense that they deliberately ignore important processes required to
shape the galaxy population.  But they are useful for demonstrating the impact
of these physical processes and how they influence the galaxy population. In
particular, we discuss the importance of our newly implemented radiative AGN
feedback scheme and demonstrate the impact of our wind metal loading scheme.

It is instructive to consider a simulation set with increasing model
complexity, starting from a physics setup close to the simulations presented
in~\cite{Vogelsberger2012} and ending at the full physics implementation
discussed above.  The different simulations of this series are summarised in
Table~\ref{table:disent_sims}. The ``plain'' simulation differs only in terms
of the IMF (Chabrier instead of Salpeter), a softer eEOS ($q=0.3$ instead of
$q=1.0$) and self-shielding corrections from the simulations presented
in~\cite{Vogelsberger2012}. We then add more physical processes on top of this
plain setup, where the parameters of the individual physics modules are chosen
according to our fiducial setup discussed above (see
Table~\ref{table:fiducial}). We start with the simulation MeGa, which includes
on top of the ``plain'' setup metal-line cooling and gas recycling. This simulation
is identical to the ``no feedback'' simulation presented in the previous section.  The last
simulation listed (MeGaWiMlQuEmRa) includes all physical effects discussed
above and is identical to the fiducial L25n256 simulation.
Table~\ref{table:disent_sims} also includes a simulation which is similar
to L25n256; i.e. our full fiducial model, but without stellar feedback
(NoWindFeed). This simulation is
useful for studying the detailed impact of stellar feedback, but still
considering all the other physical processes included in our model. We
note that MeGaWiMl explores the situation with full stellar feedback, but
without AGN feedback.

\begin{table}
\centering
\begin{tabular}{ll}
\hline
name                            & physics\\
\hline
\hline
plain                           & same as in~\cite{Vogelsberger2012}\\
                                & (except for IMF, softer eEOS, self-shielding)\\
MeGa                            & + met. line cool., gas recycl. = ``no feedback''\\
MeGaWi                          & + stellar winds\\
MeGaWiMl                        & + separate metal mass loading of winds\\
MeGaWiMlQu                      & + quasar-mode AGN feedback\\
MeGaWiMlQuEm                    & + electro-magnetic AGN feedback\\
MeGaWiMlQuEmRa                  & + radio-mode AGN = L25n256\\
\hline
NoWindFeed                      & L25n256 without stellar feedback\\
\hline
\end{tabular}
\caption{Summary of simulation series with increasing physics
  complexity, starting from a setup similar to the simulations
  presented in~\protect\cite{Vogelsberger2012} except for the IMF
  (Chabrier instead of Salpeter), the softer eEOS ($q=0.3$ instead of
  $q=1.0$) and self-shielding corrections for cooling (``plain''). The other
  simulations include additional physical processes as listed. The
  adopted parameters for these processes are the same as those of our
  fiducial model (see Table~\ref{table:fiducial}). The MeGaWiMlQuEmRa run includes
  the same physics as the L25n256 simulation presented above. We also performed one other
  simulation which is identical to L25n256 except for turning off the stellar feedback (NoWindFeed).}
\label{table:disent_sims}
\end{table}

\begin{figure*}
\centering
\includegraphics[width=0.475\textwidth]{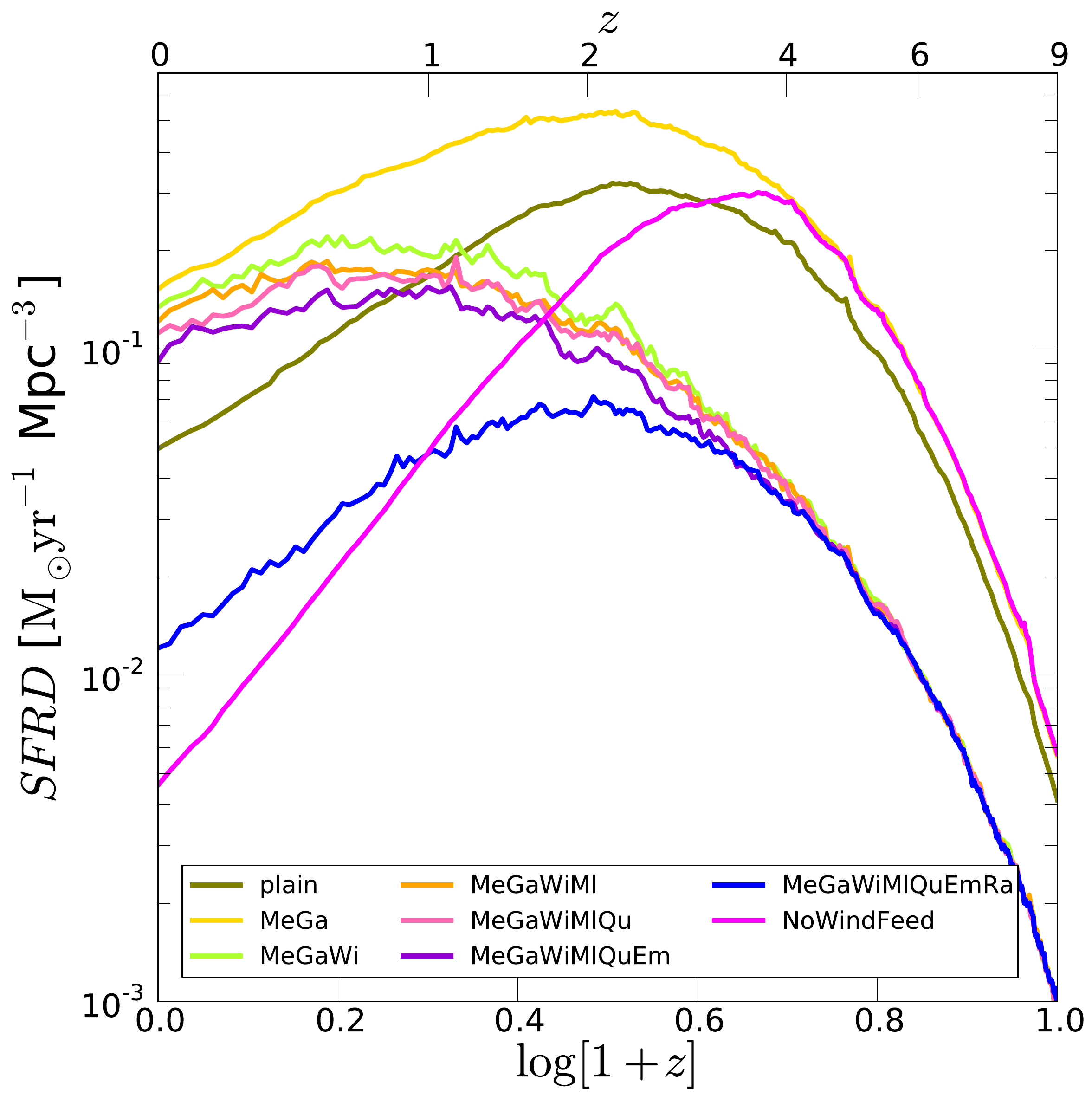}
\includegraphics[width=0.475\textwidth]{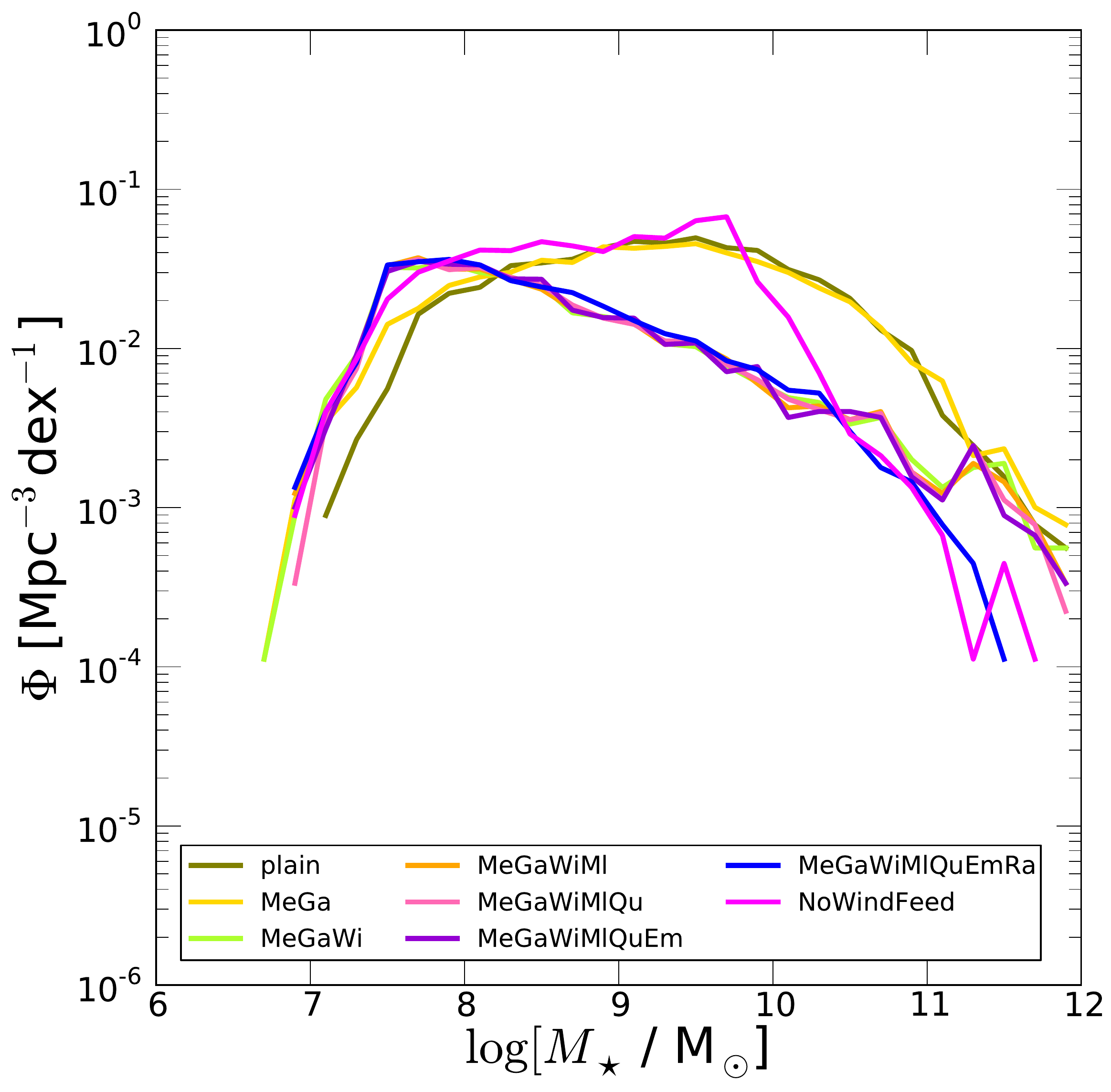}
\includegraphics[width=0.475\textwidth]{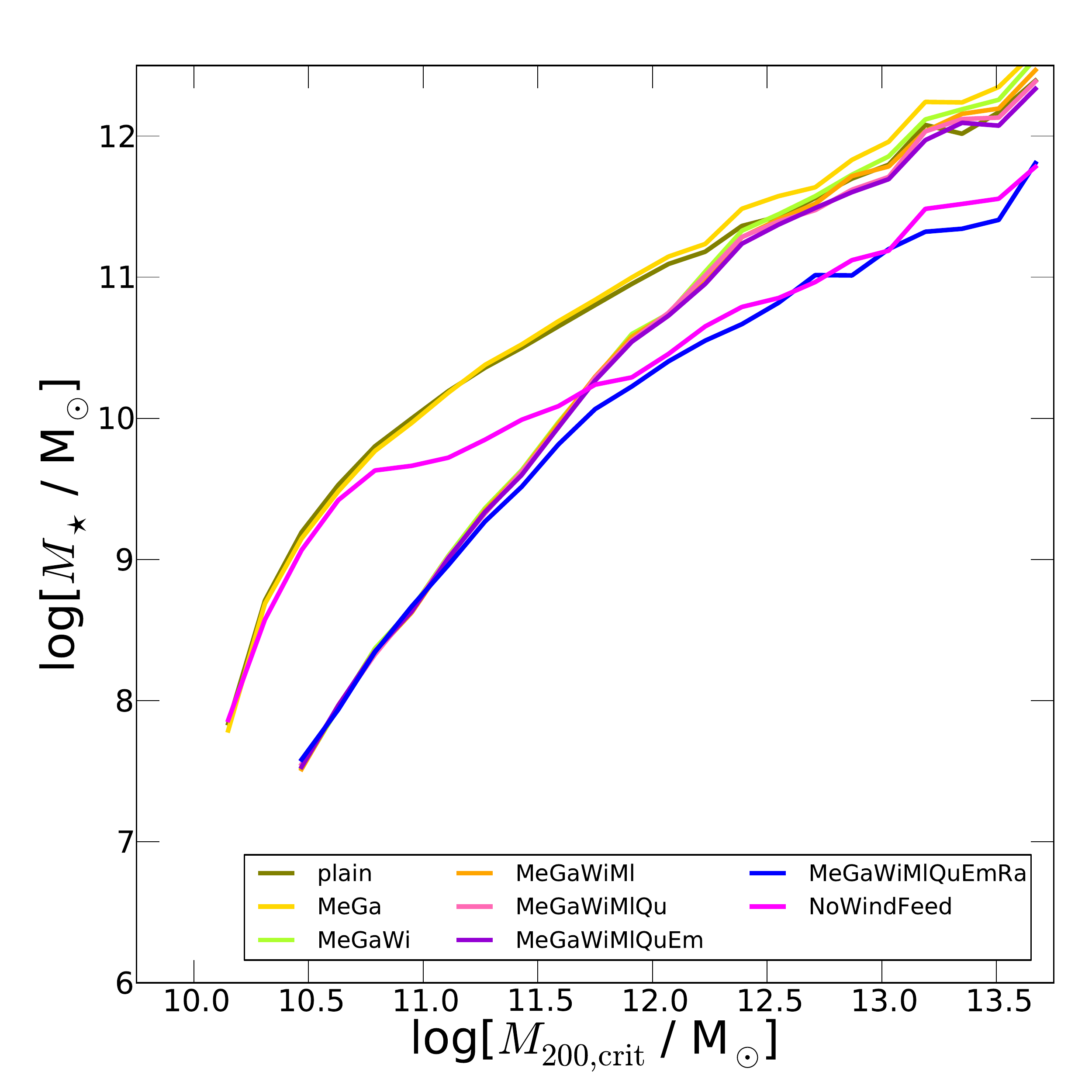}
\includegraphics[width=0.475\textwidth]{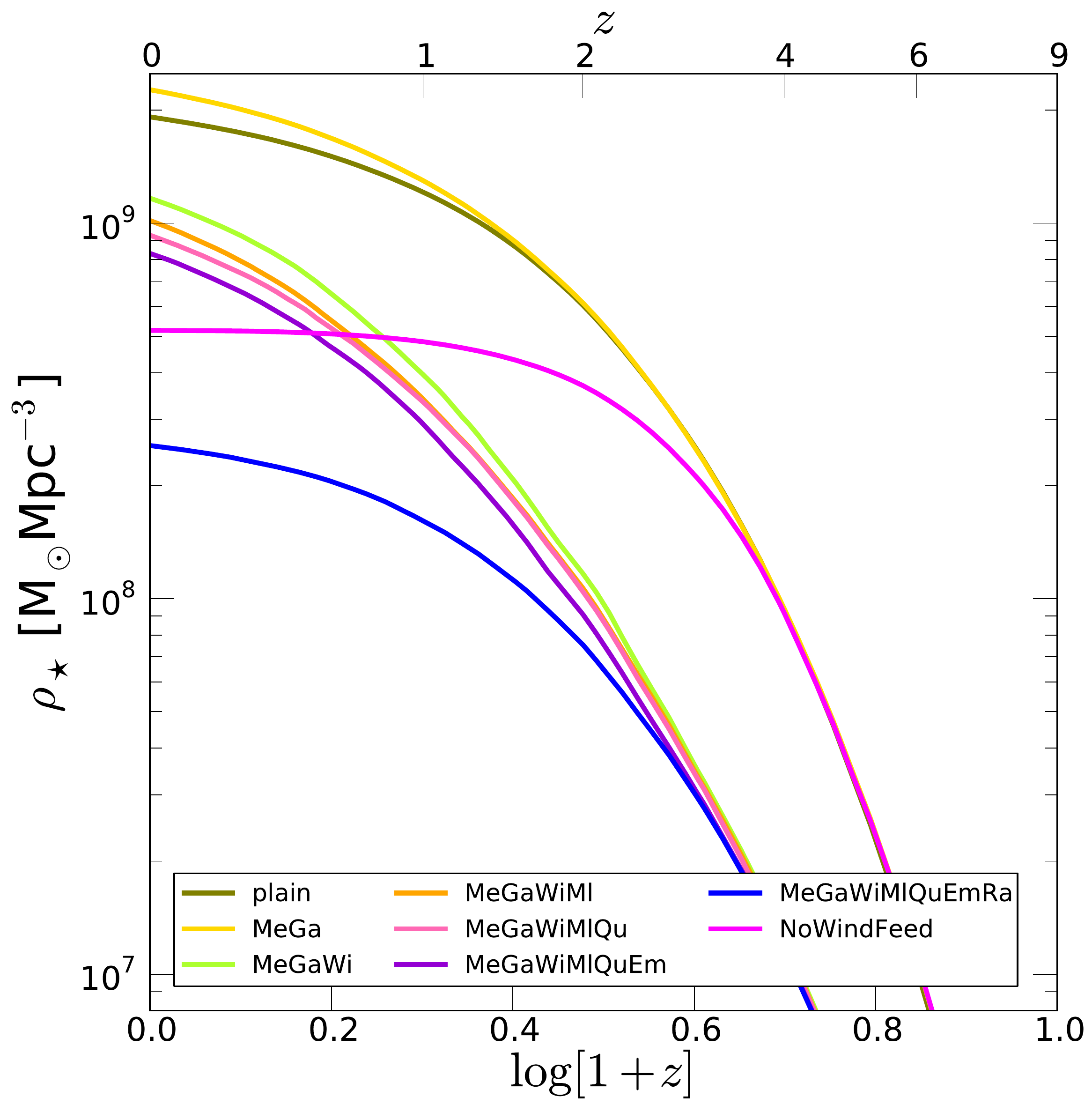}
\caption{Cosmic SFR density (upper left panel), stellar mass function (upper
right panel), stellar mass -- halo mass relation (lower left panel) and stellar
mass density (lower right panel)  of simulations with increasing modelling
complexity (see Table~\ref{table:disent_sims}). The ``plain'' setup does not
include any explicit stellar and AGN feedback. The MeGaWiMlQuEmRa simulation
includes all physical processes discussed above with parameters set as for our
fiducial model (see Table~\ref{table:fiducial}). Although most of these
simulations are unrealistic since they miss important physical ingredients, they
are helpful in disentangling the impact of various physical processes on the
galaxy population. Simulations including stellar winds, but no radio-mode AGN
feedback, lead to a significant overproduction of stars at late times. This can only
be regulated by strong radio-mode AGN feedback. Quasar-mode feedback alone has
essentially no effect on the star formation (see MeGaWiMl vs. MeGaWiMlQu).
However, radiative (electro-magnetic) AGN feedback leads to a minor reduction
of the star formation rate. Changing the metal loading of stellar winds (MeGaWi
vs. MeGaWiMl) has about the same effect on the star formation rate. As expected,
stellar feedback regulates star formation in low mass systems, whereas the stellar
content of higher mass haloes is mainly set through radio-mode AGN feedback.}
\label{fig:disent}
\end{figure*}

The upper left panel of Figure~\ref{fig:disent} shows the cosmic SFR density for the
different simulations, whereas the upper right panel presents the stellar mass
function.  We do not include observational data in these plots for graphical
clarity and because a detailed comparison to the observational data was already
presented above. Going from the ``plain'' model to the model which includes in
addition metal-line cooling and gas recycling (MeGa=``no feedback'') strongly increases the
star formation rates since net cooling rates are increased and recycled gas can
form additional stars. However, such a model still correctly produces a star
formation rate peak although at a slightly different redshift compared to the
``plain'' and full MeGaWiMlQuEmRa setup.  Including stellar winds (MeGaWi)
regulates this behaviour at early times by strongly suppressing the star
formation in low mass systems. This implies that a significant amount of gas
is not turned into stars at high redshifts in this simulation.  This gas is
then available for star formation at late times, which results in a rising SFR
density towards lower redshift. Metal line cooling and gas recycling are
so efficient then that the SFR density almost does not decline.
In fact,  a substantial decline can  be achieved to a sufficient
degree only by including radio-mode AGN feedback as demonstrated by the
MeGaWiMlQuEmRa curve.  

The MeGaWiMl simulation employs our new metal loading
scheme for outflows.  Reducing the metal loading of winds from
$\gamma_\mathrm{w}=1$ (MeGaWi) to our fiducial $\gamma_\mathrm{w}=0.4$
(MeGaWiMl)  decreases the star formation rate slightly towards lower redshifts.
This is because less enriched winds lead to less enriched gaseous haloes. As a
result, these haloes then experience less metal-line cooling yielding a lower
gas supply to central galaxies such that the star formation rate decreases.
Although this effect is clearly visible, we note however that it does not
represent a substantial change in the star formation rates, which is clearly
largely shaped by feedback processes. As we discussed above, such a metal
loading scheme is required to retain enough metals in galaxies and to obtain a
reasonable match to the galaxy mass--metallicity relation.  In fact,
$\gamma_\mathrm{w}=1$ would lead to a far too small normalisation of the
mass--metallicity relation.  

Comparing the MeGaWiMl simulation with a run that
additionally includes quasar-mode AGN feedback (MeGaWiMlQu) demonstrates that
continuous thermal AGN feedback has no significant effect on the global star
formation rates.  The MeGaWiMlQu simulation differs from the MeGaWiMlQuEm only
by radiative electro-magnetic AGN feedback, which is included in the latter.
Most importantly, this causes quasar-mode feedback to have a slightly more
noticeable effect on the star formation rates; i.e. our new radiative feedback
is stronger than the continuous thermal AGN feedback of our quasar-mode
feedback implementation alone.  Radiative AGN feedback leads to a suppression
of atomic cooling and to an increase of the heating rates, so that star
formation is further suppressed by this feedback channel. This can be
seen in the cosmic SFR density, where we find a lower star formation rate once
quasars become active.  Although the effect is visible in the star formation
rates, it is not particularly strong compared to radio-mode AGN feedback, which
is implemented in the MeGaWiMlQuEmRa simulation . The reason is that
electro-magnetic AGN feedback is effective when
the accretion rate is near Eddington in the quasar regime, 
which typically happens only for short
periods of cosmic time for any given BH. This questions previous claims
suggesting that such radiative feedback is a highly important and very
efficient feedback channel compared to thermal and mechanical AGN
feedback~\citep[e.g.,][]{Gnedin2012}.  We stress however that we focus here
only on the effect of electro-magnetic feedback on the amount of stellar mass
formed, and not do analyse the impact of the AGN radiation field on the
surrounding IGM, for example.  

Finally, the MeGaWiMlQuEmRa simulation has an
equivalent setup as the L25n256 simulation presented above. It includes all its
processes and uses the same parameterisation as our fiducial model (see
Table~\ref{table:fiducial}). Clearly, radio-mode feedback leads to a strong
suppression of star formation at late times in massive haloes.  This induces a
steep decline of the cosmic SFR density and an exponential drop of the stellar
mass function towards massive haloes. The NoWindFeed run, which does not
include winds, but is otherwise identical to L25n256, does not show any star
formation suppression at early times such that the overall normalisation of the
SFR density is significantly too high.  We conclude
that late time star formation is largely shaped by metal line cooling, gas recycling
and radio-mode AGN feedback. The high-redshift SFR density, on the other hand, is essentially
only determined through stellar feedback. Metal line cooling and gas recycling have
essentially no effect in that regime. This is expected since not many stars formed
at higher redshifts such that no significant amount of metals is available and also
mass return of stars is not important at these early times.

The upper right panel of Figure~\ref{fig:disent} demonstrates the effect of our
physics implementation on the stellar mass function. Clearly, models without
any feedback (``plain'' and MeGa) lead to a substantial overproduction of stars. This
is also true for the simulation without any explicit stellar feedback (NoWindFeed). The
incorporation of winds is the main driver for the low mass slope of the stellar
mass function. The separate metal loading of these winds has essentially no
impact on the stellar mass function (MeGaWi vs.  MeGaWiMl). The bright end is
largely shaped by radio-mode AGN feedback, which starts to dominate around
$\mathrm{M_\star} \sim 10^{11} \msun$. Also radiative (electro-magnetic) AGN feedback has only a
rather minor impact on the massive end of the stellar mass function.

The lower left panel of Figure~\ref{fig:disent} shows the stellar mass -- halo
mass relation. For clarity we only show the median relations here. As expected,
simulations without stellar feedback (``plain'', MeGa, NoWindFeed) lead to a
large overproduction of stellar mass for lower and intermediate mass systems.
The addition of stellar winds leads to about an order of magnitude suppression
of stellar mass. The inclusion of quasar-mode and radiative (electro-magnetic)
AGN feedback does not significantly affect the massive end. Only the simulation
with radio-mode AGN feedback (MeGaWiMlQuEmRa) leads to a reduction of stellar
mass for higher mass systems, resulting in the correct turnover behaviour of
the stellar mass -- halo mass relation.

The lower right panel of Figure~\ref{fig:disent} shows the stellar mass density
as a function of redshift; i.e. the integrated version of the upper left panel.
Here we find a behaviour similar to what we found for the other three
diagnostics discussed so far. The lack of stellar feedback in the ``plain'',
MeGa and NoWindFeed simulations strongly overproduces the stellar mass density
already at high redshifts. Radio-mode AGN feedback on the other hand strongly
reduces the number of stars formed at later times (MeGaWiMlQuEmRa). All other
curves follow in a narrow range between these extremes.

We conclude that the stellar content of haloes is largely dominated by stellar
feedback and radio-mode AGN feedback. Other model details, like the independent
metal loading of winds or radiative (electro-magnetic) AGN feedback lead only
to very minor changes in the stellar mass, although they can have important
consequences for other observables, such as the galaxy mass-metallicity
relation in the case of wind metal-loading.

\section{Summary and conclusions}\label{sec:conclusions}

Matching the enormous amount of low redshift observational data is a great
challenge for any galaxy formation model that tries to explain the galaxy
population from first principles. Especially in the era of ``precision
cosmology'' based on large surveys like SDSS, LSST, etc., it is crucial to have
predictive, accurate and reliable galaxy formation models to test our
understanding of structure formation by confronting these models with a
plethora of data. The most general self-consistent way to study structure
formation is through hydrodynamical simulations which follow the coupled
evolution of dark matter, dark energy and baryonic physics. This task requires
two main ingredients: (i) a highly reliable and computationally efficient scheme
to solve the basic hydrodynamical and gravity equations, and (ii) a well-motivated
and numerically meaningful model for the required physics of galaxy formation. 

\begin{figure*}
\centering
\includegraphics[width=0.475\textwidth]{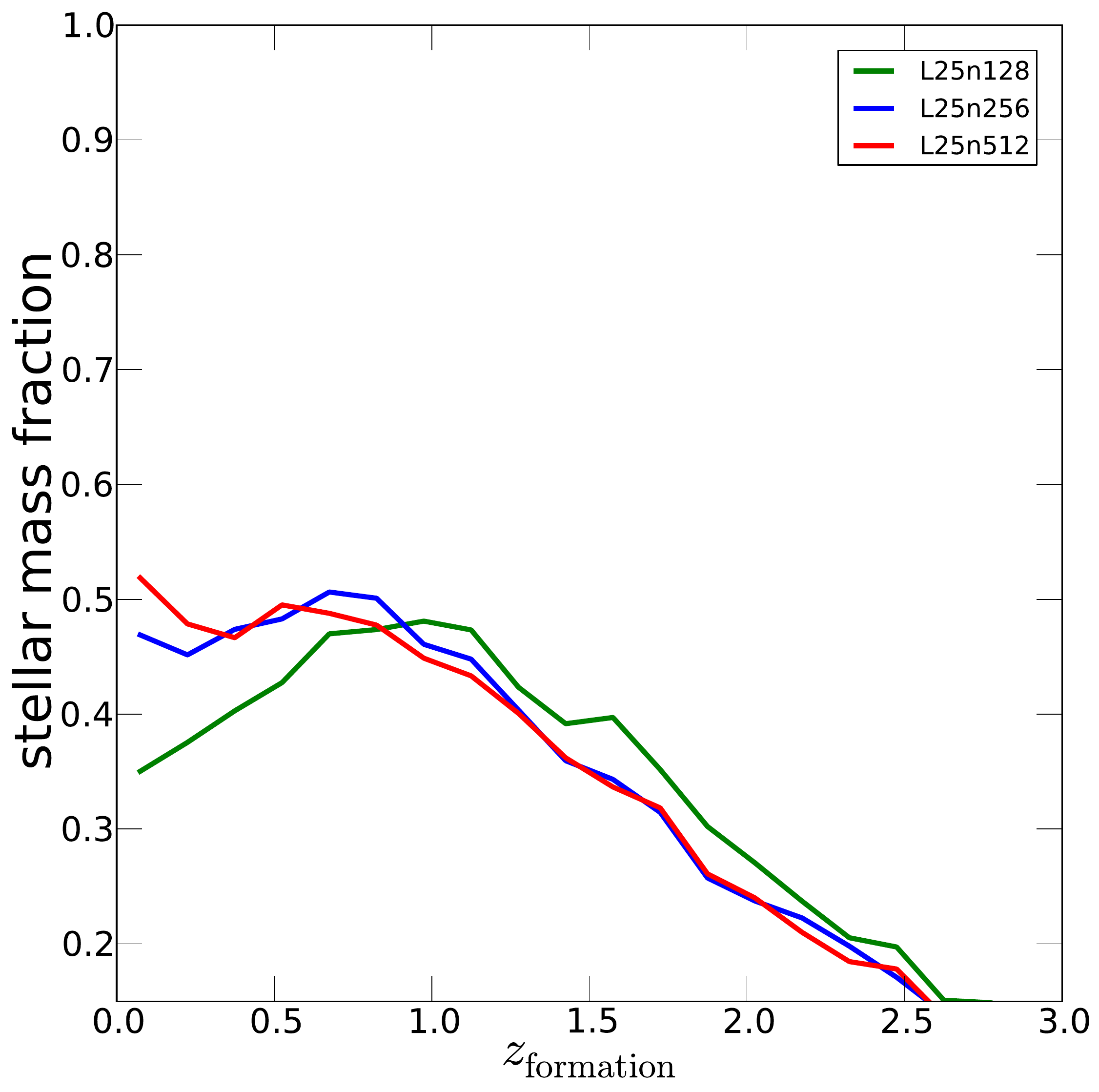}
\includegraphics[width=0.475\textwidth]{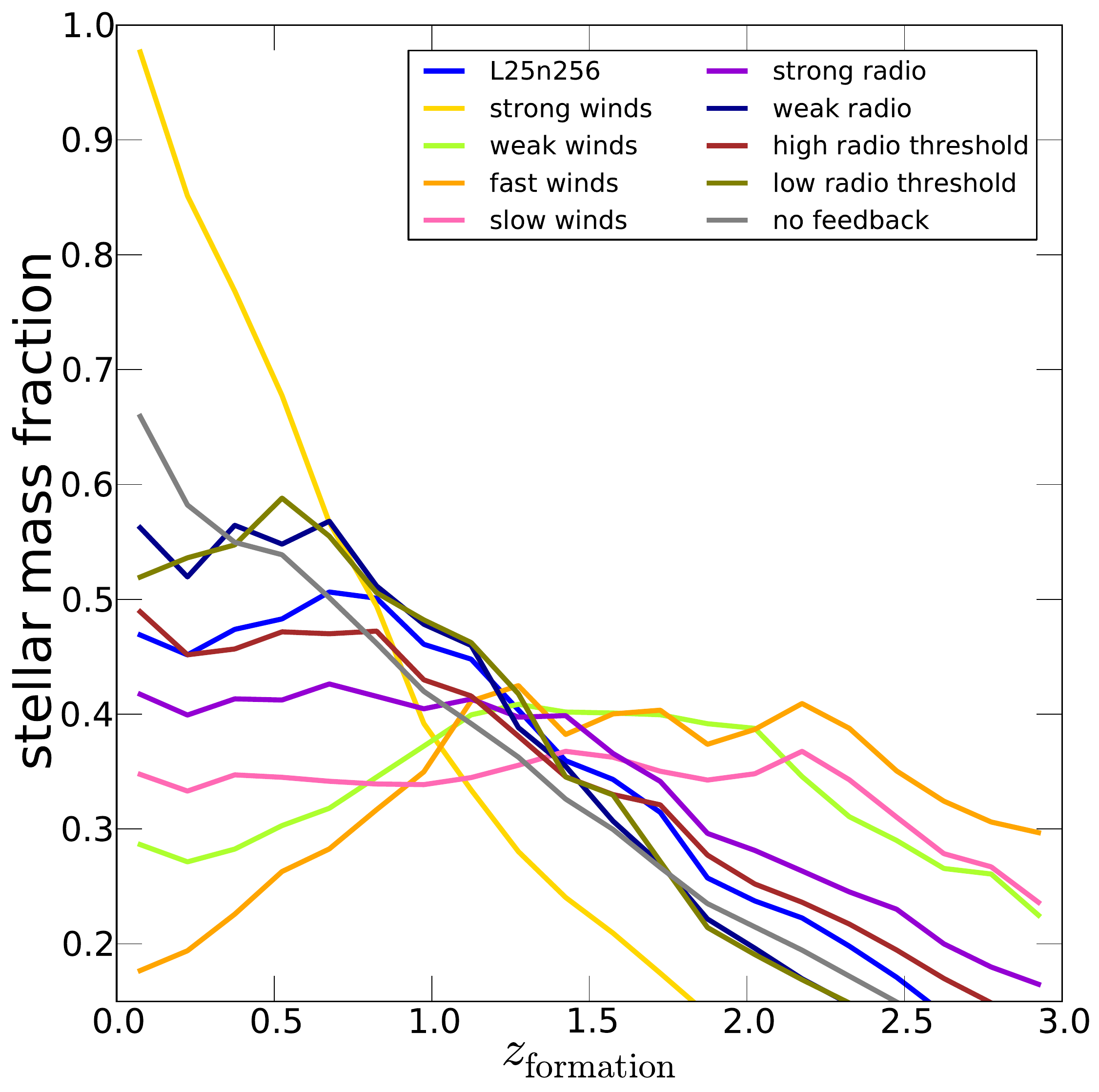}
\caption{Distribution of star formation times for all stars at $z=0$
  (left panel: resolution study; right panel: different feedback
  models). Stellar formation times are well converged. The feedback
  variations strongly affect when stars form in the simulation. The
  most extreme feedback models are the simulation with ``strong
  winds'', which leads to a great deal of late-time star formation, and the
  simulation with ``faster winds'', which yields almost no late-time
  star formation. All other feedback variations fall more or less
  between these two extreme setups.}
\label{fig:cosmo_stellar_age_dist}
\end{figure*}

We have recently demonstrated that a moving-mesh approach for solving the
underlying gas dynamics is very attractive for cosmic structure formation and
offers advantages compared to other numerical schemes for galaxy formation
simulations \citep[][]{Springel2010, Vogelsberger2012, Keres2012, Sijacki2012,
Torrey2012, Bauer2012, Nelson2013}. This approach hence addresses requirement
(i). The natural next step is to equip this new numerical scheme with the
relevant physical processes to fulfil requirement (ii), with the goal of 
carrying
out studies of galaxy formation in large volumes with higher accuracy and
fidelity to the physics than possible thus far.

Along these lines, we have presented in this paper the first galaxy formation
implementation for the moving-mesh code {\sm AREPO} specifically aimed towards
large-scale cosmological hydrodynamics simulations. Our objective has been to
construct a model which results in a realistic galaxy population, similar to
what the best semi-analytic models achieve, yet realised in a self-consistent
hydrodynamical simulation of galaxy formation. This is an important next step
in galaxy formation theory and promises a much higher predictive power from the
theoretical models.

Our star formation prescription (see Section~\ref{sec:starformation}) models
star-forming dense ISM gas with an effective equation of state, where stars
form stochastically above a certain density threshold. To reach these
densities, gas cools through primordial and metal-line cooling in the presence
of a photo-ionising UV background with self-shielding corrections (see
Section~\ref{sec:cooling}).  Primordial cooling is treated with a
self-consistent chemical network, whereas metal-line cooling rates are
tabulated based on {\sm CLOUDY} calculations.  Both processes include self-shielding
corrections from the otherwise spatially uniform UV background radiation. We
follow stellar evolution and chemical enrichment processes based on AGB, SNIa
and SNII yields and account for stellar mass loss processes (see
Section~\ref{sec:stellar_evolution_and_enrichment}).  For the chemical
enrichment we follow nine elements (H, He, C, N, O, Ne, Mg, Si, Fe). Stellar
feedback effects are based on a kinetic outflow prescription, where stellar
winds are either non-locally driven from the ISM gas directly or locally around
evolving stellar particles (see Section~\ref{sec:stellar_feedback}). Our
implementation supports energy- and momentum-driven winds or a combination of
both, although in practice we prefer energy-driven winds since they provide a
better fit to the faint-end of the stellar mass function. We introduce a novel
wind metal mass loading scheme, which allows us to decouple mass and metal
loading for gas put into the outflowing wind (see
Section~\ref{sec:wind_metal_loading}), which is required to match the
mass--metallicity relation and stellar mass function for low mass galaxies
simultaneously.  We also follow black hole growth and include quasar- and
radio-mode feedback processes (see Section~\ref{sec:BH_growth_and_AGN}).
Furthermore, we introduce a novel model for radiative (electro-magnetic) AGN
feedback (see Section~\ref{sec:EM_feedback}), which is implemented assuming an
average AGN SED and a bolometric, luminosity-dependent scaling to take into
account obscuration effects.  Finally, we introduced inlined analysis
techniques for gas tracking in the form of classical Lagrangian velocity field
tracer particles and a new Monte Carlo based scheme, and we added on-the-fly
visualisation routines to allow for direct volume rendering through a ray
casting technique while the simulation is running.

The scope of the paper is to present our galaxy formation implementation, to
identify a fiducial model with promising settings of the feedback parameters,
and to show some basic results for the newly added physics that are compared
with some key observations at $z=0$.  To this end, we have explored three
different sets of simulations: one resolution study using our fiducial feedback
parameters with a maximum resolution of $2\times512^3$ resolution elements in a
$(25\hmpc)^3$ simulation volume, a set of simulations where we slightly
modified the strength of stellar and radio-mode AGN feedback to explore changes
in the feedback prescriptions, and simulations where we systematically explored
the impact of various physical processes by specifically deactivating certain
parts of our model.

Our fiducial feedback parameters were chosen to be physically plausible and set
such that they reasonably well reproduce key observations at $z=0$.  We note
that the stellar and radio-mode AGN feedback settings of this fiducial model
differ significantly from those previously used in SPH simulations. Most
important, the AGN radio-mode feedback is more energetic than adopted
earlier.
These energetic feedback processes are required to match the
$z=0$ stellar mass function and stellar mass -- halo mass relation.  Such a
need for stronger feedback was already pointed out in~\cite{Vogelsberger2012}
and~\citet{Keres2012} where it was demonstrated that {\sm AREPO} shows 
significantly more efficient gas cooling compared to previous SPH simulations, which
is mainly due to spurious viscous heating effects in SPH and the lack of a
proper cascade of subsonic turbulent energy to
small scales.  This in turn results
in an overall increase of star formation by a factor of two towards lower
redshifts~\citep[][]{Vogelsberger2012} and also a stellar mass function which
significantly overpredicts the number of massive systems~\citep[][]{Keres2012}.
Stronger feedback needs to compensate for this increase in cooling and star
formation. The artificial heating effect of SPH is largest for high mass
systems where a quasi-hydrostatic hot atmosphere can form. SPH simulations
suffer here a form of ``numerical quenching'', reducing the amount of AGN
feedback required in this technique to regulate the amount of stars forming in
these systems. Since {\sm AREPO} does not suffer from this unphysical effect,
we need to invoke a corresponding increased radio-mode feedback strength.

The feedback variations explored in this paper are summarised in
Table~\ref{table:cosmo_sims}.  We only varied the strength of stellar and
radio-mode feedback since those influence star formation 
most significantly.  
These modifications also affect the times at which stars form in the simulation (see
Figure~\ref{fig:cosmo_stellar_age_dist}) and therefore many of the key
observables studied above are significantly altered by these feedback changes.
The most extreme feedback models, in terms of stellar formation times, are our
simulations with ``stronger winds'', which produces too much late-time star
formation, and the simulation with ``faster winds'', which leads to almost no
late-time star formation. All other feedback variations fall more or less
between these two extreme setups. ``Weaker winds'' lead, as expected, to more
star formation at higher redshifts.  Weakening the radio-mode AGN feedback
results in more late time star formation, whereas a stronger radio-mode
feedback has the opposite effect.  Interestingly, a simulation without any
feedback leads to a lower late-time star formation rate than the model with
very strong winds. The reason is that strong winds very efficiently suppress
star formation at early times such that significant quantities
of gas remain for late-time star
formation.

Before discussing how  our simulations compare to observations and how changes
in the stellar and AGN feedback strength affects this comparison, we will first briefly
elaborate on the importance of the two most novel aspects of our galaxy
formation model, namely the incorporation of radiative AGN feedback and the special
treatment of metals for stellar outflows.\\

\noindent{\bf Impact of radiative AGN feedback:} 
Our galaxy formation model includes a new phenomenological model for
radiative (electro-magnetic) AGN feedback. This feedback acts on halo gas by
changing the net cooling rates due to the radiation coming from AGN, which
is particularly strong during quasar activity. Our model assumes a fixed SED
with a simple and rather conservative prescription for obscuration effects.  We
explored for the first time the impact of this radiative AGN feedback in fully
self-consistent large-scale cosmological simulations with a focus on its
consequences for the host galaxy in terms of the stellar mass content and overall star
formation rates. We find that this form of feedback is more efficient in
regulating star formation than purely thermal, continuous quasar-mode AGN
feedback, which is mainly relevant to regulate BH growth.  However, this
radiative form of feedback is only efficient if the BH accretion rates are
close to Eddington where the radiation field is strongest. Since this happens
only for a rather short period of cosmic time, the impact of mechanical
radio-mode feedback is typically significantly stronger than radiative
feedback. However, we did not explore possible effects on the ionisation state
of the IGM, which could be more substantially influenced by radiative AGN
feedback.\\

\noindent{\bf Metal loading of stellar winds:} 
Our stellar feedback model allows the metal loading of winds to be adjusted
separately from the actual mass loading.  We found that such a treatment is
necessary to reproduce the stellar mass and metal content of galaxies
simultaneously.  Specifically, we have introduced a wind metal loading factor
which is set independently from the wind mass loading factor.  This wind metal
loading factor defines the relationship between the metallicity of newly
created wind particles and the metallicity of the ambient ISM.  We found that
the traditional wind metal loading leads to an under-production of the
gas-phase metallicity and an over enrichment of halo gas.  Both of these
directly result from the efficient ejection of metals from galaxies via winds.
Not only does this lead to poor matches to the observationally constrained
mass-metallicity relation, but the over enrichment of halo gas leads to
enhanced metal line cooling as well. We can remove this tension with our metal
loading scheme for stellar winds. In fact, such a scheme allows us to
simultaneously match the mass--metallicity relation and stellar mass function
reasonably well. While this wind treatment significantly affects the
mass--metallicity relation, it only slightly modifies other observables like
the stellar mass function. We note that metal loading factors can be constrained
empirically \citep[][]{Zahid2013}. Such studies point to rather low values for
$\gamma_\mathrm{w}$.\\

\noindent{\bf Comparison to observations:} 
Our fiducial model reproduces many key observables over a wide range of halo
masses, covering low mass and high mass systems through the combination of
strong stellar and AGN feedback. Most importantly, our model correctly
describes the transition region around $M_\mathrm{200,crit}\sim 10^{12}\msun$,
where star formation is maximally efficient. We briefly summarise our main
findings for each of the key characteristics presented in this paper.\\

\noindent{\it (i) Cosmic SFR Density:} 
Our fiducial model reproduces the cosmic SFR density as a function of redshift
reasonably well, although our highest resolution simulation is not yet fully
converged. The wind velocity of the stellar feedback has a significant impact
on the cosmic SFR density. Fast winds lead to a strong suppression of the SFR,
and such fast winds can reduce the star formation even in more massive haloes.
The impact of variations of the AGN radio mode is less dramatic and mainly
affects the late time decline of the cosmic SFR density, but no radio mode
altogether gives a far too high late time SFR.\\

\noindent{\it (ii) Stellar mass -- halo mass relation:} 
Our model predicts a stellar mass -- halo mass relation that is in good
agreement with results based on abundance matching. At higher galaxy masses
(i.e.  $M>10^9 \msun$) we find clear convergence in the stellar mass -- halo
mass relation. Most importantly, our combination of stellar feedback and
radio-mode AGN feedback leads to a shape in the stellar mass -- halo mass
relation which is in good agreement with abundance matching results, including
a correct turnover at the massive end due to efficient quenching through AGN
radio-mode feedback in massive haloes. We need strong stellar and AGN feedback
to reproduce the stellar content of low and high mass haloes.\\

\noindent{\it (iii) Stellar mass function:} 
Our model reproduces the stellar mass function well, although we find that we
have slightly too many low mass galaxies in our highest resolution simulation.
We find a similar problem for lower mass systems for the stellar mass -- halo
mass relation. Here we also slightly overproduce the amount of stars in these
systems. Both findings indicate the need for yet stronger stellar feedback,
which we partially explored in this work by increasing the SNII energy factor
by a factor of two. But we note that our fiducial simulation setup includes
already rather strong SNII feedback, and increasing this further appears
energetically problematic.  Radio-mode AGN feedback successfully produces the
sharp exponential drop-off of the stellar mass function towards larger stellar
masses. We also find that energy-driven winds are required to suppress the low
mass end of the stellar mass function sufficiently since they allow for large
mass-loading factors for low mass systems compared to momentum-driven winds.
This is essentially required to obtain a shallow slope of the stellar mass
function at the low mass end.\\

\noindent{\it (iv) Stellar mass density:} 
Our fiducial model reproduces the observed stellar mass density reasonably
well.  The ``no feedback'' simulation strongly overproduces the amount of
stellar mass, whereas the ``faster winds'' strongly underproduces it. All the
other feedback variations are quite close to the fiducial model, which provides
the best fit to the data. Changes in the radio-mode AGN feedback only affect
the late time evolution of the stellar mass density.\\

\noindent{\it (v) Tully-Fisher relation:} 
We find good convergence for the Tully-Fisher relation, and our highest
resolution fiducial model agrees very well with observations both in terms of
normalisation and slope. This is an important result of our model since it
implies that our galaxies, in general, have the correct dynamical structure.
Variations of the feedback model parameter choices lead to only negligible
changes in the resulting Tully-Fisher relation, implying that the Tully-Fisher
relation is a rather stable outcome of our model.  The exception is the ``fast
wind'' model which gives rise to the largest change in the Tully-Fisher
relation by increasing the rotational velocity at a fixed stellar mass. Also a model
without any feedback does not reproduce the Tully-Fisher relation.\\

\noindent{\it (vi) Mass-metallicity relation:} 
Among all the relations that we examined in this paper, the mass--metallicity
relation shows the largest tension between observations and our simulations. Although
we find a positive correlation between stellar mass and [O/H], the overall
amplitude of this relation is not fully correctly reproduced. But given the
observational uncertainties we do not consider this a very severe problem at
this point.  The main issue is that the simulations overpredict the metallicity
of higher mass systems. Here our simulations do not predict the turnover in the
mass--metallicity relation correctly, and therefore lead to too many metals in
massive galaxies. For lower to intermediate mass systems we can only achieve a
reasonable agreement with the observations through our novel wind metal loading
scheme, which decouples the actual mass loading from the metal loading of
winds. Without such a scheme, low mass galaxies particularly would lose far too
many metals owing to the high mass loadings of our energy-driven wind scalings.
However, such high mass loadings are required to reduce star formation
sufficiently in these low mass systems.\\

\noindent{\it (vii) Luminosity function:} 
We used stellar population synthesis models to derive SDSS-band luminosity
functions (g-, r-, i-, z-bands).  Our highest resolution simulations produce
luminosity functions at $z=0$ which are in reasonable agreement with
observations.  Especially the faint end slope and the exponential drop towards
brighter systems are well reproduced. Our highest resolution simulation has a
slightly too high faint-end normalisation compared to the observations. Also
the exponential drop happens at slightly too high halo masses. Interestingly,
this effect is strongest for the g-band luminosity function, whereas the
agreement with the other bands is typically better. Feedback models with faster
winds strongly disagree with the observations.  Weakening the radio-mode AGN
feedback leads to a significant overshoot of the luminosity function at the
bright end while weak winds cause an overproduction of faint galaxies.\\

\noindent{\it (viii) Black hole mass -- stellar mass relation:}
Black hole growth is mainly regulated by quasar-mode feedback in our model,
which results together with BH accretion in a tightly self-regulated feedback
loop. Our simulated BH mass -- stellar mass relation is  well converged and
agrees with the observations both in slope and in normalisation. This
demonstrates that our modifications to the BH accretion procedure do not affect
BH growth in any significant way.  Furthermore, the slope of the BH mass --
stellar mass relation does not change strongly if we alter the stellar or
radio-mode AGN feedback parameters.  Normalisation changes in the relation are
mainly due to changes in the stellar mass caused by the different feedback
choices. \\

Overall, we find that our fiducial model reproduces a significant number of key
observations of the low redshift Universe.  As demonstrated in \cite{Torrey2013} our model also reproduces many high redshift observables.
Furthermore, \cite{Marinacci2013} shows that our galaxy formation
implementation also produces close analogs of Milky Way-like disk galaxies in
high resolution simulations of Milky Way-sized DM haloes.  We therefore
conclude that we have implemented a successful galaxy formation model in the
new moving-mesh code {\sm AREPO}, which is suitable for large-scale galaxy
formation simulations.  Clearly, the simulations presented in this paper are
not sufficient to fully explore all regimes of galaxy formation.  Most
important, our simulation volume is too small to sample very high-mass systems
and ideally we would like to have slightly higher mass and spatial resolution.
The convergence studies presented in this paper show that our model requires a
mass resolution of at least $10^6\hmsun$ for the baryonic component to achieve
reasonably converged results. Therefore, predictive large-scale structure
simulations with our model require a simulation volume of the order of
$(100\hmpc)^3$ with a mass resolution of the order of $10^6\hmsun$, which
implies that the simulation has to follow the evolution of many billions of
resolution elements. In forthcoming work we will present such simulations based
on the methodology presented here (Vogelsberger et al., in prep).

\section*{Acknowledgements}

The simulations in this paper were run on the Odyssey cluster supported by the
FAS Science Division Research Computing Group at Harvard University, the
Stampede supercomputer at the Texas Advanced Computing Center, the Magny
Cluster at HITS, and the CURIE supercomputer at CEA/France as part of PRACE
project RA0844.  We thank Simeon Bird, Laura Blecha, Charlie Conroy, Daniel
Eisenstein, Claude-Andre Faucher-Giguere, Chris Hayward, Dusan Keres, Lisa
Kewley, Bence Kocsis, Federico Marinacci, Diego Munoz, Dylan Nelson,
R\"{u}diger Pakmor, Christoph Pfrommer, Ewald Puchwein, Vicente
Rodriguez-Gomez, Laura Sales, Rob Simcoe, Gregory Snyder, Joshua Suresh, Rob
Wiersma, Dandan Xu, and Jabran Zahid for useful discussions.  MV acknowledges
support from NASA through Hubble Fellowship grant HST-HF-51317.01. VS
acknowledges support through SFB 881 'The Milky Way System' of the DFG, and
through the European Research Council under ERC-StG grant EXAGAL-308037.

\bibliography{paper}

\begin{thebibliography}{241}
\expandafter\ifx\csname natexlab\endcsname\relax\def\natexlab#1{#1}\fi

\bibitem[{{Abazajian} {et~al.}(2009){Abazajian}, {Adelman-McCarthy},
  {Ag{\"u}eros}, {Allam}, {Allende Prieto}, {An}, {Anderson}, {Anderson},
  {Annis}, {Bahcall}, \& et~al.}]{Abazajian2009}
{Abazajian}, K.~N., {Adelman-McCarthy}, J.~K., {Ag{\"u}eros}, M.~A., {et~al.}
  2009, \apjs, 182, 543

\bibitem[{{Abel}(2011)}]{Abel2011}
{Abel}, T. 2011, \mnras, 413, 271

\bibitem[{{Abel} \& {Wandelt}(2002)}]{Abel2002}
{Abel}, T., \& {Wandelt}, B.~D. 2002, \mnras, 330, L53

\bibitem[{{Agertz} {et~al.}(2011){Agertz}, {Teyssier}, \& {Moore}}]{Agertz2011}
{Agertz}, O., {Teyssier}, R., \& {Moore}, B. 2011, \mnras, 410, 1391

\bibitem[{{Agertz} {et~al.}(2007){Agertz}, {Moore}, {Stadel}, {Potter},
  {Miniati}, {Read}, {Mayer}, {Gawryszczak}, {Kravtsov}, {Nordlund}, {Pearce},
  {Quilis}, {Rudd}, {Springel}, {Stone}, {Tasker}, {Teyssier}, {Wadsley}, \&
  {Walder}}]{Agertz2007}
{Agertz}, O., {Moore}, B., {Stadel}, J., {et~al.} 2007, \mnras, 380, 963

\bibitem[{{Ascasibar} {et~al.}(2002){Ascasibar}, {Yepes}, {Gottl{\"o}ber}, \&
  {M{\"u}ller}}]{Ascasibar2002}
{Ascasibar}, Y., {Yepes}, G., {Gottl{\"o}ber}, S., \& {M{\"u}ller}, V. 2002,
  \aap, 387, 396

\bibitem[{{Aumer} {et~al.}(2013){Aumer}, {White}, {Naab}, \&
  {Scannapieco}}]{Aumer2013}
{Aumer}, M., {White}, S., {Naab}, T., \& {Scannapieco}, C. 2013, ArXiv e-prints

\bibitem[{{Baldry} {et~al.}(2008){Baldry}, {Glazebrook}, \&
  {Driver}}]{Bladry2008}
{Baldry}, I.~K., {Glazebrook}, K., \& {Driver}, S.~P. 2008, \mnras, 388, 945

\bibitem[{{Bauer} \& {Springel}(2012)}]{Bauer2012}
{Bauer}, A., \& {Springel}, V. 2012, \mnras, 423, 2558

\bibitem[{{Baugh}(2006)}]{Baugh2006}
{Baugh}, C.~M. 2006, Reports on Progress in Physics, 69, 3101

\bibitem[{{Behroozi} {et~al.}(2013){Behroozi}, {Wechsler}, \&
  {Conroy}}]{Behroozi2012}
{Behroozi}, P.~S., {Wechsler}, R.~H., \& {Conroy}, C. 2013, \apjl, 762, L31

\bibitem[{{Bell} \& {de Jong}(2001)}]{Bell2001}
{Bell}, E.~F., \& {de Jong}, R.~S. 2001, \apj, 550, 212

\bibitem[{{Benson}(2012)}]{Benson2012}
{Benson}, A.~J. 2012, \na, 17, 175

\bibitem[{{Berger} \& {Colella}(1989)}]{Berger1989}
{Berger}, M.~J., \& {Colella}, P. 1989, Journal of Computational Physics, 82,
  64

\bibitem[{{Bird} {et~al.}(2011){Bird}, {Peiris}, {Viel}, \& {Verde}}]{Bird2011}
{Bird}, S., {Peiris}, H.~V., {Viel}, M., \& {Verde}, L. 2011, \mnras, 413, 1717

\bibitem[{{Bird} {et~al.}(2013){Bird}, {Vogelsberger}, {Sijacki},
  {Zaldarriaga}, {Springel}, \& {Hernquist}}]{Bird2013}
{Bird}, S., {Vogelsberger}, M., {Sijacki}, D., {et~al.} 2013, \mnras, 429, 3341

\bibitem[{{Blanton} {et~al.}(2005){Blanton}, {Lupton}, {Schlegel}, {Strauss},
  {Brinkmann}, {Fukugita}, \& {Loveday}}]{Blanton2005}
{Blanton}, M.~R., {Lupton}, R.~H., {Schlegel}, D.~J., {et~al.} 2005, \apj, 631,
  208

\bibitem[{{Blumenthal} {et~al.}(1984){Blumenthal}, {Faber}, {Primack}, \&
  {Rees}}]{Blumenthal1984}
{Blumenthal}, G.~R., {Faber}, S.~M., {Primack}, J.~R., \& {Rees}, M.~J. 1984,
  \nat, 311, 517

\bibitem[{{Booth} \& {Schaye}(2009)}]{Booth2009}
{Booth}, C.~M., \& {Schaye}, J. 2009, \mnras, 398, 53

\bibitem[{{Bouwens} {et~al.}(2008){Bouwens}, {Illingworth}, {Franx}, \&
  {Ford}}]{Bouwens2008}
{Bouwens}, R.~J., {Illingworth}, G.~D., {Franx}, M., \& {Ford}, H. 2008, \apj,
  686, 230

\bibitem[{{Bower} {et~al.}(2012){Bower}, {Benson}, \& {Crain}}]{Bower2012}
{Bower}, R.~G., {Benson}, A.~J., \& {Crain}, R.~A. 2012, \mnras, 422, 2816

\bibitem[{{Bower} {et~al.}(2006){Bower}, {Benson}, {Malbon}, {Helly}, {Frenk},
  {Baugh}, {Cole}, \& {Lacey}}]{Bower2006}
{Bower}, R.~G., {Benson}, A.~J., {Malbon}, R., {et~al.} 2006, \mnras, 370, 645

\bibitem[{{Boylan-Kolchin} {et~al.}(2010){Boylan-Kolchin}, {Springel}, {White},
  \& {Jenkins}}]{BoylanKolchin2010}
{Boylan-Kolchin}, M., {Springel}, V., {White}, S.~D.~M., \& {Jenkins}, A. 2010,
  \mnras, 406, 896

\bibitem[{{Brook} {et~al.}(2004){Brook}, {Kawata}, {Gibson}, \&
  {Flynn}}]{Brook2004}
{Brook}, C.~B., {Kawata}, D., {Gibson}, B.~K., \& {Flynn}, C. 2004, \mnras,
  349, 52

\bibitem[{{Bruzual} \& {Charlot}(2003)}]{BC03}
{Bruzual}, G., \& {Charlot}, S. 2003, \mnras, 344, 1000

\bibitem[{{Cantalupo} \& {Porciani}(2011)}]{Cantalupo2011}
{Cantalupo}, S., \& {Porciani}, C. 2011, \mnras, 411, 1678

\bibitem[{{Cappellari} {et~al.}(2006){Cappellari}, {Bacon}, {Bureau}, {Damen},
  {Davies}, {de Zeeuw}, {Emsellem}, {Falc{\'o}n-Barroso}, {Krajnovi{\'c}},
  {Kuntschner}, {McDermid}, {Peletier}, {Sarzi}, {van den Bosch}, \& {van de
  Ven}}]{Cappellari2006}
{Cappellari}, M., {Bacon}, R., {Bureau}, M., {et~al.} 2006, \mnras, 366, 1126

\bibitem[{{Caputi} {et~al.}(2011){Caputi}, {Cirasuolo}, {Dunlop}, {McLure},
  {Farrah}, \& {Almaini}}]{Caputi2011}
{Caputi}, K.~I., {Cirasuolo}, M., {Dunlop}, J.~S., {et~al.} 2011, \mnras, 413,
  162

\bibitem[{{Cattaneo} {et~al.}(2006){Cattaneo}, {Dekel}, {Devriendt},
  {Guiderdoni}, \& {Blaizot}}]{Cattaneo2006}
{Cattaneo}, A., {Dekel}, A., {Devriendt}, J., {Guiderdoni}, B., \& {Blaizot},
  J. 2006, \mnras, 370, 1651

\bibitem[{{Cen}(1992)}]{Cen1992}
{Cen}, R. 1992, \apjs, 78, 341

\bibitem[{{Chabrier}(2003)}]{Chabrier2003}
{Chabrier}, G. 2003, \pasp, 115, 763

\bibitem[{{Choi} {et~al.}(2012){Choi}, {Ostriker}, {Naab}, \&
  {Johansson}}]{Choi2012}
{Choi}, E., {Ostriker}, J.~P., {Naab}, T., \& {Johansson}, P.~H. 2012, \apj,
  754, 125

\bibitem[{{Ciotti} \& {Ostriker}(2007)}]{Ciotti2007}
{Ciotti}, L., \& {Ostriker}, J.~P. 2007, \apj, 665, 1038

\bibitem[{{Ciotti} {et~al.}(2009){Ciotti}, {Ostriker}, \& {Proga}}]{Ciotti2009}
{Ciotti}, L., {Ostriker}, J.~P., \& {Proga}, D. 2009, \apj, 699, 89

\bibitem[{{Collins} {et~al.}(2010){Collins}, {Xu}, {Norman}, {Li}, \&
  {Li}}]{Collins2010}
{Collins}, D.~C., {Xu}, H., {Norman}, M.~L., {Li}, H., \& {Li}, S. 2010, \apjs,
  186, 308

\bibitem[{{Conroy} \& {van Dokkum}(2012)}]{Conroy2012}
{Conroy}, C., \& {van Dokkum}, P.~G. 2012, \apj, 760, 71

\bibitem[{{Conroy} \& {Wechsler}(2009)}]{Conroy2009}
{Conroy}, C., \& {Wechsler}, R.~H. 2009, \apj, 696, 620

\bibitem[{{Conroy} {et~al.}(2006){Conroy}, {Wechsler}, \&
  {Kravtsov}}]{Conroy2006}
{Conroy}, C., {Wechsler}, R.~H., \& {Kravtsov}, A.~V. 2006, \apj, 647, 201

\bibitem[{{Crain} {et~al.}(2009){Crain}, {Theuns}, {Dalla Vecchia}, {Eke},
  {Frenk}, {Jenkins}, {Kay}, {Peacock}, {Pearce}, {Schaye}, {Springel},
  {Thomas}, {White}, \& {Wiersma}}]{Crain2009}
{Crain}, R.~A., {Theuns}, T., {Dalla Vecchia}, C., {et~al.} 2009, \mnras, 399,
  1773

\bibitem[{{Croft} {et~al.}(2009){Croft}, {Di Matteo}, {Springel}, \&
  {Hernquist}}]{Croft2009}
{Croft}, R.~A.~C., {Di Matteo}, T., {Springel}, V., \& {Hernquist}, L. 2009,
  \mnras, 400, 43

\bibitem[{{Croton} {et~al.}(2006){Croton}, {Springel}, {White}, {De Lucia},
  {Frenk}, {Gao}, {Jenkins}, {Kauffmann}, {Navarro}, \& {Yoshida}}]{Croton2006}
{Croton}, D.~J., {Springel}, V., {White}, S.~D.~M., {et~al.} 2006, \mnras, 365,
  11

\bibitem[{{Cullen} \& {Dehnen}(2010)}]{Cullen2010}
{Cullen}, L., \& {Dehnen}, W. 2010, \mnras, 408, 669

\bibitem[{{Dahlen} {et~al.}(2004){Dahlen}, {Strolger}, {Riess}, {Mobasher},
  {Chary}, {Conselice}, {Ferguson}, {Fruchter}, {Giavalisco}, {Livio}, {Madau},
  {Panagia}, \& {Tonry}}]{Dahlen2004}
{Dahlen}, T., {Strolger}, L.-G., {Riess}, A.~G., {et~al.} 2004, \apj, 613, 189

\bibitem[{{Dalla Vecchia} \& {Schaye}(2008)}]{DallaVecchia2008}
{Dalla Vecchia}, C., \& {Schaye}, J. 2008, \mnras, 387, 1431

\bibitem[{{Dalla Vecchia} \& {Schaye}(2012)}]{DallaVecchia2012}
---. 2012, \mnras, 426, 140

\bibitem[{{Dav{\'e}} {et~al.}(2011{\natexlab{a}}){Dav{\'e}}, {Finlator}, \&
  {Oppenheimer}}]{Dave2011b}
{Dav{\'e}}, R., {Finlator}, K., \& {Oppenheimer}, B.~D. 2011{\natexlab{a}},
  \mnras, 416, 1354

\bibitem[{{Dav{\'e}} {et~al.}(2011{\natexlab{b}}){Dav{\'e}}, {Oppenheimer}, \&
  {Finlator}}]{Dave2011}
{Dav{\'e}}, R., {Oppenheimer}, B.~D., \& {Finlator}, K. 2011{\natexlab{b}},
  \mnras, 415, 11

\bibitem[{{Debuhr} {et~al.}(2011){Debuhr}, {Quataert}, \& {Ma}}]{Debuhr2011}
{Debuhr}, J., {Quataert}, E., \& {Ma}, C.-P. 2011, \mnras, 412, 1341

\bibitem[{{Dekel} \& {Silk}(1986)}]{Dekel1986}
{Dekel}, A., \& {Silk}, J. 1986, \apj, 303, 39

\bibitem[{{Di Matteo} {et~al.}(2008){Di Matteo}, {Colberg}, {Springel},
  {Hernquist}, \& {Sijacki}}]{DiMatteo2008}
{Di Matteo}, T., {Colberg}, J., {Springel}, V., {Hernquist}, L., \& {Sijacki},
  D. 2008, \apj, 676, 33

\bibitem[{{Di Matteo} {et~al.}(2005){Di Matteo}, {Springel}, \&
  {Hernquist}}]{DiMatteo2005}
{Di Matteo}, T., {Springel}, V., \& {Hernquist}, L. 2005, \nat, 433, 604

\bibitem[{{Dickinson} {et~al.}(2003){Dickinson}, {Papovich}, {Ferguson}, \&
  {Budav{\'a}ri}}]{Dickinson2003}
{Dickinson}, M., {Papovich}, C., {Ferguson}, H.~C., \& {Budav{\'a}ri}, T. 2003,
  \apj, 587, 25

\bibitem[{{Dolag} {et~al.}(2009){Dolag}, {Borgani}, {Murante}, \&
  {Springel}}]{Dolag2009b}
{Dolag}, K., {Borgani}, S., {Murante}, G., \& {Springel}, V. 2009, \mnras, 399,
  497

\bibitem[{{Dolag} {et~al.}(2004){Dolag}, {Jubelgas}, {Springel}, {Borgani}, \&
  {Rasia}}]{Dolag2004}
{Dolag}, K., {Jubelgas}, M., {Springel}, V., {Borgani}, S., \& {Rasia}, E.
  2004, \apjl, 606, L97

\bibitem[{{Dolag} {et~al.}(2011){Dolag}, {Reinecke}, {Gheller}, {Rivi},
  {Krokos}, \& {Jin}}]{Dolag2011}
{Dolag}, K., {Reinecke}, M., {Gheller}, C., {et~al.} 2011, in Astrophysics
  Source Code Library, record ascl:1103.005, 3005

\bibitem[{{Dolag} \& {Stasyszyn}(2009)}]{Dolag2009}
{Dolag}, K., \& {Stasyszyn}, F. 2009, \mnras, 398, 1678

\bibitem[{{Dubois} {et~al.}(2012){Dubois}, {Devriendt}, {Slyz}, \&
  {Teyssier}}]{Dubois2012}
{Dubois}, Y., {Devriendt}, J., {Slyz}, A., \& {Teyssier}, R. 2012, \mnras, 420,
  2662

\bibitem[{{Dubois} {et~al.}(2013){Dubois}, {Gavazzi}, {Peirani}, \&
  {Silk}}]{Dubois2013}
{Dubois}, Y., {Gavazzi}, R., {Peirani}, S., \& {Silk}, J. 2013, ArXiv e-prints,
  1301.3092

\bibitem[{{Dubois} \& {Teyssier}(2008)}]{Dubois2008}
{Dubois}, Y., \& {Teyssier}, R. 2008, \aap, 477, 79

\bibitem[{{Efstathiou}(1992)}]{Efstathiou1992}
{Efstathiou}, G. 1992, \mnras, 256, 43P

\bibitem[{{Elvis} {et~al.}(1994){Elvis}, {Wilkes}, {McDowell}, {Green},
  {Bechtold}, {Willner}, {Oey}, {Polomski}, \& {Cutri}}]{Elivs1994}
{Elvis}, M., {Wilkes}, B.~J., {McDowell}, J.~C., {et~al.} 1994, \apjs, 95, 1

\bibitem[{{Faucher-Gigu{\`e}re}
  {et~al.}(2008{\natexlab{a}}){Faucher-Gigu{\`e}re}, {Lidz}, {Hernquist}, \&
  {Zaldarriaga}}]{Faucher2008b}
{Faucher-Gigu{\`e}re}, C.-A., {Lidz}, A., {Hernquist}, L., \& {Zaldarriaga}, M.
  2008{\natexlab{a}}, \apj, 688, 85

\bibitem[{{Faucher-Gigu{\`e}re} {et~al.}(2009){Faucher-Gigu{\`e}re}, {Lidz},
  {Zaldarriaga}, \& {Hernquist}}]{Faucher2009}
{Faucher-Gigu{\`e}re}, C.-A., {Lidz}, A., {Zaldarriaga}, M., \& {Hernquist}, L.
  2009, \apj, 703, 1416

\bibitem[{{Faucher-Gigu{\`e}re}
  {et~al.}(2008{\natexlab{b}}){Faucher-Gigu{\`e}re}, {Prochaska}, {Lidz},
  {Hernquist}, \& {Zaldarriaga}}]{Faucher2008}
{Faucher-Gigu{\`e}re}, C.-A., {Prochaska}, J.~X., {Lidz}, A., {Hernquist}, L.,
  \& {Zaldarriaga}, M. 2008{\natexlab{b}}, \apj, 681, 831

\bibitem[{{Few} {et~al.}(2012){Few}, {Courty}, {Gibson}, {Kawata}, {Calura}, \&
  {Teyssier}}]{Few2012}
{Few}, C.~G., {Courty}, S., {Gibson}, B.~K., {et~al.} 2012, \mnras, 424, L11

\bibitem[{{Fontana} {et~al.}(2006){Fontana}, {Salimbeni}, {Grazian},
  {Giallongo}, {Pentericci}, {Nonino}, {Fontanot}, {Menci}, {Monaco},
  {Cristiani}, {Vanzella}, {de Santis}, \& {Gallozzi}}]{Fontana2006}
{Fontana}, A., {Salimbeni}, S., {Grazian}, A., {et~al.} 2006, \aap, 459, 745

\bibitem[{{Francis} {et~al.}(1993){Francis}, {Hooper}, \&
  {Impey}}]{Francis1993}
{Francis}, P.~J., {Hooper}, E.~J., \& {Impey}, C.~D. 1993, \aj, 106, 417

\bibitem[{{Frenk} {et~al.}(1999){Frenk}, {White}, {Bode}, {Bond}, {Bryan},
  {Cen}, {Couchman}, {Evrard}, {Gnedin}, {Jenkins}, {Khokhlov}, {Klypin},
  {Navarro}, {Norman}, {Ostriker}, {Owen}, {Pearce}, {Pen}, {Steinmetz},
  {Thomas}, {Villumsen}, {Wadsley}, {Warren}, {Xu}, \& {Yepes}}]{Frenk1999}
{Frenk}, C.~S., {White}, S.~D.~M., {Bode}, P., {et~al.} 1999, \apj, 525, 554

\bibitem[{{Genel} {et~al.}(2013){Genel}, {Vogelsberger}, {Nelson}, {Sijacki},
  {Springel}, \& {Hernquist}}]{Genel2013}
{Genel}, S., {Vogelsberger}, M., {Nelson}, D., {et~al.} 2013, ArXiv e-prints

\bibitem[{{Genel} {et~al.}(2012)}]{Genel2012}
{Genel}, S., {et~al.} 2012, \mnras, 11

\bibitem[{{Gerritsen} \& {Icke}(1997)}]{Gerritsen1997}
{Gerritsen}, J.~P.~E., \& {Icke}, V. 1997, \aap, 325, 972

\bibitem[{{Gingold} \& {Monaghan}(1977)}]{Gingold1977}
{Gingold}, R.~A., \& {Monaghan}, J.~J. 1977, \mnras, 181, 375

\bibitem[{{Gnedin}(1995)}]{Gnedin1995}
{Gnedin}, N.~Y. 1995, \apjs, 97, 231

\bibitem[{{Gnedin} \& {Hollon}(2012)}]{Gnedin2012}
{Gnedin}, N.~Y., \& {Hollon}, N. 2012, \apjs, 202, 13

\bibitem[{{Gonz{\'a}lez} {et~al.}(2011){Gonz{\'a}lez}, {Labb{\'e}}, {Bouwens},
  {Illingworth}, {Franx}, \& {Kriek}}]{Gonzalez2011}
{Gonz{\'a}lez}, V., {Labb{\'e}}, I., {Bouwens}, R.~J., {et~al.} 2011, \apjl,
  735, L34

\bibitem[{{Governato} {et~al.}(2010){Governato}, {Brook}, {Mayer}, {Brooks},
  {Rhee}, {Wadsley}, {Jonsson}, {Willman}, {Stinson}, {Quinn}, \&
  {Madau}}]{Governato2010}
{Governato}, F., {Brook}, C., {Mayer}, L., {et~al.} 2010, \nat, 463, 203

\bibitem[{{Greggio}(2005)}]{Greggio2005}
{Greggio}, L. 2005, \aap, 441, 1055

\bibitem[{{Guedes} {et~al.}(2011){Guedes}, {Callegari}, {Madau}, \&
  {Mayer}}]{Guedes2011}
{Guedes}, J., {Callegari}, S., {Madau}, P., \& {Mayer}, L. 2011, \apj, 742, 76

\bibitem[{{Guo} {et~al.}(2010){Guo}, {White}, {Li}, \&
  {Boylan-Kolchin}}]{Guo2010}
{Guo}, Q., {White}, S., {Li}, C., \& {Boylan-Kolchin}, M. 2010, \mnras, 404,
  1111

\bibitem[{{Guo} {et~al.}(2011){Guo}, {White}, {Boylan-Kolchin}, {De Lucia},
  {Kauffmann}, {Lemson}, {Li}, {Springel}, \& {Weinmann}}]{Guo2011}
{Guo}, Q., {White}, S., {Boylan-Kolchin}, M., {et~al.} 2011, \mnras, 413, 101

\bibitem[{{Haas} {et~al.}(2012{\natexlab{a}}){Haas}, {Schaye}, {Booth}, {Dalla
  Vecchia}, {Springel}, {Theuns}, \& {Wiersma}}]{Haas2012a}
{Haas}, M.~R., {Schaye}, J., {Booth}, C.~M., {et~al.} 2012{\natexlab{a}}, ArXiv
  e-prints, 1211.1021

\bibitem[{{Haas} {et~al.}(2012{\natexlab{b}}){Haas}, {Schaye}, {Booth}, {Dalla
  Vecchia}, {Springel}, {Theuns}, \& {Wiersma}}]{Haas2012b}
---. 2012{\natexlab{b}}, ArXiv e-prints, 1211.3120

\bibitem[{{Hambrick} {et~al.}(2009){Hambrick}, {Ostriker}, {Naab}, \&
  {Johansson}}]{Hambrick2009}
{Hambrick}, D.~C., {Ostriker}, J.~P., {Naab}, T., \& {Johansson}, P.~H. 2009,
  \apj, 705, 1566

\bibitem[{{Hambrick} {et~al.}(2011){Hambrick}, {Ostriker}, {Naab}, \&
  {Johansson}}]{Hambrick2011}
---. 2011, \apj, 738, 16

\bibitem[{{H{\"a}ring} \& {Rix}(2004)}]{Haering2004}
{H{\"a}ring}, N., \& {Rix}, H.-W. 2004, \apjl, 604, L89

\bibitem[{{Hernquist}(1993)}]{Hernquist1993}
{Hernquist}, L. 1993, \apj, 404, 717

\bibitem[{{Hopkins} \& {Beacom}(2006{\natexlab{a}})}]{Hopkins2006}
{Hopkins}, A.~M., \& {Beacom}, J.~F. 2006{\natexlab{a}}, \apj, 651, 142

\bibitem[{{Hopkins} \& {Beacom}(2006{\natexlab{b}})}]{HopkinsBeacom2006}
---. 2006{\natexlab{b}}, \apj, 651, 142

\bibitem[{{Hopkins}(2013)}]{Hopkins2013}
{Hopkins}, P.~F. 2013, \mnras, 428, 2840

\bibitem[{{Hopkins} {et~al.}(2009{\natexlab{a}}){Hopkins}, {Cox}, {Dutta},
  {Hernquist}, {Kormendy}, \& {Lauer}}]{Hopkins2009}
{Hopkins}, P.~F., {Cox}, T.~J., {Dutta}, S.~N., {et~al.} 2009{\natexlab{a}},
  \apjs, 181, 135

\bibitem[{{Hopkins} {et~al.}(2013{\natexlab{a}}){Hopkins}, {Cox}, {Hernquist},
  {Narayanan}, {Hayward}, \& {Murray}}]{Hopkins2013b}
{Hopkins}, P.~F., {Cox}, T.~J., {Hernquist}, L., {et~al.} 2013{\natexlab{a}},
  \mnras, 430, 1901

\bibitem[{{Hopkins} {et~al.}(2008){Hopkins}, {Hernquist}, {Cox}, {Dutta}, \&
  {Rothberg}}]{Hopkins2008}
{Hopkins}, P.~F., {Hernquist}, L., {Cox}, T.~J., {Dutta}, S.~N., \& {Rothberg},
  B. 2008, \apj, 679, 156

\bibitem[{{Hopkins} {et~al.}(2013{\natexlab{b}}){Hopkins}, {Keres}, \&
  {Murray}}]{Hopkins2013c}
{Hopkins}, P.~F., {Keres}, D., \& {Murray}, N. 2013{\natexlab{b}}, ArXiv
  e-prints

\bibitem[{{Hopkins} {et~al.}(2012{\natexlab{a}}){Hopkins}, {Kere{\v s}},
  {Murray}, {Quataert}, \& {Hernquist}}]{Hopkins2012d}
{Hopkins}, P.~F., {Kere{\v s}}, D., {Murray}, N., {Quataert}, E., \&
  {Hernquist}, L. 2012{\natexlab{a}}, \mnras, 427, 968

\bibitem[{{Hopkins} {et~al.}(2009{\natexlab{b}}){Hopkins}, {Lauer}, {Cox},
  {Hernquist}, \& {Kormendy}}]{Hopkins2009b}
{Hopkins}, P.~F., {Lauer}, T.~R., {Cox}, T.~J., {Hernquist}, L., \& {Kormendy},
  J. 2009{\natexlab{b}}, \apjs, 181, 486

\bibitem[{{Hopkins} \& {Quataert}(2010)}]{Hopkins2010}
{Hopkins}, P.~F., \& {Quataert}, E. 2010, \mnras, 407, 1529

\bibitem[{{Hopkins} {et~al.}(2012{\natexlab{b}}){Hopkins}, {Quataert}, \&
  {Murray}}]{Hopkins2012b}
{Hopkins}, P.~F., {Quataert}, E., \& {Murray}, N. 2012{\natexlab{b}}, \mnras,
  421, 3522

\bibitem[{{Hopkins} {et~al.}(2012{\natexlab{c}}){Hopkins}, {Quataert}, \&
  {Murray}}]{Hopkins2012}
---. 2012{\natexlab{c}}, \mnras, 421, 3522

\bibitem[{{Hopkins} {et~al.}(2012{\natexlab{d}}){Hopkins}, {Quataert}, \&
  {Murray}}]{Hopkins2012a}
---. 2012{\natexlab{d}}, \mnras, 421, 3488

\bibitem[{{Hopkins} {et~al.}(2007){Hopkins}, {Richards}, \&
  {Hernquist}}]{Hopkins2007}
{Hopkins}, P.~F., {Richards}, G.~T., \& {Hernquist}, L. 2007, \apj, 654, 731

\bibitem[{{Hummels} \& {Bryan}(2012)}]{Hummels2012}
{Hummels}, C.~B., \& {Bryan}, G.~L. 2012, \apj, 749, 140

\bibitem[{{Ikeuchi} \& {Ostriker}(1986)}]{Ikeuchi1986}
{Ikeuchi}, S., \& {Ostriker}, J.~P. 1986, \apj, 301, 522

\bibitem[{{Jubelgas} {et~al.}(2004){Jubelgas}, {Springel}, \&
  {Dolag}}]{Jubelgas2004}
{Jubelgas}, M., {Springel}, V., \& {Dolag}, K. 2004, \mnras, 351, 423

\bibitem[{{Jubelgas} {et~al.}(2008){Jubelgas}, {Springel}, {En{\ss}lin}, \&
  {Pfrommer}}]{Jubelgas2008}
{Jubelgas}, M., {Springel}, V., {En{\ss}lin}, T., \& {Pfrommer}, C. 2008, \aap,
  481, 33

\bibitem[{{Kannan} {et~al.}(2013){Kannan}, {Stinson}, {Macci{\`o}}, {Brook},
  {Weinmann}, {Wadsley}, \& {Couchman}}]{Kannan2013}
{Kannan}, R., {Stinson}, G.~S., {Macci{\`o}}, A.~V., {et~al.} 2013, ArXiv
  e-prints, 1302.2618

\bibitem[{{Karakas}(2010)}]{Karakas2010}
{Karakas}, A.~I. 2010, \mnras, 403, 1413

\bibitem[{{Katz} {et~al.}(1992){Katz}, {Hernquist}, \& {Weinberg}}]{Katz1992}
{Katz}, N., {Hernquist}, L., \& {Weinberg}, D.~H. 1992, \apjl, 399, L109

\bibitem[{{Katz} {et~al.}(1996){Katz}, {Weinberg}, \& {Hernquist}}]{Katz1996}
{Katz}, N., {Weinberg}, D.~H., \& {Hernquist}, L. 1996, \apjs, 105, 19

\bibitem[{{Kauffmann} {et~al.}(1993){Kauffmann}, {White}, \&
  {Guiderdoni}}]{Kauffmann1993}
{Kauffmann}, G., {White}, S.~D.~M., \& {Guiderdoni}, B. 1993, \mnras, 264, 201

\bibitem[{{Kawata} \& {Gibson}(2003)}]{Kawata2003}
{Kawata}, D., \& {Gibson}, B.~K. 2003, \mnras, 340, 908

\bibitem[{{Kawata} \& {Gibson}(2005)}]{Kawata2005}
---. 2005, \mnras, 358, L16

\bibitem[{{Kay} {et~al.}(2002){Kay}, {Pearce}, {Frenk}, \& {Jenkins}}]{Kay2002}
{Kay}, S.~T., {Pearce}, F.~R., {Frenk}, C.~S., \& {Jenkins}, A. 2002, \mnras,
  330, 113

\bibitem[{{Kere{\v s}} {et~al.}(2012){Kere{\v s}}, {Vogelsberger}, {Sijacki},
  {Springel}, \& {Hernquist}}]{Keres2012}
{Kere{\v s}}, D., {Vogelsberger}, M., {Sijacki}, D., {Springel}, V., \&
  {Hernquist}, L. 2012, \mnras, 425, 2027

\bibitem[{{Kewley} \& {Ellison}(2008)}]{KewleyEllison2008}
{Kewley}, L.~J., \& {Ellison}, S.~L. 2008, \apj, 681, 1183

\bibitem[{{Kim} {et~al.}(2011){Kim}, {Wise}, {Alvarez}, \& {Abel}}]{Kim2011}
{Kim}, J.-h., {Wise}, J.~H., {Alvarez}, M.~A., \& {Abel}, T. 2011, \apj, 738,
  54

\bibitem[{{Klypin} {et~al.}(2011){Klypin}, {Trujillo-Gomez}, \&
  {Primack}}]{Klypin2011}
{Klypin}, A.~A., {Trujillo-Gomez}, S., \& {Primack}, J. 2011, \apj, 740, 102

\bibitem[{{Kobayashi}(2004)}]{Kobayashi2004}
{Kobayashi}, C. 2004, \mnras, 347, 740

\bibitem[{{Kobulnicky} \& {Kewley}(2004)}]{KK04}
{Kobulnicky}, H.~A., \& {Kewley}, L.~J. 2004, \apj, 617, 240

\bibitem[{{Korista} {et~al.}(1997){Korista}, {Baldwin}, {Ferland}, \&
  {Verner}}]{Korista1997}
{Korista}, K., {Baldwin}, J., {Ferland}, G., \& {Verner}, D. 1997, \apjs, 108,
  401

\bibitem[{{Kurosawa} \& {Proga}(2009)}]{Kurosawa2009}
{Kurosawa}, R., \& {Proga}, D. 2009, \mnras, 397, 1791

\bibitem[{{Le Borgne} {et~al.}(2004){Le Borgne}, {Rocca-Volmerange},
  {Prugniel}, {Lan{\c c}on}, {Fioc}, \& {Soubiran}}]{Pegase}
{Le Borgne}, D., {Rocca-Volmerange}, B., {Prugniel}, P., {et~al.} 2004, \aap,
  425, 881

\bibitem[{{Leitherer} {et~al.}(1999){Leitherer}, {Schaerer}, {Goldader},
  {Gonz{\'a}lez Delgado}, {Robert}, {Kune}, {de Mello}, {Devost}, \&
  {Heckman}}]{Starburst99}
{Leitherer}, C., {Schaerer}, D., {Goldader}, J.~D., {et~al.} 1999, \apjs, 123,
  3

\bibitem[{{Leitner} \& {Kravtsov}(2011)}]{Leitner2011}
{Leitner}, S.~N., \& {Kravtsov}, A.~V. 2011, \apj, 734, 48

\bibitem[{{Lia} {et~al.}(2002){Lia}, {Portinari}, \& {Carraro}}]{Lia2002}
{Lia}, C., {Portinari}, L., \& {Carraro}, G. 2002, \mnras, 330, 821

\bibitem[{{Lilly} {et~al.}(1996){Lilly}, {Le Fevre}, {Hammer}, \&
  {Crampton}}]{Lilly1996}
{Lilly}, S.~J., {Le Fevre}, O., {Hammer}, F., \& {Crampton}, D. 1996, \apjl,
  460, L1

\bibitem[{{Ling} {et~al.}(2010){Ling}, {Nezri}, {Athanassoula}, \&
  {Teyssier}}]{Ling2010}
{Ling}, F.-S., {Nezri}, E., {Athanassoula}, E., \& {Teyssier}, R. 2010, \jcap,
  2, 12

\bibitem[{{Lucy}(1977)}]{Lucy1977}
{Lucy}, L.~B. 1977, \aj, 82, 1013

\bibitem[{{Madau} {et~al.}(1998){Madau}, {Pozzetti}, \&
  {Dickinson}}]{Madau1998}
{Madau}, P., {Pozzetti}, L., \& {Dickinson}, M. 1998, \apj, 498, 106

\bibitem[{{Mannucci} {et~al.}(2006){Mannucci}, {Della Valle}, \&
  {Panagia}}]{Mannucci2006}
{Mannucci}, F., {Della Valle}, M., \& {Panagia}, N. 2006, \mnras, 370, 773

\bibitem[{{Maoz} {et~al.}(2012){Maoz}, {Mannucci}, \& {Brandt}}]{Maoz2012}
{Maoz}, D., {Mannucci}, F., \& {Brandt}, T.~D. 2012, \mnras, 426, 3282

\bibitem[{{Marchesini} {et~al.}(2009){Marchesini}, {van Dokkum}, {F{\"o}rster
  Schreiber}, {Franx}, {Labb{\'e}}, \& {Wuyts}}]{Marchesini2009}
{Marchesini}, D., {van Dokkum}, P.~G., {F{\"o}rster Schreiber}, N.~M., {et~al.}
  2009, \apj, 701, 1765

\bibitem[{{Marinacci} {et~al.}(2013){Marinacci}, {Pakmor}, \&
  {Springel}}]{Marinacci2013}
{Marinacci}, F., {Pakmor}, R., \& {Springel}, V. 2013, ArXiv e-prints

\bibitem[{{Martin}(2005)}]{Martin2005}
{Martin}, C.~L. 2005, \apj, 621, 227

\bibitem[{{Matteucci} {et~al.}(2006){Matteucci}, {Panagia}, {Pipino},
  {Mannucci}, {Recchi}, \& {Della Valle}}]{Matteucci2006}
{Matteucci}, F., {Panagia}, N., {Pipino}, A., {et~al.} 2006, \mnras, 372, 265

\bibitem[{{McCarthy} {et~al.}(2012){McCarthy}, {Schaye}, {Font}, {Theuns},
  {Frenk}, {Crain}, \& {Dalla Vecchia}}]{McCarthy2012}
{McCarthy}, I.~G., {Schaye}, J., {Font}, A.~S., {et~al.} 2012, \mnras, 427, 379

\bibitem[{{McCarthy} {et~al.}(2010){McCarthy}, {Schaye}, {Ponman}, {Bower},
  {Booth}, {Dalla Vecchia}, {Crain}, {Springel}, {Theuns}, \&
  {Wiersma}}]{McCarthy2010}
{McCarthy}, I.~G., {Schaye}, J., {Ponman}, T.~J., {et~al.} 2010, \mnras, 406,
  822

\bibitem[{{McQuinn} {et~al.}(2009){McQuinn}, {Lidz}, {Zaldarriaga},
  {Hernquist}, {Hopkins}, {Dutta}, \& {Faucher-Gigu{\`e}re}}]{McQuinn2009}
{McQuinn}, M., {Lidz}, A., {Zaldarriaga}, M., {et~al.} 2009, \apj, 694, 842

\bibitem[{{Mihos} \& {Hernquist}(1994)}]{Mihos1994}
{Mihos}, J.~C., \& {Hernquist}, L. 1994, \apj, 437, 611

\bibitem[{{Mitchell} {et~al.}(2009){Mitchell}, {McCarthy}, {Bower}, {Theuns},
  \& {Crain}}]{Mitchell2009}
{Mitchell}, N.~L., {McCarthy}, I.~G., {Bower}, R.~G., {Theuns}, T., \& {Crain},
  R.~A. 2009, \mnras, 395, 180

\bibitem[{{Monaghan}(1992)}]{Monaghan1992}
{Monaghan}, J.~J. 1992, \araa, 30, 543

\bibitem[{{Monaghan}(2005)}]{Monaghan2005}
---. 2005, Reports on Progress in Physics, 68, 1703

\bibitem[{{Mortlock} {et~al.}(2011){Mortlock}, {Conselice}, {Bluck}, {Bauer},
  {Gr{\"u}tzbauch}, {Buitrago}, \& {Ownsworth}}]{Mortlock2011}
{Mortlock}, A., {Conselice}, C.~J., {Bluck}, A.~F.~L., {et~al.} 2011, \mnras,
  413, 2845

\bibitem[{{Mosconi} {et~al.}(2001){Mosconi}, {Tissera}, {Lambas}, \&
  {Cora}}]{Mosconi2001}
{Mosconi}, M.~B., {Tissera}, P.~B., {Lambas}, D.~G., \& {Cora}, S.~A. 2001,
  \mnras, 325, 34

\bibitem[{{Moster} {et~al.}(2013){Moster}, {Naab}, \& {White}}]{Moster2013}
{Moster}, B.~P., {Naab}, T., \& {White}, S.~D.~M. 2013, \mnras, 428, 3121

\bibitem[{{Moster} {et~al.}(2010){Moster}, {Somerville}, {Maulbetsch}, {van den
  Bosch}, {Macci{\`o}}, {Naab}, \& {Oser}}]{Moster2010}
{Moster}, B.~P., {Somerville}, R.~S., {Maulbetsch}, C., {et~al.} 2010, \apj,
  710, 903

\bibitem[{{Murray} {et~al.}(2005){Murray}, {Quataert}, \&
  {Thompson}}]{Murray2005}
{Murray}, N., {Quataert}, E., \& {Thompson}, T.~A. 2005, \apj, 618, 569

\bibitem[{{Narayan} \& {Yi}(1994)}]{Narayan1994}
{Narayan}, R., \& {Yi}, I. 1994, \apjl, 428, L13

\bibitem[{{Navarro} \& {White}(1993)}]{Navarro1993}
{Navarro}, J.~F., \& {White}, S.~D.~M. 1993, \mnras, 265, 271

\bibitem[{{Nelson} {et~al.}(2013){Nelson}, {Vogelsberger}, {Genel}, {Sijacki},
  {Kere{\v s}}, {Springel}, \& {Hernquist}}]{Nelson2013}
{Nelson}, D., {Vogelsberger}, M., {Genel}, S., {et~al.} 2013, \mnras, 429, 3353

\bibitem[{{Ocvirk} {et~al.}(2008){Ocvirk}, {Pichon}, \&
  {Teyssier}}]{Ocvirk2008}
{Ocvirk}, P., {Pichon}, C., \& {Teyssier}, R. 2008, \mnras, 390, 1326

\bibitem[{{Okamoto} {et~al.}(2010){Okamoto}, {Frenk}, {Jenkins}, \&
  {Theuns}}]{Okamoto2010}
{Okamoto}, T., {Frenk}, C.~S., {Jenkins}, A., \& {Theuns}, T. 2010, \mnras,
  406, 208

\bibitem[{{Okamoto} {et~al.}(2008){Okamoto}, {Nemmen}, \&
  {Bower}}]{Okamoto2008}
{Okamoto}, T., {Nemmen}, R.~S., \& {Bower}, R.~G. 2008, \mnras, 385, 161

\bibitem[{{Oppenheimer} \& {Dav{\'e}}(2006)}]{Oppenheimer2006}
{Oppenheimer}, B.~D., \& {Dav{\'e}}, R. 2006, \mnras, 373, 1265

\bibitem[{{Oppenheimer} \& {Dav{\'e}}(2008)}]{Oppenheimer2008}
---. 2008, \mnras, 387, 577

\bibitem[{{Oppenheimer} {et~al.}(2010){Oppenheimer}, {Dav{\'e}}, {Kere{\v s}},
  {Fardal}, {Katz}, {Kollmeier}, \& {Weinberg}}]{Oppenheimer2010}
{Oppenheimer}, B.~D., {Dav{\'e}}, R., {Kere{\v s}}, D., {et~al.} 2010, \mnras,
  406, 2325

\bibitem[{{O'Shea} {et~al.}(2004){O'Shea}, {Bryan}, {Bordner}, {Norman},
  {Abel}, {Harkness}, \& {Kritsuk}}]{OShea2004}
{O'Shea}, B.~W., {Bryan}, G., {Bordner}, J., {et~al.} 2004, ArXiv Astrophysics
  e-prints

\bibitem[{{Pakmor} {et~al.}(2011){Pakmor}, {Bauer}, \& {Springel}}]{Pakmor2011}
{Pakmor}, R., {Bauer}, A., \& {Springel}, V. 2011, \mnras, 418, 1392

\bibitem[{{Pakmor} \& {Springel}(2012)}]{Pakmor2012}
{Pakmor}, R., \& {Springel}, V. 2012, ArXiv e-prints, 1212.1452

\bibitem[{{Pawlik} \& {Schaye}(2008)}]{Pawlink2008}
{Pawlik}, A.~H., \& {Schaye}, J. 2008, \mnras, 389, 651

\bibitem[{{Pawlik} \& {Schaye}(2011)}]{Pawlink2011}
---. 2011, \mnras, 412, 1943

\bibitem[{{Peeples} \& {Shankar}(2011)}]{Peeples2011}
{Peeples}, M.~S., \& {Shankar}, F. 2011, \mnras, 417, 2962

\bibitem[{{Pen}(1998)}]{Pen1998}
{Pen}, U. 1998, \apjs, 115, 19

\bibitem[{{P{\'e}rez-Gonz{\'a}lez} {et~al.}(2008){P{\'e}rez-Gonz{\'a}lez},
  {Rieke}, {Villar}, {Barro}, {Blaylock}, {Egami}, {Gallego}, {Gil de Paz},
  {Pascual}, {Zamorano}, \& {Donley}}]{PerezGonzalez2008}
{P{\'e}rez-Gonz{\'a}lez}, P.~G., {Rieke}, G.~H., {Villar}, V., {et~al.} 2008,
  \apj, 675, 234

\bibitem[{{Perlmutter} {et~al.}(1999)}]{Perlmutter1999}
{Perlmutter}, S., {et~al.} 1999, \apj, 517, 565

\bibitem[{{Petkova} \& {Springel}(2010)}]{Petkova2010}
{Petkova}, M., \& {Springel}, V. 2010, \mnras, 1851

\bibitem[{{Pettini} \& {Pagel}(2004)}]{PP04}
{Pettini}, M., \& {Pagel}, B.~E.~J. 2004, \mnras, 348, L59

\bibitem[{{Piontek} \& {Steinmetz}(2011)}]{Piontek2011}
{Piontek}, F., \& {Steinmetz}, M. 2011, \mnras, 410, 2625

\bibitem[{{Portinari} {et~al.}(1998){Portinari}, {Chiosi}, \&
  {Bressan}}]{Portinari1998}
{Portinari}, L., {Chiosi}, C., \& {Bressan}, A. 1998, \aap, 334, 505

\bibitem[{{Pozzetti} {et~al.}(2007){Pozzetti}, {Bolzonella}, {Lamareille},
  {Zamorani}, {Franzetti}, {Le F{\`e}vre}, {Iovino}, {Temporin}, {Ilbert},
  {Arnouts}, {Charlot}, {Brinchmann}, {Zucca}, {Tresse}, {Scodeggio}, {Guzzo},
  {Bottini}, {Garilli}, {Le Brun}, {Maccagni}, {Picat}, {Scaramella},
  {Vettolani}, {Zanichelli}, {Adami}, {Bardelli}, {Cappi}, {Ciliegi},
  {Contini}, {Foucaud}, {Gavignaud}, {McCracken}, {Marano}, {Marinoni},
  {Mazure}, {Meneux}, {Merighi}, {Paltani}, {Pell{\`o}}, {Pollo}, {Radovich},
  {Bondi}, {Bongiorno}, {Cucciati}, {de la Torre}, {Gregorini}, {Mellier},
  {Merluzzi}, {Vergani}, \& {Walcher}}]{Pozzetti2007}
{Pozzetti}, L., {Bolzonella}, M., {Lamareille}, F., {et~al.} 2007, \aap, 474,
  443

\bibitem[{{Price}(2007)}]{Price2007}
{Price}, D.~J. 2007, PASA, 24, 159

\bibitem[{{Price}(2008)}]{Price2008}
---. 2008, Journal of Computational Physics, 227, 10040

\bibitem[{{Price} \& {Federrath}(2010)}]{Price2010}
{Price}, D.~J., \& {Federrath}, C. 2010, \mnras, 406, 1659

\bibitem[{{Puchwein} {et~al.}(2008){Puchwein}, {Sijacki}, \&
  {Springel}}]{Puchwein2008}
{Puchwein}, E., {Sijacki}, D., \& {Springel}, V. 2008, \apjl, 687, L53

\bibitem[{{Puchwein} \& {Springel}(2013)}]{Puchwein2013}
{Puchwein}, E., \& {Springel}, V. 2013, \mnras, 428, 2966

\bibitem[{{Rahmati} {et~al.}(2013){Rahmati}, {Pawlik}, {Rai{\v c}evi{\'c}}, \&
  {Schaye}}]{Rahmati2013}
{Rahmati}, A., {Pawlik}, A.~H., {Rai{\v c}evi{\'c}}, M., \& {Schaye}, J. 2013,
  \mnras, 765

\bibitem[{{Read} \& {Hayfield}(2012)}]{Read2012}
{Read}, J.~I., \& {Hayfield}, T. 2012, \mnras, 422, 3037

\bibitem[{{Rees}(1986)}]{Rees1986}
{Rees}, M.~J. 1986, \mnras, 218, 25P

\bibitem[{{Rees} \& {Ostriker}(1977)}]{Rees1977}
{Rees}, M.~J., \& {Ostriker}, J.~P. 1977, \mnras, 179, 541

\bibitem[{{Reyes} {et~al.}(2012){Reyes}, {Mandelbaum}, {Gunn}, {Nakajima},
  {Seljak}, \& {Hirata}}]{Reyes2011}
{Reyes}, R., {Mandelbaum}, R., {Gunn}, J.~E., {et~al.} 2012, \mnras, 425, 2610

\bibitem[{{Riess} {et~al.}(1999){Riess}, {Filippenko}, {Li}, {Treffers},
  {Schmidt}, {Qiu}, {Hu}, {Armstrong}, {Faranda}, {Thouvenot}, \&
  {Buil}}]{Riess1999}
{Riess}, A.~G., {Filippenko}, A.~V., {Li}, W., {et~al.} 1999, \aj, 118, 2675

\bibitem[{{Robertson} {et~al.}(2006){Robertson}, {Cox}, {Hernquist}, {Franx},
  {Hopkins}, {Martini}, \& {Springel}}]{Robertson2006}
{Robertson}, B., {Cox}, T.~J., {Hernquist}, L., {et~al.} 2006, \apj, 641, 21

\bibitem[{{Robertson} {et~al.}(2004){Robertson}, {Yoshida}, {Springel}, \&
  {Hernquist}}]{Robertson2004}
{Robertson}, B., {Yoshida}, N., {Springel}, V., \& {Hernquist}, L. 2004, \apj,
  606, 32

\bibitem[{{Saitoh} \& {Makino}(2013)}]{Saitoh2013}
{Saitoh}, T.~R., \& {Makino}, J. 2013, \apj, 768, 44

\bibitem[{{Sales} {et~al.}(2012){Sales}, {Navarro}, {Theuns}, {Schaye},
  {White}, {Frenk}, {Crain}, \& {Dalla Vecchia}}]{Sales2012}
{Sales}, L.~V., {Navarro}, J.~F., {Theuns}, T., {et~al.} 2012, \mnras, 423,
  1544

\bibitem[{{Sazonov} {et~al.}(2005){Sazonov}, {Ostriker}, {Ciotti}, \&
  {Sunyaev}}]{Sazonov2005}
{Sazonov}, S.~Y., {Ostriker}, J.~P., {Ciotti}, L., \& {Sunyaev}, R.~A. 2005,
  \mnras, 358, 168

\bibitem[{{Scannapieco} {et~al.}(2005){Scannapieco}, {Tissera}, {White}, \&
  {Springel}}]{Scannapieco2005}
{Scannapieco}, C., {Tissera}, P.~B., {White}, S.~D.~M., \& {Springel}, V. 2005,
  \mnras, 364, 552

\bibitem[{{Scannapieco} {et~al.}(2012){Scannapieco}, {Wadepuhl}, {Parry},
  {Navarro}, {Jenkins}, {Springel}, {Teyssier}, {Carlson}, {Couchman}, {Crain},
  {Dalla Vecchia}, {Frenk}, {Kobayashi}, {Monaco}, {Murante}, {Okamoto},
  {Quinn}, {Schaye}, {Stinson}, {Theuns}, {Wadsley}, {White}, \&
  {Woods}}]{Scannapieco2012}
{Scannapieco}, C., {Wadepuhl}, M., {Parry}, O.~H., {et~al.} 2012, \mnras, 423,
  1726

\bibitem[{{Schaye} \& {Dalla Vecchia}(2008)}]{Schaye2008}
{Schaye}, J., \& {Dalla Vecchia}, C. 2008, \mnras, 383, 1210

\bibitem[{{Schaye} {et~al.}(2010){Schaye}, {Dalla Vecchia}, {Booth}, {Wiersma},
  {Theuns}, {Haas}, {Bertone}, {Duffy}, {McCarthy}, \& {van de
  Voort}}]{Schaye2010}
{Schaye}, J., {Dalla Vecchia}, C., {Booth}, C.~M., {et~al.} 2010, \mnras, 402,
  1536

\bibitem[{{Schiminovich} \& et~al.(2005)}]{Schiminovich2005}
{Schiminovich}, D., \& et~al. 2005, \apjl, 619, L47

\bibitem[{{Shakura} \& {Sunyaev}(1973)}]{Shakura1973}
{Shakura}, N.~I., \& {Sunyaev}, R.~A. 1973, \aap, 24, 337

\bibitem[{{Sijacki} \& {Springel}(2006)}]{Sijacki2006}
{Sijacki}, D., \& {Springel}, V. 2006, \mnras, 366, 397

\bibitem[{{Sijacki} {et~al.}(2007){Sijacki}, {Springel}, {Di Matteo}, \&
  {Hernquist}}]{Sijacki2007}
{Sijacki}, D., {Springel}, V., {Di Matteo}, T., \& {Hernquist}, L. 2007,
  \mnras, 380, 877

\bibitem[{{Sijacki} {et~al.}(2012){Sijacki}, {Vogelsberger}, {Kere{\v s}},
  {Springel}, \& {Hernquist}}]{Sijacki2012}
{Sijacki}, D., {Vogelsberger}, M., {Kere{\v s}}, D., {Springel}, V., \&
  {Hernquist}, L. 2012, \mnras, 424, 2999

\bibitem[{{Silk}(1977)}]{Silk1977}
{Silk}, J. 1977, \apj, 211, 638

\bibitem[{{Smith} {et~al.}(2008){Smith}, {Sigurdsson}, \& {Abel}}]{Smith2008}
{Smith}, B., {Sigurdsson}, S., \& {Abel}, T. 2008, \mnras, 385, 1443

\bibitem[{{Somerville} {et~al.}(2008){Somerville}, {Hopkins}, {Cox},
  {Robertson}, \& {Hernquist}}]{Somerville2008}
{Somerville}, R.~S., {Hopkins}, P.~F., {Cox}, T.~J., {Robertson}, B.~E., \&
  {Hernquist}, L. 2008, \mnras, 391, 481

\bibitem[{{Sommer-Larsen} {et~al.}(2003){Sommer-Larsen}, {G{\"o}tz}, \&
  {Portinari}}]{SommerLarsen2003}
{Sommer-Larsen}, J., {G{\"o}tz}, M., \& {Portinari}, L. 2003, \apj, 596, 47

\bibitem[{{Springel}(2010)}]{Springel2010}
{Springel}, V. 2010, \mnras, 401, 791

\bibitem[{{Springel} {et~al.}(2005{\natexlab{a}}){Springel}, {Di Matteo}, \&
  {Hernquist}}]{Springel2005c}
{Springel}, V., {Di Matteo}, T., \& {Hernquist}, L. 2005{\natexlab{a}}, \apjl,
  620, L79

\bibitem[{{Springel} {et~al.}(2005{\natexlab{b}}){Springel}, {Di Matteo}, \&
  {Hernquist}}]{Springel2005}
---. 2005{\natexlab{b}}, \mnras, 361, 776

\bibitem[{{Springel} \& {Hernquist}(2002)}]{Springel2002}
{Springel}, V., \& {Hernquist}, L. 2002, \mnras, 333, 649

\bibitem[{{Springel} \& {Hernquist}(2003{\natexlab{a}})}]{Springel2003a}
---. 2003{\natexlab{a}}, \mnras, 339, 289

\bibitem[{{Springel} \& {Hernquist}(2003{\natexlab{b}})}]{Springel2003b}
---. 2003{\natexlab{b}}, \mnras, 339, 312

\bibitem[{{Springel} {et~al.}(2001){Springel}, {White}, {Tormen}, \&
  {Kauffmann}}]{Springel2001}
{Springel}, V., {White}, S.~D.~M., {Tormen}, G., \& {Kauffmann}, G. 2001,
  \mnras, 328, 726

\bibitem[{{Springel} {et~al.}(2005{\natexlab{c}}){Springel}, {White},
  {Jenkins}, {Frenk}, {Yoshida}, {Gao}, {Navarro}, {Thacker}, {Croton},
  {Helly}, {Peacock}, {Cole}, {Thomas}, {Couchman}, {Evrard}, {Colberg}, \&
  {Pearce}}]{SpringelWhite2005}
{Springel}, V., {White}, S.~D.~M., {Jenkins}, A., {et~al.} 2005{\natexlab{c}},
  \nat, 435, 629

\bibitem[{{Springel} {et~al.}(2008){Springel}, {Wang}, {Vogelsberger},
  {Ludlow}, {Jenkins}, {Helmi}, {Navarro}, {Frenk}, \& {White}}]{Springel2008}
{Springel}, V., {Wang}, J., {Vogelsberger}, M., {et~al.} 2008, \mnras, 391,
  1685

\bibitem[{{Steinmetz} \& {Mueller}(1994)}]{Steinmetz1994}
{Steinmetz}, M., \& {Mueller}, E. 1994, \aap, 281, L97

\bibitem[{{Steinmetz} \& {Navarro}(1999)}]{Steinmetz1999}
{Steinmetz}, M., \& {Navarro}, J.~F. 1999, \apj, 513, 555

\bibitem[{{Stinson} {et~al.}(2006){Stinson}, {Seth}, {Katz}, {Wadsley},
  {Governato}, \& {Quinn}}]{Stinson2006}
{Stinson}, G., {Seth}, A., {Katz}, N., {et~al.} 2006, \mnras, 373, 1074

\bibitem[{{Stinson} {et~al.}(2013){Stinson}, {Brook}, {Macci{\`o}}, {Wadsley},
  {Quinn}, \& {Couchman}}]{Stinson2013}
{Stinson}, G.~S., {Brook}, C., {Macci{\`o}}, A.~V., {et~al.} 2013, \mnras, 428,
  129

\bibitem[{{Stone} \& {Norman}(1992)}]{Stone1992}
{Stone}, J.~M., \& {Norman}, M.~L. 1992, \apjs, 80, 753

\bibitem[{{Strolger}(2004)}]{Strolger2004}
{Strolger}, L.-G.~a. 2004, \apj, 613, 200

\bibitem[{{Sutherland} \& {Dopita}(1993)}]{Sutherland1993}
{Sutherland}, R.~S., \& {Dopita}, M.~A. 1993, \apjs, 88, 253

\bibitem[{{Teyssier}(2002)}]{Teyssier2002}
{Teyssier}, R. 2002, \aap, 385, 337

\bibitem[{{Teyssier} {et~al.}(2006){Teyssier}, {Fromang}, \&
  {Dormy}}]{Teyssier2006}
{Teyssier}, R., {Fromang}, S., \& {Dormy}, E. 2006, Journal of Computational
  Physics, 218, 44

\bibitem[{{Teyssier} {et~al.}(2011){Teyssier}, {Moore}, {Martizzi}, {Dubois},
  \& {Mayer}}]{Teyssier2011}
{Teyssier}, R., {Moore}, B., {Martizzi}, D., {Dubois}, Y., \& {Mayer}, L. 2011,
  \mnras, 414, 195

\bibitem[{{Thacker} \& {Couchman}(2000)}]{Thacker2000}
{Thacker}, R.~J., \& {Couchman}, H.~M.~P. 2000, \apj, 545, 728

\bibitem[{{Thacker} {et~al.}(2006){Thacker}, {Scannapieco}, \&
  {Couchman}}]{Thacker2006}
{Thacker}, R.~J., {Scannapieco}, E., \& {Couchman}, H.~M.~P. 2006, \apj, 653,
  86

\bibitem[{{Thielemann} {et~al.}(2003){Thielemann}, {Argast}, {Brachwitz},
  {Hix}, {H{\"o}flich}, {Liebend{\"o}rfer}, {Martinez-Pinedo}, {Mezzacappa},
  {Nomoto}, \& {Panov}}]{Thielemann2003}
{Thielemann}, F.-K., {Argast}, D., {Brachwitz}, F., {et~al.} 2003, in From
  Twilight to Highlight: The Physics of Supernovae, ed. {W.~Hillebrandt \&
  B.~Leibundgut}, 331

\bibitem[{{Tornatore} {et~al.}(2007){Tornatore}, {Borgani}, {Dolag}, \&
  {Matteucci}}]{Tornatore2007}
{Tornatore}, L., {Borgani}, S., {Dolag}, K., \& {Matteucci}, F. 2007, \mnras,
  382, 1050

\bibitem[{{Torrey} {et~al.}(2013){Torrey}, {Vogelsberger}, {Genel}, {Sijacki},
  {Springel}, \& {Hernquist}}]{Torrey2013}
{Torrey}, P., {Vogelsberger}, M., {Genel}, S., {et~al.} 2013, ArXiv e-prints

\bibitem[{{Torrey} {et~al.}(2012){Torrey}, {Vogelsberger}, {Sijacki},
  {Springel}, \& {Hernquist}}]{Torrey2012}
{Torrey}, P., {Vogelsberger}, M., {Sijacki}, D., {Springel}, V., \&
  {Hernquist}, L. 2012, \mnras, 427, 2224

\bibitem[{{Travaglio} {et~al.}(2004){Travaglio}, {Hillebrandt}, {Reinecke}, \&
  {Thielemann}}]{Travaglio2004}
{Travaglio}, C., {Hillebrandt}, W., {Reinecke}, M., \& {Thielemann}, F.-K.
  2004, \aap, 425, 1029

\bibitem[{{Tully} \& {Fisher}(1977)}]{Tully1977}
{Tully}, R.~B., \& {Fisher}, J.~R. 1977, \aap, 54, 661

\bibitem[{{Turk} {et~al.}(2011){Turk}, {Smith}, {Oishi}, {Skory}, {Skillman},
  {Abel}, \& {Norman}}]{Turk2011}
{Turk}, M.~J., {Smith}, B.~D., {Oishi}, J.~S., {et~al.} 2011, \apjs, 192, 9

\bibitem[{{van de Voort} {et~al.}(2011){van de Voort}, {Schaye}, {Booth}, \&
  {Dalla Vecchia}}]{vandeVoort2011}
{van de Voort}, F., {Schaye}, J., {Booth}, C.~M., \& {Dalla Vecchia}, C. 2011,
  \mnras, 415, 2782

\bibitem[{{Vogelsberger} {et~al.}(2012){Vogelsberger}, {Sijacki}, {Kere{\v s}},
  {Springel}, \& {Hernquist}}]{Vogelsberger2012}
{Vogelsberger}, M., {Sijacki}, D., {Kere{\v s}}, D., {Springel}, V., \&
  {Hernquist}, L. 2012, \mnras, 425, 3024

\bibitem[{{Vogelsberger} {et~al.}(2009){Vogelsberger}, {Helmi}, {Springel},
  {White}, {Wang}, {Frenk}, {Jenkins}, {Ludlow}, \&
  {Navarro}}]{Vogelsberger2009}
{Vogelsberger}, M., {Helmi}, A., {Springel}, V., {et~al.} 2009, \mnras, 395,
  797

\bibitem[{{Wadsley} {et~al.}(2008){Wadsley}, {Veeravalli}, \&
  {Couchman}}]{Wadsley2008}
{Wadsley}, J.~W., {Veeravalli}, G., \& {Couchman}, H.~M.~P. 2008, \mnras, 387,
  427

\bibitem[{{Weiner} {et~al.}(2009){Weiner}, {Coil}, {Prochaska}, {Newman},
  {Cooper}, {Bundy}, {Conselice}, {Dutton}, {Faber}, {Koo}, {Lotz}, {Rieke}, \&
  {Rubin}}]{Weiner2009}
{Weiner}, B.~J., {Coil}, A.~L., {Prochaska}, J.~X., {et~al.} 2009, \apj, 692,
  187

\bibitem[{{White} \& {Frenk}(1991)}]{White1991}
{White}, S.~D.~M., \& {Frenk}, C.~S. 1991, \apj, 379, 52

\bibitem[{{White} \& {Rees}(1978)}]{White1978}
{White}, S.~D.~M., \& {Rees}, M.~J. 1978, \mnras, 183, 341

\bibitem[{{Wiersma} {et~al.}(2009{\natexlab{a}}){Wiersma}, {Schaye}, \&
  {Smith}}]{Wiersma2009a}
{Wiersma}, R.~P.~C., {Schaye}, J., \& {Smith}, B.~D. 2009{\natexlab{a}},
  \mnras, 393, 99

\bibitem[{{Wiersma} {et~al.}(2009{\natexlab{b}}){Wiersma}, {Schaye}, {Theuns},
  {Dalla Vecchia}, \& {Tornatore}}]{Wiersma2009b}
{Wiersma}, R.~P.~C., {Schaye}, J., {Theuns}, T., {Dalla Vecchia}, C., \&
  {Tornatore}, L. 2009{\natexlab{b}}, \mnras, 399, 574

\bibitem[{{Woosley} \& {Weaver}(1995)}]{Woosley1995}
{Woosley}, S.~E., \& {Weaver}, T.~A. 1995, \apjs, 101, 181

\bibitem[{{Wu} {et~al.}(2013){Wu}, {Hahn}, {Wechsler}, {Mao}, \&
  {Behroozi}}]{Wu2013}
{Wu}, H.-Y., {Hahn}, O., {Wechsler}, R.~H., {Mao}, Y.-Y., \& {Behroozi}, P.~S.
  2013, \apj, 763, 70

\bibitem[{{Yu} \& {Tremaine}(2002)}]{YuTremaine2002}
{Yu}, Q., \& {Tremaine}, S. 2002, \mnras, 335, 965

\bibitem[{{Zahid} {et~al.}(2012){Zahid}, {Dima}, {Kewley}, {Erb}, \&
  {Dav{\'e}}}]{Zahid2012}
{Zahid}, H.~J., {Dima}, G.~I., {Kewley}, L.~J., {Erb}, D.~K., \& {Dav{\'e}}, R.
  2012, \apj, 757, 54

\bibitem[{{Zahid} {et~al.}(2013){Zahid}, {Torrey}, {Vogelsberger}, {Hernquist},
  {Kewley}, \& {Dave}}]{Zahid2013}
{Zahid}, H.~J., {Torrey}, P., {Vogelsberger}, M., {et~al.} 2013, ArXiv e-prints

\bibitem[{{Zamorani} {et~al.}(1981){Zamorani}, {Henry}, {Maccacaro},
  {Tananbaum}, {Soltan}, {Avni}, {Liebert}, {Stocke}, {Strittmatter},
  {Weymann}, {Smith}, \& {Condon}}]{Zamorani1981}
{Zamorani}, G., {Henry}, J.~P., {Maccacaro}, T., {et~al.} 1981, \apj, 245, 357

\end{thebibliography}

\label{lastpage}

\end{document}